\pdfoutput=1

\documentclass[11pt,twoside,a4paper,cmspaper,final,collab]{cms-tdr}
\def\svnVersion{326779}\def\cmsCernNo{2013-037}\def\cmsCernDate{\today}\def\cmsMessage{Published in the Journal of Instrumentation as \href{http://dx.doi.org/10.1088/1748-0221/11/01/P01019}{\doi{10.1088/1748-0221/11/01/P01019.}}}

\begin{document}\cmsNoteHeader{TAU-14-001}

\hyphenation{had-ron-i-za-tion}
\hyphenation{cal-or-i-me-ter}
\hyphenation{de-vices}
\RCS$Revision: 318416 $
\RCS$HeadURL: svn+ssh://svn.cern.ch/reps/tdr2/papers/TAU-14-001/trunk/TAU-14-001.tex $
\RCS$Id: TAU-14-001.tex 318416 2016-01-19 16:48:06Z veelken $
\newlength{\figWidth}
\setlength\figWidth{0.38\textwidth}
\newlength\figWidthScale
\setlength\figWidthScale{0.19\textwidth}

\providecommand{\NA}{\text{---}\xspace}
\providecommand{\mT}{\ensuremath{m_{\mathrm{T}}}\xspace}
\newcommand{\vecMET}{\ptvecmiss\xspace}
\newcommand{\tauh}{\ensuremath{\Pgt_{\mathrm{h}}}\xspace}
\newcommand{\Plepton}{\ensuremath{\ell}}
\newcommand{\PLepton}{\ensuremath{\ell}}
\newcommand{\Ppizero}{\ensuremath{\Pgpz}}
\newcommand{\Pnut}{\ensuremath{\PGn_{\Pgt}}}
\newcommand{\APnue}{\ensuremath{\PAGn_{\Pe}}}
\newcommand{\APnum}{\ensuremath{\PAGn_{\Pgm}}}
\newcommand{\APnut}{\ensuremath{\PAGn_{\Pgt}}}
\renewcommand{\PJgy}{\ensuremath{\cmsSymbolFace{J}/\Psi}\xspace}
\renewcommand{\Pgg}{\ensuremath{\PGg}\xspace}
\newcommand{\Pggx}{\ensuremath{\PGg^{*}}\xspace}
\newcommand{\PWprime}{\ensuremath{\PW'}\xspace}
\newcommand{\PZprime}{\ensuremath{\cPZ'}\xspace}
\newcommand{\PHiggs}{\ensuremath{\PH}}
\newcommand{\Pquark}{\ensuremath{\PQq}}
\newcommand{\APquark}{\ensuremath{\PAQq}}
\newcommand{\Pgamma}{\ensuremath{\PGg}}
\newcommand{\Pbottom}{\ensuremath{\PQb}}
\newcommand{\mVis}{\ensuremath{m_{\text{vis}}}\xspace}
\newcommand{\Ntracks}{\ensuremath{N_{\text{tracks}}}\xspace}
\newcommand{\Nvtx}{\ensuremath{N_{\text{vtx}}}\xspace}
\newcommand{\oneProngZeroPizero}{\ensuremath{\mathrm{h}^{\pm}}\xspace}
\newcommand{\oneProngOnePizero}{\ensuremath{\mathrm{h}^{\pm}\Ppizero}\xspace}
\newcommand{\oneProngTwoPizero}{\ensuremath{\mathrm{h}^{\pm}\Ppizero\Ppizero}\xspace}
\newcommand{\oneProngPizeros}{\ensuremath{\mathrm{h}^{\pm}\Ppizero\text{s}}\xspace}
\newcommand{\threeProngZeroPizero}{\ensuremath{\mathrm{h}^{\pm}\mathrm{h}^{\mp}\mathrm{h}^{\pm}}\xspace}

\cmsNoteHeader{TAU-14-001}
\title{Reconstruction and identification of \texorpdfstring{$\Pgt$}{tau} lepton decays to hadrons and \texorpdfstring{$\Pnut$}{tau neutrino} at CMS}

\date{\today}

\abstract{
  This paper describes the algorithms used by the CMS experiment to reconstruct and identify $\Pgt \to \text{hadrons} + \Pnut$ decays during Run 1 of the LHC.
  The performance of the algorithms is studied in proton-proton collisions recorded at a centre-of-mass energy of 8\TeV,
  corresponding to an integrated luminosity of 19.7\fbinv.
  The algorithms achieve an identification efficiency of 50--60\%,
  with misidentification rates for quark and gluon jets, electrons, and muons between per mille and per cent levels.
}

\hypersetup{%
pdfauthor={CMS Collaboration},%
pdftitle={Reconstruction and identification of tau lepton decays to hadrons and tau neutrino at CMS},%
pdfsubject={CMS},%
pdfkeywords={CMS, physics, tau lepton}}

\maketitle

\section{Introduction}
\label{sec:introduction}

Decays of $\Pgt$ leptons provide an important experimental signature for analyses at the CERN LHC.
Evidence for decays of the standard model (SM) Higgs boson
($\PHiggs$) into $\Pgt\Pgt$ has been reported~\cite{HIG-13-004,Aad:2015vsa},
as have searches for neutral and charged Higgs bosons in decays to
$\Pgt$ leptons that have special interest
in the context of the minimal supersymmetric extension of the SM (MSSM)~\cite{HIG-10-002,Aad:2011rv,HIG-13-021,Aad:2014vgg,HIG-11-019,Aad:2014kga}.
The CMS collaboration has published analyses of Drell--Yan ($\Pquark\APquark \to \cPZ/\Pggx \to \Pgt\Pgt$)
and top quark pair production~\cite{EWK-10-013,TOP-11-006,TOP-12-026}
in final states with $\Pgt$ leptons.
Searches for supersymmetry, leptoquarks, $\PWprime$ and $\PZprime$ bosons,
as well as other non-SM Higgs bosons~\cite{SUS-12-004,SUS-13-002,SUS-13-003,EXO-12-002,EXO-11-031,HIG-12-005}
benefit from the high performance $\Pgt$ reconstruction and identification capabilities of the CMS detector.

With a mass of $m_{\Pgt} = 1.777$\GeV~\cite{PDG}, the $\Pgt$ is the only lepton heavy enough to decay into hadrons ($\mathrm{h}$),
and it does so in about two thirds of the cases,
typically into either one or three charged pions or kaons and up to two neutral pions ($\Ppizero$), and one neutrino ($\Pnut$).
The $\Ppizero$ meson decays almost exclusively into $\Pgamma\Pgamma$.
In about 35\% of the cases, $\Pgt$ leptons decay into an electron or muon and two neutrinos.
The branching fractions for the main $\Pgt$ decay modes are given in Table~\ref{tab:tauDecayModes}.
The decays $\Pgt^{-} \to \mathrm{h}^{-} \, \Ppizero \, \Pnut$, $\Pgt^{-} \to \mathrm{h}^{-} \, \Ppizero \, \Ppizero \, \Pnut$,
and $\Pgt^{-} \to \mathrm{h}^{-} \, \mathrm{h}^+ \, \mathrm{h}^{-} \,
\Pnut$ (with corresponding channels for $\Pgt^{+}$) proceed via
intermediate $\Pgr$ and $\mathrm{a}_{1}(1260)$ meson resonances.
The electrons and muons originating from $\Pgt$ decays are difficult to distinguish from electrons and muons produced directly in the primary proton-proton ($\Pp\Pp$) interaction,
and are handled using the standard CMS algorithms for electron and muon reconstruction and identification.
The algorithms for $\Pgt$ reconstruction and identification presented in this paper focus on
$\Pgt$ lepton decays to $\text{hadrons} + \Pnut$, that we refer to as ``hadronic'' $\Pgt$ decays and denote by $\tauh$.
The algorithms provide the means for reconstructing individually the dominant $\tauh$ decay modes.
In comparing the energies of reconstructed $\tauh$ candidates to their true energies,
we refer to the charged hadrons and neutral pions produced in the $\Pgt$ decay as ``visible'' $\Pgt$ decay products,
and ignore the $\Pnut$.

\begin{table}[htbp]
\centering
\topcaption{
  Approximate branching fractions ($\mathcal{B}$) of different $\Pgt$ decay modes~\cite{PDG}.
  The generic symbol $\mathrm{h}^{-}$ represents a charged hadron (either a pion or a kaon).
  Charge conjugation invariance is assumed in this paper.
}
\label{tab:tauDecayModes}
\newcolumntype{.}{D{.}{.}{-1}}
\begin{tabular}{lc.}
\hline
Decay mode & Meson resonance    & \multicolumn{1}{c}{$\mathcal{B}$\,[\%]} \\
\hline
$\Pgt^{-} \to \Pem \, \Pagne \, \Pnut$     &                         & 17.8 \\
$\Pgt^{-} \to \Pgm^{-} \, \Pagngm \, \Pnut$     &                         & 17.4 \\
\hline
$\Pgt^{-} \to \mathrm{h}^{-} \, \Pnut$                             &             & 11.5 \\
$\Pgt^{-} \to \mathrm{h}^{-} \, \Ppizero \, \Pnut$                    & $\Pgr$ & 26.0 \\
$\Pgt^{-} \to \mathrm{h}^{-} \, \Ppizero \, \Ppizero \, \Pnut$            & $\mathrm{a}_1(1260)$            & 9.5 \\
$\Pgt^{-} \to \mathrm{h}^{-} \, \mathrm{h}^+ \, \mathrm{h}^{-} \, \Pnut$            & $\mathrm{a}_1(1260)$            & 9.8 \\
$\Pgt^{-} \to \mathrm{h}^{-} \, \mathrm{h}^+ \, \mathrm{h}^{-} \, \Ppizero \, \Pnut$ &                         & 4.8 \\
Other modes with hadrons &                         & 3.2 \\
\hline
All modes containing hadrons &                         & 64.8 \\
\hline
\end{tabular}
\end{table}

The mean lifetime of $\Pgt$ leptons at rest is $290 \times 10^{-15}$\unit{s}~\cite{PDG}.
The distances that $\Pgt$ leptons travel between their production and decay are small,
but nevertheless significant compared to the transverse impact parameter and secondary-vertex resolution of the CMS tracking detector~\cite{Chatrchyan:2008zzk}.
Energetic $\Pgt$ leptons originating from $\cPZ$ or SM Higgs boson decays typically traverse distances of a few millimetres before decaying.

The main challenge in identifying hadronic $\Pgt$ decays is
distinguishing them from quark and gluon jet background.
The cross section for multijet production from perturbative quantum
chromodynamical (QCD) calculations exceeds by many orders of magnitude
the rate at which $\Pgt$ leptons are produced at the LHC.
To reduce the background arising from quark and gluon jets,
we exploit the fact that hadronic $\Pgt$ decays result in a lower
particle multiplicity, and are more collimated and isolated relative to other particles in the event.
In some analyses, the misidentification of electrons or muons as $\tauh$ candidates
may constitute a sizeable problem, and dedicated
algorithms have been developed to reduce this type of background.

The performance of $\tauh$ reconstruction and identification algorithms has been validated using the first LHC data recorded at $\sqrt{s} = 7$\TeV~\cite{TAU-11-001}.
Since then, the algorithms have been further developed, especially to improve their performance in dealing with
additional inelastic $\Pp\Pp$ interactions (pileup) that occur in the same bunch crossing as the hard scattering of interest.
Moreover, the rejection of backgrounds arising from misidentification of jets, electrons, and muons as $\tauh$
has improved significantly through the introduction of multivariate analysis (MVA) techniques.
In this paper, we report on the performance of the improved algorithms
used to analyze the 8\TeV $\Pp\Pp$ data at CMS,
corresponding to an integrated luminosity of 19.7\fbinv.

The paper is organized as follows.
The CMS detector is described briefly in Section~\ref{sec:detector}.
Section~\ref{sec:datasamples_and_MonteCarloSimulation} describes the data and the Monte Carlo (MC) simulations used for studying the performance of $\tauh$ reconstruction and identification.
The reconstruction of electrons, muons, and jets, along with various kinematic quantities is described in Section~\ref{sec:eventReconstruction}.
The algorithms used for reconstruction and identification of $\tauh$ decays are detailed in Section~\ref{sec:tauId}.
The performance of the algorithms in simulated events is presented in Section~\ref{sec:expectedPerformance}.
Sections~\ref{sec:validation}--\ref{sec:e_and_muToTauFakeRates} detail the validation of the algorithms with data.
The results are summarized in Section~\ref{sec:summary}.

\section{CMS detector}
\label{sec:detector}

The central feature of the CMS detector is a superconducting solenoid of 6\unit{m} internal diameter, providing a magnetic field of 3.8\unit{T}.
A silicon pixel and strip tracker,
a lead tungstate crystal electromagnetic calorimeter (ECAL), and a brass and scintillator hadron calorimeter (HCAL),
each composed of a barrel and two endcap sections,
are positioned within the solenoid volume.
Muons are measured and identified in gas-ionization detectors embedded in the steel flux-return yoke outside the solenoid.
Extensive forward calorimetry complements the coverage provided by the barrel and endcap detectors.

The CMS tracker is a cylindrical detector of 5.5\unit{m} length and 2.5\unit{m} diameter, constructed entirely of silicon modules.
It provides an active sensor area of about 200\unit{m$^2$} to reconstruct charged particles
within the pseudorapidity range $\abs{\eta} < 2.5$.
The innermost region around the interaction point, subject to the highest particle flux, is instrumented with silicon pixel sensors.
The central part of the pixel detector consists of three cylindrical layers, installed at transverse radii of $r = 4.4$, 7.3, and 10.2\unit{cm},
which extend over a total length of 53\unit{cm}.
The central part is complemented by two forward endcap disks of radius $6 < r < 15$\unit{cm},
located at longitudinal distances $\abs{z} = 34.5$ and 46.5\unit{cm} on either side of the interaction point.
The central part of the silicon strip detector consists of
ten cylindrical layers and twelve endcap disks that surround the pixel detector volume.
The cylindrical layers cover radial distances of up to 108\unit{cm} and $\abs{z} < 109$\unit{cm},
and the disks cover up to $r < 113$\unit{cm} and $\abs{z} < 280$\unit{cm}.
Tracks of charged hadrons are reconstructed with an efficiency of 75--95\% that depends on the transverse momentum \pt and $\eta$~\cite{Chatrchyan:2014fea}.

\begin{figure}[htb]
\centering
\includegraphics[width=0.50\textwidth]{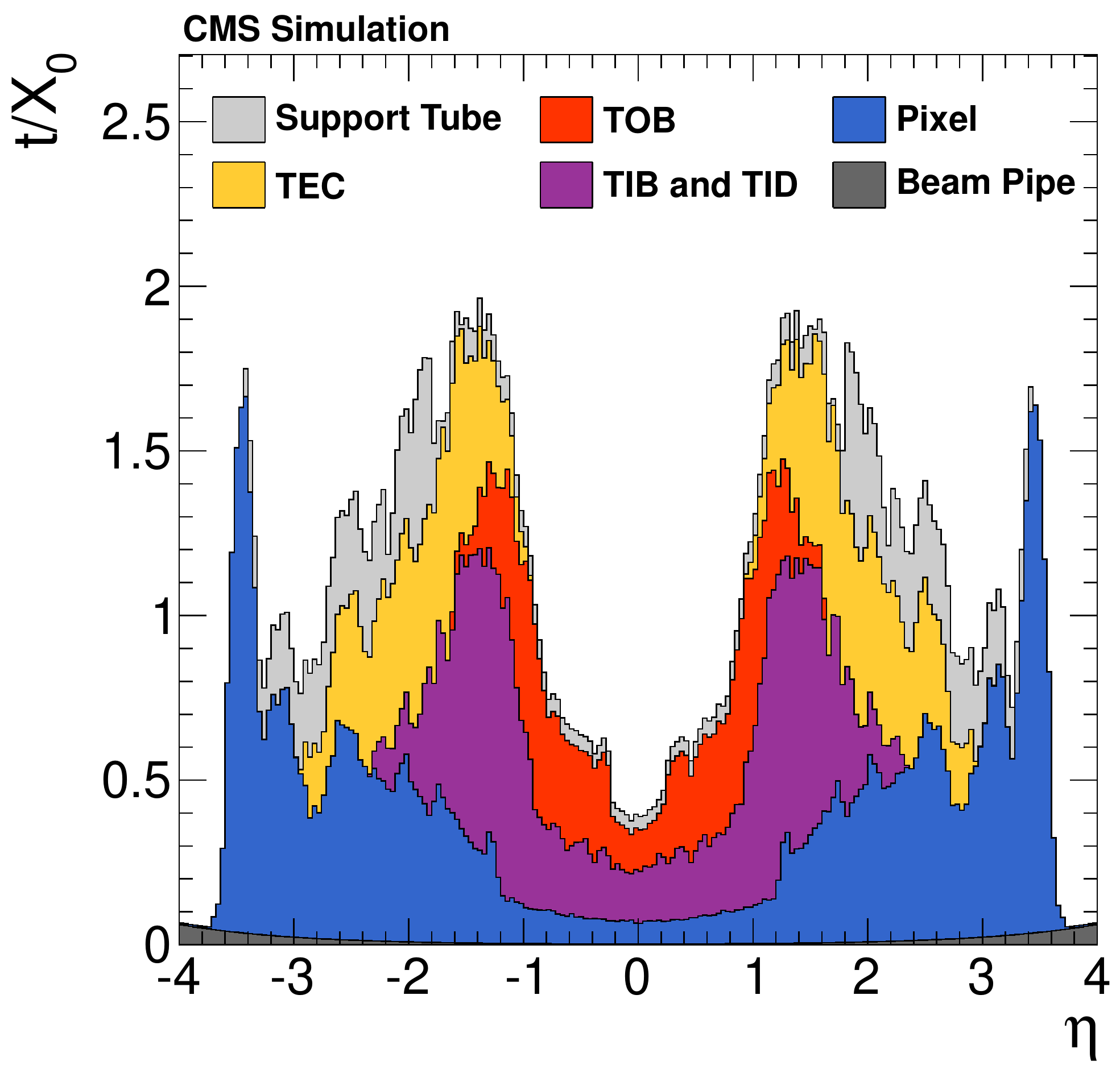}
\caption{
  The total material thickness ($t$) in units of radiation length $X_{0}$,
  as a function of $\eta$, that a particle produced at the interaction point must traverse before it reaches the ECAL.
  The material used for sensors, readout electronics, mechanical structures, cooling, and services
  is given separately for the silicon pixel detector and for individual components of the silicon strip detector (``TEC'', ``TOB'', ``TIB and TID'')~\cite{Chatrchyan:2014fea}.
  The material used for the beam pipe and for the support tube that separates the tracker from the ECAL is also shown separately.
}
\label{fig:Material}
\end{figure}

The silicon tracker adds a significant amount of material in front of the ECAL, mainly because of the mechanical structure, the services, and the cooling system.
Figure~\ref{fig:Material} shows, as a function of $\eta$, the number of radiation lengths ($X_{0}$) of material
that particles produced at the interaction point must traverse before they reach the ECAL.
This rises from about $0.4 X_{0}$ at $\abs{\eta} \approx 0$ to about $2.0 X_{0}$ at $\abs{\eta} \approx 1.4$,
and decreases to about $1.3  X_{0}$ at $\abs{\eta} \approx 2.5$.
As a result, photons originating from $\Ppizero \to \Pgg\Pgg$ decays have a high probability for converting to $\Pep\Pem$ pairs within the volume of the tracking detector.

The ECAL is a homogeneous and hermetic calorimeter made of PbWO$_4$ scintillating crystals.
It is composed of a central barrel, covering $\abs{\eta} < 1.48$,
and two endcaps covering $1.48 < \abs{\eta} < 3.0$.
The barrel is made of $61\,200$ trapezoidal crystals of front-face transverse section $22{\times}22\unit{mm}^2$,
giving a granularity of $0.0174{\times}0.0174$ in $\eta$ and azimuth $\phi$,
and a length of 230\unit{mm} ($25.8 X_0$).
The crystals are organized in 36 supermodules, 18 on each side of $\eta = 0$.
Each supermodule contains 1700 crystals, covers $\pi/9$ radians in $\phi$, and is made of four modules along $\eta$.
This structure has a few thin uninstrumented regions between the modules in $\eta$
(at $\abs{\eta} = 0$, 0.435, 0.783, 1.131, and 1.479),
and between the supermodules in $\phi$ (every $\pi/9$ radians).
The crystals are installed with a quasi-projective geometry, tilted by an angle of 3$^{\circ}$
relative to the projective axis that passes through the centre of CMS (the nominal interaction point), to minimize the passage of electrons or photons
through uninstrumented regions.
The endcaps consist of a total of $14\,648$ trapezoidal crystals with front-face transverse sections of $28.62{\times}28.62$\unit{mm$^2$},
and lengths of 220\unit{mm} ($24.7 X_0$).
The small radiation length ($X_{0} = 0.89$\unit{cm}) and small Moli\`{e}re radius (2.3\unit{cm}) of the PbWO$_4$ crystals
provide a compact calorimeter with excellent two-shower separation.

The HCAL is a sampling calorimeter, with brass as passive absorber, and plastic scintillator tiles serving as active material,
and provides coverage for $\abs{\eta} < 2.9$.
The calorimeter cells are grouped in projective towers of approximate size
$0.087{\times}0.087$ in $\eta \times \phi$ in the barrel
and $0.17{\times}0.17$ in the endcaps.

The muon system is composed of a cylindrical barrel section, and two planar endcaps
that surround the solenoid with about $25\,000$\unit{m$^{2}$} of detection planes.
Drift tube (DT) and cathode strip chamber (CSC) layers provide muon reconstruction, identification, and trigger capability within $\abs{\eta} < 2.4$.
The muon system consists of four muon stations, located at different distances from the centre of CMS,
and separated by layers of steel plates.
Drift tubes are installed in the barrel region $\abs{\eta} < 1.2$, where the muon rate is low and the magnetic field in the return yoke is uniform.
Each DT station contains eight layers of tubes that measure the position in the transverse plane ($r$-$\phi$),
and four layers that provide position information in the $r$-$z$ plane,
except for the outermost station, which contains only eight $r$-$\phi$ layers.
In the endcaps, where the muon rates as well as the background from neutron radiation are higher and the magnetic field is non-uniform,
CSC detectors cover the region $0.9 < \abs{\eta} < 2.4$.
Each CSC station contains six layers of anode wires and cathode planes to measure the position in the bending plane (precise in $\phi$, coarse in $r$).
The combination of DT and CSC detectors covers the pseudorapidity interval $\abs{\eta} < 2.4$ without any gaps in acceptance.
The DT and CSC systems are complemented by a system of resistive-plate chambers (RPC)
that provide precise timing signals for triggering on muons within the region $\abs{\eta} < 1.6$.
Particles produced at the nominal interaction point must traverse more than 10 and 15 interaction lengths ($\lambda$) of absorber material before they reach their respective innermost and outermost detection planes.
This greatly reduces the contribution from punch-through particles.

The first level of the CMS trigger system, based on special hardware processors,
uses information from calorimeters and muon detectors to select the most interesting events in a fixed time interval of $<$4\mus.
The high-level trigger processor farm further decreases the event rate from $<$100\unit{kHz} to $\approx$400\unit{Hz}, before data storage.

A more detailed description of the CMS detector and of the kinematic variables used in the analysis can be found in Ref.~\cite{Chatrchyan:2008zzk}.
\section{Data samples and Monte Carlo simulation}
\label{sec:datasamples_and_MonteCarloSimulation}

The $\Pgt$ reconstruction and identification performance in the data
is compared with MC simulations,
using samples of $\cPZ/\Pggx \to \PLepton\PLepton$ ($\PLepton$ corresponds to $\Pe$, $\Pgm$, and $\Pgt$),
$\PW$+jets, $\cPqt\cPaqt$, single top quark, diboson ($\PW\PW$, $\PW\cPZ$, and $\cPZ\cPZ$), and QCD multijet events.
The $\PW$+jets, $\cPqt\cPaqt$, and diboson samples are generated using the leading-order (LO) \MADGRAPH 5.1 program~\cite{MadGraph}, and
single top quark events with the next-to-leading-order (NLO) program \POWHEG 1.0~\cite{POWHEG1,POWHEG2,POWHEG3}.
The $\cPZ/\Pggx \to \PLepton\PLepton$ samples are generated using \MADGRAPH and \POWHEG.
The QCD multijet samples are produced using the LO generator \PYTHIA 6.4~\cite{pythia6_4} with the Z2* tune.
In fact, \PYTHIA with the Z2* tune is also used to model parton shower and hadronization processes for all MC event samples.
The \PYTHIA Z2* tune is obtained from the Z1 tune~\cite{PYTHIA_Z1tune_CMS},
which uses the \textsc{CTEQ5L} parton distribution functions (PDF),
whereas Z2* adopts CTEQ6L~\cite{CTEQ6}.
The decays of $\Pgt$ leptons, including polarization effects, are
modelled with \TAUOLA~\cite{tauola}.
The samples produced by \PYTHIA and \MADGRAPH are based on the CTEQ6L1 set of PDFs,
while the samples produced by \POWHEG use
CTEQ6M~\cite{CTEQ6}.
The $\cPZ/\Pggx \to \PLepton\PLepton$ and $\PW$+jets events are normalized
to cross sections computed at next-to-next-to-leading-order accuracy~\cite{FEWZ}.
The $\cPqt\cPaqt$ production cross section measured by CMS~\cite{TOP-12-007} is used to normalize the $\cPqt\cPaqt$ sample.
A reweighting is applied to MC-generated $\cPqt\cPaqt$ events to improve the modelling of the \pt spectrum of the top quark relative to data~\cite{Chatrchyan:2012saa,TOP-12-028}.
The cross sections for single top quark and diboson production are computed at NLO accuracy~\cite{MCFMdiBosonXsection}.

Simulated samples of hypothetical heavy Higgs bosons and heavy charged ($\PWprime$) and neutral ($\PZprime$) gauge bosons
are used to train MVA-based $\Pgt$ identification discriminators.
The heavy $\PHiggs$, $\PWprime$, and $\PZprime$ boson events are generated
using the \PYTHIA program
and increase the size of the training sample with $\Pgt$ leptons of
high \pt,
for which the SM production rate is very small.
The Higgs boson samples are produced in the mass range 80--1000\GeV,
the $\PWprime$ and $\PZprime$ samples in the mass range
900--4000\GeV and 750--2500\GeV, respectively.
The list of training samples is complemented by SM $\PHiggs \to \Pgt\Pgt$ events, generated using \POWHEG.
The QCD samples used for the MVA training extend up to a scale of $\hat{p}_\mathrm{T} = 3000$\GeV.

\begin{figure}[htb]
\centering
\includegraphics[width=0.32\textwidth]{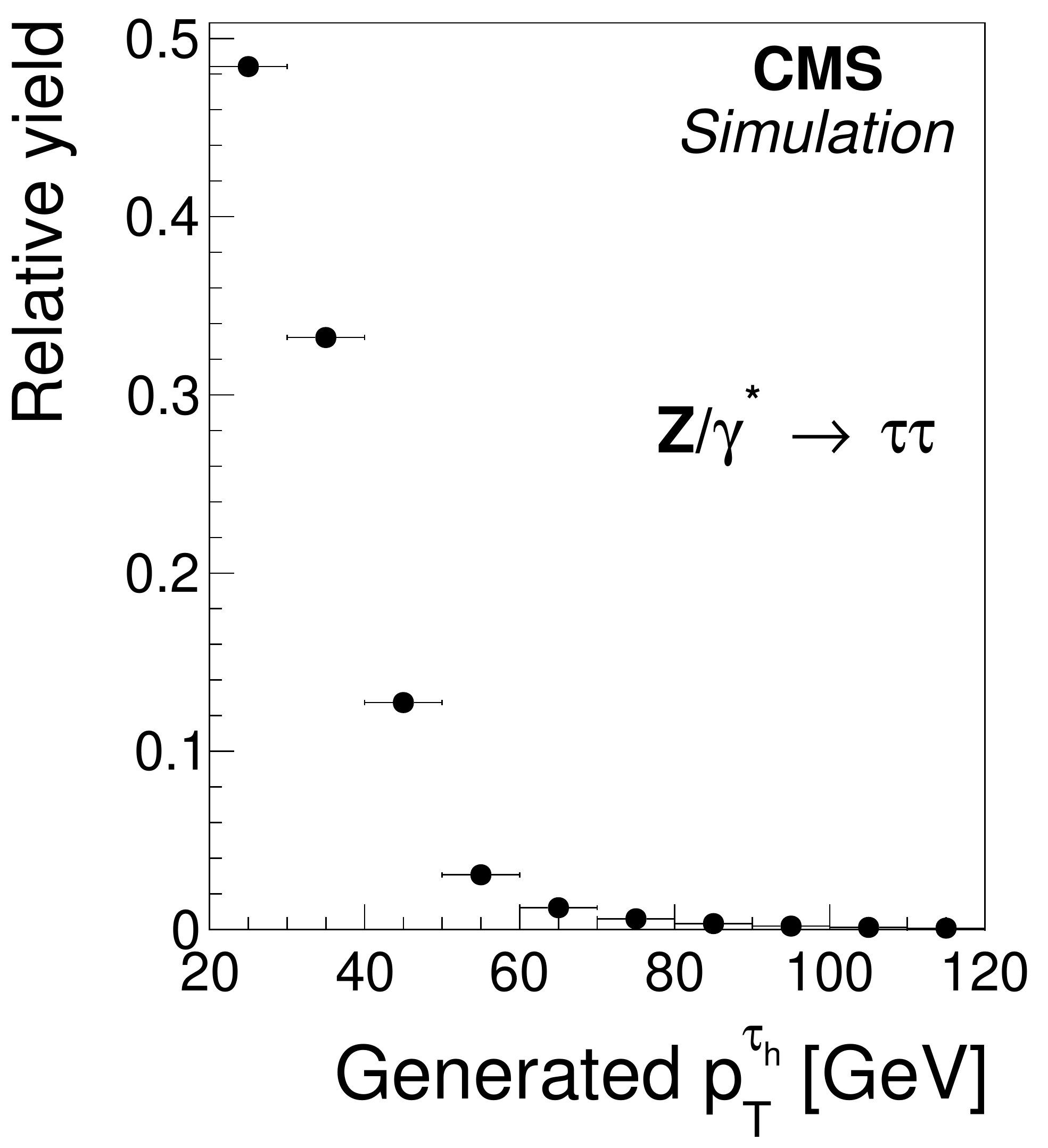}
\includegraphics[width=0.32\textwidth]{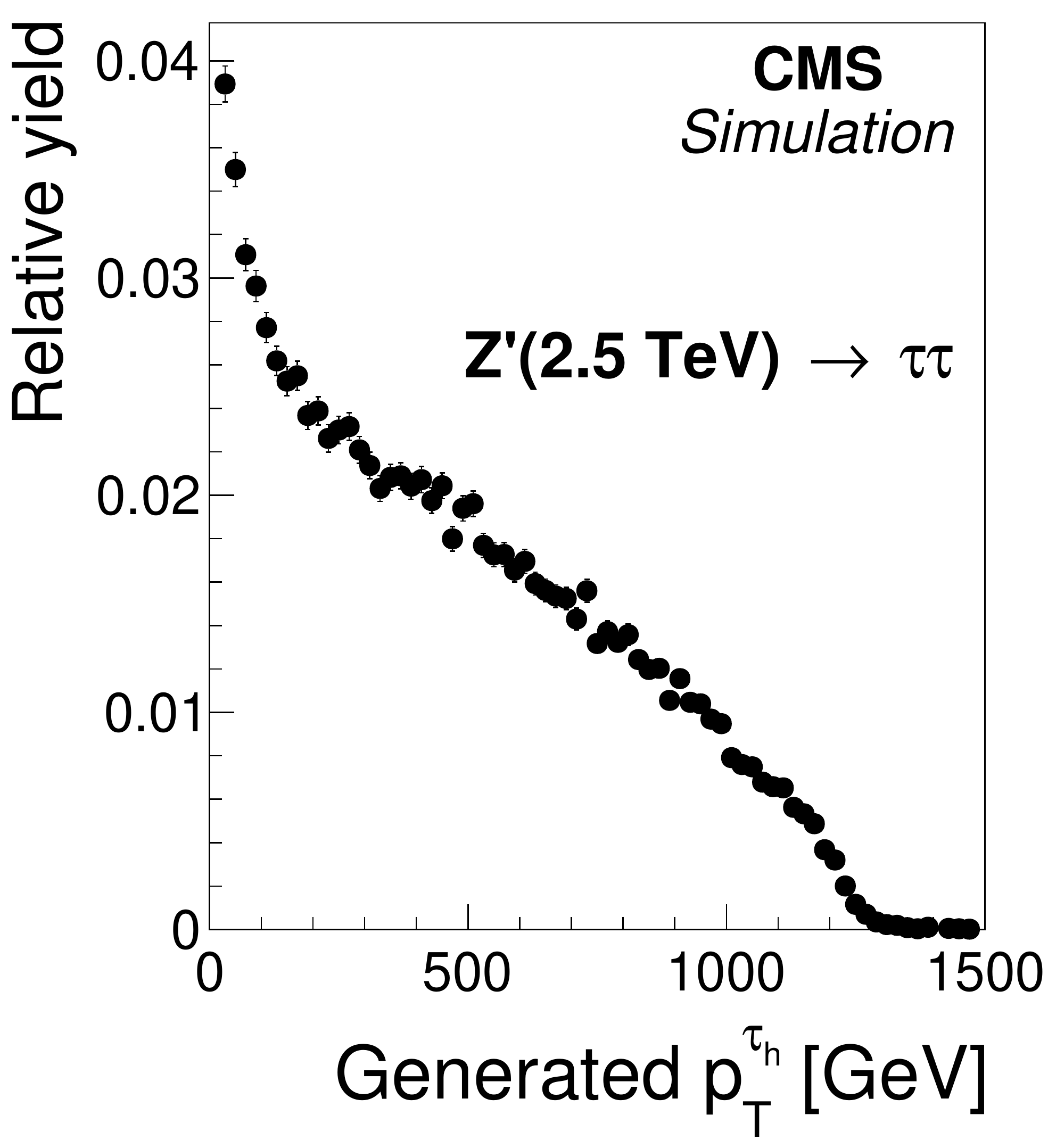}
\includegraphics[width=0.32\textwidth]{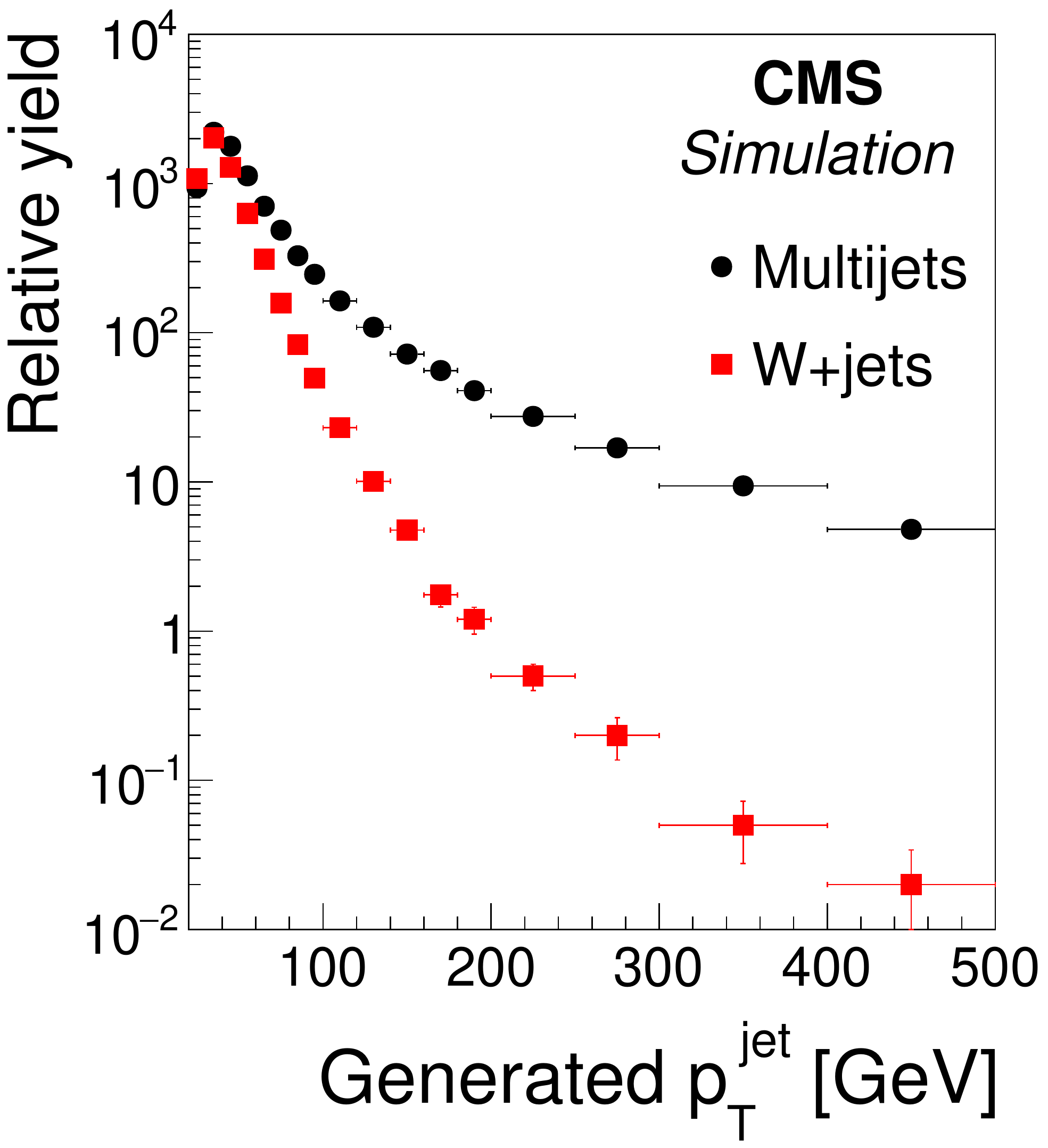}
\caption{
  Transverse momentum distributions of the visible decay products of $\tauh$ decays,
  in (left) simulated $\cPZ/\Pggx \to \Pgt\Pgt$ events, (middle) $\PZprime(2.5\TeV) \to \Pgt\Pgt$ events,
  and (right) of quark and gluon jets in simulated $\PW$+jets and multijet events,
  at the generator level.
}
\label{fig:expectedPerformance_pt}
\end{figure}

The transverse momentum distribution of the visible $\Pgt$ decay products in simulated $\cPZ/\Pggx \to \Pgt\Pgt$ and $\PZprime \to \Pgt\Pgt$ events
is shown in Fig.~\ref{fig:expectedPerformance_pt}.
The $\PZprime$ sample is generated for a mass of $m_{\PZprime} = 2.5$\TeV, and used to study the efficiency to identify $\tauh$ decays at high \pt.
The \pt distribution of generator level quark and gluon jets in simulated $\PW$+jets and QCD multijet events is also shown in the figure.
The jets are constructed using the anti-\kt algorithm~\cite{Cacciari:2008gp} with a distance parameter of 0.5.

On average, 21 inelastic $\Pp\Pp$ interactions occur per LHC bunch crossing.
Minimum bias events generated with \PYTHIA are overlaid on all simulated events,
according to the lu\-mi\-no\-si\-ty profile of the analyzed data.

All generated events are passed through a detailed simulation of the CMS apparatus, based on \GEANTfour~\cite{geant},
and are reconstructed using the same version of the CMS event reconstruction software as used for data.

Small differences between data and MC simulation are observed in selection efficiencies and in energy and momentum measurements of electrons
and muons, as well as in the efficiencies for electron, muon, and
$\tauh$ final states to pass the trigger requirements.
These differences are corrected by applying suitably-chosen weights to
simulated events.
The corrections are determined by comparing $\cPZ/\Pggx \to
\PLepton\PLepton$ events in simulation and data.
Differences in response and resolution of the missing transverse momentum in data and simulation are corrected as described in Ref.~\cite{JME-13-003}.
\section{Event reconstruction}
\label{sec:eventReconstruction}

The information available from all CMS subdetectors is employed in the particle-flow (PF) algorithm~\cite{PFT-09-001,PFT-10-001,PFT-10-002,PFT-10-003}
to identify and reconstruct individual particles in the event, namely muons, electrons, photons, and charged and neutral hadrons.
These particles are used to reconstruct jets, $\tauh$ candidates, and the vector imbalance in transverse momentum in the event, referred to as $\vecMET$,
as well as to quantify the isolation of leptons.

Electrons are reconstructed by matching tracks in the inner detector with energy depositions in the ECAL~\cite{PFT-09-001,Baffioni:2006cd}.
The tracks of electron candidates are reconstructed using a Gaussian-sum filter (GSF)~\cite{Adam:2003kg} algorithm,
which accounts for the emission of bremsstrahlung photons along the electron trajectory.
Energy loss in bremsstrahlung is reconstructed by searching for energy depositions in the ECAL located in directions tangential to the electron track.
A multivariate approach based on boosted decision trees (BDT)~\cite{TMVA}
is employed for electron identification~\cite{EGM-13-001}.
Observables that quantify the quality of the electron track,
the compactness of the electron cluster in directions transverse and longitudinal to the electron track,
and the compatibility between the track momentum and the energy depositions in the ECAL are used as inputs to the BDT.
Additional requirements are applied to reject electrons originating from photon conversions to $\Pep\Pem$ pairs in detector material.

The identification of muons is based on
linking track segments reconstructed in the silicon tracking detector and in the muon system~\cite{Chatrchyan:2012xi}.
The matching between track segments is done outside-in, starting from a track in the muon system, and inside-out,
starting from a track reconstructed in the inner detector.
In case a link can be established, the track parameters are refitted using the combined hits in the inner and outer detectors,
with the resulting track referred to as a global muon track.
Quality criteria are applied on the multiplicity of hits, on the number of matched segments, and on the fit quality of the global muon track, quantified through a $\chi^{2}$.

Electrons and muons originating from decays of $\PW$ and $\cPZ$ bosons are expected to be isolated,
while leptons from heavy flavour (charm and bottom quark) decays,
as well as from in-flight decays of pions and kaons,
are often reconstructed within jets.
The signal is distinguished from multijet background through the sum of scalar \pt values of charged particles, neutral hadrons,
and photons, reconstructed within a cone of size $\Delta R = \sqrt{\smash[b]{(\Delta\eta)^{2} + (\Delta\phi)^{2}}}$ of 0.4, centred around the lepton direction, using the PF algorithm.
Neutral hadrons and photons within
the innermost region of the cone are excluded from the sum,
to prevent the footprint of the lepton in ECAL and HCAL from causing the lepton to fail isolation criteria.
Charged particles close to the direction of electrons are also excluded from the computation,
to avoid counting tracks from converted photons emitted by bremsstrahlung.
Efficiency loss due to pileup is kept minimal
by considering only charged particles originating from the lepton production vertex in the isolation sum.
The contribution of the neutral component of pileup to the isolation of the lepton is taken into account by means of so-called $\Delta\beta$ corrections:
\begin{equation}
I_{\ell} = \sum_{\text{charged}} \pt + \max \left\{ 0, \sum_{\text{neutrals}} \pt - \Delta\beta \right\},
\label{eq:lepIsolationDeltaBeta}
\end{equation}
where $\Plepton$ corresponds to either $\Pe$ or $\Pgm$, and the sums extend over, respectively,
the charged particles that originate from the lepton production vertex and the neutral particles.
Charged and neutral particles are required to be
within a cone of size $\Delta R = 0.4$ around the lepton direction.
The $\Delta\beta$ corrections are computed by summing the scalar \pt of charged particles
that are within a cone of size $\Delta R = 0.4$ around the lepton direction and do not originate from the lepton production vertex,
and scaling this sum down by a factor of two:
\begin{equation}
\Delta\beta = 0.5 \sum_{\text{charged, pileup}} \pt.
\end{equation}
The factor of 0.5 approximates the phenomenological ratio of neutral-to-charged hadron production in the hadronization of inelastic $\Pp\Pp$ collisions.

Collision vertices are reconstructed using a deterministic annealing algorithm~\cite{Chabanat:2005zz,Fruhwirth:2007hz}.
The reconstructed vertex position is required to be compatible with the location of the LHC beam in the $x$-$y$ plane.
The primary collision vertex (PV) is taken to be the vertex that maximizes $\sum_{\text{tracks}} \pt^{2}$.
The sum extends over all tracks associated with a given vertex.

Jets within the range $\abs{\eta} < 4.7$ are reconstructed
using the anti-\kt algorithm~\cite{Cacciari:2008gp} with a distance parameter of 0.5.
As mentioned previously, the particles reconstructed by the PF algorithm are used as input to the jet reconstruction.
Reconstructed jets are required not to overlap with identified electrons, muons, or $\tauh$ within $\Delta R < 0.5$,
and to pass two levels of jet identification criteria:
(i) misidentified jets, mainly arising from calorimeter noise, are rejected by requiring reconstructed jets
to pass a set of loose jet identification criteria~\cite{JME-10-003} and
(ii) jets originating from pileup interactions are rejected through an MVA-based jet identification discriminant,
relying on information about the vertex and energy distribution within the jet~\cite{mvaJetId}.
The energy of reconstructed jets is calibrated as a function of jet \pt and $\eta$~\cite{Chatrchyan:2011ds}.
The contribution of pileup to the energy of jets originating from the hard scattering
is compensated by determining a median transverse momentum density ($\rho$) for each event,
and subtracting the product of $\rho$ times the area of the jet,
computed in the $\eta-\phi$ plane, from the reconstructed jet \pt~\cite{Cacciari:2008gn, Cacciari:2007fd}.
Jets originating from the hadronization of $\Pbottom$ quarks
are identified through the combined secondary vertex (CSV) algorithm~\cite{Chatrchyan:2012jua},
which exploits observables related to the long lifetime of $\Pbottom$ hadrons and the higher particle multiplicity and mass
of $\Pbottom$ jets compared to light-quark and gluon jets.

Two algorithms are used to reconstruct $\vecMET$, the imbalance in transverse momentum in the event,
whose magnitude is referred to as $\MET$.
The standard algorithm computes the negative vectorial sum of all particle momenta reconstructed using the PF algorithm.
In addition, a multivariate regression algorithm~\cite{JME-13-003} has been developed to reduce the effect of pileup on the resolution in $\MET$.
The algorithm utilizes the fact that pileup predominantly produces jets of low \pt,
while leptons and high-\pt jets are produced almost exclusively in the hard-scatter.

The transverse mass, $\mT$, of the system constituted by an electron or a muon and $\MET$ is
used to either select or remove events that are due to $\PW$+jets and $\cPqt\cPaqt$ production.
It is defined by:
\begin{equation}
\mT = \sqrt{2 \pt^{\ell}  \MET  \left( 1 - \cos \Delta\phi \right)},
\label{eq:MtDefinition}
\end{equation}
where the symbol $\ell$ refers to electron or muon
and $\Delta\phi$ denotes the difference in azimuthal angle between the lepton momentum and the $\vecMET$ vector.

\section{Algorithm for \texorpdfstring{$\tauh$}{hadronic tau} reconstruction and identification}
\label{sec:tauId}

The $\tauh$ decays are reconstructed and identified using the hadrons-plus-strips (HPS) algorithm~\cite{TAU-11-001}.
The algorithm is designed to reconstruct individual decay modes of the $\Pgt$ lepton,
taking advantage of the excellent performance of the PF algorithm
in reconstructing individual charged and neutral particles.

The reconstruction and identification of $\tauh$ decays in the HPS algorithm is performed in two steps:
\begin{enumerate}
\item \textbf{Reconstruction:}
  combinations of charged and neutral particles reconstructed by the PF algorithm
  that are compatible with specific $\tauh$ decays are constructed,
  and the four-momentum, expressed in terms of (\pt, $\eta$, $\phi$, and mass) of $\tauh$ candidates, is computed.
\item \textbf{Identification:}
  discriminators that separate $\tauh$ decays from quark and gluon jets, and from electrons and muons, are computed.
  This provides a reduction in the jet $\to \tauh$, $\Pe \to \tauh$, and $\Pgm \to \tauh$ misidentification rates.
\end{enumerate}
The HPS algorithm is seeded by jets of $\pt > 14$\GeV and $\abs{\eta} < 2.5$,
reconstructed using the anti-\kt algorithm~\cite{Cacciari:2008gp} with a distance parameter of 0.5.
The \pt criterion is applied on the jet momentum given by the vectorial sum of all particle constituents of the jet,
before the jet energy calibration and pileup corrections described in Section~\ref{sec:eventReconstruction} are taken into account.

\subsection{Identification of decay modes}
\label{sec:decay_mode_reconstruction}

Reconstruction of specific $\tauh$ decay modes requires reconstruction of neutral pions that are present in most of the hadronic $\Pgt$ decays.
The high probability for photons originating from $\Ppizero \to \Pgg\Pgg$ decays to convert to $\Pep\Pem$ pairs within the volume of the CMS tracking detector
is taken into account by clustering the photon and electron constituents of the $\Pgt$-seeding jet into ``strips'' in the $\eta-\phi$ plane.
The clustering of electrons and photons of $\pt > 0.5$\GeV into strips proceeds via an iterative procedure.
The electron or photon of highest \pt not yet included into any strip is used to seed a new strip.
The initial position of the strip in the $\eta-\phi$ plane is set according to the $\eta$ and $\phi$ of the seed $\Pe$ or $\Pgg$.
The $\Pe$ or $\Pgg$ of next-highest \pt that is within an $\eta \times \phi$ window centred on the strip location
is merged into the strip.
The strip position is then recomputed as an energy-weighted average of all electrons and photons contained in the strip:
\begin{equation*}\begin{aligned}
\eta_{\text{strip}} & = \frac{1}{\pt^{\text{strip}}}  \sum \pt^{\Pgg}  \eta_{\Pgg}\\
\phi_{\text{strip}} & = \frac{1}{\pt^{\text{strip}}}  \sum \pt^{\Pgg}  \phi_{\Pgg},
\end{aligned}\end{equation*}
with $\pt^{\text{strip}} = \sum \pt^{\Pgg}$.
The construction of the strip ends when no additional electrons or photons are found within an $\eta \times \phi$ window of size $0.05 \times 0.20$.
In which case the clustering proceeds by constructing a new strip, which is seeded by the $\Pe$ or $\Pgg$ with next highest \pt.
The size of the window is enlarged in the $\phi$ direction to account for the bending of $\Pep$ and $\Pem$
from photon conversions in the 3.8\unit{T} magnetic field.
Strips with \pt sums of electrons and photons in the strip of $>$2.5\GeV are kept as $\Ppizero$ candidates.

Hadronic $\Pgt$ candidates are formed by combining the strips with the charged-particle constituents of the jet.
The charged particles are required to satisfy the condition $\pt > 0.5$\GeV.
The distance of closest approach between their tracks and the hypothetical production vertex of the $\tauh$ candidate,
taken to be the vertex closest to the charged particle of highest \pt within the jet,
is required to be less than 0.4\cm in the $z$ direction and $<$0.03\cm in the transverse plane.
The requirements for tracks to be compatible with the production vertex of the $\Pgt$ removes spurious tracks and significantly reduces the effect of pileup,
while being sufficiently loose so as not to lose efficiency
because of the small distances that $\Pgt$ leptons traverse between their production and decay.

A combinatorial approach is taken for constructing hadronic $\Pgt$ candidates.
Multiple $\tauh$ hypotheses, corresponding to combinations of either one or three charged particles and up to two strips,
are constructed for each jet. To reduce computing time, the set of
input objects is restricted to the 6 charged particles and the 6
strips with highest \pt.

The four-momentum of each $\tauh$ candidate hypothesis (\pt, $\eta$, $\phi$, and mass)
is given by the four-momentum sum of the charged particles and strips.
In a few per cent of the cases, the charged particles included in the $\tauh$ candidates are identified as electrons or muons,
and are assigned their respective electron or muon masses by the PF algorithm.
The HPS algorithm sets the mass of all charged particles included in $\tauh$ candidates to that of the charged pion,
except for electron constituents of strips, which are treated as massless.
The charge of $\tauh$ candidates is reconstructed by summing the charges
of all particles included in the construction of the $\tauh$ candidate,
except for the electrons contained in strips.
The probability for misreconstructing the $\tauh$ charge is $\approx$1\%,
with a moderate dependence on \pt and $\eta$,
for taus from $\cPZ$ decays.

{\sloppy
The following criteria are applied to assure the compatibility of each hypothesis with the
signatures expected for the different $\tauh$ decays in Table~\ref{tab:tauDecayModes}:
\begin{enumerate}
\item $\threeProngZeroPizero$:
  Combination of three charged particles with mass $0.8 < m_{\tauh} < 1.5$\GeV.
  The tracks are required to originate within $\Delta z < 0.4$\cm of the same event vertex,
  and to have a total charge of one.
\item $\oneProngTwoPizero$:
  Combination of a single charged particle with two strips.
  The mass of the $\tauh$ candidate is required to satisfy the
  condition $0.4 < m_{\tauh} < 1.2\sqrt{\pt\,[\GeVns{}]/100}$\GeV.
  The size of the mass window is enlarged for $\tauh$ candidates of high \pt to account for resolution effects.
  The upper limit on the mass window is constrained to be at least 1.2 and at most 4.0\GeV.
\item $\oneProngOnePizero$:
  Combination of one charged particle and one strip with mass $0.3 < m_{\tauh} <1.3 \sqrt{\pt\,[\GeVns{}]/100}\GeV.$ The upper limit on the mass window is constrained to be at least 1.3 and at most 4.2\GeV.
\item $\oneProngZeroPizero$:
  A single charged particle without any strips.
\end{enumerate}
The combinations of charged particles and strips considered by the HPS algorithm
represent all hadronic $\Pgt$ decay modes in Table~\ref{tab:tauDecayModes},
except $\Pgt^{-} \to \mathrm{h}^{-} \mathrm{h}^{+} \mathrm{h}^{-}\Ppizero\Pnut$.
The latter corresponds to a branching fraction of 4.8\%, and is not considered
in the present version of the algorithm, because of its contamination by jets.
The $\oneProngOnePizero$ and $\oneProngTwoPizero$ decays are analyzed together,
and referred to as $\oneProngPizeros$.
}

Hypotheses that fail the mass window selection for the corresponding decay mode are discarded,
as are hypotheses that have a charge different from unity, or hypotheses that
include any charged hadron or strip outside of a signal cone of $\Delta R = 3.0/\pt$\,[\GeVns{}]
of the axis given by the momentum vector of the $\tauh$ candidate.
The size of the cone takes into account the fact that decay products of energetic $\Pgt$ leptons are more collimated.
When $\Delta R$ is smaller than 0.05 or exceeds 0.10,
a cone of size $\Delta R = 0.05$ or $\Delta R = 0.10$ is used as the
limit, respectively.

When multiple combinations of charged hadrons and strips pass the mass window and the signal cone requirements,
the hypothesis for the candidate with largest \pt is retained. All other combinations are discarded,
resulting in a unique $\tauh$ candidate to be associated to each jet.

\begin{figure}[htb]
\centering
\includegraphics*[width=0.48\textwidth]{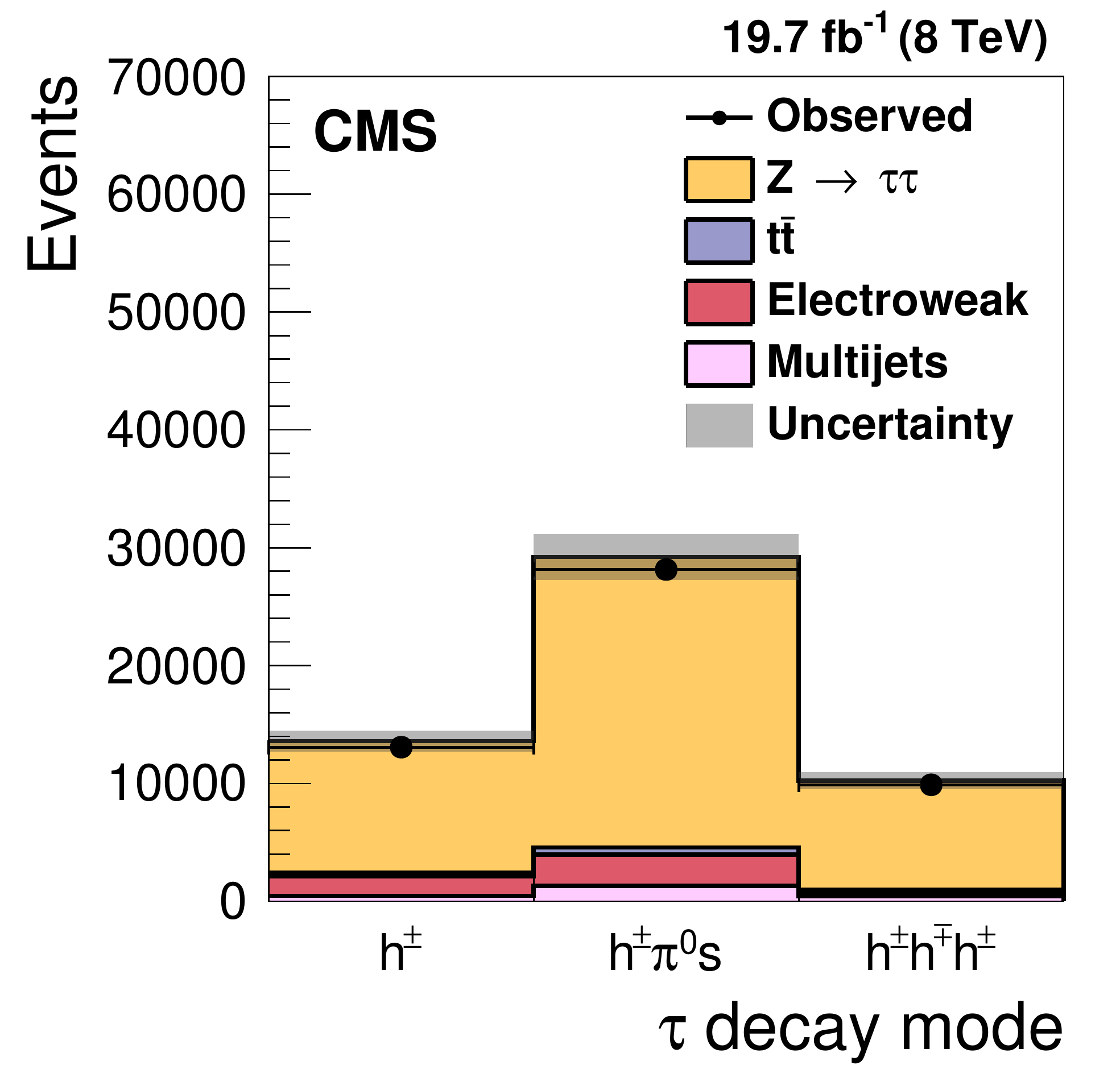}
\includegraphics*[width=0.48\textwidth]{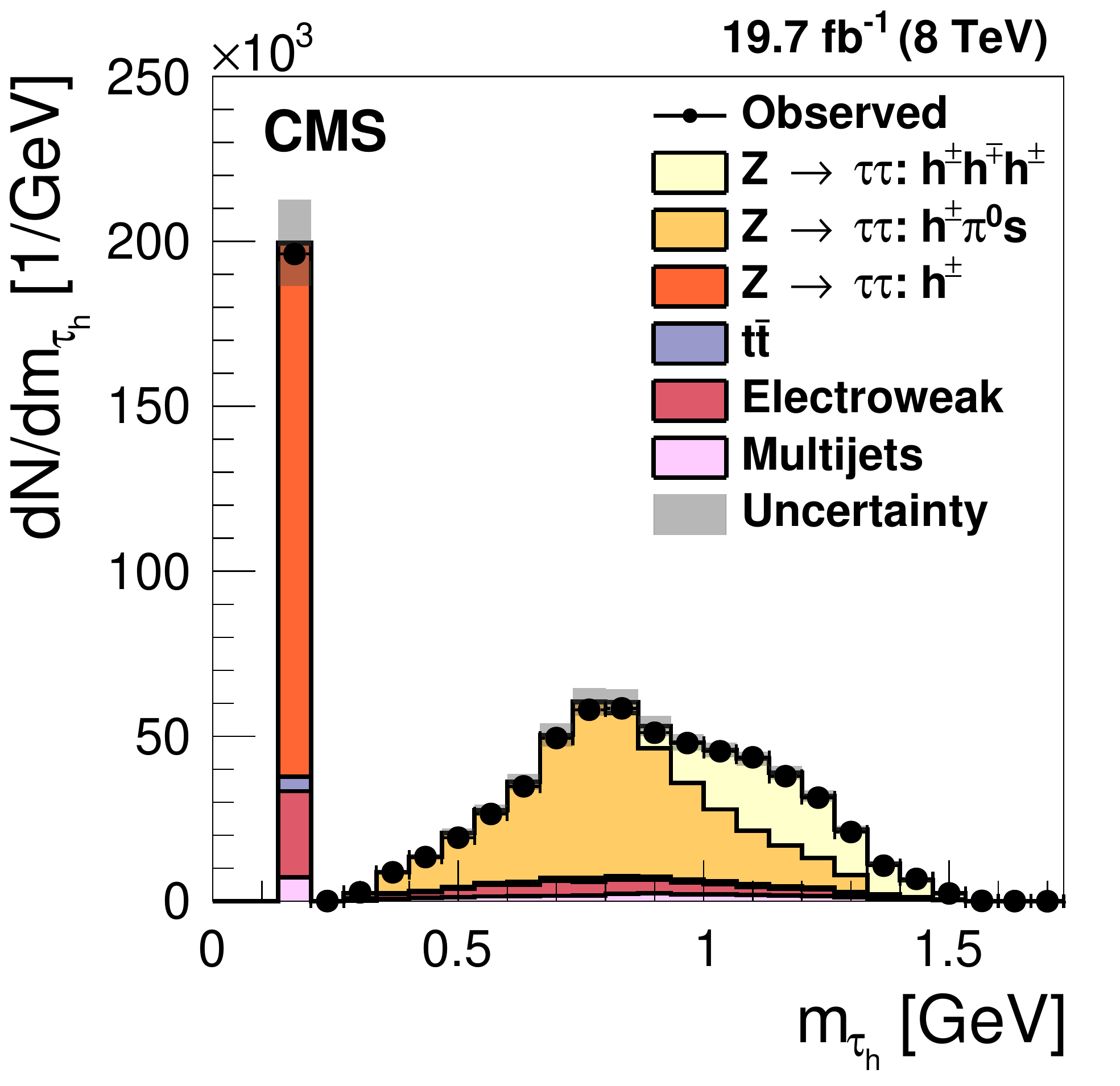}
\caption{
  Distributions in (left) reconstructed $\tauh$ decay modes and (right) $\tauh$ candidate masses
  in $\cPZ/\Pggx \to \Pgt\Pgt$ events selected in data, compared to MC expectations.
  The $\cPZ/\Pggx \to \Pgt\Pgt$ events are selected in the decay channel of muon and $\tauh$,
  as described in Section~\ref{sec:validation_eventSelection_ZTT}.
  The $\tauh$ are required to pass the medium working point of the MVA-based $\tauh$ isolation discriminant.
  The mass of $\tauh$ candidates reconstructed in simulated $\cPZ/\Pggx \to \Pgt\Pgt$ events is corrected for small data/MC differences in the $\tauh$ energy scale,
  discussed in Section~\ref{sec:Tau_energy_scale}.
  The electroweak background is dominated by $\PW$+jets production, with minor contributions arising from single top quark and diboson production.
  The shaded uncertainty band represents the sum of systematic and statistical uncertainties on the MC simulation.
}
\label{fig:tauIdAlgorithm_ZTT_dm_and_mTau}
\end{figure}

The distributions in the decay modes and in the mass of $\tauh$ candidates in $\cPZ/\Pggx \to \Pgt\Pgt$ events are shown in Fig.~\ref{fig:tauIdAlgorithm_ZTT_dm_and_mTau}.
The contribution of the $\cPZ/\Pggx \to \Pgt\Pgt$ signal is split according to the reconstructed $\tauh$ mode,
as shown in the legend.
For $\tauh$ candidates reconstructed in the $\oneProngPizeros$ and $\threeProngZeroPizero$ modes,
the $m_{\tauh}$ distribution peaks near the intermediate $\Pgr$ and $\text{a}_1(1260)$ meson resonances (cf. Table~\ref{tab:tauDecayModes}),
as expected.
The narrow peak at the charged pion mass is due to $\tauh$ candidates reconstructed in the $\oneProngZeroPizero$ mode.

\subsection{Tau-isolation discriminants}
\label{sec:tauIdDiscriminators}

Requiring reconstructed $\tauh$ candidates to pass strict isolation requirements
constitutes the main handle for reducing the large multijet background.
Tau leptons are usually isolated relative to other particles in the event,
and so are their decay products, in contrast to quark and gluon jets.
Two types of $\tauh$ isolation discriminants have been developed,
using simple cutoff-based selections and an MVA approach.
An overview of the discriminants, with their respective efficiencies and misidentification rates, is given in Table~\ref{tab:tauIdPerformance}.

\subsubsection{Cutoff-based discriminants}
\label{sec:tauIdDiscrCutBased}

The isolation of $\tauh$ candidates is computed
by summing the scalar values of \pt of charged particles and photons with $\pt > 0.5$\GeV,
reconstructed with the PF algorithm,
within an isolation cone of size $\Delta R = 0.5$, centred on the $\tauh$ direction.
The effect of pileup is reduced by requiring the tracks associated to charged particles considered in the isolation sum
to be compatible with originating from the production vertex of the $\tauh$ candidate
within a distance of $\Delta z < 0.2$\cm and $\Delta r < 0.03$\cm.
Charged hadrons used to form the $\tauh$ candidate are excluded from the isolation sum,
as are electrons and photons used to construct any of the strips.
The effect of pileup on photon isolation is compensated on a statistical basis
through the modified $\Delta\beta$ corrections:
\begin{equation}
I_{\Pgt} = \sum_{\text{charged},\Delta z < 0.2\cm} \pt + \max \left\{ 0, \sum_{\Pgg} \pt - \Delta\beta \right\},
\label{eq:tauIsolation}
\end{equation}
where the $\Delta\beta$ are computed by summing the \pt of charged particles that are within a cone of size $\Delta R = 0.8$ around the $\tauh$ direction,
and are associated to tracks that have a distance to the $\tauh$ production vertex of more than 0.2\cm in $z$.
The sum is scaled by a factor 0.46, chosen to make the $\tauh$ identification efficiency insensitive to pileup:
\begin{equation}
\Delta\beta = 0.46\sum_{\text{charged},\Delta z > 0.2\cm} \pt.
\label{eq:tauIsolationDB}
\end{equation}

Loose, medium, and tight working points (WP) are defined for the cutoff-based $\tauh$ isolation discriminants
by requiring the \pt sum defined by Eq.~(\ref{eq:tauIsolation}) not to exceed thresholds of 2.0, 1.0, and 0.8\GeV, respectively.

\subsubsection{MVA-based  discriminants}
\label{sec:tauIdDiscrMVABased}

\begin{figure}[htbp]\centering
\includegraphics[width=0.32\textwidth]{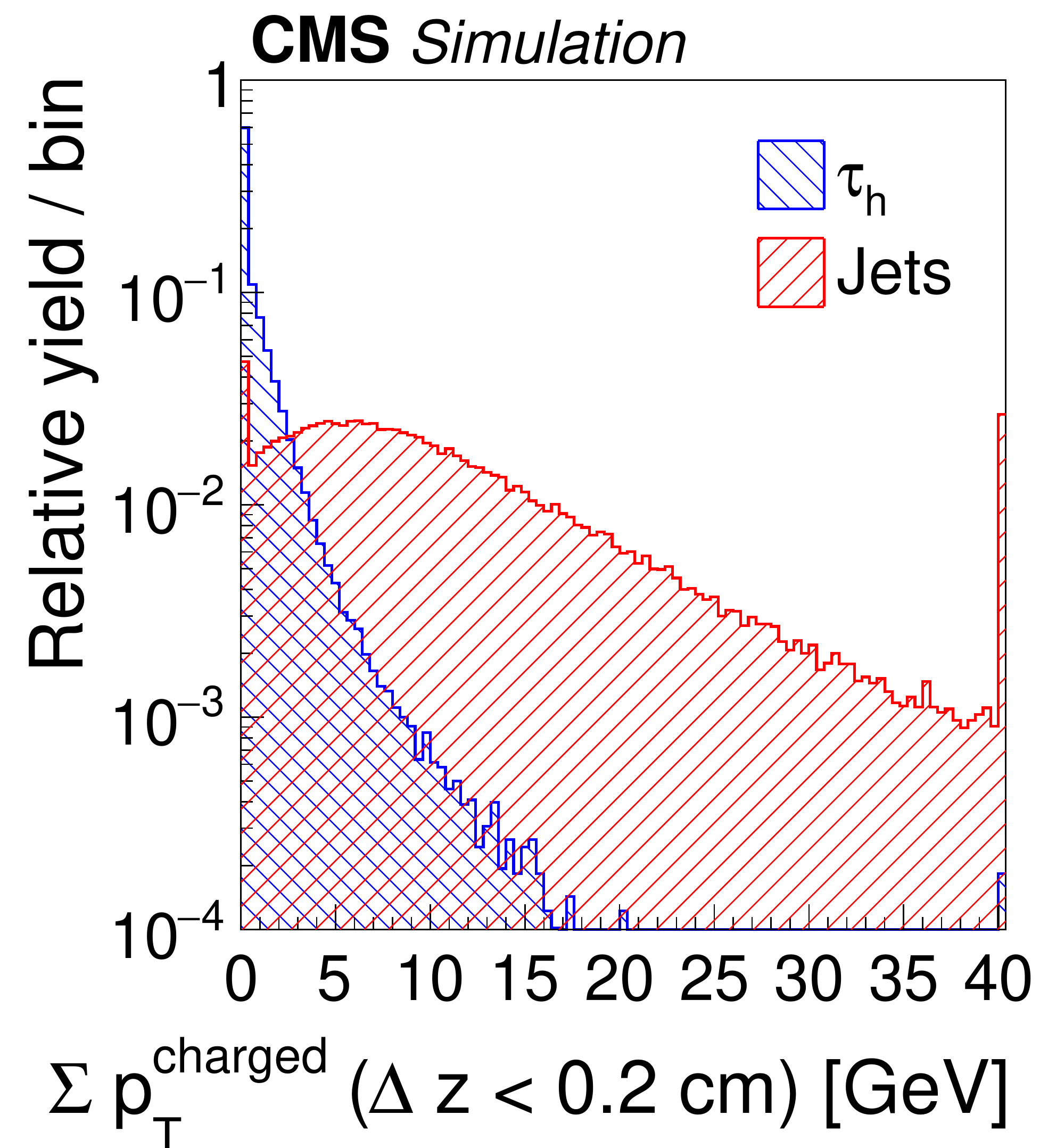}
\includegraphics[width=0.32\textwidth]{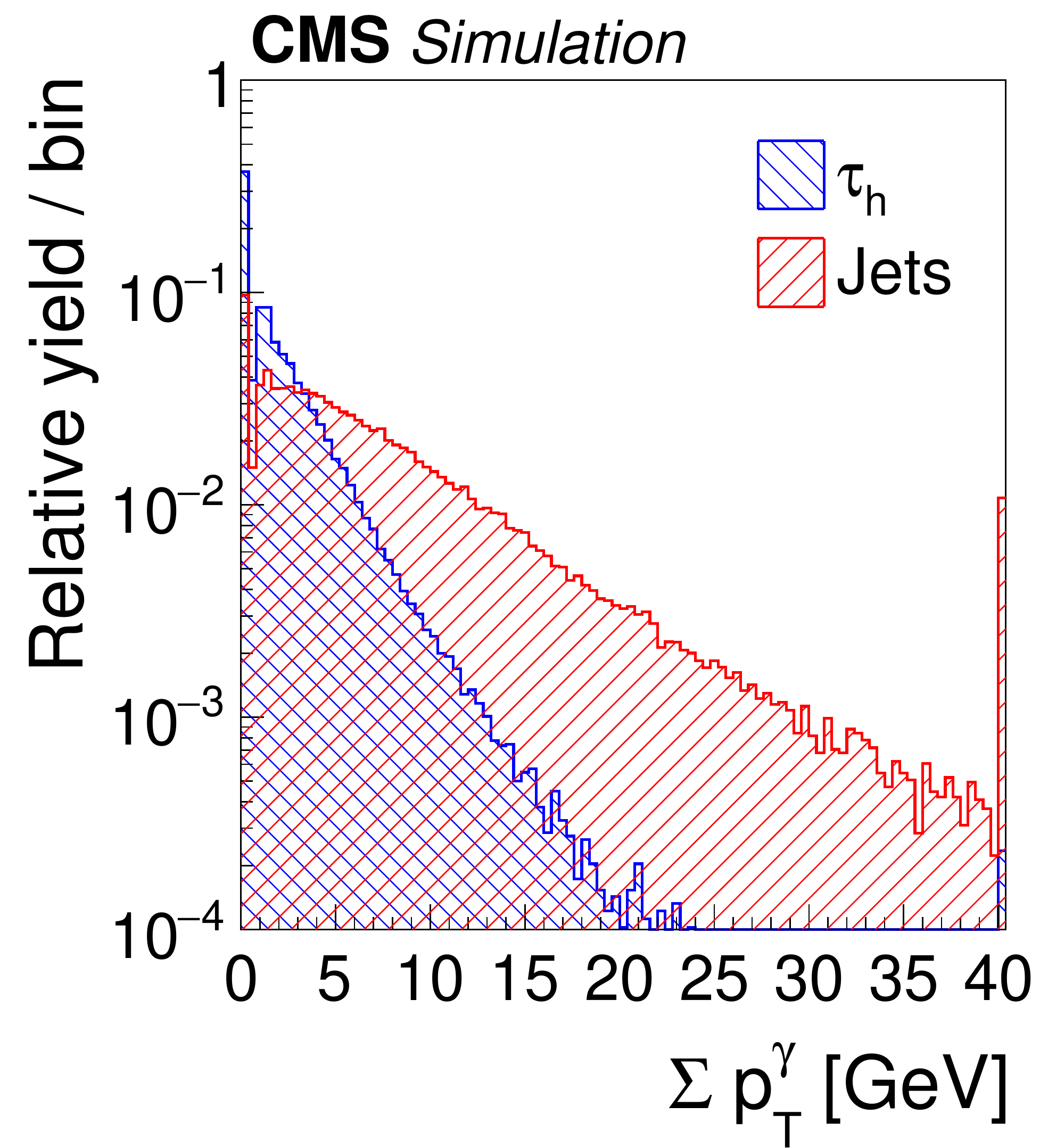}
\includegraphics[width=0.32\textwidth]{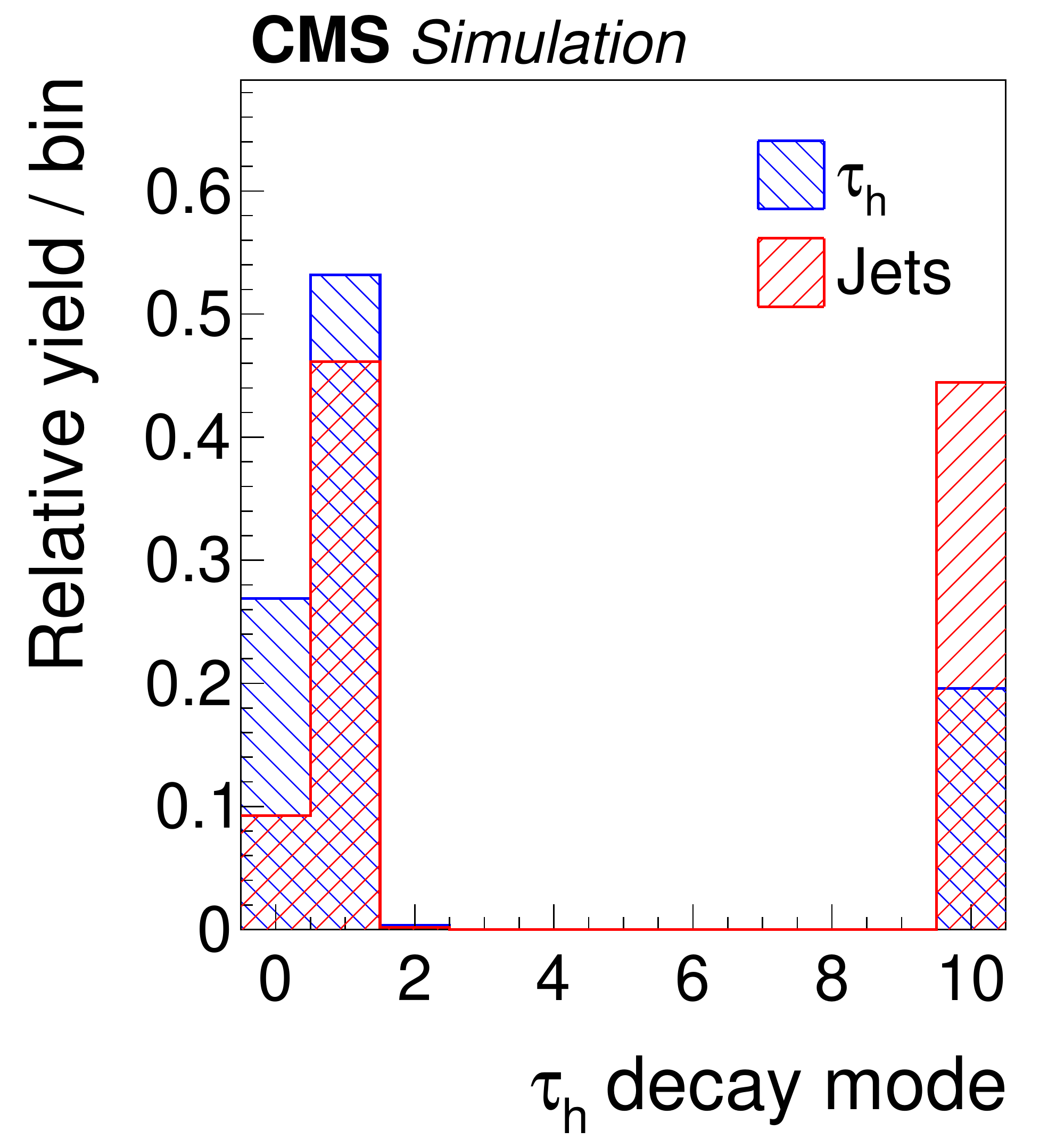}
\includegraphics[width=0.32\textwidth]{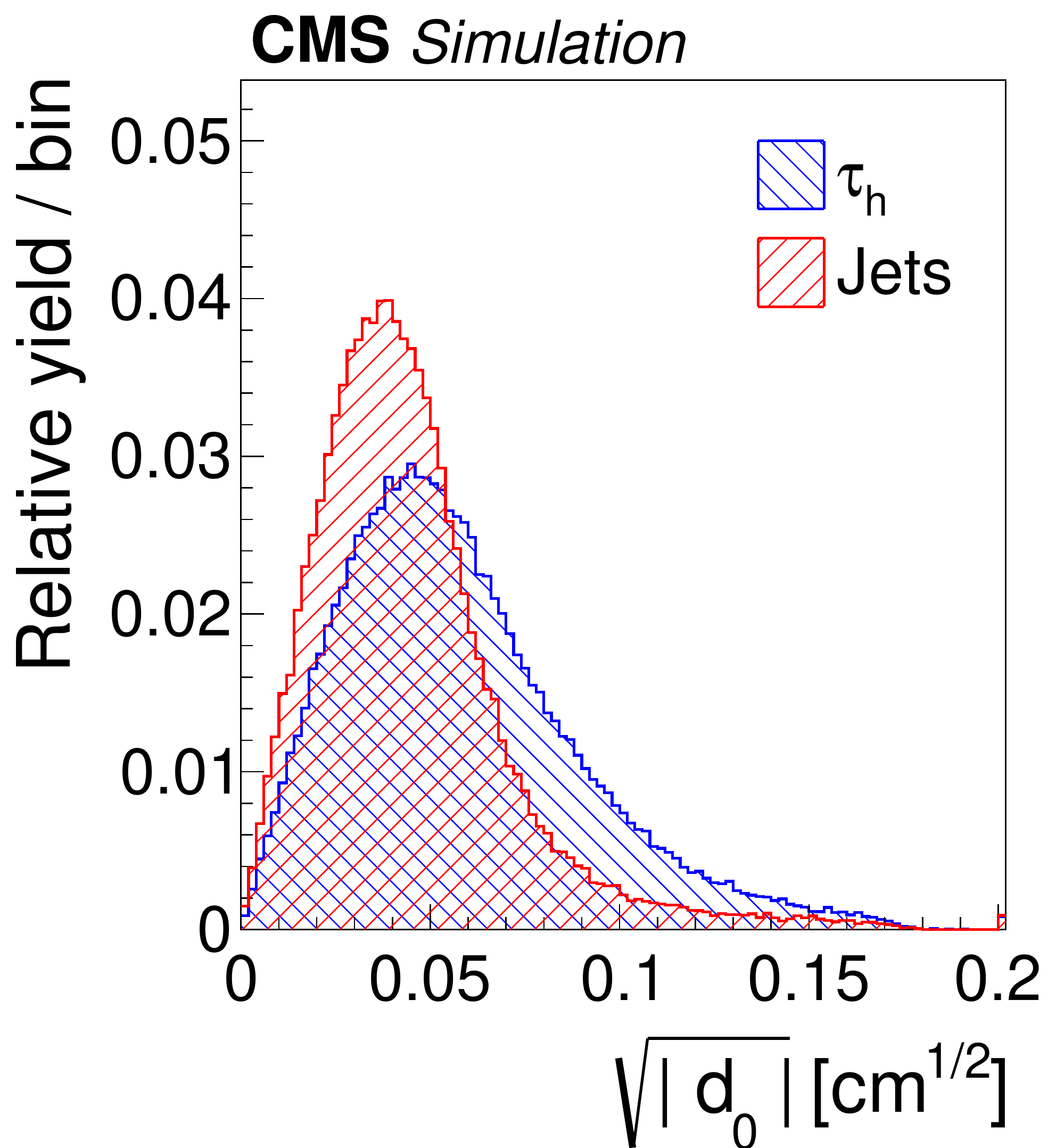}
\includegraphics[width=0.32\textwidth]{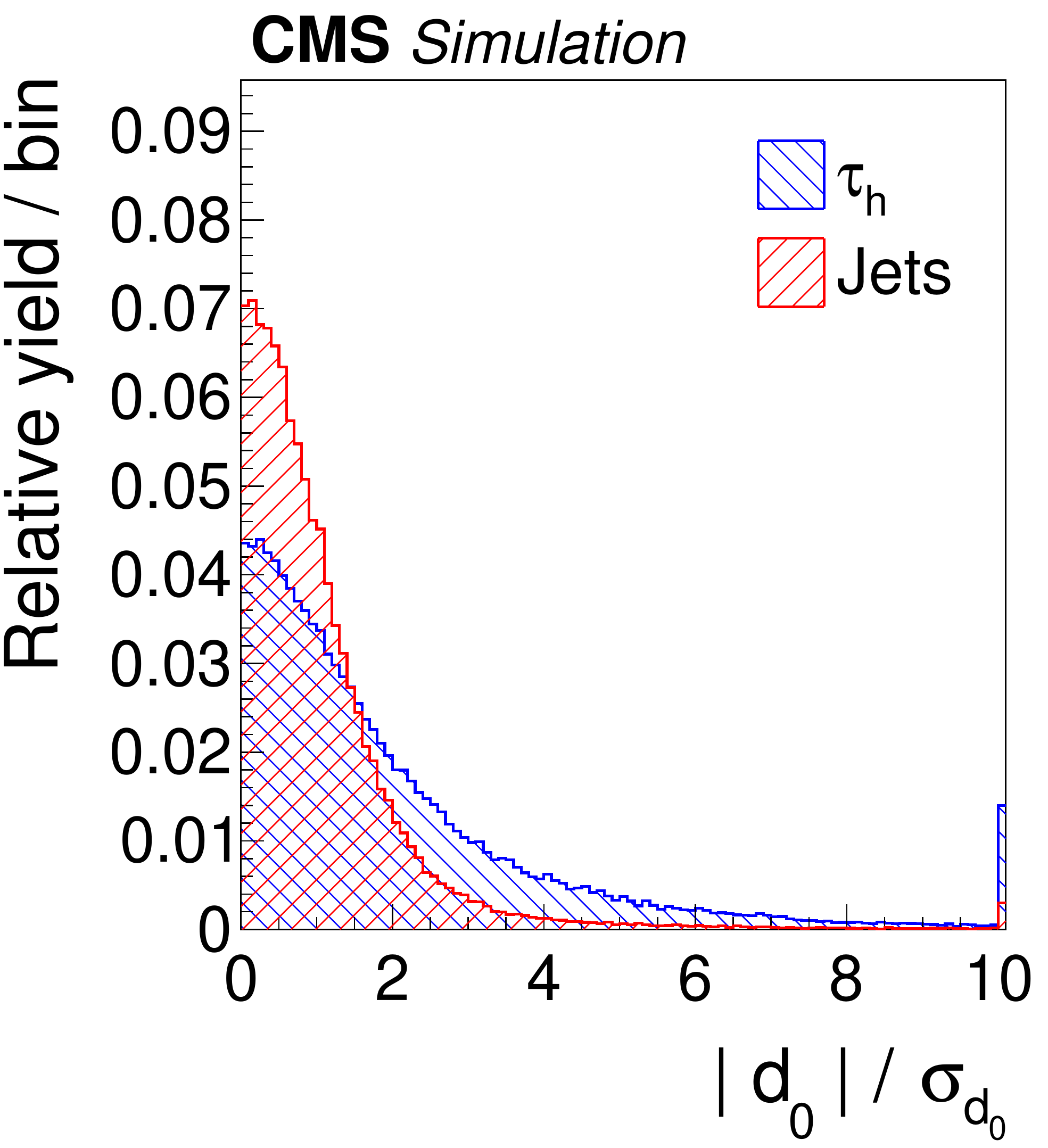}\\
\includegraphics[width=0.32\textwidth]{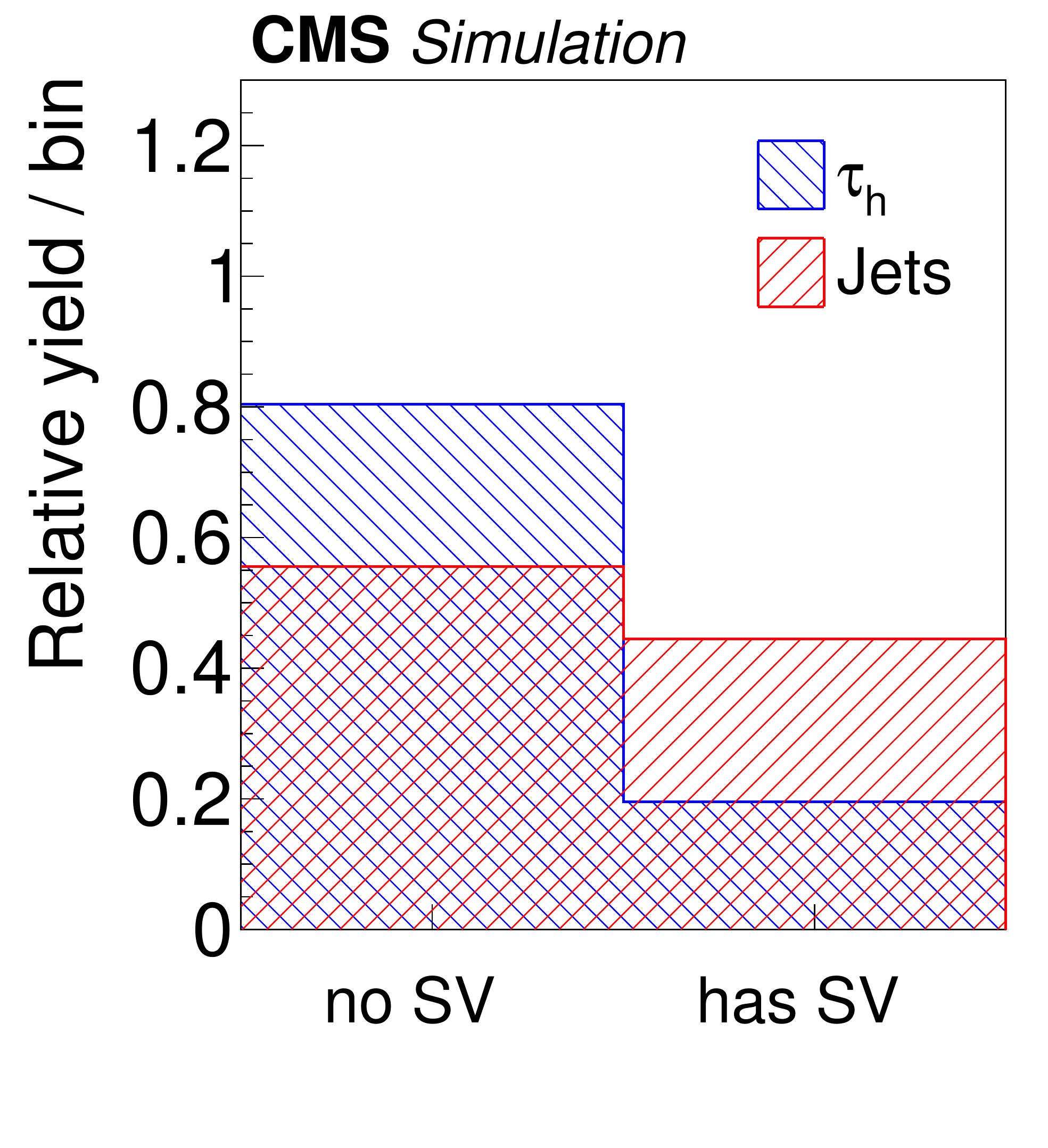}
\includegraphics[width=0.32\textwidth]{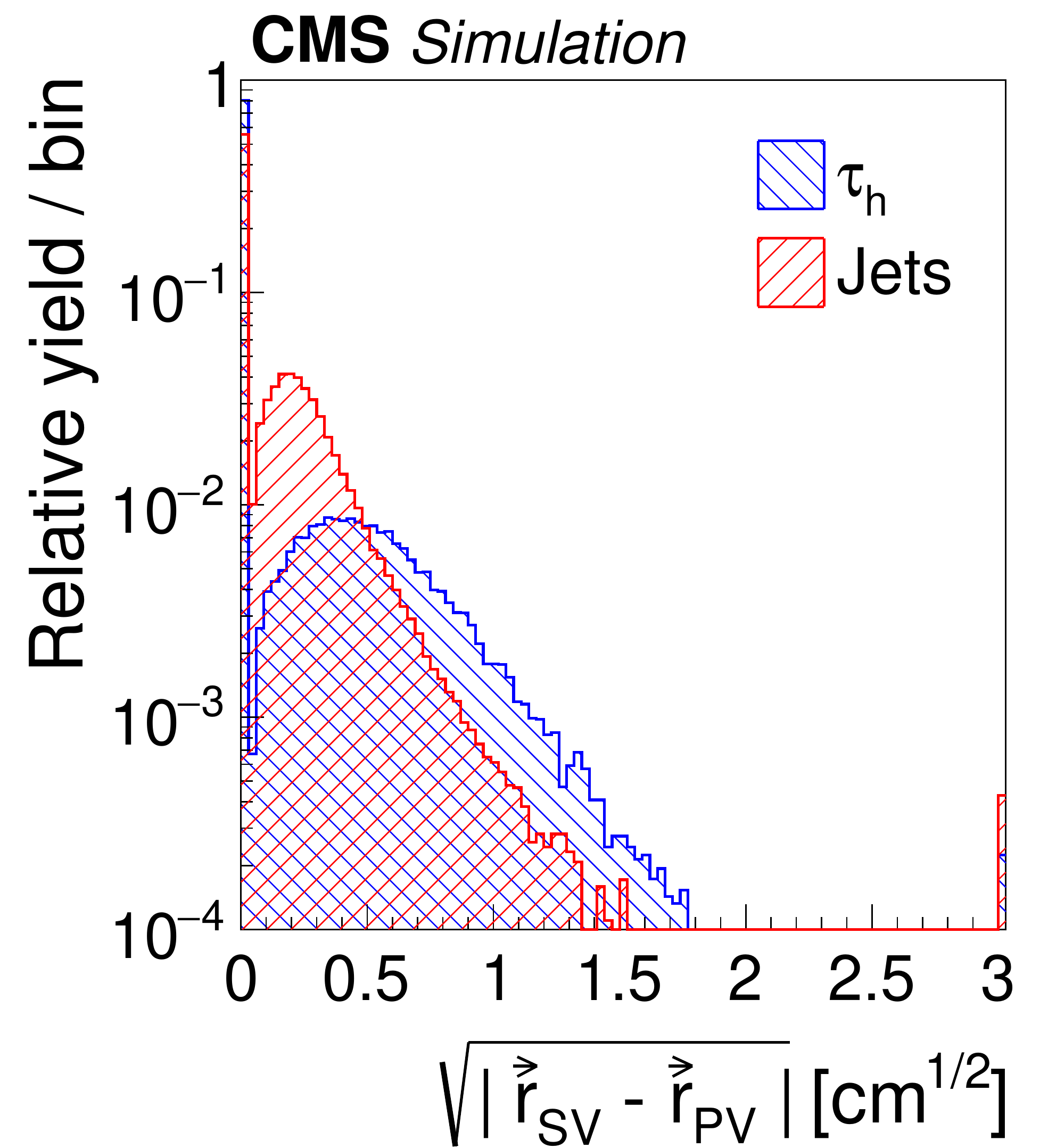}
\includegraphics[width=0.32\textwidth]{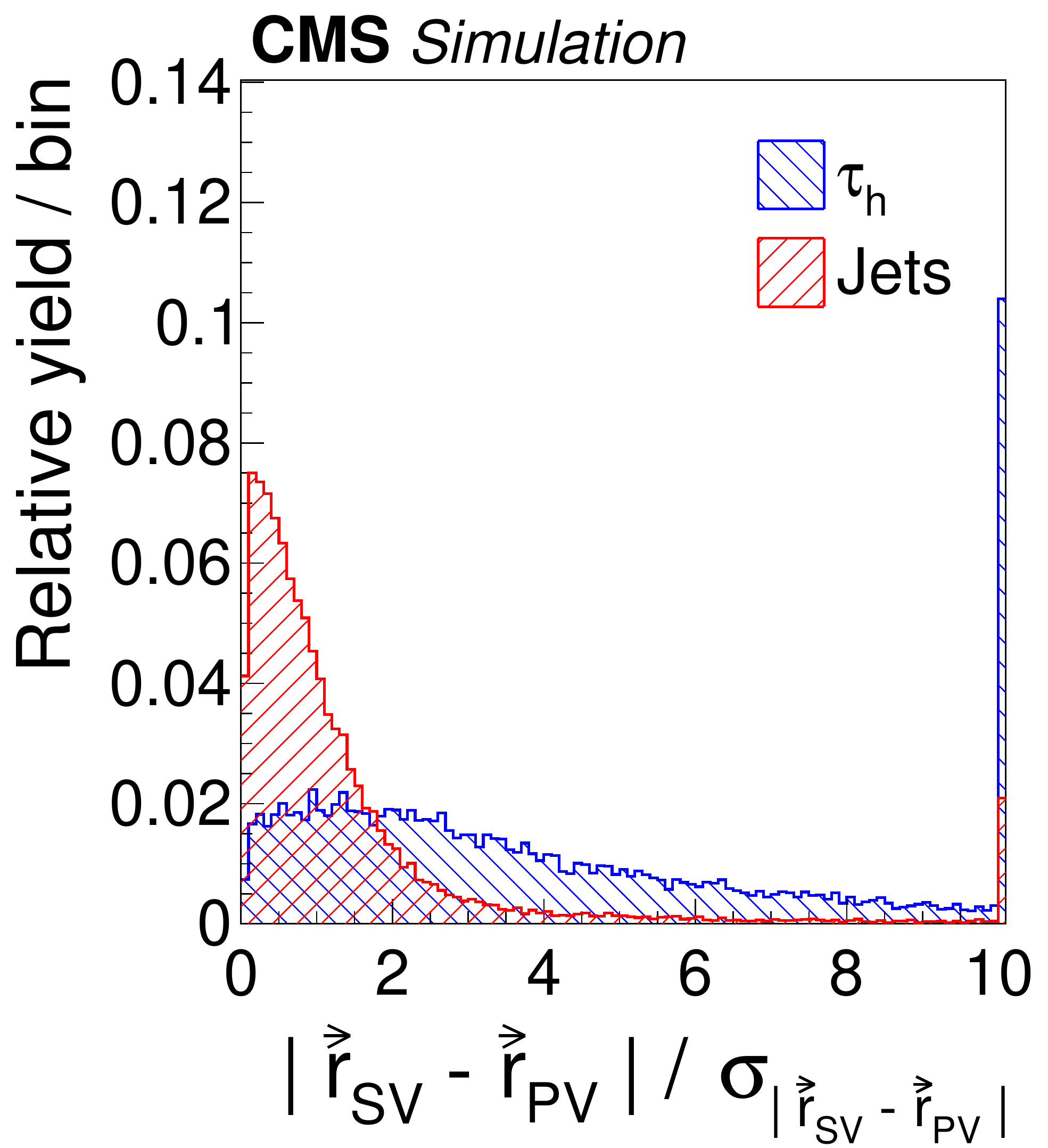}
\caption{
  Distributions, normalized to unity, in observables used as input variables to the MVA-based isolation discriminant,
  for hadronic $\Pgt$ decays in simulated $\cPZ/\Pggx \to \Pgt\Pgt$ (blue), and jets in simulated $\PW$+jets (red) events.
  The $\tauh$ candidates must have $\pt > 20$\GeV and $\abs{\eta} < 2.3$,
  and be reconstructed in one of the decay modes $\oneProngZeroPizero$, $\oneProngOnePizero$, $\oneProngTwoPizero$, or $\threeProngZeroPizero$.
  In the plot of the $\tauh$ decay mode on the upper right, an entry at 0 represents the decay mode $\oneProngZeroPizero$,
  1 and 2 represent the decay modes $\oneProngOnePizero$ and $\oneProngTwoPizero$, respectively,
  and entry 10 represents the $\threeProngZeroPizero$ decay mode.
}
\label{fig:tauIdMVAInputVariableDistributions}
\end{figure}

In order to minimize the jet $\to \tauh$ background,
the MVA-based $\tauh$ identification discriminant utilizes the transverse impact parameter of the ``leading'' (highest \pt) track of the $\tauh$ candidate,
defined as the distance of closest approach in the transverse plane of
the track to the $\tauh$ production vertex.
It also uses, for $\tauh$ candidates reconstructed in the $\threeProngZeroPizero$ decay mode,
the distance between the $\Pgt$ production point and the decay vertex.
A BDT is used to discriminate $\tauh$ decays (``signal'') from quark and gluon jets (``background'').
The variables used as inputs to the BDT are:
\begin{enumerate}
\item The charged- and neutral-particle isolation sums defined in Eq.~(\ref{eq:tauIsolation}) as separate inputs.
\item The reconstructed $\tauh$ decay mode, represented by an integer that takes the value of
      0 for $\tauh$ candidates reconstructed in the $\oneProngZeroPizero$ decay mode,
      as 1 and 2 for candidates reconstructed in the $\oneProngOnePizero$ and $\oneProngTwoPizero$ decay modes, respectively,
      and 10 for candidates reconstructed in the $\threeProngZeroPizero$ decay mode.
\item The transverse impact parameter $d_{0}$ of the leading track of the $\tauh$ candidate,
      and its value divided by its uncertainty, which corresponds to its significance $d_{0}/\sigma_{d_{0}}$.
\item The distance between the $\Pgt$ production and decay vertices, $\vert \vec{r}_{\mathrm{SV}} - \vec{r}_{\text{PV}} \vert$,
      and its significance, $\vert \vec{r}_{\mathrm{SV}} - \vec{r}_{\text{PV}} \vert/\sigma_{\vert \vec{r}_{\mathrm{SV}} - \vec{r}_{\text{PV}} \vert}$,
      and a flag indicating whether a decay vertex has successfully been reconstructed for a given $\tauh$ candidate.
      The positions of the vertices, $\vec{r}_{\mathrm{SV}}$ and $\vec{r}_\text{{PV}}$,
      are reconstructed using the adaptive vertex fitter algorithm~\cite{Fruhwirth:2007hz}.
\end{enumerate}
The position of the primary event vertex is refitted after excluding the tracks associated with the $\tauh$ candidate.
The discrimination power of individual input variables is illustrated in Fig.~\ref{fig:tauIdMVAInputVariableDistributions}.

The inputs are complemented by the \pt and $\eta$ of the $\tauh$ candidate
and by the $\Delta\beta$ correction defined in Eqs.~(\ref{eq:tauIsolation}) and~(\ref{eq:tauIsolationDB}).
The purpose of the \pt and $\eta$ variables is to parameterize possible dependences of the other input variables
on \pt and $\eta$.
The events used for the training of the BDT are reweighted such that the two-dimensional \pt and $\eta$ distribution of the $\tauh$ candidates
for signal and background are identical,
which makes the MVA result independent of event kinematics.
The $\Delta\beta$ correction parameterizes the dependence on pileup, in particular, the \pt sum of the neutral particles.

\begin{figure}[htb]
\centering
\includegraphics[width=0.65\textwidth]{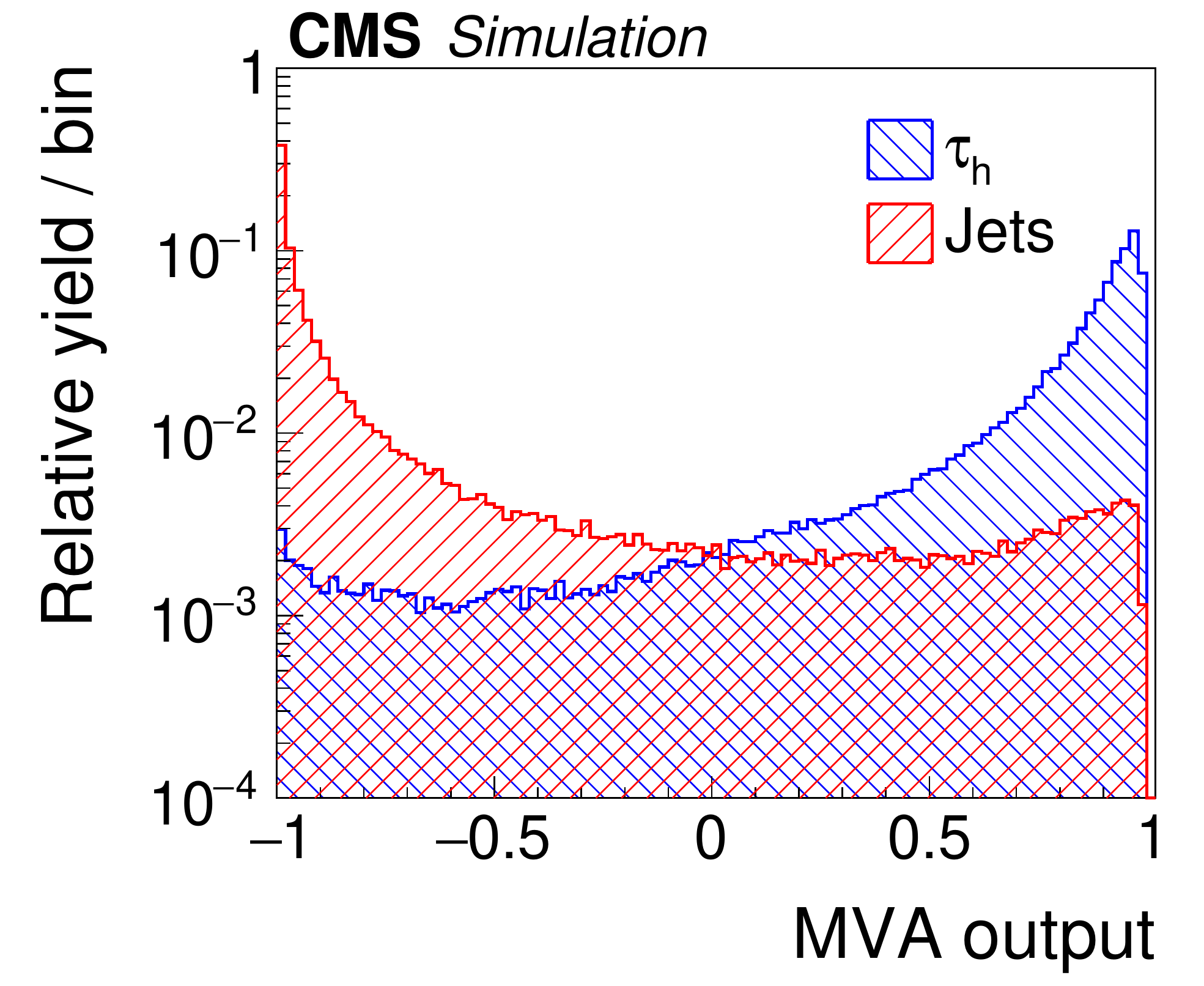}
\caption{
  Distribution of MVA output for the $\tauh$ identification discriminant that includes lifetime information
  for hadronic $\Pgt$ decays in simulated $\cPZ/\Pggx \to \Pgt\Pgt$ (blue), and jets in simulated $\PW$+jets (red) events.
}
\label{fig:tauIdMVAOutputDistribution}
\end{figure}

The BDT is trained on event samples produced using MC simulation.
Samples of $\cPZ/\Pggx \to \Pgt\Pgt$, $\PHiggs \to \Pgt\Pgt$, $\PZprime \to \Pgt\Pgt$, and $\PWprime \to \Pgt\APnut$ events are used for the ``signal'' category.
Reconstructed $\tauh$ candidates are required to match $\tauh$ decays within $\Delta R < 0.3$ at the generator level.
Multijet and $\PW$+jets events are used for the ``background'' category.
The $\tauh$ candidates that match leptons originating from the $\PW$ boson decays are excluded from the training.
The samples contain $\approx 10^{7}$ events in total,
and cover the range 20--2000\GeV in $\tauh$ candidate \pt.
Half of the available events are used for training, the other half for evaluating the MVA performance,
and conducting overtraining checks.
The distribution in MVA output is shown in Fig.~\ref{fig:tauIdMVAOutputDistribution}.

Different working points, corresponding to different $\tauh$ identification efficiencies and jet $\to \tauh$ misidentification rates,
are defined by changing the selections on the MVA output.
The thresholds are adjusted as function of the \pt of the $\tauh$ candidate,
such that the $\tauh$ identification efficiency for each WP is constant as function of \pt.

\subsection{Discriminants against electrons and muons}
\label{sec:discrAgainstElectronsAndMuons}

Electrons and muons have a sizeable probability to get reconstructed in the $\oneProngZeroPizero$ decay mode.
Electrons radiating a bremsstrahlung photon that subsequently converts may also get reconstructed in the $\oneProngOnePizero$ decay mode.
In particular, electrons and muons originating from decays of $\PW$ and $\cPZ$ bosons, which are produced with cross sections of $\approx$100\unit{nb} at the LHC at $\sqrt{s} = 8$\TeV,
have a high chance to pass isolation-based $\tauh$ identification criteria.
Dedicated discriminants have been developed to separate $\tauh$ from electrons and muons.
The separation of $\tauh$ from electrons is based on an MVA approach.
A cutoff-based and an MVA based discriminant are used to separate $\tauh$ from muons.

\subsubsection{MVA-based electron discriminant}
\label{sec:antiElectronDiscrMVABased}

\begin{figure}[htbp]
\centering
\includegraphics[width=0.32\textwidth]{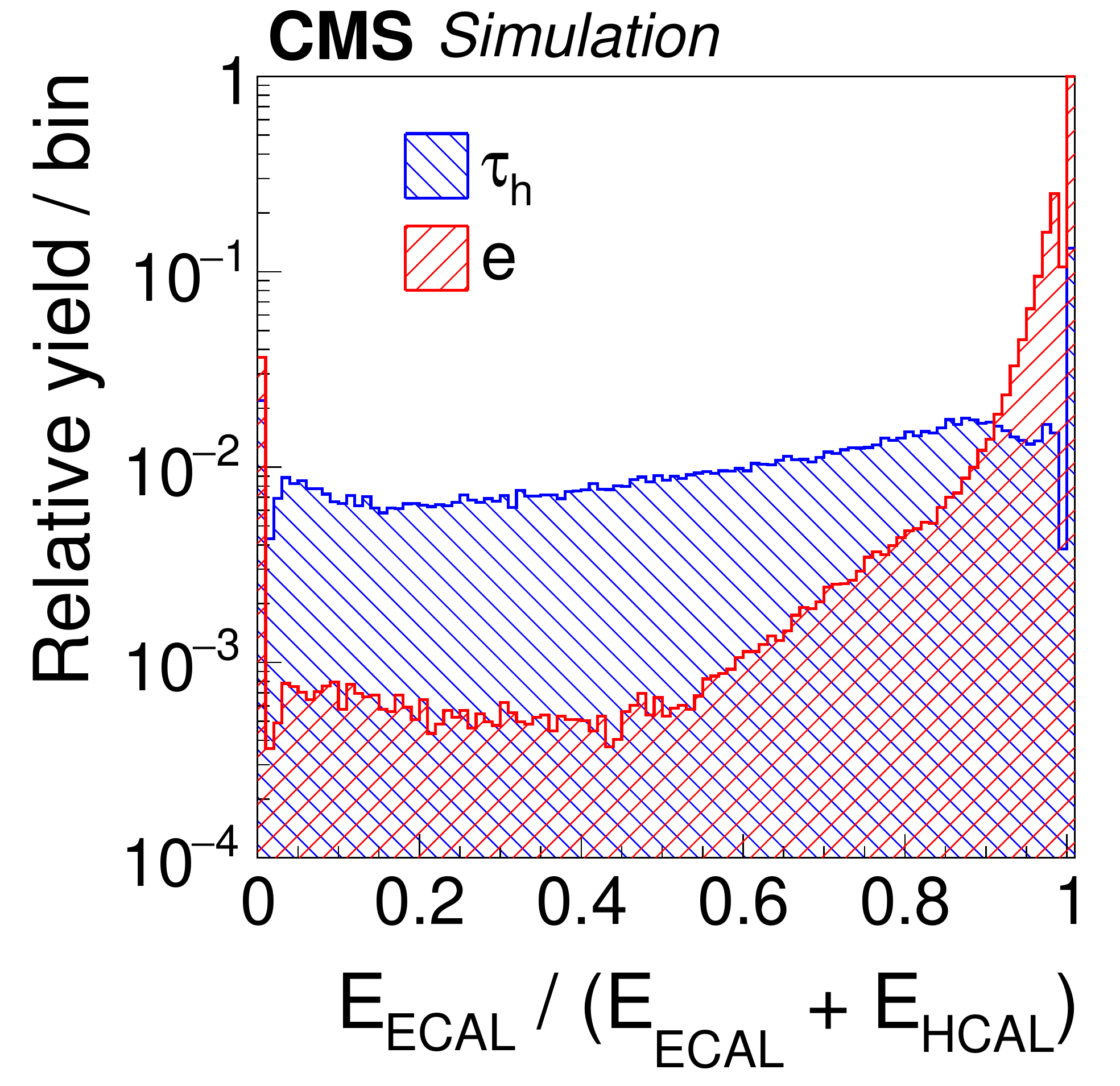}
\includegraphics[width=0.32\textwidth]{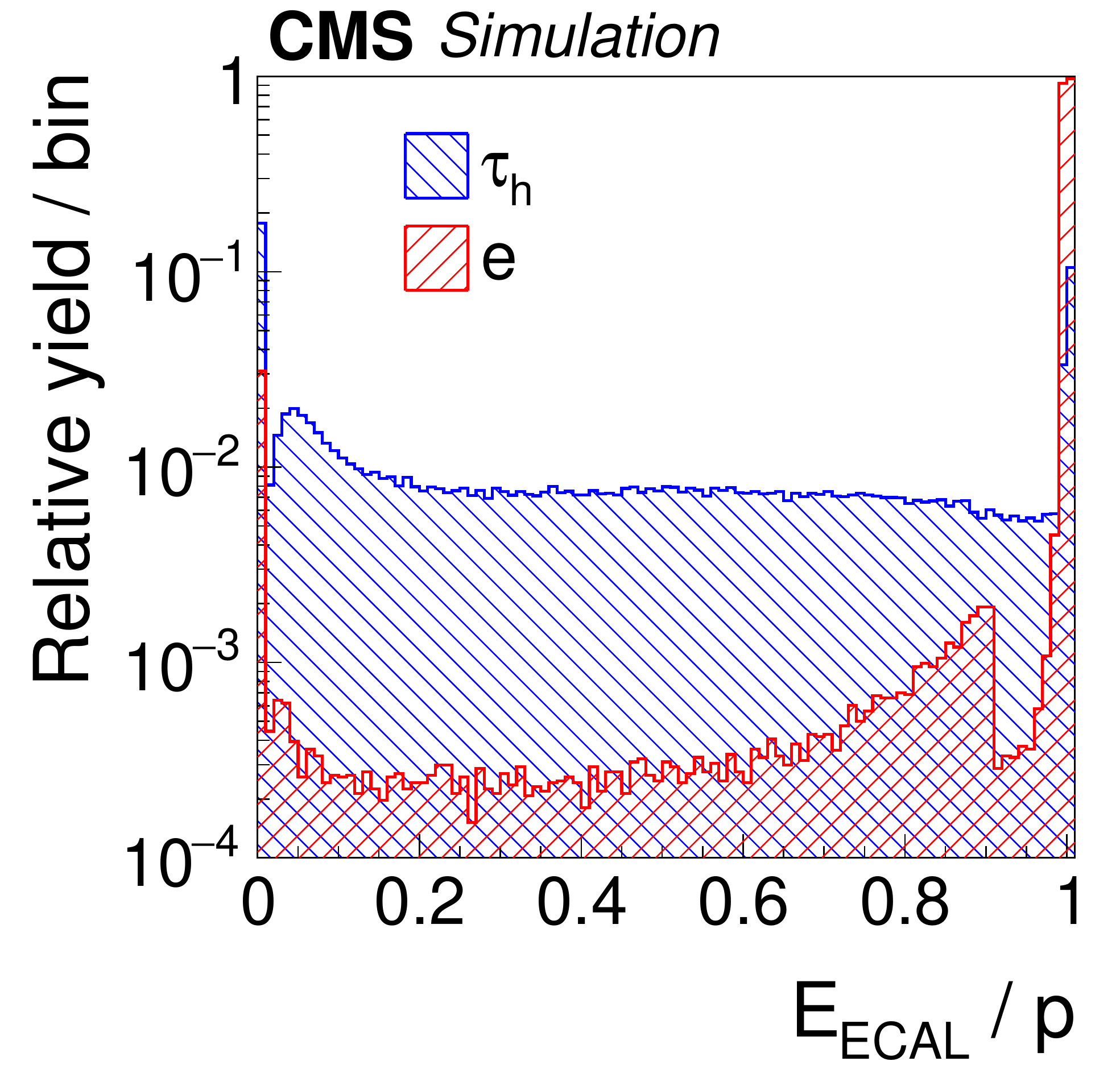}
\includegraphics[width=0.32\textwidth]{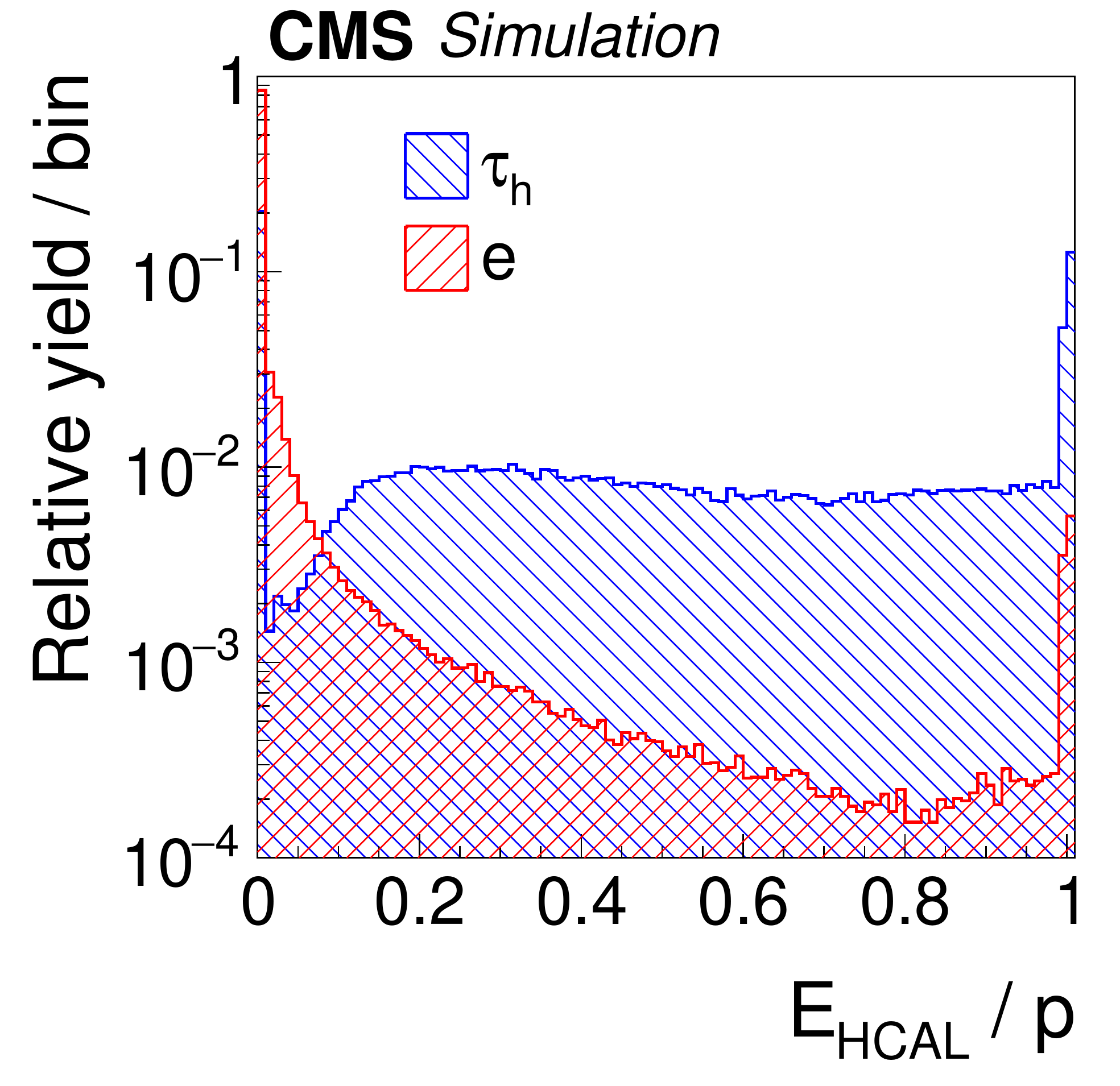}
\includegraphics[width=0.32\textwidth]{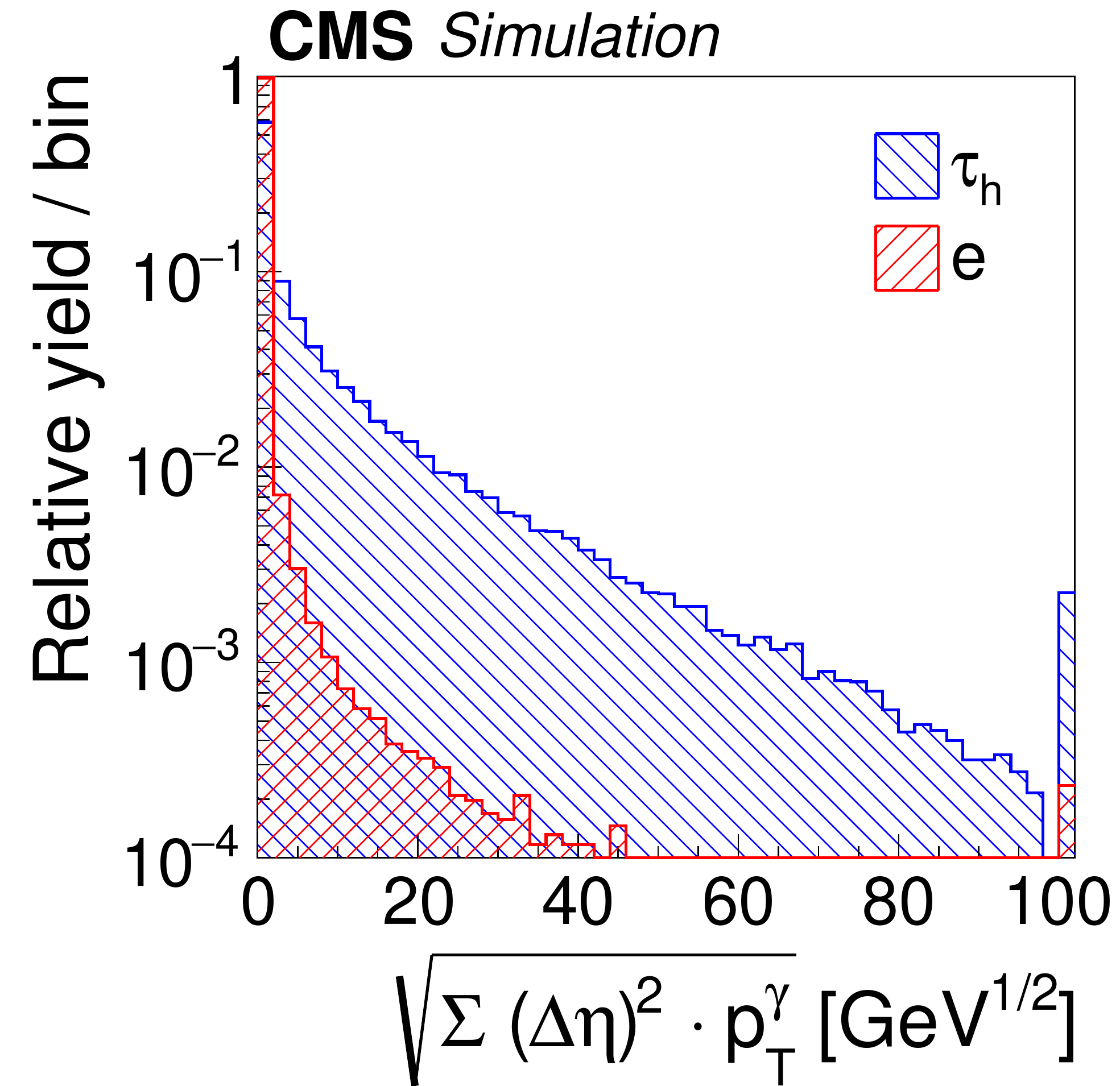}
\includegraphics[width=0.32\textwidth]{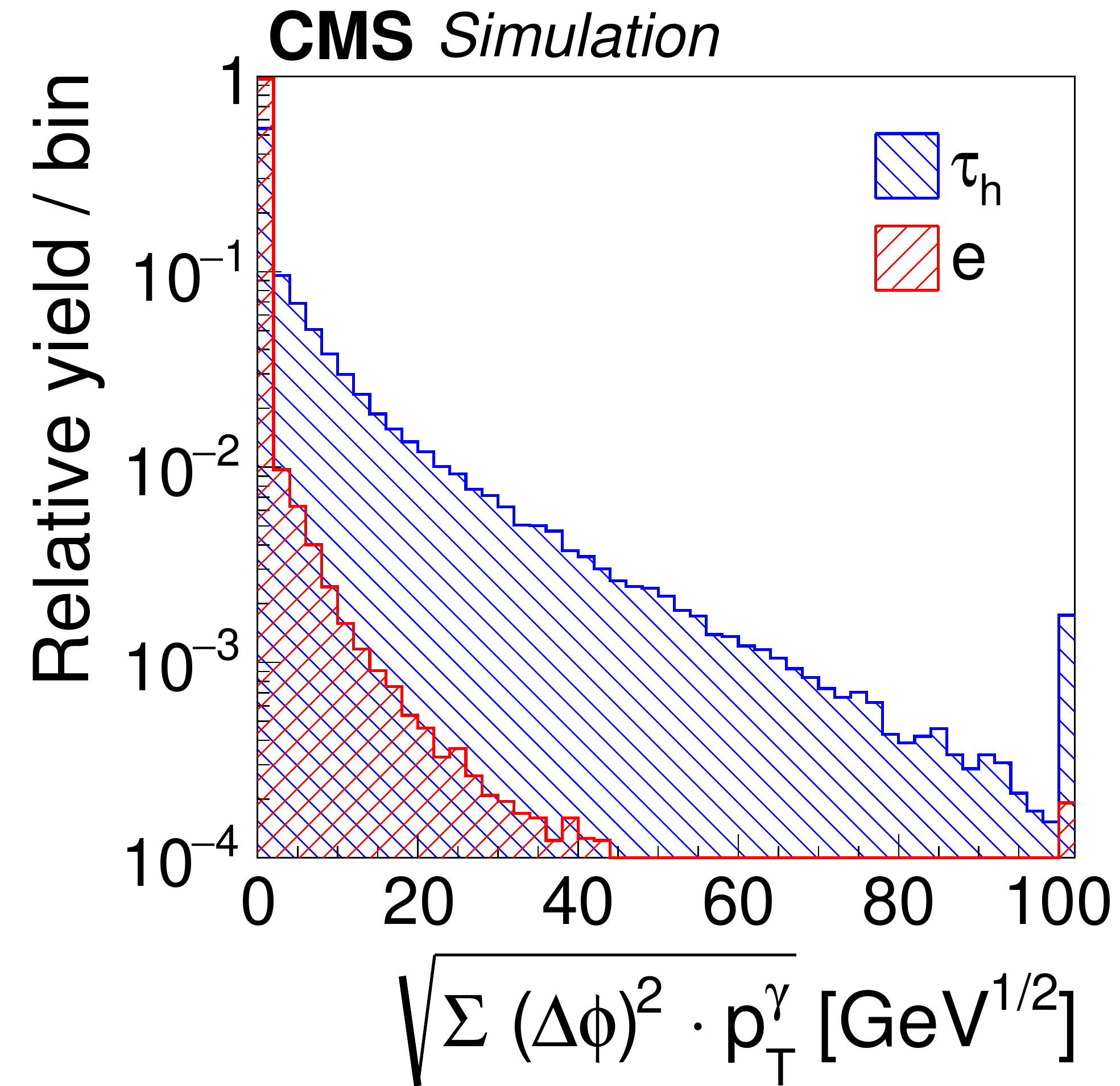}
\includegraphics[width=0.32\textwidth]{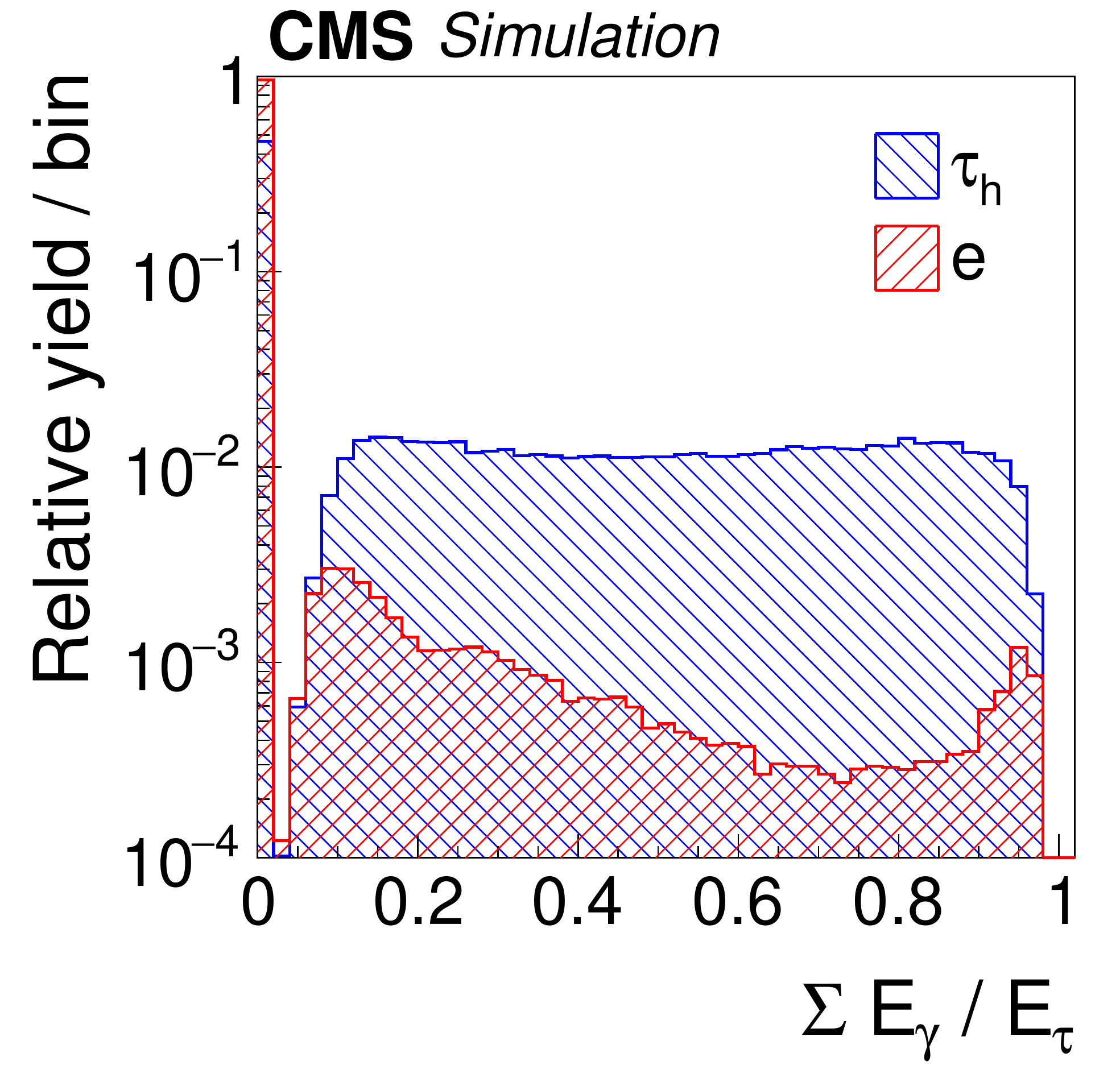}
\includegraphics[width=0.32\textwidth]{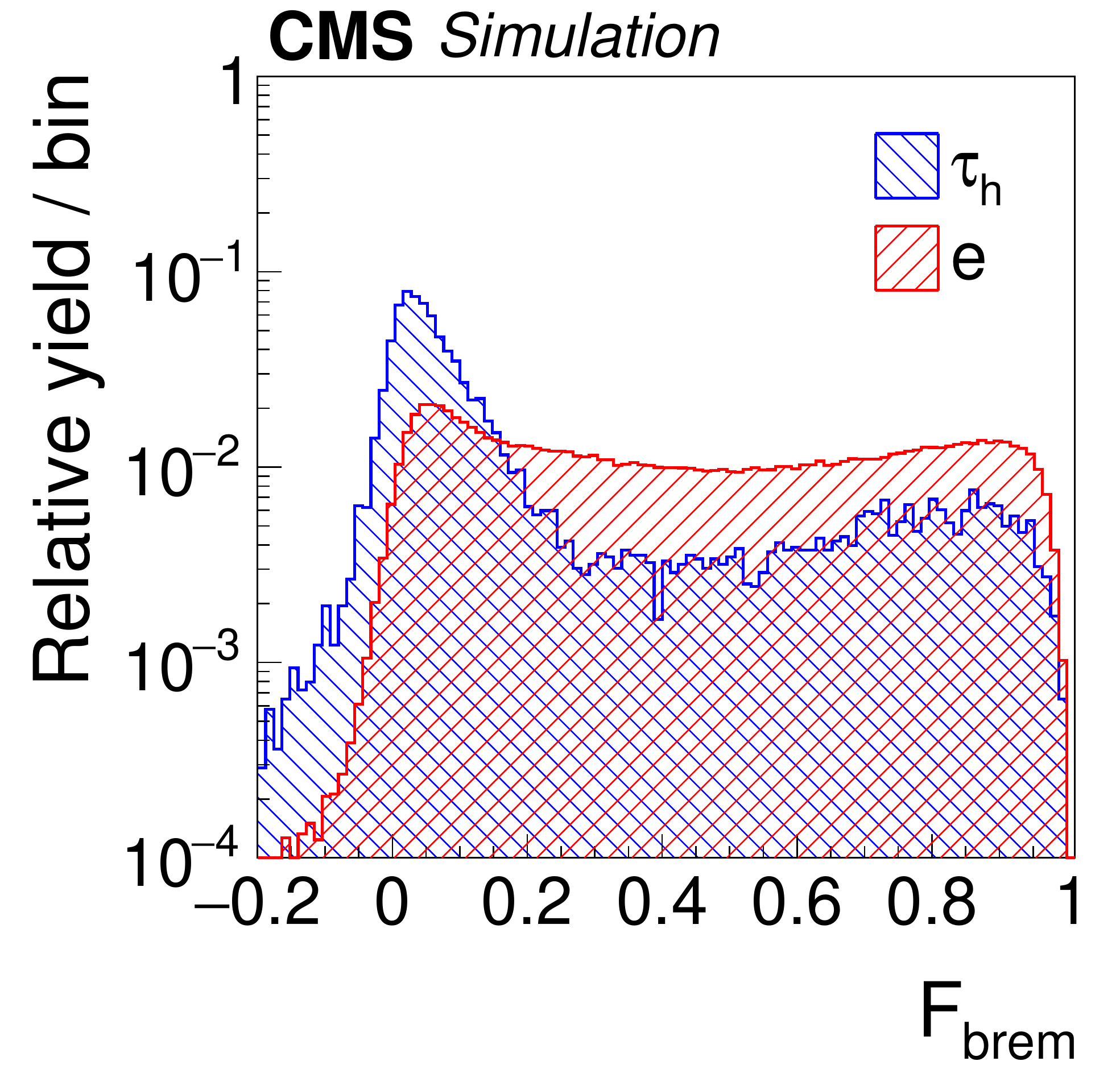}
\includegraphics[width=0.32\textwidth]{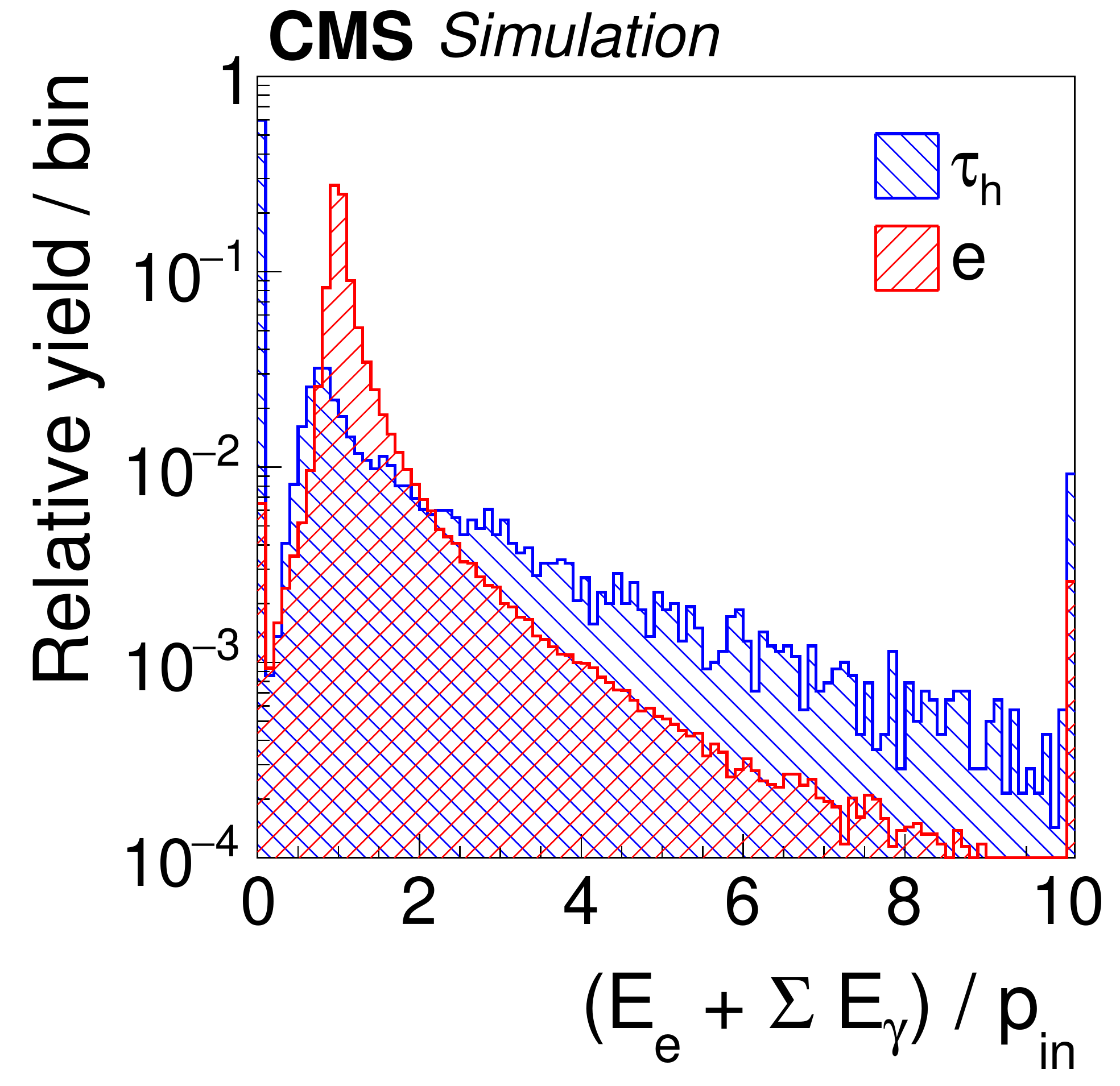}
\includegraphics[width=0.32\textwidth]{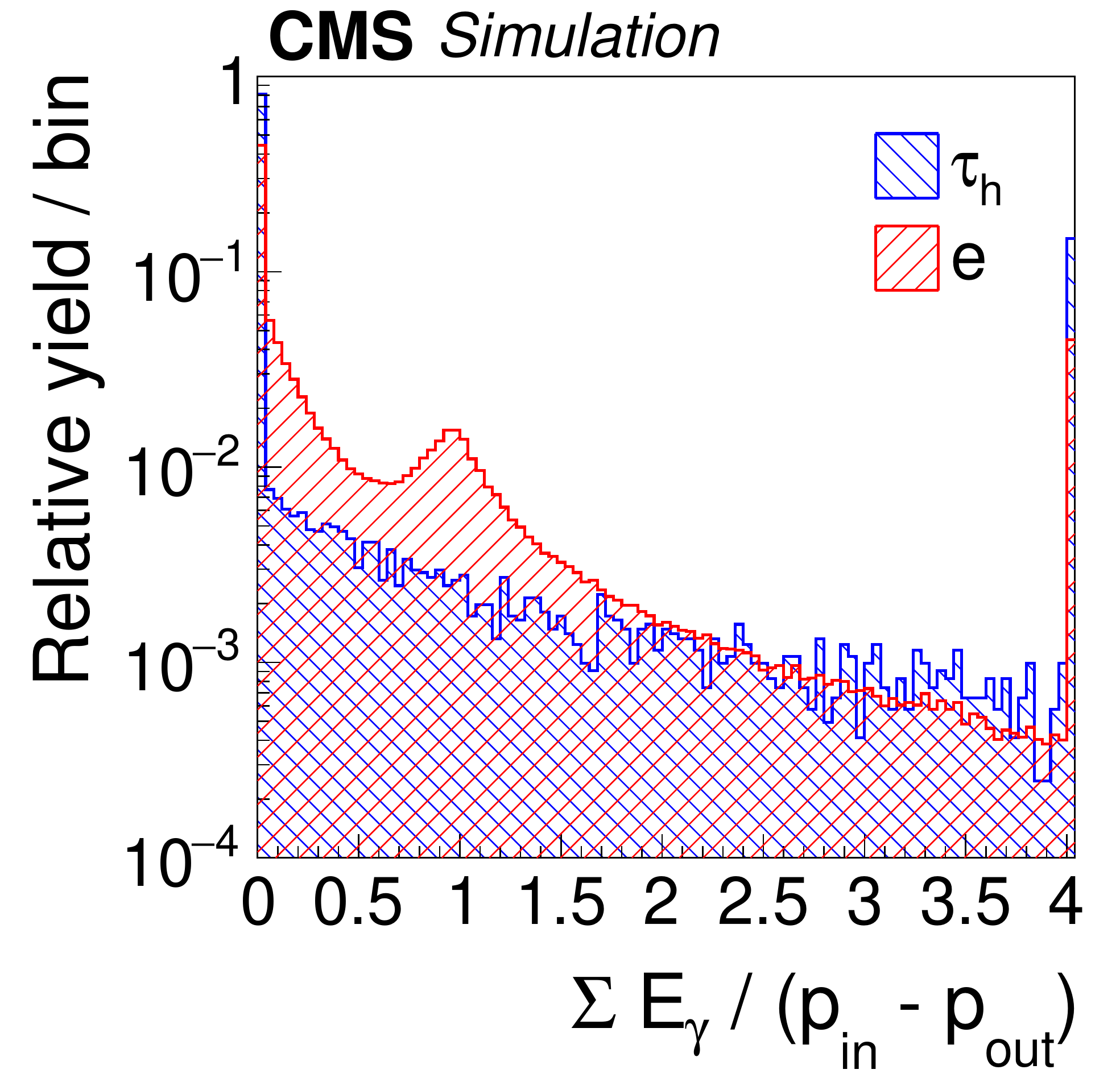}
\includegraphics[width=0.32\textwidth]{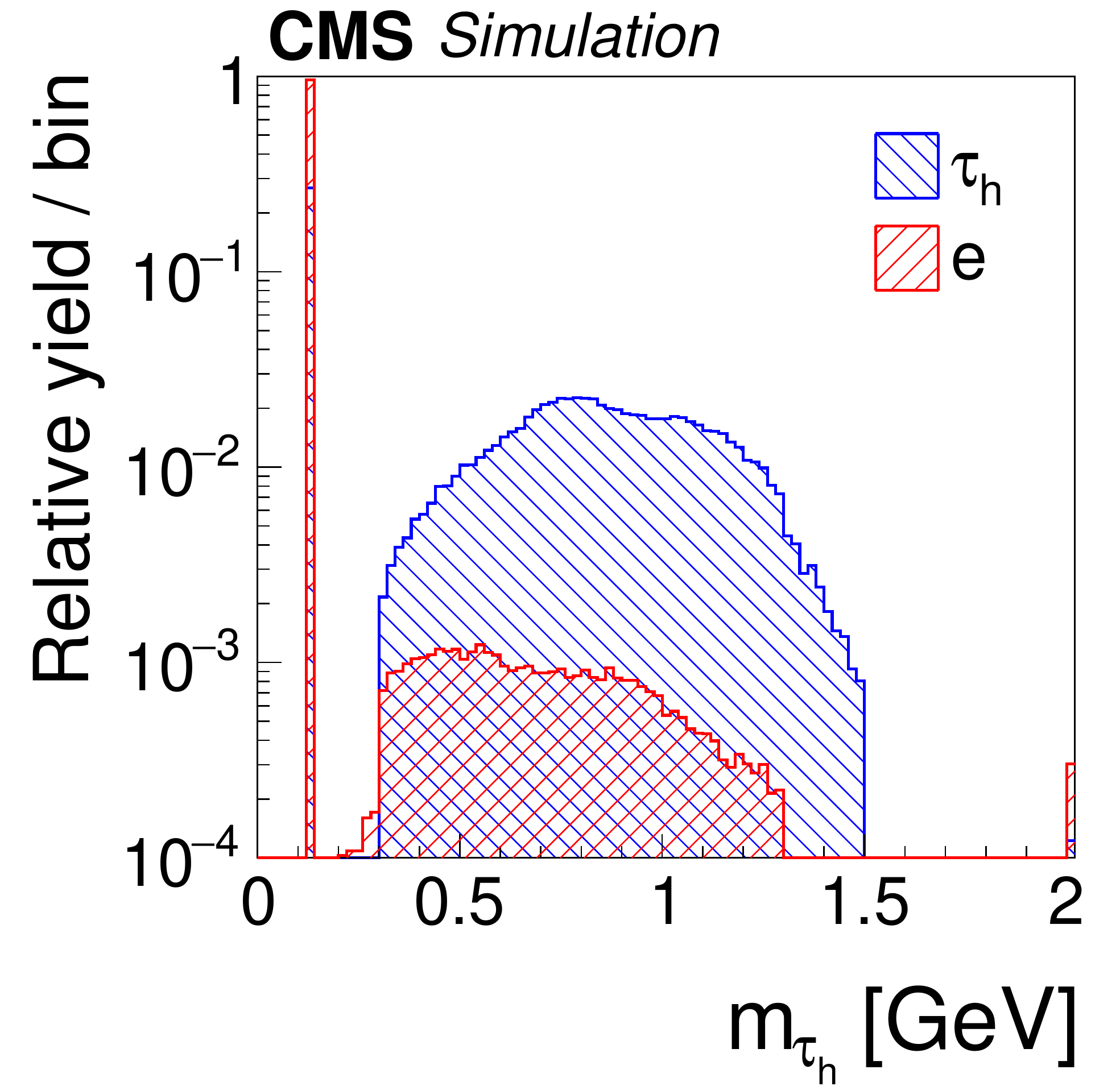}
\includegraphics[width=0.32\textwidth]{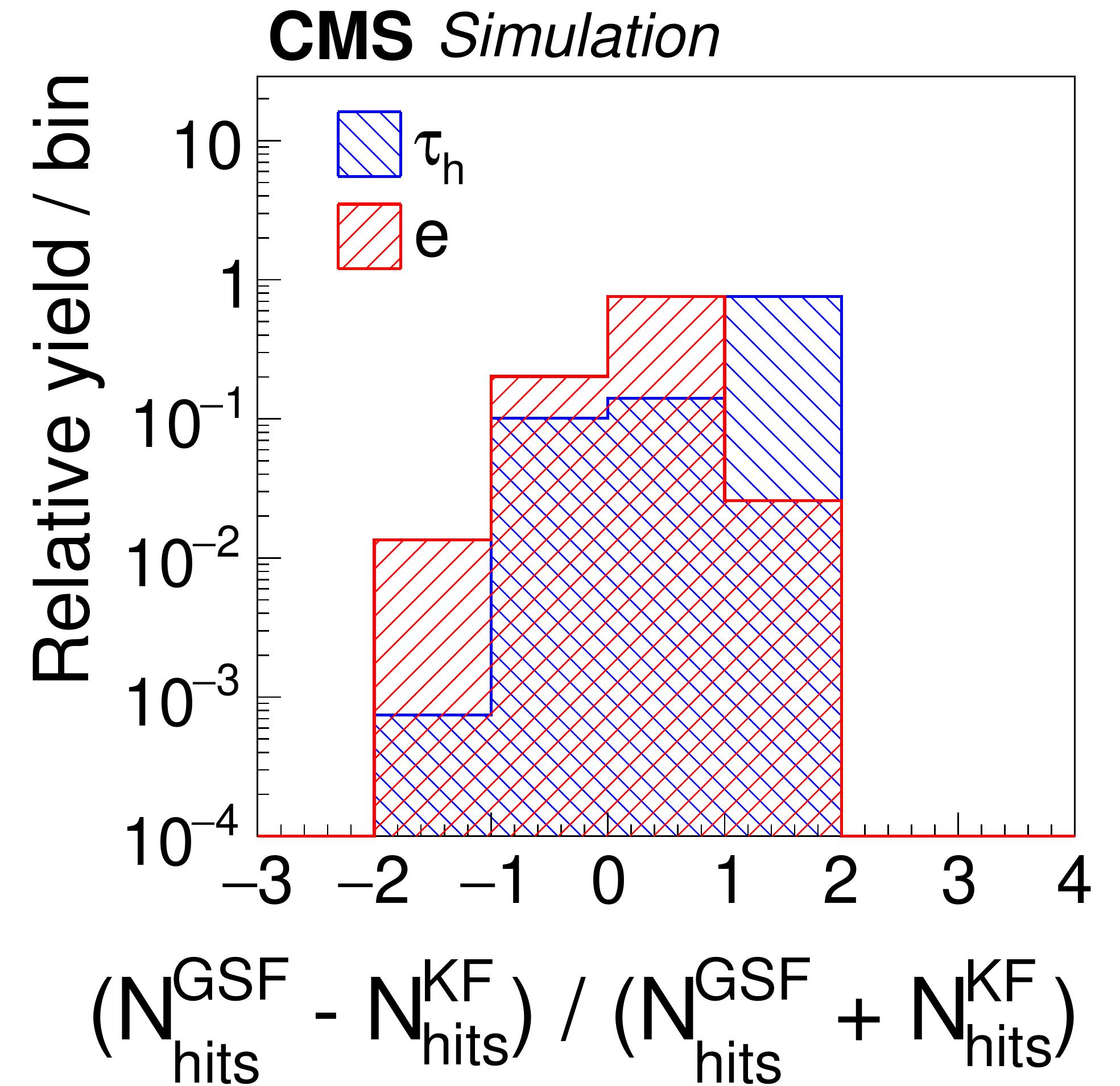}
\includegraphics[width=0.32\textwidth]{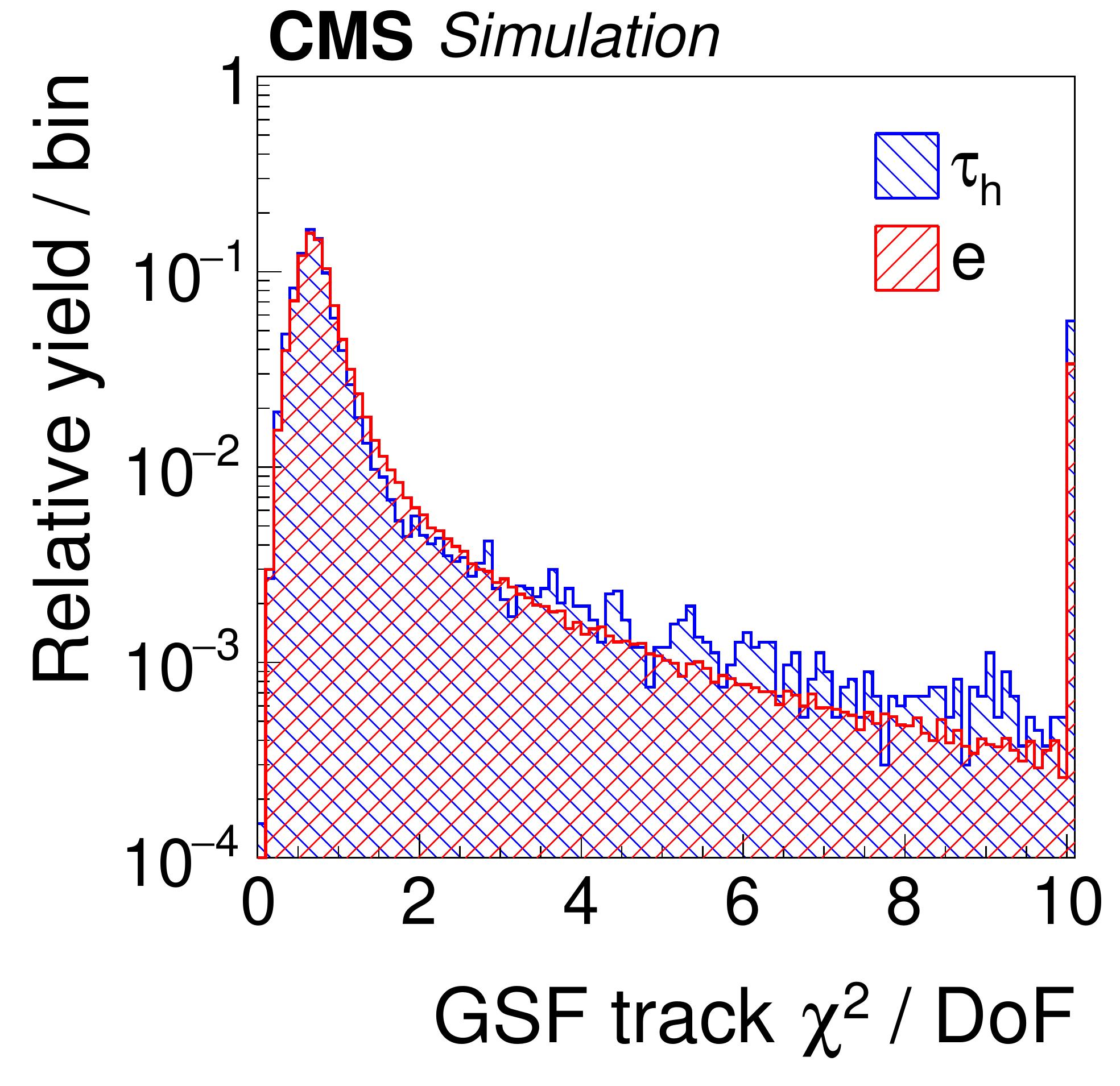}
\caption{
  Distributions, normalized to unity, in observables that are used as inputs to the MVA-based electron discriminant,
  for hadronic $\Pgt$ decays in simulated $\cPZ/\Pggx \to \Pgt\Pgt$ (blue), and electrons in simulated $\cPZ/\Pggx \to \Pe\Pe$ (red) events.
  The $\tauh$ candidates must have $\pt > 20$\GeV and $\abs{\eta} < 2.3$,
  and be reconstructed in one of the decay modes $\oneProngZeroPizero$, $\oneProngOnePizero$, $\oneProngTwoPizero$, or $\threeProngZeroPizero$.
  The rightmost bin of the distributions is used as overflow bin.
}
\label{fig:antiEMVAInputVariableDistributions}
\end{figure}

A BDT discriminant is trained to separate $\tauh$ decays from
electrons.
The algorithm utilizes observables that quantify the distribution in energy depositions in the ECAL,
in combination with observables sensitive to the amount of bremsstrahlung emitted along the leading track,
and observables sensitive to the overall particle multiplicity,
to distinguish electromagnetic from hadronic showers.
More specifically, the following variables are used as inputs to the BDT:
\begin{enumerate}
\item Electromagnetic energy fraction,
  $E_{\text{ECAL}}/(E_{\text{ECAL}}+E_{\text{HCAL}})$, defined
  as the ratio of energy depositions in the ECAL
  to the sum of energy in the ECAL and HCAL,
  associated with the charged particles and photons that constitute the $\tauh$ candidate.
\item $E_{\text{ECAL}}/p$ and $E_{\text{HCAL}}/p$, defined as ratios of ECAL and HCAL energies relative to the momentum of the leading charged-particle track of the $\tauh$ candidate.
\item $\sqrt{\sum (\Delta\eta)^{2}\pt^{\Pgg}}$ and $\sqrt{\sum (\Delta\phi)^{2}  \pt^{\Pgg}}$,
  the respective \pt-weighted (in \GeV) root-mean-square distances in $\eta$ and $\phi$ between the photons in any strip and the leading charged particle.
\item $\sum E_{\Pgg}/E_{\Pgt}$, the fraction of $\tauh$ energy carried by photons.
\item $F_{\text{brem}}=(p_{\text{in}} - p_{\text{out}})/p_{\text{in}}$, where $p_{\text{in}}$ and $p_{\text{out}}$
  are measured by the curvature of the leading track, reconstructed using the GSF algorithm, at the innermost and outermost positions of the tracker.
\item $(E_{\Pe} + \sum E_{\Pgg})/p_{\text{in}}$, the ratio between the total ECAL energy and the inner track momentum.
  The quantities $E_{\Pe}$ and $\sum E_{\Pgg}$ represent the energies
  of the electron cluster and of bremsstrahlung photons, respectively.
  $\sum E_{\Pgg}$ is reconstructed by summing the energy depositions
  in ECAL clusters located along the tangent to the GSF track.
\item $\sum E_{\Pgg}/(p_{\text{in}} - p_{\text{out}})$, the ratio of energies of the bremsstrahlung photons measured in the ECAL and in the tracker.
\item $m_{\tauh}$, the mass of the $\tauh$ candidate.
\item $(N_{\text{hits}}^{\text{GSF}} - N_{\text{hits}}^{\text{KF}})/(N_{\text{hits}}^{\text{GSF}} + N_{\text{hits}}^{\text{KF}})$,
  with $N_{\text{hits}}^{\text{GSF}}$ and
  $N_{\text{hits}}^{\text{KF}}$ representing, respectively, the number of hits in the silicon pixel and strip tracking detector
  associated with the track reconstructed using, respectively, the GSF and Kalman
  filter (KF) track reconstruction algorithms.
  The KF algorithm is the standard algorithm for track reconstruction at CMS~\cite{Chatrchyan:2014fea}.
  The number of hits associated with GSF and KF track is sensitive to the emission of hard bremsstrahlung photons.
\item $\chi^{2}$ per degree-of-freedom (DoF) of the GSF track.
\end{enumerate}
The discriminating power of these variables is illustrated in Fig.~\ref{fig:antiEMVAInputVariableDistributions}.

The inputs are complemented by the \pt and $\eta$ of the $\tauh$ candidate,
the \pt, $\sigma_{\pt} / \pt$, and $\eta$ of the GSF track,
and by the distances in $\eta$ and in $\phi$ of the GSF track to the nearest boundary between ECAL modules.
These variables are used to parameterize the dependence of the other input variables.
Electrons entering the boundaries between ECAL modules are more difficult to discriminate from $\tauh$ decays,
as their electromagnetic showers are often not well reconstructed,
and the probability to reach the hadron calorimeter increases in these regions.

Samples of simulated $\cPZ/\Pggx \to \Pgt\Pgt$, $\cPZ/\Pggx \to \Pe\Pe$, $\PW \to \Pgt\APnut$, $\PW \to \Pe\APnue$, $\cPqt\cPaqt$, $\PHiggs \to \Pgt\Pgt$,
$\PZprime \to \Pgt\Pgt$, $\PZprime \to \Pe\Pe$, $\PWprime \to \Pgt\APnut$, and $\PWprime \to \Pe\APnue$ events have been used to train the BDT.
Reconstructed $\tauh$ candidates are considered as signal or background when
they are matched, respectively, within $\Delta R < 0.3$ to a hadronic $\Pgt$ decay or to an electron at the generator level.

Different WP are defined by changing the cutoff on the BDT output.
The $\tauh$ candidates reconstructed in the uninstrumented region between ECAL barrel and endcap,
$1.45 < \eta < 1.56$, are rejected in all cases.

\subsubsection{Cutoff-based muon discriminant}
\label{sec:antiMuonDiscrCutBased}

The cutoff-based discriminant against muons vetoes $\tauh$ candidates
when signals in the muon system are found near the $\tauh$ direction.
Two working points are provided:
\begin{enumerate}
\item \textbf{Loose}:
  $\tauh$ candidates pass the cutoff on this discriminant, except when
  track segments are found in at least two muon stations within a cone of size $\Delta R = 0.3$ centred on the $\tauh$ direction,
  or when the sum of the energies in the ECAL and HCAL
  corresponds to $< 0.2$ of the momentum of the leading track of the $\tauh$ candidate.
\item \textbf{Tight}:
  $\tauh$ candidates pass this discriminant restriction when
  they pass the loose WP, and no hits are present
  within a cone of $\Delta R = 0.3$ around the $\tauh$ direction
  in the CSC, DT, and RPC detectors located in the two outermost muon stations.
\end{enumerate}

\subsubsection{MVA-based muon discriminant}
\label{sec:antiMuonDiscrMVABased}

A multivariate BDT discriminant has also been trained to separate $\tauh$ decays from muons.
The following variables are used as BDT inputs:
\begin{enumerate}
\item The calorimeter energy associated with the leading charged particle of the $\tauh$ candidate,
  with separate energy sums computed for ECAL and HCAL.
\item The calorimeter energy associated in the PF algorithm with any charged particle or photon constituting the $\tauh$ candidate, again, with separate energy sums computed for ECAL and HCAL.
\item The fraction of \pt carried by the charged particle with highest \pt.
\item The number of track segments in the muon system reconstructed within a cone of size $\Delta R = 0.5$ around the $\tauh$ direction.
\item The number of muon stations with at least one hit detected within a cone of size $\Delta R = 0.5$ centred on the $\tauh$ direction, computed separately for DT, CSC, and RPC detectors.
\end{enumerate}

The inputs are complemented by the $\eta$ of the $\tauh$ candidate,
to parameterize the dependence of the input variables on the DT, CSC, and RPC muon acceptance,
and on the path length of muons traversed in the ECAL and HCAL.

The BDT is trained using
samples of simulated $\cPZ/\Pggx \to \Pgt\Pgt$, $\cPZ/\Pggx \to \Pgm\Pgm$, $\PW \to \Pgt\APnut$, $\PW \to \Pgm\APnum$, $\cPqt\cPaqt$, $\PHiggs \to \Pgt\Pgt$,
$\PZprime \to \Pgt\Pgt$, $\PZprime \to \Pgm\Pgm$, $\PWprime \to \Pgt\APnut$, and $\PWprime \to \Pgm\APnum$ events.
Reconstructed $\tauh$ candidates are considered as signal or background when they are matched, respectively, to generator-level hadronic tau decays or muons
within $\Delta R < 0.3$.

Different WP are defined by changing the cutoff on the MVA output.

\section{Expected performance}
\label{sec:expectedPerformance}

The expected performance of the HPS $\tauh$ identification algorithm
is studied in terms of decay modes and energy reconstruction, $\tauh$ identification efficiency,
and misidentification rates for jets, electrons, and muons
using simulated samples of
$\cPZ/\Pggx \to \PLepton\PLepton$ ($\PLepton = \Pe$, $\Pgm$, $\Pgt$), $\PZprime \to \Pgt\Pgt$, $\PW$+jets, and multijet events.

Tau identification efficiencies and misidentification rates in MC simulated events,
averaged over \pt and $\eta$, for pileup conditions characteristic of the data-taking period,
are given in Table~\ref{tab:tauIdPerformance}.

\begin{table}
\centering
\topcaption{
  Expected efficiencies and misidentification rates of various $\tauh$ identification discriminants,
  averaged over \pt and $\eta$, for pileup conditions characteristic of the LHC Run 1 data-taking period.
  The DM-finding criterion refers to the requirement that the $\tauh$ candidate be reconstructed
in one of the decay modes $\oneProngZeroPizero$, $\oneProngOnePizero$, $\oneProngTwoPizero$, or $\threeProngZeroPizero$ (cf. Section~\ref{sec:decay_mode_reconstruction}).
}
\label{tab:tauIdPerformance}
\begin{tabular}{lcc|cc}
\hline
\multicolumn{5}{c}{DM-finding and $\tauh$ isolation discriminants} \\
\hline
\multirow{2}{10mm}{WP} & \multicolumn{2}{c|}{Efficiency} & \multicolumn{2}{c}{Jet $\to \tauh$ misidentification rate} \\
\cline{2-5}
 & $\cPZ/\Pggx \to \Pgt\Pgt$ & $\PZprime(2.5\TeV) \to \Pgt\Pgt$ & $\PW$+jets & Multijet \\
\hline
\multicolumn{5}{c}{Cutoff-based} \\
\hline
Loose      & 49.0\% & 58.9\% & $9.09 \times 10^{-3}$ & $3.86 \times 10^{-3}$ \\
Medium     & 40.8\% & 50.8\% & $5.13 \times 10^{-3}$ & $2.06 \times 10^{-3}$ \\
Tight      & 38.1\% & 48.1\% & $4.38 \times 10^{-3}$ & $1.75 \times 10^{-3}$ \\
\hline
\multicolumn{5}{c}{MVA-based} \\
\hline
Very loose & 55.9\% & 71.2\% & $1.29 \times 10^{-2}$ & $6.21 \times 10^{-3}$ \\
Loose      & 50.7\% & 64.3\% & $7.38 \times 10^{-3}$ & $3.21 \times 10^{-3}$ \\
Medium     & 39.6\% & 50.7\% & $3.32 \times 10^{-3}$ & $1.30 \times 10^{-3}$ \\
Tight      & 27.3\% & 36.4\% & $1.56 \times 10^{-3}$ & $4.43 \times 10^{-4}$ \\
\hline
\end{tabular}
\\[1em]
\begin{tabular}{lcc|c}
\hline
\multicolumn{4}{c}{Discriminant against electrons} \\
\hline
\multirow{2}{10mm}{WP} & \multicolumn{2}{c|}{Efficiency} & $\Pe \to \tauh$ misidentification rate \\
\cline{2-3}
 & $\cPZ/\Pggx \to \Pgt\Pgt$ & $\PZprime(2.5$\TeV$) \to \Pgt\Pgt$ & $\cPZ/\Pggx \to \Pe\Pe$ \\
\hline
Very loose & 94.3\% & 89.6\% & $2.38 \times 10^{-2}$ \\
Loose      & 90.6\% & 81.5\% & $4.43 \times 10^{-3}$ \\
Medium     & 84.8\% & 73.2\% & $1.38 \times 10^{-3}$ \\
Tight      & 78.3\% & 65.1\% & $6.21 \times 10^{-4}$ \\
Very tight & 72.1\% & 60.0\% & $3.54 \times 10^{-4}$ \\
\hline
\end{tabular}
\\[1em]
\begin{tabular}{lcc|c}
\hline
\multicolumn{4}{c}{Discriminant against muons} \\
\hline
\multirow{2}{10mm}{WP} & \multicolumn{2}{c|}{Efficiency} & $\Pgm \to \tauh$ misidentification rate \\
\cline{2-3}
 & $\cPZ/\Pggx \to \Pgt\Pgt$ & $\PZprime(2.5\TeV) \to \Pgt\Pgt$ & $\cPZ/\Pggx \to \Pgm\Pgm$ \\
\hline
\multicolumn{4}{c}{Cutoff-based} \\
\hline
Loose  & 99.3\% & 96.4\% & $1.77 \times 10^{-3}$ \\
Tight  & 99.1\% & 95.0\% & $7.74 \times 10^{-4}$ \\
\hline
\multicolumn{4}{c}{MVA-based} \\
\hline
Loose  & 99.5\% & 99.4\% & $5.20 \times 10^{-4}$ \\
Medium & 99.0\% & 98.8\% & $3.67 \times 10^{-4}$ \\
Tight  & 98.0\% & 97.7\% & $3.18 \times 10^{-4}$ \\
\hline
\end{tabular}
\end{table}

\subsection{Decay modes and energy reconstruction}
\label{sec:expectedPerformance_decaymode_and_energy_reconstruction}

\begin{figure}[htb]
\centering
\includegraphics[width=0.48\textwidth]{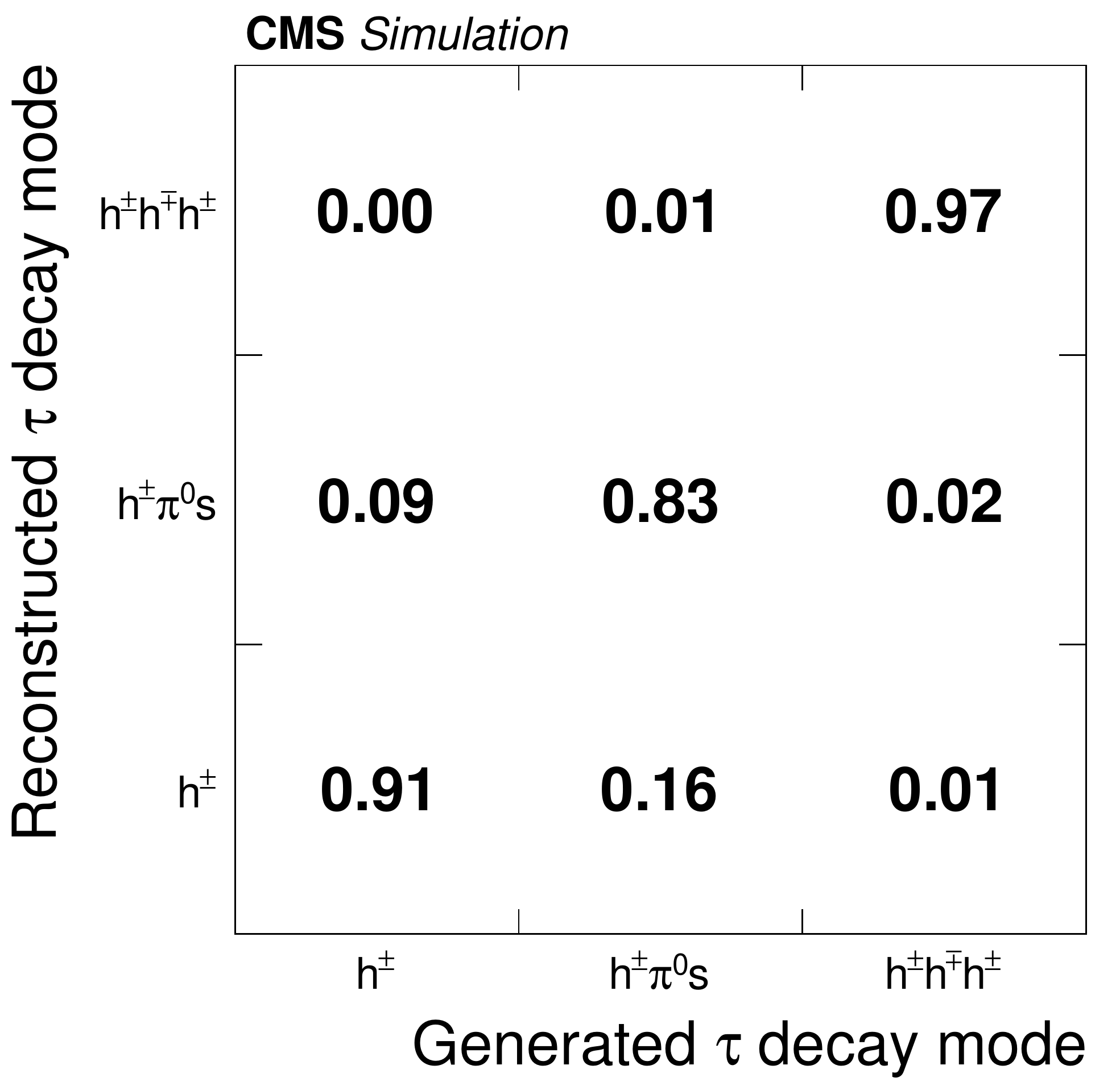}
\includegraphics[width=0.48\textwidth]{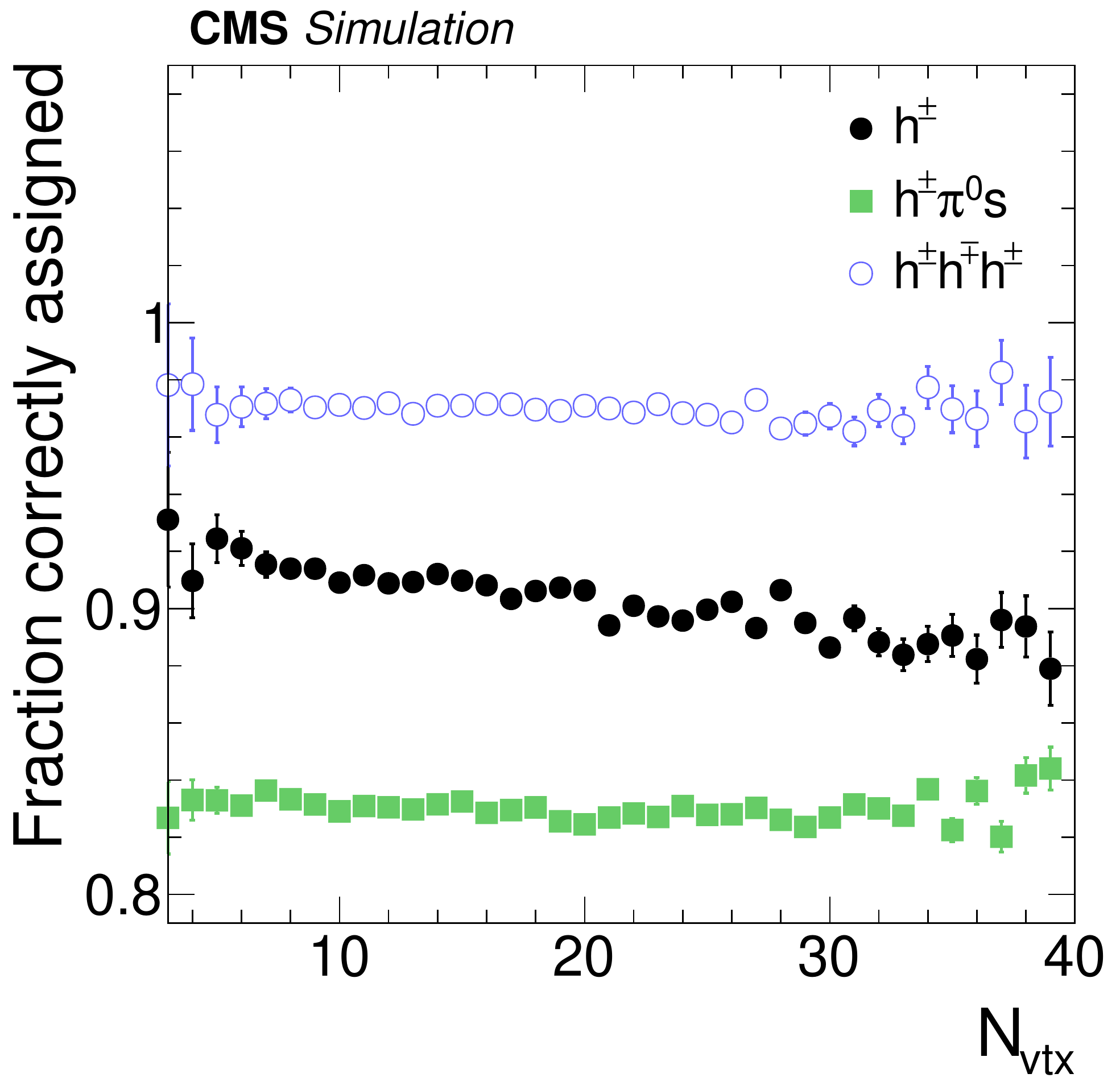}
\caption{
  Left: Correlation between generated and reconstructed $\tauh$ decay modes for $\tauh$ decays in $\cPZ/\Pggx \to \Pgt\Pgt$ events,
  simulated for pileup conditions characteristic of the LHC Run 1 data-taking period.
  Right: Fraction of generated $\tauh$ reconstructed in the correct decay mode as function of $\Nvtx$.
  Reconstructed $\tauh$ candidates are required to be matched to hadronic $\Pgt$ decays at the generator-level within $\Delta R < 0.3$,
  to be reconstructed in one of the decay modes $\oneProngZeroPizero$, $\oneProngOnePizero$, $\oneProngTwoPizero$, or $\threeProngZeroPizero$,
  and pass $\pt > 20$\GeV, $\abs{\eta} < 2.3$, and the loose WP of the cutoff-based $\tauh$ isolation discriminant.
}
\label{fig:recVsGenTauDecayMode_simZtautau}
\end{figure}

The $\tauh$ decay mode reconstruction
is studied in simulated $\cPZ/\Pggx \to \Pgt\Pgt$ events.
The performance is quantified by the correlation between reconstructed and generator-level $\tauh$ decay modes.
Figure~\ref{fig:recVsGenTauDecayMode_simZtautau} demonstrates that the true $\Pgt$ decay mode is reconstructed in about 90\% of the cases,
irrespective of pileup conditions, represented by the number of reconstructed vertices ($\Nvtx$).
The few per cent decrease in the fraction of $\Pgt$ leptons decaying to a single charged hadron that are reconstructed in the true decay mode
is due to events in which particles from pileup deposit energy in the ECAL near the $\Pgt$,
causing the $\Pgt$ to be reconstructed in the $\oneProngOnePizero$ or $\oneProngTwoPizero$ decay modes.

\begin{figure}[htb]
\centering
\includegraphics[width=0.48\textwidth]{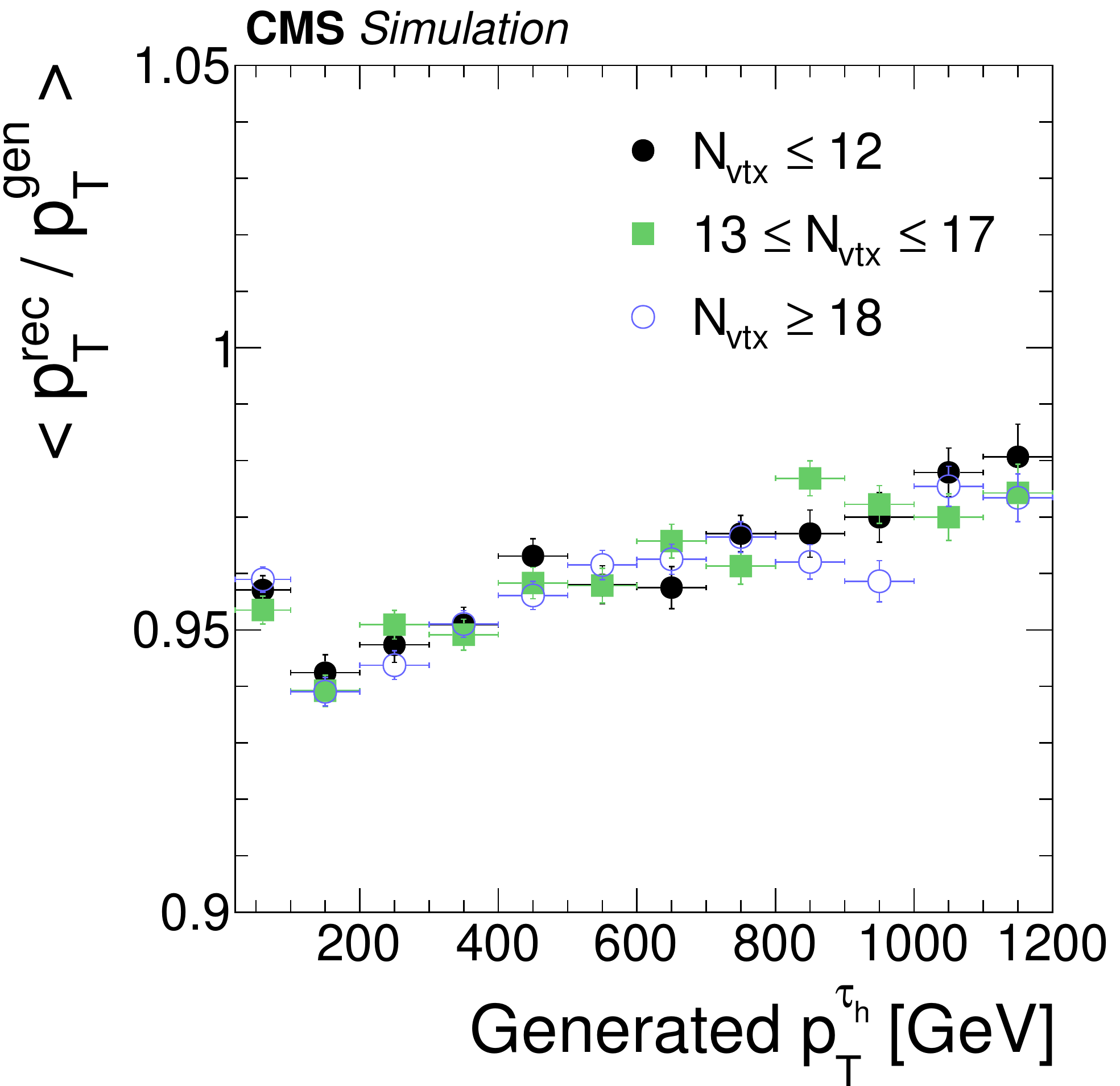}
\includegraphics[width=0.48\textwidth]{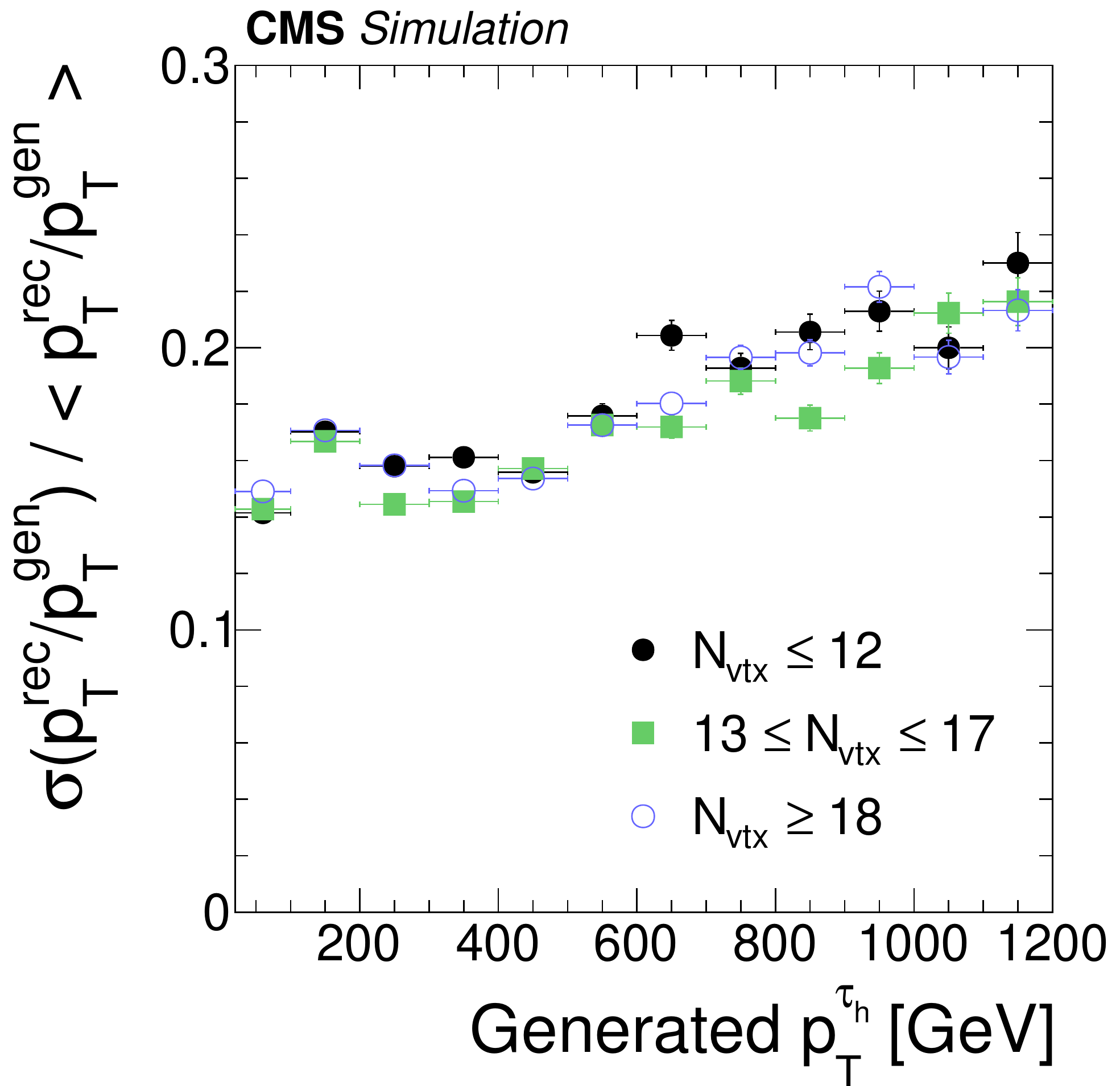}
\caption{
  The $\tauh$ energy response (left) and relative resolution (right) as function of generator-level visible $\Pgt$ \pt
  in simulated $\PZprime \to \Pgt\Pgt$ events for different pileup conditions:
  $\Nvtx \leq 12$, $13 \leq \Nvtx \leq 17$, and $\Nvtx \geq 18$.
  Reconstructed $\tauh$ candidates are required to be matched to hadronic $\Pgt$ decays at the generator-level within $\Delta R < 0.3$,
  to be reconstructed in one of the decay modes $\oneProngZeroPizero$, $\oneProngOnePizero$, $\oneProngTwoPizero$ or $\threeProngZeroPizero$,
  and to pass $\pt > 20$\GeV, $\abs{\eta} < 2.3$, and the loose WP of the cutoff-based $\tauh$ isolation discriminant.
}
\label{fig:recDivGenVisTauPt}
\end{figure}

The performance of energy reconstruction is studied in simulated $\cPZ/\Pggx \to \Pgt\Pgt$ and $\PZprime \to \Pgt\Pgt$ events,
and quantified in terms of response and resolution,
defined as the mean and standard deviation of the reconstructed momentum distribution relative to the generator-level momentum of the visible $\Pgt$ decay products.
The distributions for $\tauh$ decays in simulated $\cPZ/\Pggx \to \Pgt\Pgt$ and $\PZprime \to \Pgt\Pgt$ events
are shown in Fig.~\ref{fig:recDivGenVisTauPt}.
The average response is below 1.0, because of an asymmetry of the $\langle p_{T}^{\text{rec}}/p_{T}^{\text{gen}} \rangle$ distribution, where $\pt^{\text{rec}}$ and $\pt^{\text{gen}}$ refer, respectively,
to the \pt of the reconstructed $\tauh$ candidate and to the \pt of the vectorial momentum sum of the visible $\Pgt$ decay products at the generator level.
The most probable value of the ratio $\langle p_{T}^{\text{rec}}/p_{T}^{\text{gen}} \rangle$ is close to 1.0.
The effect of pileup on $\Pgt$ reconstruction is small.

\subsection{The \texorpdfstring{$\tauh$}{hadronic tau} identification efficiency}
\label{sec:expectedPerformance_efficiency}

The efficiency to pass the decay mode reconstruction
and the different $\tauh$ identification discriminants
is determined for hadronic $\Pgt$ decays with visible decay products
that satisfy the conditions $\pt > 20$\GeV and $\abs{\eta} < 2.3$ at the generator level.
More specifically, the efficiency is defined by the percentage of $\tauh$ candidates that satisfy:
\begin{equation}
\varepsilon_{\Pgt} = \frac{\pt^{\text{rec}} > 20\GeV, \, \abs{\eta_{\text{rec}}} < 2.3, \,  \text{DM-finding}, \, \text{$\tauh$ ID discriminant}}{\pt^{\text{gen}} > 20\GeV, \,  \abs{\eta_{\text{gen}}} < 2.3},
\label{eq:tauIdEffIsolationDiscriminators}
\end{equation}
where $\eta_{\text{rec}}$ and $\eta_{\text{gen}}$ refer, respectively, to the $\eta$
of the reconstructed $\tauh$ candidate and to the $\eta$ of the vectorial momentum sum of the visible $\Pgt$ decay products at the generator level.
The DM-finding criterion refers to the requirement that the $\tauh$ candidate be reconstructed
in one of the decay modes $\oneProngZeroPizero$, $\oneProngOnePizero$, $\oneProngTwoPizero$, or $\threeProngZeroPizero$ (cf. Section~\ref{sec:decay_mode_reconstruction}),
and $\tauh$ ID refers to the $\tauh$ identification discriminant used in the analysis.
The $\pt^{\text{gen}}$ and $\eta_{\text{gen}}$ selection criteria in the denominator are also applied in the numerator.
Only those $\tauh$ candidates matched to generator-level hadronic $\Pgt$ decays within $\Delta R < 0.3$ are considered in the numerator.

The efficiencies of the discriminants against electrons and muons are determined
for $\tauh$ candidates
matched to generator-level $\tauh$ decays within $\Delta R < 0.3$,
passing $\pt^{\text{rec}} > 20$\GeV, $\lvert \eta_{\text{rec}} \rvert < 2.3$,
reconstructed in one of the decay modes $\oneProngZeroPizero$, $\oneProngOnePizero$, $\oneProngTwoPizero$, or $\threeProngZeroPizero$,
and satisfying the loose WP of the cutoff-based $\tauh$ isolation discriminant:
\begin{equation}
\varepsilon_{\Pgt} = \frac{\text{lepton discriminant}}{\pt^{\text{rec}} > 20\GeV, \, \abs{\eta_{\text{rec}}} < 2.3, \, \text{DM-finding}, \, \text{loose cutoff-based isolation}}.
\label{eq:tauIdEffDiscriminatorsAgainstElectronsAndMuons}
\end{equation}

The selection criteria in the denominators of Eqs.~(\ref{eq:tauIdEffIsolationDiscriminators}) and~(\ref{eq:tauIdEffDiscriminatorsAgainstElectronsAndMuons})
are also applied in the numerators.

\begin{figure}[htbp]
\centering
\includegraphics[width=0.48\textwidth]{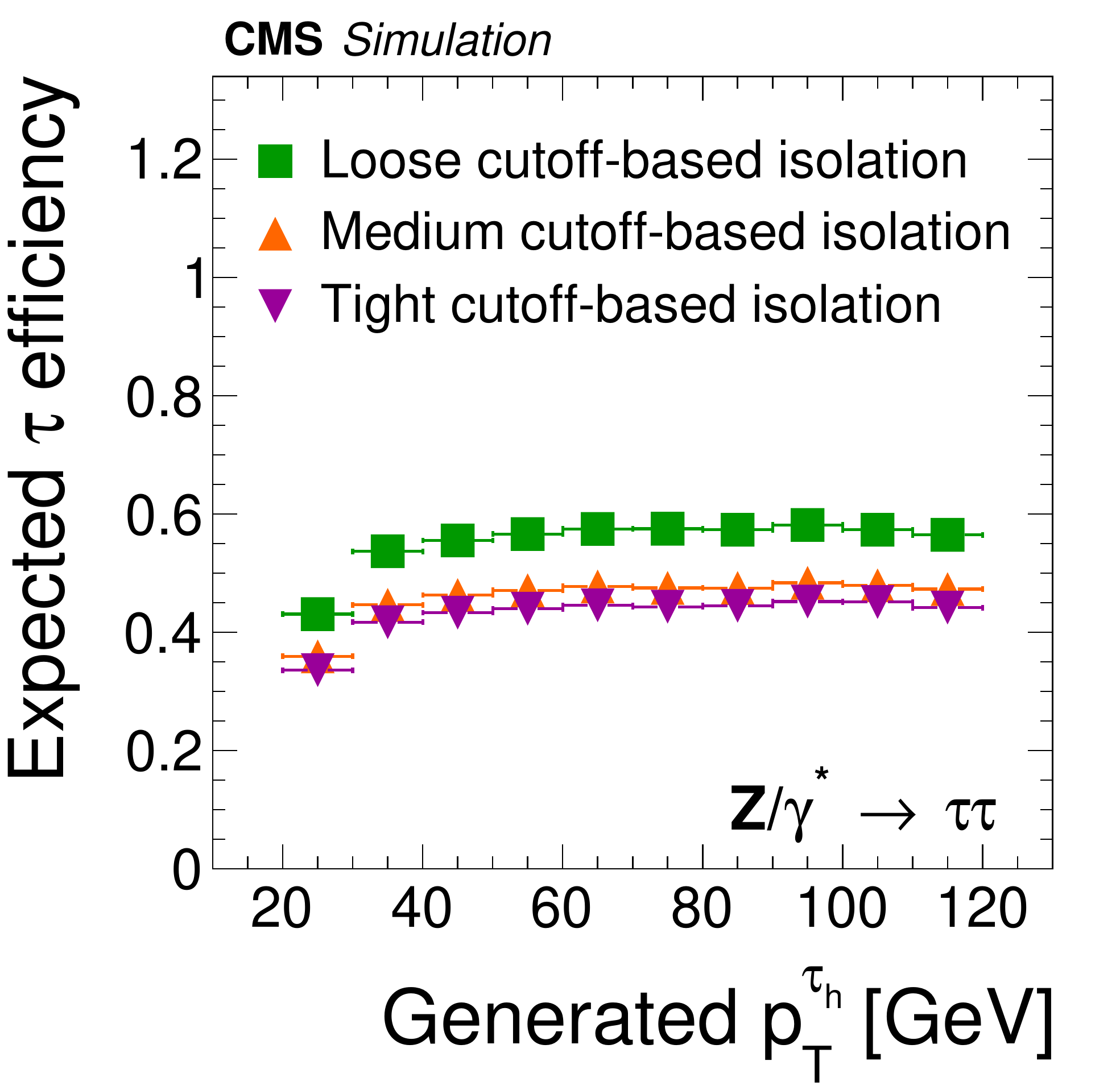}
\includegraphics[width=0.48\textwidth]{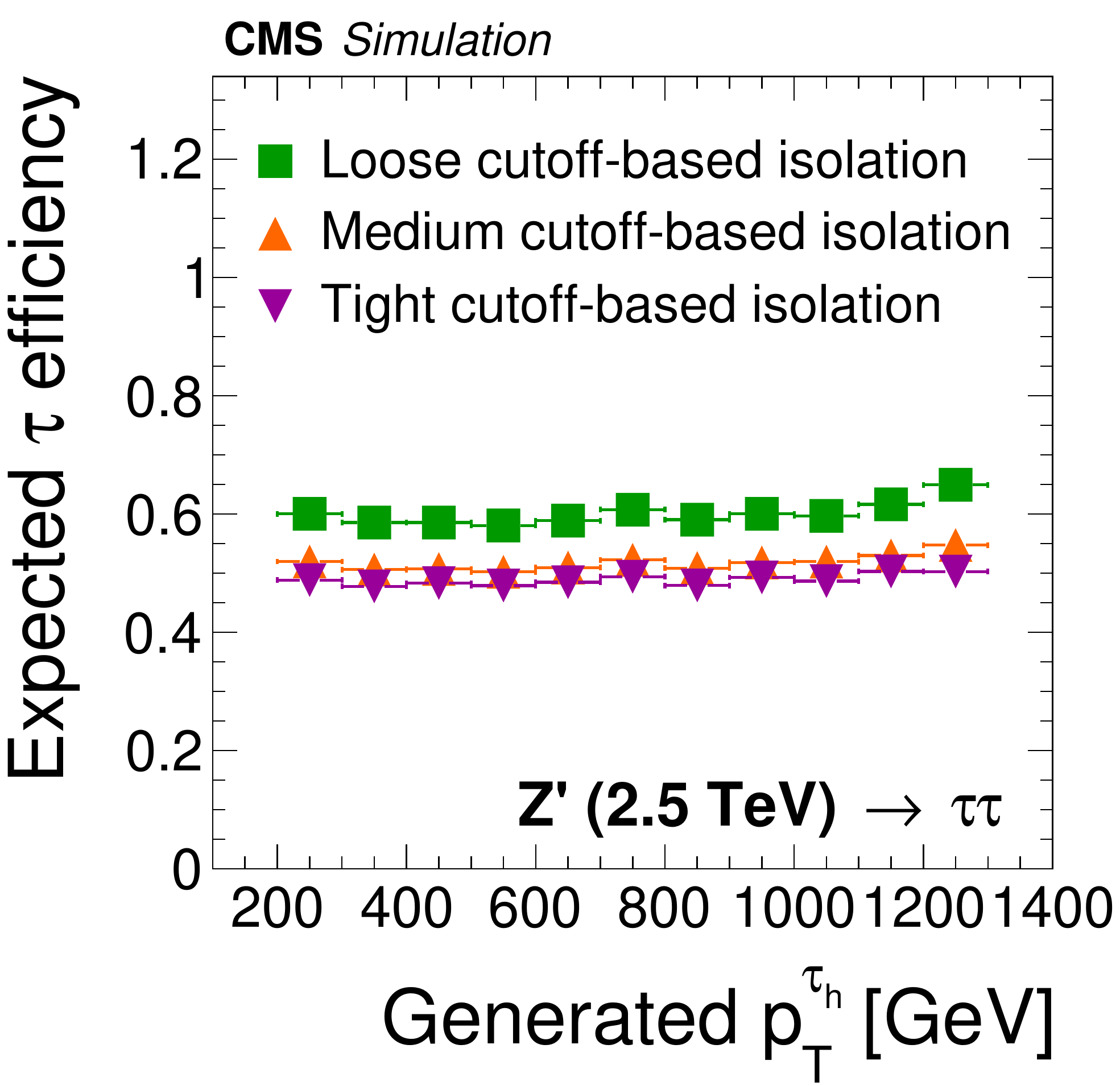}
\includegraphics[width=0.48\textwidth]{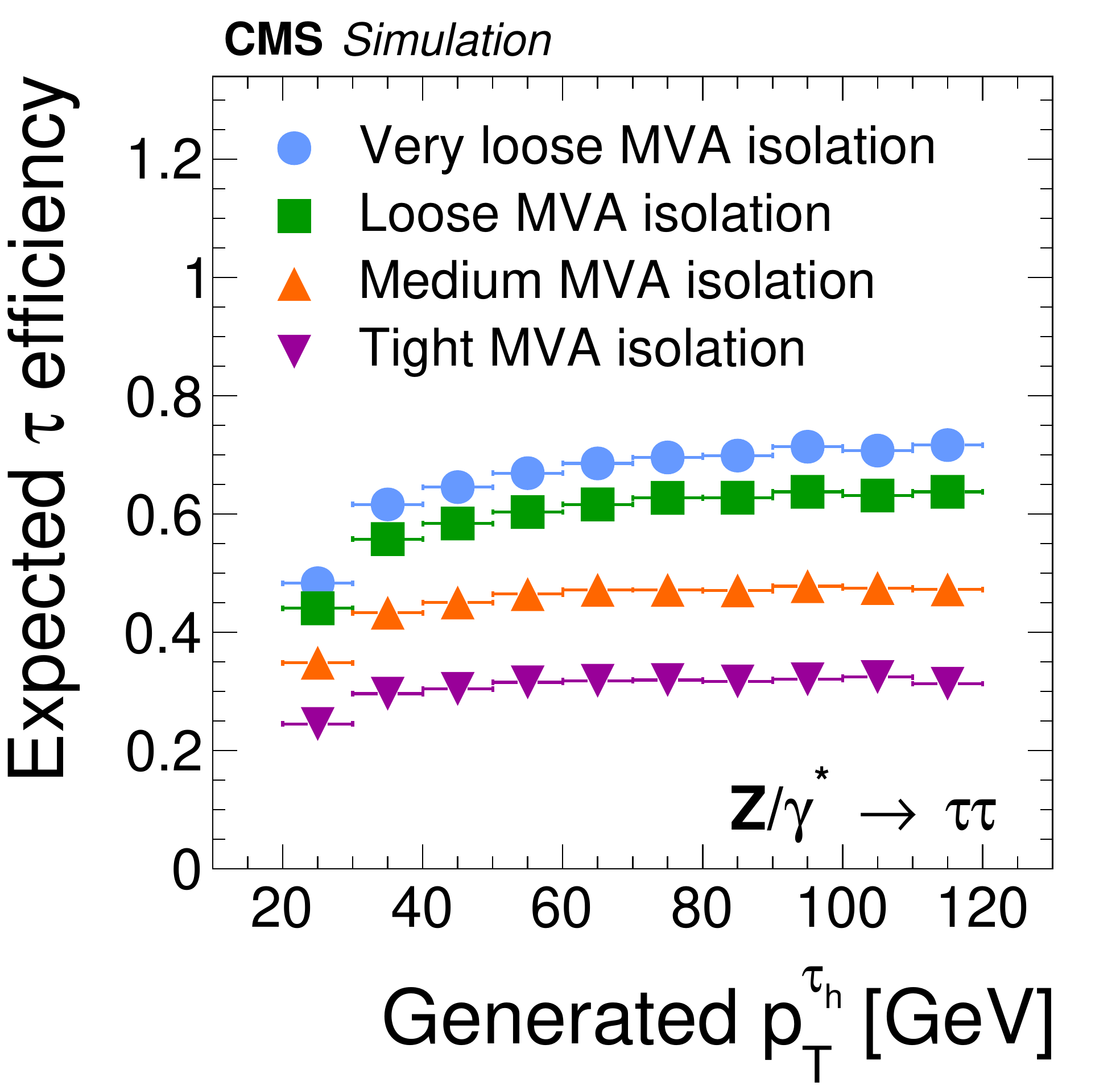}
\includegraphics[width=0.48\textwidth]{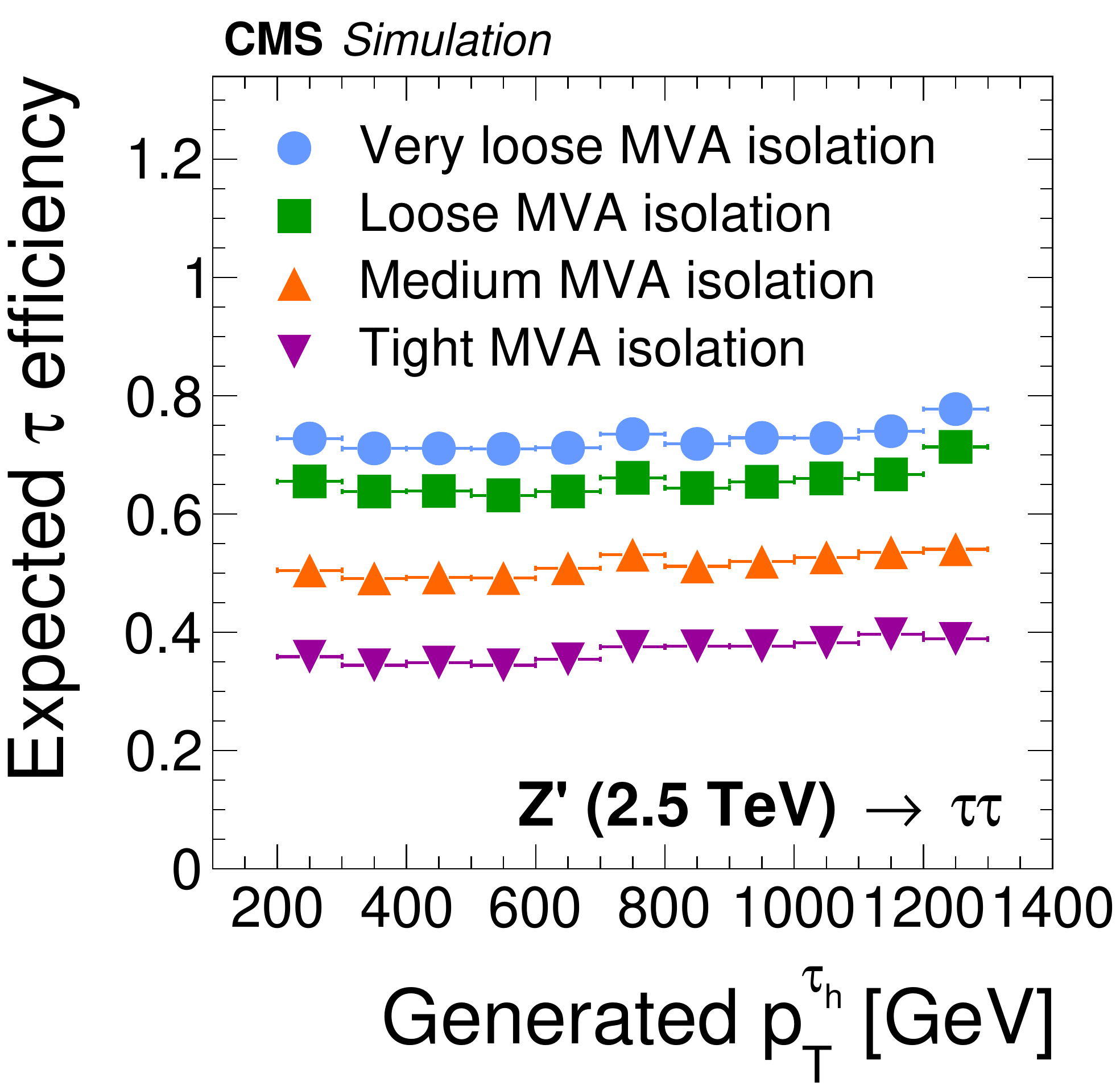}
\caption{
  Efficiency for $\tauh$ decays in simulated $\cPZ/\Pggx \to \Pgt\Pgt$ (left) and $\PZprime \to \Pgt\Pgt$ (right) events
  to be reconstructed in one of the decay modes $\oneProngZeroPizero$, $\oneProngOnePizero$, $\oneProngTwoPizero$, or $\threeProngZeroPizero$,
  to satisfy the conditions $\pt > 20$\GeV and $\abs{\eta} < 2.3$, and to pass:
  the loose, medium and tight WP of the cutoff-based $\tauh$ isolation discriminant (top)
  and the very loose, loose, medium and tight WP of the MVA-based tau isolation discriminant (bottom).
  The efficiency is shown as a function of the generator-level \pt of the visible $\Pgt$ decay products
  in $\tauh$ decays that are within $\abs{\eta} < 2.3$.
}
\label{fig:expEffHPScombIso3Hit_and_MVAisoOldDMwLT}
\end{figure}

The efficiency for $\tauh$ decays to pass the cutoff-based and MVA-based $\tauh$ identification discriminants
are shown for simulated $\cPZ/\Pggx \to \Pgt\Pgt$ and $\PZprime \to \Pgt\Pgt$ events in Fig.~\ref{fig:expEffHPScombIso3Hit_and_MVAisoOldDMwLT}.

The efficiencies are higher in $\PZprime \to \Pgt\Pgt$ than in SM $\cPZ/\Pggx \to \Pgt\Pgt$ events,
as the $\Pgt$ leptons have larger \pt in the former case.
The expected efficiencies of the isolation discriminants range between 40\% and 70\%,
depending on whether tight or loose criteria are applied.
The discrimination against electrons and against muons have respective efficiencies between 60\% and 95\%, and between 95\% and 99\%.

\subsection{Misidentification rate for jets}
\label{sec:expectedPerformance_fakerate_jets}

The rate at which quark and gluon jets are reconstructed as $\tauh$ candidates passing $\Pgt$ identification
is computed for jets with $\pt^{\text{jet}} > 20$\GeV and $\lvert \eta_{\text{jet}} \rvert < 2.3$ as follows:
\begin{equation}
P_{\text{misid}} = \frac{\pt^{\tauh} > 20\GeV, \, \abs{\eta_{\tauh}} < 2.3, \, \text{DM-finding}, \, \text{$\tauh$ ID discriminant}}{\pt^{\text{jet}} > 20\GeV , \,  \abs{ \eta_{\text{jet}}} < 2.3}.
\label{eq:jetToTauFakeRate}
\end{equation}
The $\pt^{\text{jet}}$ and $\eta_{\text{jet}}$ selection criteria of the denominator are also applied in the numerator.
Note that \pt and $\eta$ are different in the numerator and denominator,
because $\pt^{\text{jet}}$ and $\eta_{\text{jet}}$ are computed by summing the momenta of all the particle constituents of the jet,
while $\pt^{\tauh}$ and $\eta_{\tauh}$ refer to only the charged particles and photons included in the decay mode reconstruction of the $\tauh$ candidate.
Besides, jet energies are calibrated~\cite{Chatrchyan:2011ds} and corrected for pileup effects~\cite{Cacciari:2008gn, Cacciari:2007fd},
whereas no energy calibration or pileup correction is applied to $\tauh$ candidates.

The rates of jet $\to \tauh$ misidentification range from a few $10^{-4}$ to $10^{-2}$.
They differ for $\PW$+jets and multijet events,
because of the different fractions of quark and gluon jets in the two samples, and because of
differences in jet \pt spectra, which are relevant due to the dependence of the jet $\to \tauh$ misidentification rates on jet \pt
(cf. Section~\ref{sec:jetToTauFakeRate}).

The MVA-based $\tauh$ identification discriminants that include lifetime information
reduce the jet $\to \tauh$ misidentification rate by about 40\% relative to cutoff-based discriminants,
while the $\tauh$ identification efficiencies are very similar.

\subsection{Misidentification rate for electrons and muons}
\label{sec:expectedPerformance_fakerate_electrons_and_muons}

The misidentification rates for $\Pe \to \tauh$ and $\Pgm \to \tauh$ are determined for electrons and muons with $\pt^{\Plepton} > 20$\GeV and $\lvert \eta_{\Plepton} \rvert < 2.3$,
and can be written as follows:
\begin{equation}
P_{\text{misid}} = \frac{\pt^{\tauh} > 20\GeV, \, \abs{\eta_{\tauh}}, < 2.3, \,  \text{DM-finding}, \, \text{loose cutoff-based isolation}, \, \text{lepton discriminant}}{\pt^{\Plepton} > 20\GeV, \,  \abs{\eta_{\Plepton}} < 2.3}.
\label{eq:lepToTauFakeRate}
\end{equation}
Only $\tauh$ candidates reconstructed within $\Delta R < 0.3$ of a generator-level electron or muon trajectory are considered for the numerator.
The $\pt^{\Plepton}$ and $\eta_{\Plepton}$ symbols refer to the generator-level \pt and $\eta$ of the electron or muon.

Typical $\Pe \to \tauh$ misidentification rates range from a few per mille to a few per cent.
The rates for $\Pgm \to \tauh$ misidentification are at or below the per mille level.

\section{Validation with data}
\label{sec:validation}

Different kinds of events are used to evaluate the $\tauh$ reconstruction and identification in data.
The $\tauh$ identification efficiency and energy scale are validated using $\cPZ/\Pggx \to \Pgt\Pgt$ events.
The efficiency to reconstruct and identify $\tauh$ of higher \pt in more dense hadronic environments is measured using $\cPqt\cPaqt$ events.
Samples of $\PW$+jets and multijet events are used to validate the rates with which quark and gluon jets are misidentified as $\tauh$ candidates.
The misidentification rates for electrons and muons are measured using $\cPZ/\Pggx \to \Pe\Pe$ and $\cPZ/\Pggx \to \Pgm\Pgm$ events.

The selection of event samples is described in Section~\ref{sec:validation_eventSelection}.
Systematic uncertainties relevant to the validation of the $\tauh$ reconstruction and identification
are detailed in Section~\ref{sec:validation_systematicUncertainties}.
The measurement of $\tauh$ identification efficiency,
as well as of the rates at which electrons and muons are misidentified as $\tauh$ candidates,
is based on determining the yield of signal and background processes, for which we use fits of simulated distributions (templates) to data, as described in Section~\ref{sec:validation_templateFits}.

\subsection{Event selection}
\label{sec:validation_eventSelection}

\subsubsection{\texorpdfstring{$\cPZ/\Pggx \to \Pgt\Pgt$}{Z/gamma* to tau tau} events}
\label{sec:validation_eventSelection_ZTT}

The sample of $\cPZ/\Pggx \to \Pgt\Pgt$ events is selected in decay channels of $\Pgt$ leptons to muon and $\tauh$ final states.
Except for extracting the $\tauh$ identification efficiency,
$\cPZ/\Pggx \to \Pgt\Pgt \to \Pgm\tauh$ events are recorded using a trigger that demands the presence of a muon and $\tauh$~\cite{EWK-10-013}.
The events used for the $\tauh$ identification efficiency measurement are recorded using a single-muon trigger~\cite{Chatrchyan:2012xi},
to avoid potential bias that may arise from requiring a $\tauh$ at the trigger level.
The reconstructed muon is required to satisfy the conditions $\pt > 20$\GeV and $\abs{\eta} < 2.1$,
to pass tight identification criteria,
and to be isolated relative to other particles in the event by $I_{\Pgm} < 0.10\pt^{\Pgm}$,
computed according to Eq.~(\ref{eq:lepIsolationDeltaBeta}).
The $\tauh$ candidates are required to be reconstructed in one of the decay modes described in Section~\ref{sec:decay_mode_reconstruction},
to satisfy the conditions $\pt > 20$\GeV and $\abs{\eta} < 2.3$,
and to pass the loose WP of the cutoff-based $\tauh$ isolation discriminant,
the tight WP of the cutoff-based discriminant against muons, and the loose WP of the discriminant against electrons.
The muon and $\tauh$ candidate are required to be compatible with originating from the primary collision vertex and be of opposite charge.
In case multiple combinations of muon and $\tauh$ exist in an event,
the combination with the highest sum in scalar \pt is chosen.
Background arising from $\PW$+jets production is removed
by requiring the transverse mass computed in
Eq.~(\ref{eq:MtDefinition}) to satisfy the condition $\mT < 40$\GeV.
Events containing a second muon of $\pt > 15$\GeV and $\lvert \eta_{\mu} \rvert < 2.4$,
passing loose identification and isolation criteria, are rejected to suppress $\cPZ/\Pggx \to \Pgm\Pgm$ Drell--Yan (DY) background.

\begin{figure}[htbp]
\centering
\includegraphics[width=0.48\textwidth]{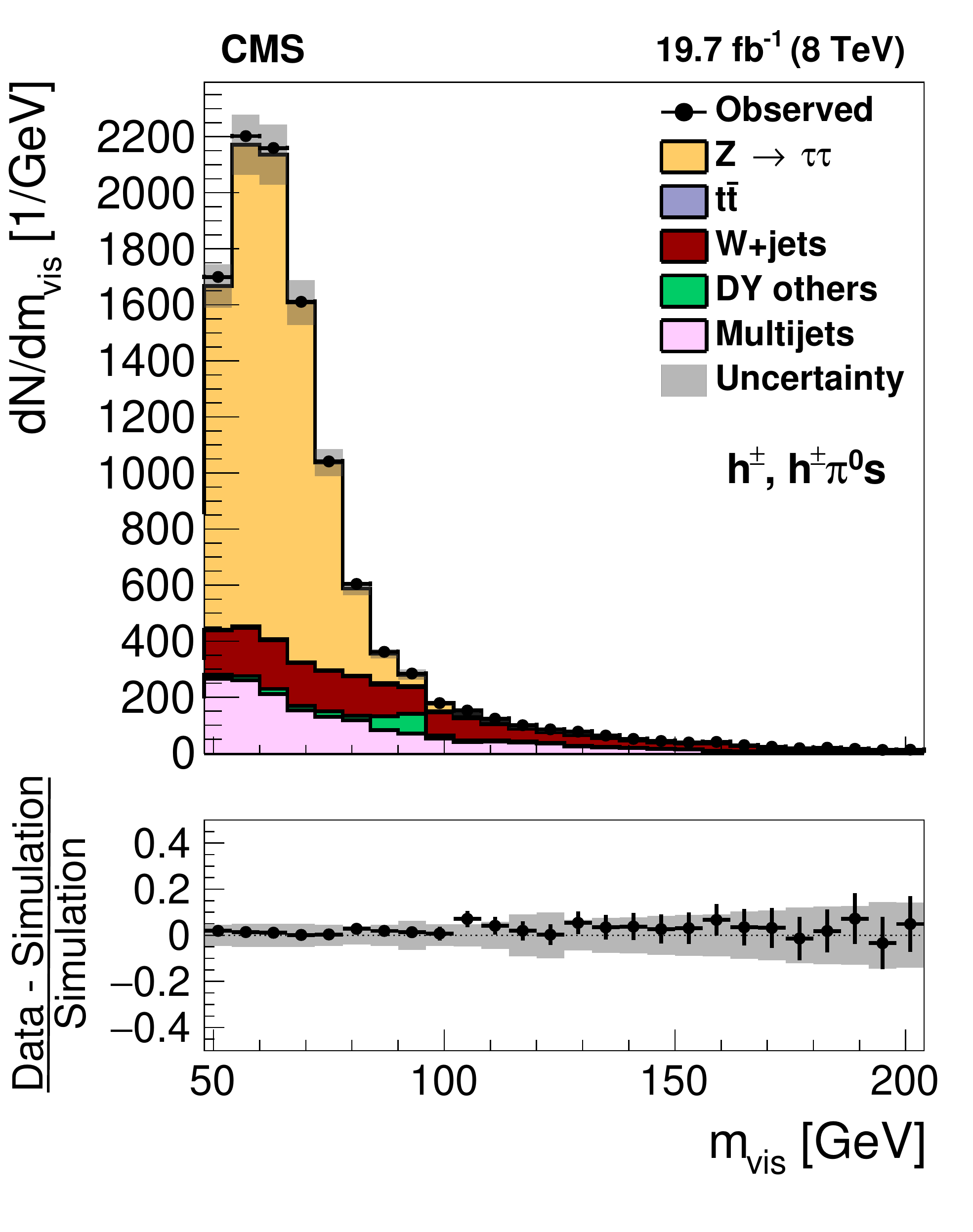}
\includegraphics[width=0.48\textwidth]{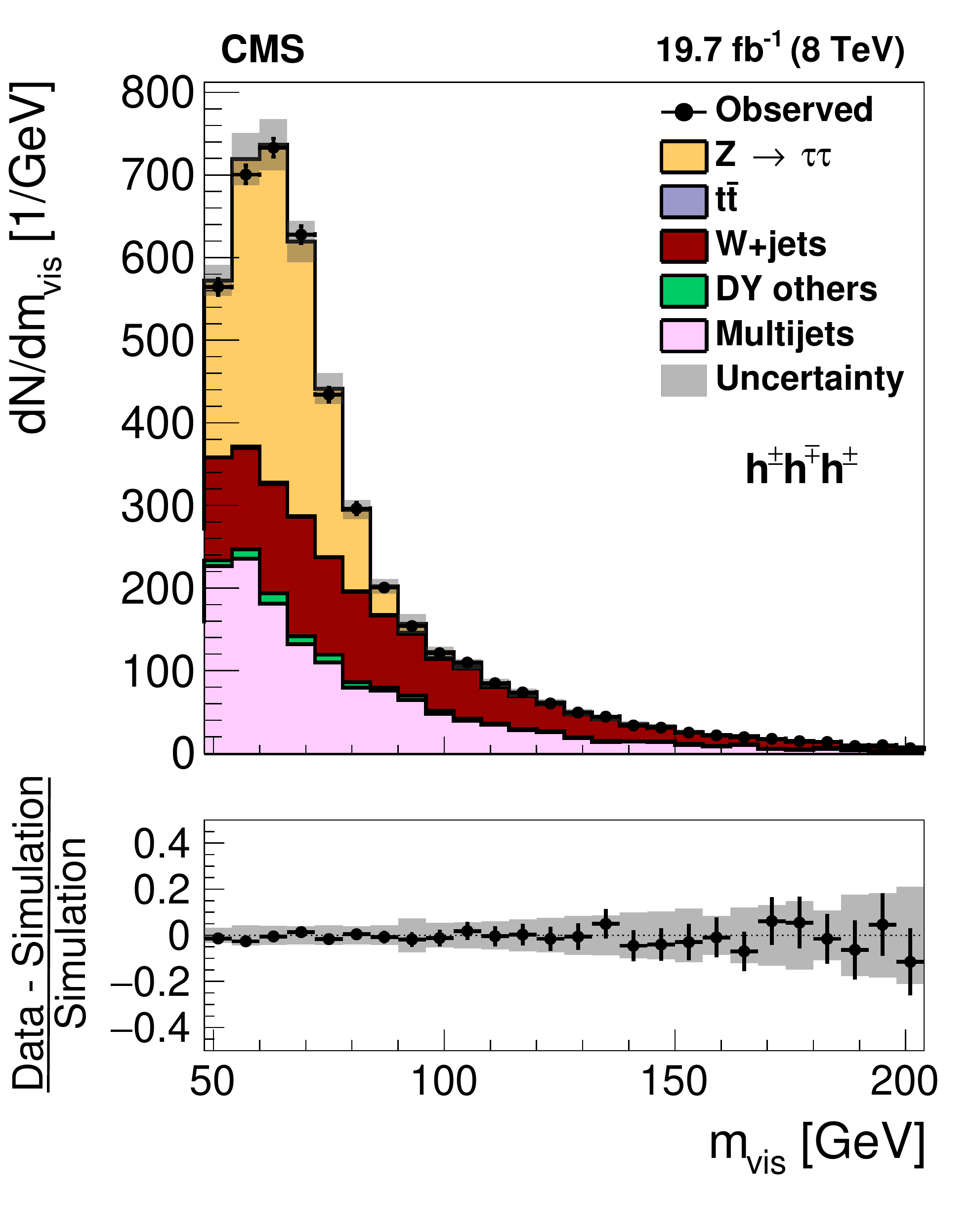}
\includegraphics[width=0.48\textwidth]{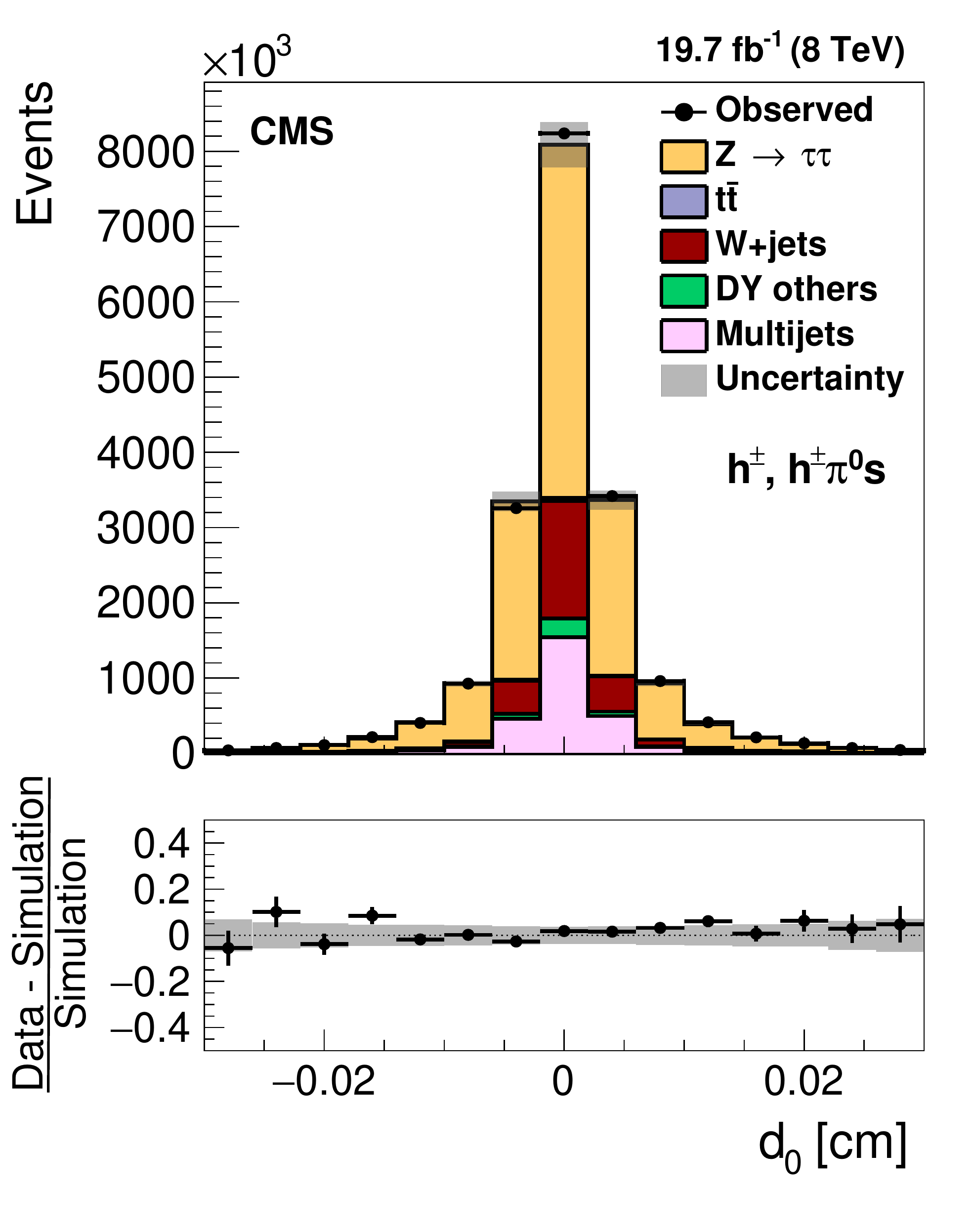}
\includegraphics[width=0.48\textwidth]{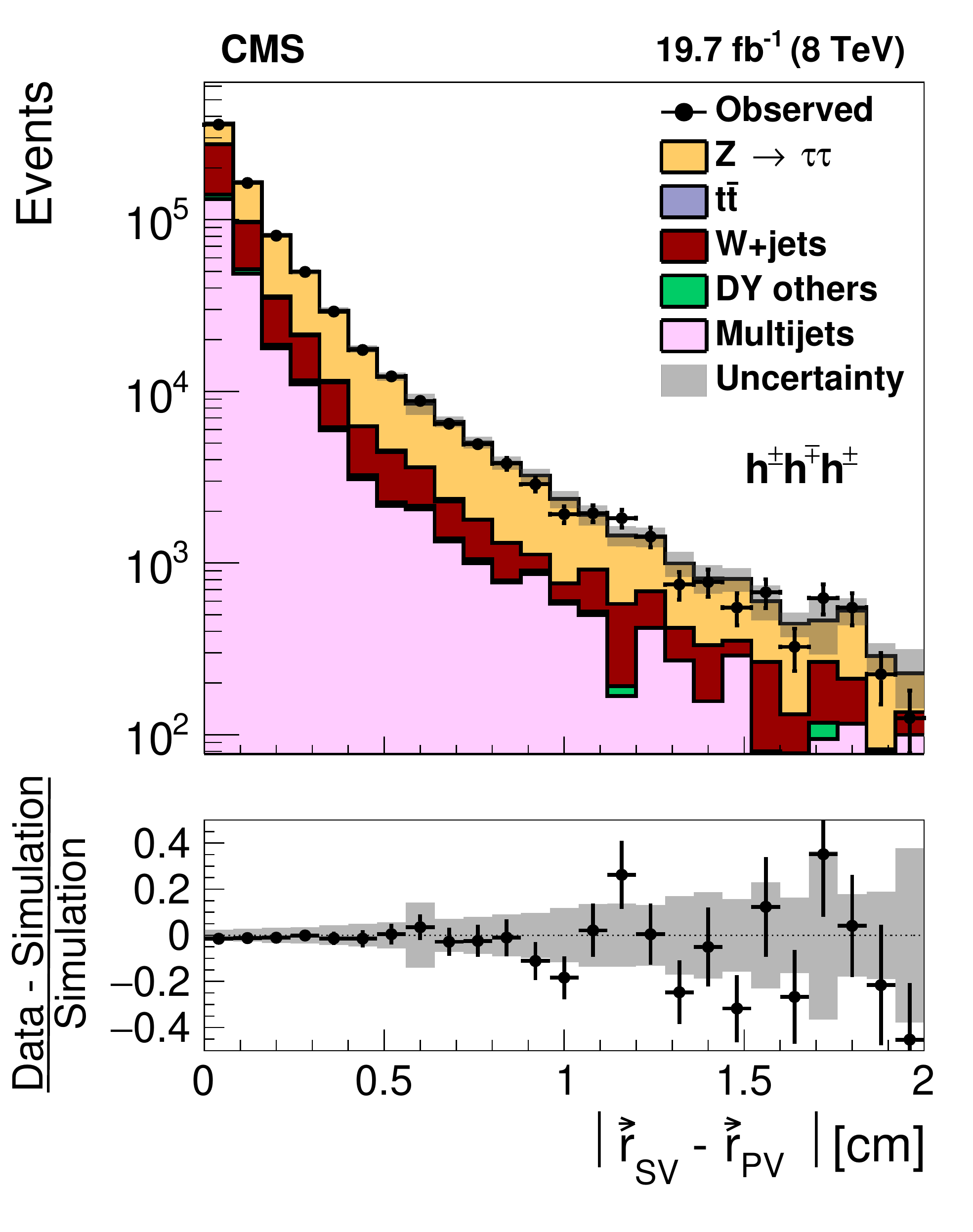}
\caption{
  Top: Distribution in the visible mass of $\cPZ/\Pggx \to \Pgt\Pgt \to \Pgm\tauh$ candidate events,
  in which the reconstructed $\tauh$ candidate contains (upper left) a single or (upper right) three charged particles.
  Bottom: Distribution in (lower left) transverse impact parameter for events in which the $\tauh$ candidate contains one charged particle
  and (lower right) in the distance between the $\Pgt$ production and decay vertex for events in which the $\tauh$ candidate contains three charged particles.
  The $\cPZ/\Pggx \to \PLepton\PLepton$ ($\PLepton = \Pe$, $\Pgm$, $\Pgt$) events, in which
  either the reconstructed muon or the reconstructed $\tauh$ candidate are misidentified, are denoted by  ``DY others''.
}
\label{fig:validation_controlPlots_ZTT}
\end{figure}

The transverse impact parameter $d_{0}$ and the distance $\abs{\vec{r}_{\mathrm{SV}} - \vec{r}_{\text{PV}}}$ between the $\Pgt$ production and decay vertices
in selected $\cPZ/\Pggx \to \Pgt\Pgt$ events are shown in Fig.~\ref{fig:validation_controlPlots_ZTT}.
The normalization of the $\cPZ/\Pggx \to \Pgt\Pgt \to \Pgm\tauh$ signal and of background processes
is determined through a template fit to the data,
as described in Section~\ref{sec:validation_templateFits},
using the visible mass of the muon and $\tauh$ ($\mVis$) as observable in the fit.
Separate fits are performed for events with $\tauh$ candidates containing one and three charged particles.
The fitted $\mVis$ spectra are also shown in Fig.~\ref{fig:validation_controlPlots_ZTT}.
The shaded areas represent the sum of statistical uncertainties of the MC samples
and systematic uncertainties, added in quadrature, as discussed in Section~\ref{sec:validation_systematicUncertainties}.
All distributions agree well with their respective MC simulations.

\subsubsection{$\cPqt\cPaqt$ events}
\label{sec:validation_eventSelection_ttbar}

A sample of $\cPqt\cPaqt$ events is also selected in the $\mu\tauh$ channel.
The $\cPqt\cPaqt \to \cPqb\cPaqb\Pgm\tauh$ events are required to pass a single-muon trigger
and to contain a muon with $\pt > 25$\GeV and $\abs{\eta} < 2.1$.
The muon is required to pass tight identification criteria and to be isolated at the level of $I_{\Pgm} < 0.10\pt^{\Pgm}$.
The $\tauh$ candidate is required to be reconstructed in one of the decay modes described in Section~\ref{sec:decay_mode_reconstruction},
to satisfy the conditions $\pt > 20$\GeV and $\abs{\eta} < 2.3$,
to pass the loose WP of the cutoff-based $\tauh$ isolation discriminant,
and to be separated from the muon by $\Delta R > 0.5$.
The event is also required to contain two jets of $\pt > 30$\GeV and $\abs{\eta} < 2.5$,
separated from the muon and the $\tauh$ candidate by $\Delta R > 0.5$.
At least one of the jets is required to meet the $\Pbottom$ tagging criteria~\cite{Chatrchyan:2012jua,BTV-13-001}.
Background from $\cPZ/\Pggx \to \PLepton\PLepton$ ($\PLepton = \Pe$, $\Pgm$, $\Pgt$) events is reduced
by requiring $\MET > 40$\GeV.
Events containing an electron of $\pt > 15$\GeV and $\abs{\eta} < 2.3$,
or a second muon of $\pt > 10$\GeV and $\abs{\eta} < 2.4$ that pass loose identification and isolation criteria are rejected.

\begin{figure}[htb]
\centering
\includegraphics[width=0.48\textwidth]{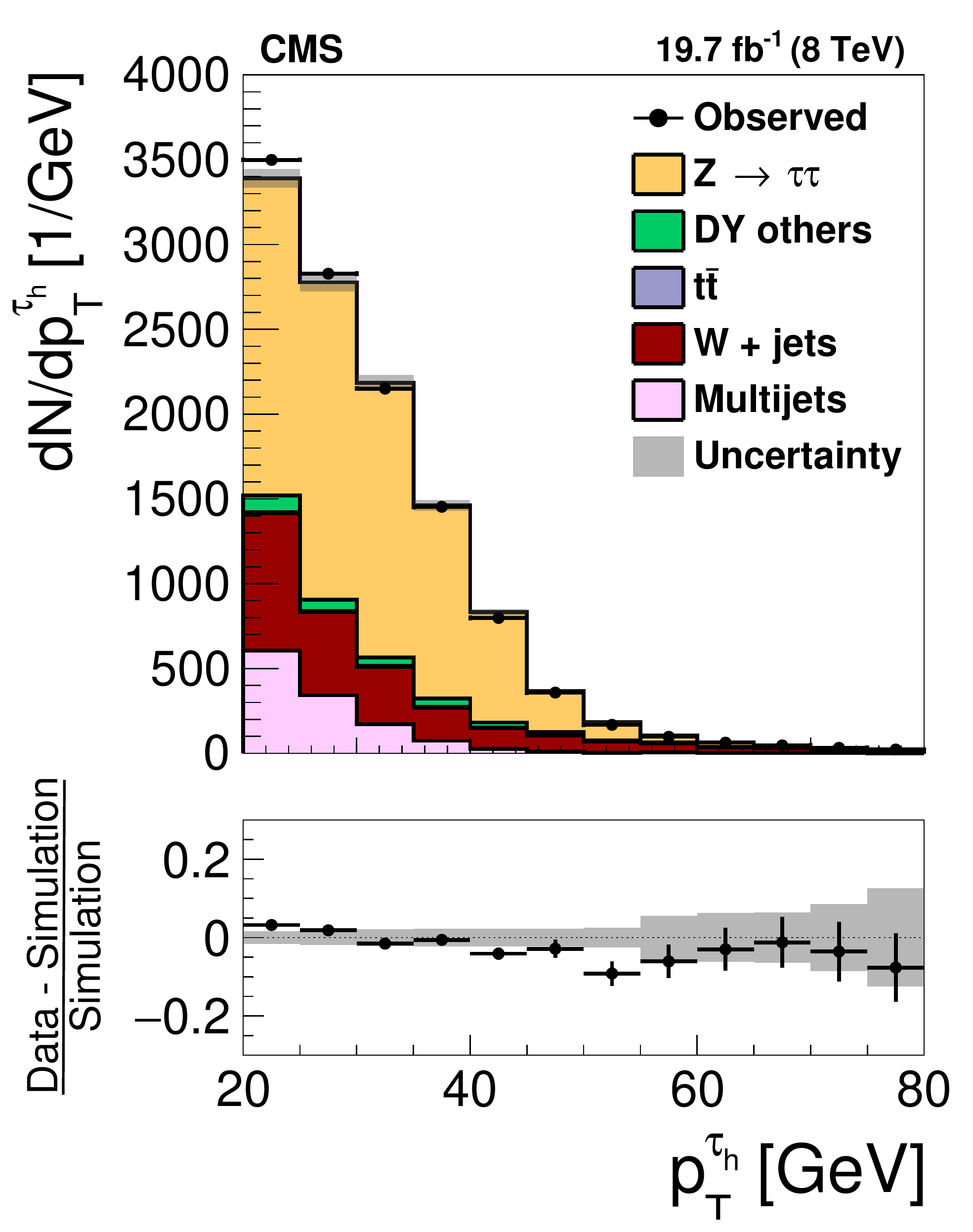}
\includegraphics[width=0.48\textwidth]{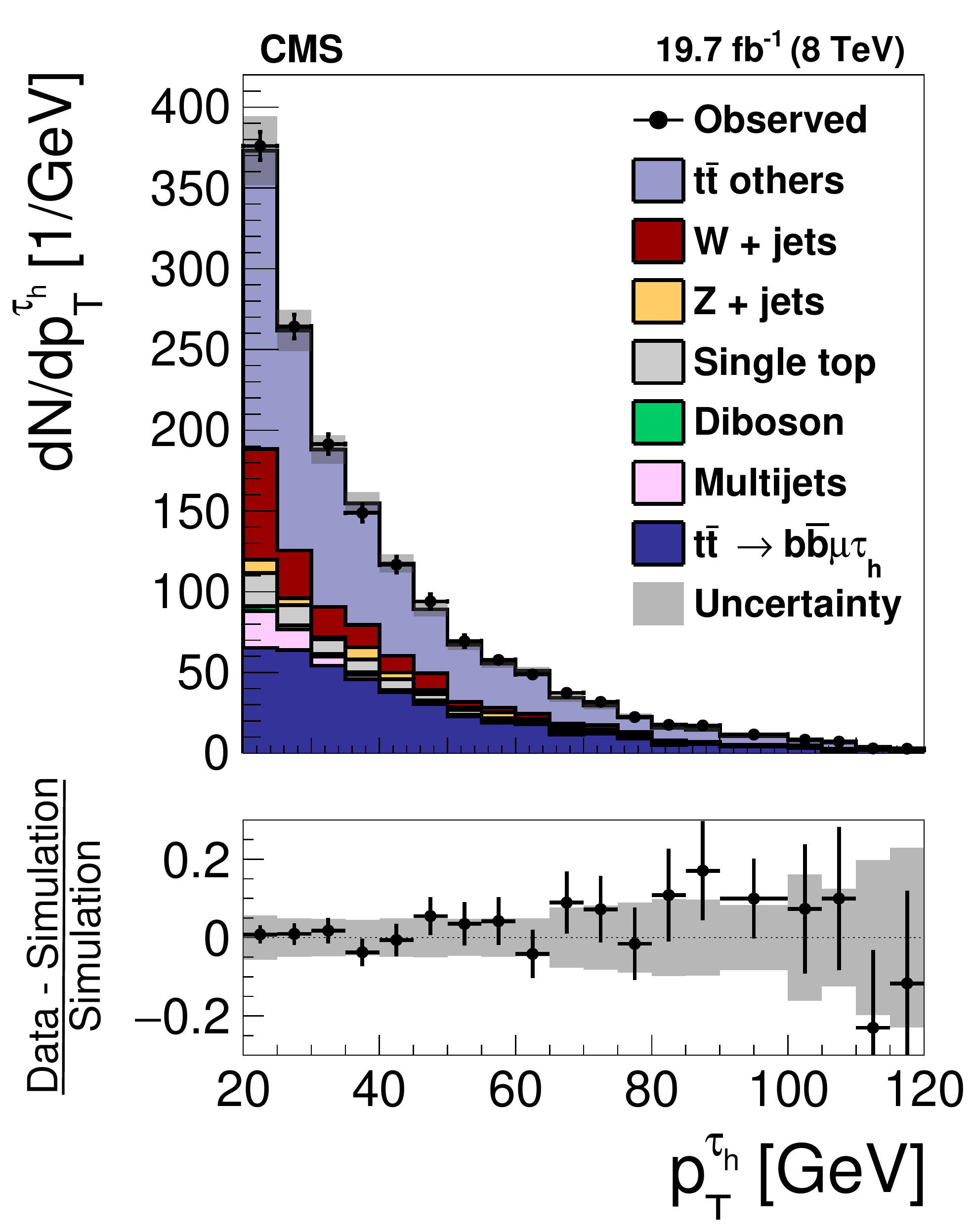}
\caption{
  Distribution in the \pt of $\tauh$ candidates in (left) $\cPZ/\Pggx \to \Pgt\Pgt$ and (right) $\cPqt\cPaqt$ events in data and in simulations.
  The $\cPZ/\Pggx \to \PLepton\PLepton$ ($\PLepton = \Pe$, $\Pgm$, $\Pgt$) and $\cPqt\cPaqt$ events, in which
  either the reconstructed muon or the reconstructed $\tauh$ candidate is misidentified, are denoted in the MC simulation
  by  ``DY others'' and ``$\cPqt\cPaqt$ others'', respectively.
}
\label{fig:validation_taupt_ZTT_and_TT}
\end{figure}

The \pt distribution of $\tauh$ candidates in the $\cPqt\cPaqt$ sample is compared to the $\cPZ/\Pggx \to \Pgt\Pgt$ sample in Fig.~\ref{fig:validation_taupt_ZTT_and_TT}.

\subsubsection{The $\PW$+jets sample}
\label{sec:validation_eventSelection_Wjets}

Events selected for the $\PW$+jets sample are required to pass the single-muon trigger
and to contain a muon with $\pt > 25$\GeV and $\abs{\eta} < 2.1$,
passing tight identification and isolation criteria $I_{\Pgm} < 0.10\pt^{\Pgm}$.
The muon and $\MET$ transverse mass, computed according to
Eq.~(\ref{eq:MtDefinition}), is required to satisfy the condition $\mT > 50$\GeV.
Selected $\PW$+jets candidate events
are further required to contain at least one jet with $\pt > 20$\GeV and $\abs{\eta} < 2.3$
that is separated from the muon by $\Delta R > 0.5$.

\subsubsection{Multijet sample}
\label{sec:validation_eventSelection_QCD}

The sample of multijet events is selected by requiring the events to pass a single-jet trigger
with the \pt threshold of 320\GeV.
The trigger was not prescaled during the whole data-taking period.
The jet that passes the trigger is required to satisfy the conditions $\pt > 350$\GeV and $\abs{\eta} < 2.5$.
In order to measure the jet $\to \tauh$ misidentification rate for jets unbiased by the trigger selection,
the following procedure is used:
If only one jet in the event passes the trigger requirement,
that jet is excluded from the computation of the jet $\to \tauh$ misidentification rate, and the other jets
with $\pt > 20$\GeV and $\abs{\eta} < 2.3$ in the event are used instead.
When two or more jets in the event pass the trigger requirement,
all jets with $\pt > 20$\GeV and $\abs{\eta} < 2.3$ are included in the computation of the misidentification rate.
Each jet is unbiased relative to the trigger selection,
because the event would have been triggered by another jet regardless of the rest of the objects in the event.

\begin{figure}[htb]
\centering
\includegraphics[width=0.48\textwidth]{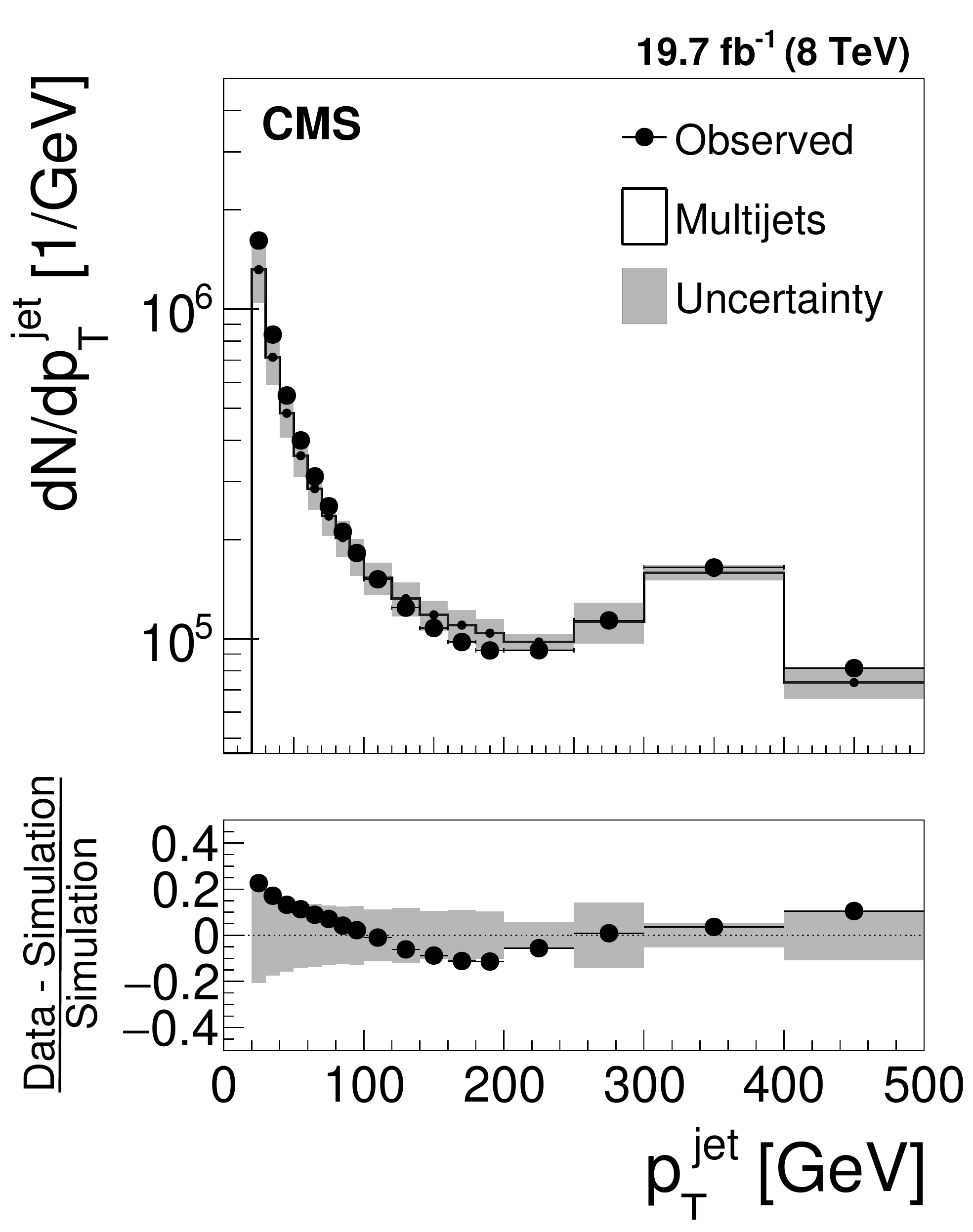}
\includegraphics[width=0.48\textwidth]{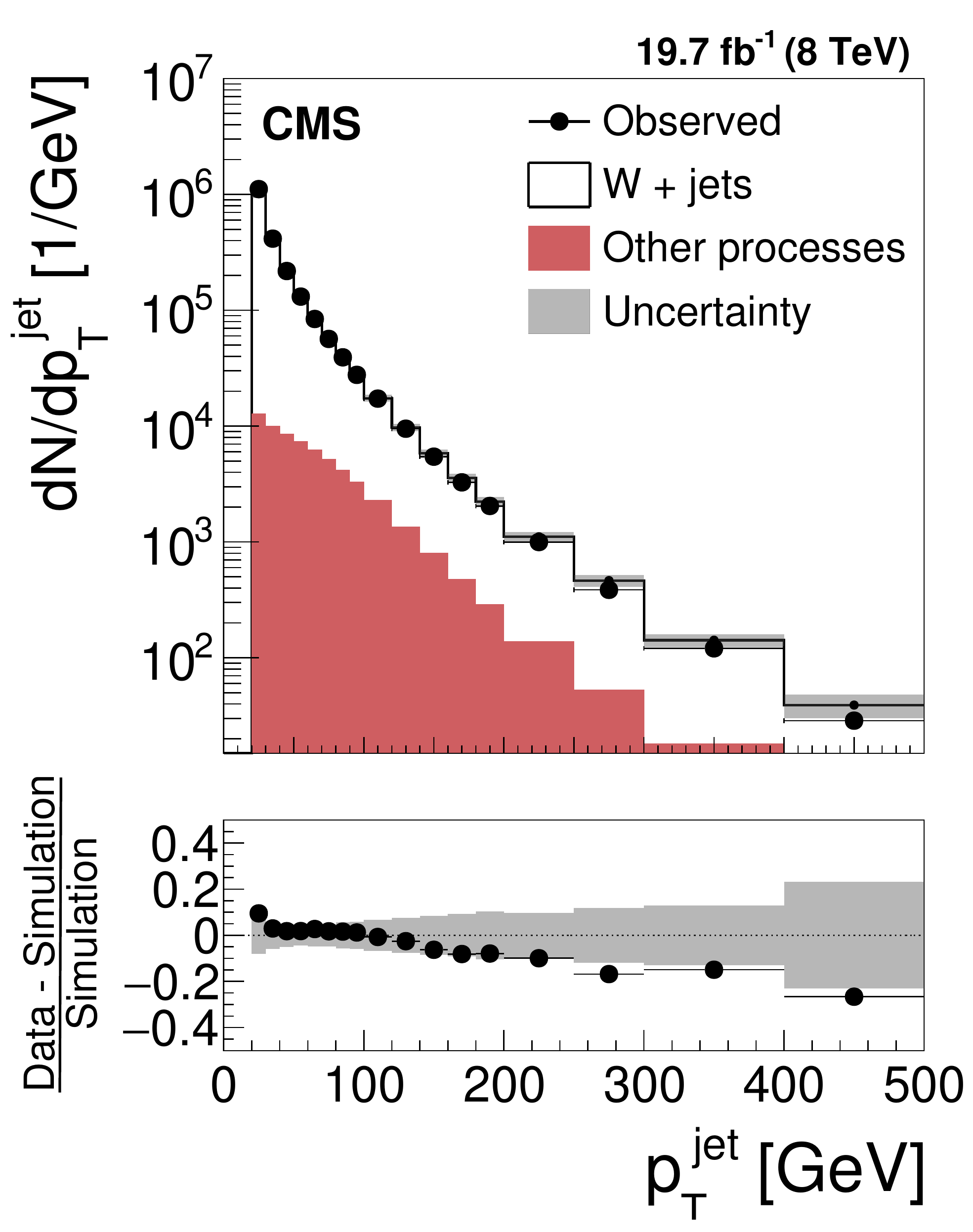}
\caption{
  Jet \pt distribution in (left) multijet and (right) $\PW$+jets events observed in data, compared to the MC expectation.
  The uncertainty in the MC expectation is dominated by the uncertainty in the jet energy scale.
}
\label{fig:validation_controlPlots_QCD}
\end{figure}

The \pt distribution of jets considered for the computation of the jet $\to \tauh$ misidentification rate
is compared for $\PW$+jets and multijet samples in Fig.~\ref{fig:validation_controlPlots_QCD}.
The multijet sample provides more jets with large \pt.
Since the single-jet trigger used to select the multijet events requires at least one jet with \pt greater than 320\GeV,
the sample is enriched with events containing high \pt jets that are likely recoiling against each other.
This is the reason for the increase in the jet \pt spectrum in bin 300--400\GeV.
The distributions observed in data agree with the MC expectation within uncertainties.

\subsubsection{The $\cPZ/\Pggx \to \Pe\Pe$ and $\cPZ/\Pggx \to \Pgm\Pgm$ events}
\label{sec:validation_eventSelection_Zll}

A high-purity sample of $\cPZ/\Pggx \to \Pe\Pe$ and $\cPZ/\Pggx \to \Pgm\Pgm$ events is selected
by requiring candidate events to contain at least one well-identified and isolated electron or muon, referred to as the ``tag'',
and one $\tauh$ candidate that passes loose preselection criteria, referred to as the ``probe''.
The $\Pe \to \tauh$ and $\Pgm \to \tauh$ misidentification rates are given by the fraction of probes
that pass the $\Pgt$ isolation criteria, as well as one of the dedicated discriminants for vetoing electrons or muons.

Tag electrons are required to pass a single-electron trigger,
to satisfy the conditions $\pt > 30$\GeV and $\abs{\eta} < 2.1$,
pass tight electron identification criteria, and isolation, with $I_{\Pe} < 0.10\pt^{\Pe}$.
Tag electrons reconstructed in the transition region between ECAL barrel and endcap, $1.46 < \abs{\eta} < 1.56$, are discarded. Similarly, tag
muons are required to pass a single-muon trigger,
to satisfy the conditions $\pt > 25$\GeV and $\abs{\eta} < 2.1$,
pass tight muon identification criteria, and isolation, with $I_{\Pgm} < 0.10\pt^{\Pgm}$.

The probe is required to be reconstructed in one of the decay modes $\oneProngZeroPizero$, $\oneProngOnePizero$, $\oneProngTwoPizero$, or $\threeProngZeroPizero$,
to satisfy the conditions $\pt > 20$\GeV and $\abs{\eta} < 2.3$, and to be separated from the tag electron or muon by $\Delta R > 0.5$.
The \pt and $\eta$ requirements are applied on the values reconstructed using the HPS algorithm.
When an event contains two electron or muon candidates that pass the tight selection criteria and qualify as tags,
the event is selected if it contains at least one combination of tag and probe leptons that are separated by $\Delta R > 0.5$.
In this case, all combinations of tag and probe leptons separated by $\Delta R > 0.5$ are considered in the analysis.

The contribution of $\PW$+jets and $\cPqt\cPaqt$ backgrounds is reduced
by requiring selected events to pass a requirement on the transverse mass of a tag electron or muon and $\MET$,
computed according to Eq.~(\ref{eq:MtDefinition}), respectively, of $\mT < 25$\GeV or $< 40$\GeV.
The contribution of the $\PW$+jets and $\cPqt\cPaqt$ background to the $\cPZ/\Pggx \to \Pe\Pe$ event sample
is further suppressed by requiring $\MET < 25$\GeV.

\subsection{Systematic uncertainties}
\label{sec:validation_systematicUncertainties}

Various imprecisely known or simulated effects can affect the level of agreement between data and simulation,
irrespective of $\tauh$ reconstruction and identification.

Electron and muon trigger, identification, and isolation efficiencies
are measured using $\cPZ/\Pggx \to \Pe\Pe$ and $\cPZ/\Pggx \to \Pgm\Pgm$ events via the ``tag-and-probe'' method~\cite{cmsTnP}
with a precision of 2\%~\cite{EGM-13-001,Chatrchyan:2012xi}.
The uncertainty on the $\tauh$ trigger efficiency is $\approx$3\%.

The jet energy scale (JES) is validated using $\Pgg$+jets, $\cPZ$+jets, and dijet events~\cite{Chatrchyan:2011ds}.
The uncertainty in JES ranges from 1\% to 10\%, depending on \pt and $\eta$ of the jet.
The effect of uncertainty in energy resolution is found to be small and is not considered in this analysis.
The efficiency for $\Pbottom$ jets to pass the medium WP of the CSV $\Pbottom$ tagging algorithm,
and the mistag rates for light-quark and gluon jets are measured using $\cPqt\cPaqt$ and multijet events,
and are in the ranges of 2--7\% and 10--20\%, respectively~\cite{Chatrchyan:2012jua,BTV-13-001}.

The uncertainty in the integrated luminosity is 2.6\%~\cite{LUM-13-001}.

The theoretical uncertainties on the production cross sections are
5\% for $\cPZ/\Pggx \to \PLepton\PLepton$ ($\PLepton = \Pe$, $\Pgm$, $\Pgt$) DY production,
and 15\% for the  $\cPqt\cPaqt$, diboson, and single top quark backgrounds.
These include the uncertainties in PDF,
estimated following the recommendation in Refs.~\cite{Alekhin:2011sk,Botje:2011sn},
and on the modelling of parton showers and of the underlying event.

The uncertainty in reweighting of simulated $\cPqt\cPaqt$ MC events,
described in Section~\ref{sec:datasamples_and_MonteCarloSimulation},
is estimated by changing the weights from their nominal values to the square of the nominal values and to no reweighting.

The energy scale of electrons and muons is calibrated using $\PJgy \to \Plepton\Plepton$, $\Upsilon \to \Plepton\Plepton$, and $\cPZ \to \Plepton\Plepton$ ($\Plepton$ = $\Pe$, $\Pgm$) events,
and is known to an uncertainty of 1\%~\cite{EGM-13-001,Chatrchyan:2012xi}.

The $\MET$ scale and resolution is known to a few per cent uncertainty from studies performed in $\cPZ/\Pggx \to \Pgm\Pgm$, $\cPZ/\Pggx \to \Pe\Pe$, and $\Pgg$+jets events~\cite{JME-13-003}.

\subsection{Template fits}
\label{sec:validation_templateFits}

The measurements of the $\tauh$ identification efficiency, of the $\tauh$ energy scale, and of the misidentification rates for electrons and muons
are based on fitting the distribution of some observable in data with templates representing signal and background processes.

The likelihood function $\mathcal{L}$ used in the fit is given by the product of Poisson probabilities to observe $n_{i}$ events in each bin $i$
of the distribution,
given a number $\nu_{i}$ events expected from signal and background processes in that bin:
\begin{equation}
\mathcal{L} \left( \mu, \theta \right)
  = \mathcal{P} \left(\text{data}\vert \mu, \theta \right) \, p( \tilde{\theta} \vert \theta )
  = \prod_{i} \frac{\nu_{i}^{n_{i}}}{n_{i}!} \exp ( -\nu_{i} ) \, p( \tilde{\theta} \vert \theta ).
\label{eq:likelihoodFunction}
\end{equation}
The number of expected events depends on the parameter of interest (POI) $\mu$ that we wish to measure,
such as the $\tauh$ identification efficiency, the energy scale, or the misidentification rates for electrons and muons,
and on the values of ``nuisance'' parameters $\theta$ that represent the systematic uncertainties discussed in the previous section.

The function $p( \tilde{\theta} \vert \theta )$ represents the probability to
observe a value $\tilde{\theta}$ in an auxiliary measurement of the nuisance parameter,
given that the true value is $\theta$.
The nuisance parameters are treated via the frequentist paradigm,
as described in Refs.~\cite{ATL-PHYS-PUB-2011-011,Chatrchyan:2012tx}.
Constraints on nuisance parameters that affect the normalization, but not the shape of the distribution
are represented by log-normal probability density functions.
Systematic uncertainties that affect the distribution as well as the normalization
are incorporated into the likelihood fit via the technique detailed in Ref.~\cite{Conway:2011in}
and constrained by Gaussian probability density functions.

Statistical uncertainties on the templates are accounted for
by introducing additional nuisance parameters into the likelihood fit
that provide uncorrelated single-bin fluctuations of the background expectation,
following the method described in Ref.~\cite{Barlow:1993dm}.

The value of $\mu$ that maximizes the likelihood function $\mathcal{L}$ in Eq.~(\ref{eq:likelihoodFunction})
is taken as the best-fit estimate for the parameter of interest, referred to as $\mu^{\text{obs}}$.
The uncertainty in the measured value $\mu^{\text{obs}}$ is obtained by determining lower and upper bounds, $\mu^{\text{min}}$ and $\mu^{\text{max}}$,
for which the negative logarithm of the likelihood function exceeds the maximum by half a unit:
\begin{equation}
-\ln \mathcal{L} \left( \mu^{\text{min}}, \hat{\theta}_{\mu^{\text{min}}} \right) = -\ln \mathcal{L} \left( \mu^{\text{obs}}, \hat{\theta}_{\mu^{\text{obs}}} \right) + 0.5,
\end{equation}
and similarly for $\mu^{\text{max}}$.
The nuisance parameters are profiled,
that is, the values $\hat{\theta}_{\mu^{\text{min}}}$ and $\hat{\theta}_{\mu^{\text{max}}}$ are chosen such
that the likelihood function reaches its local maximum,
subject to the constraint that the POI value equals $\mu^{\text{min}}$ and $\mu^{\text{max}}$, respectively.

The best-fit value of the POI that we obtain from one measurement, e.g. of the $\tauh$ identification efficiency,
can depend on the POI of another measurement, e.g. of the $\tauh$ energy scale.
Correlations of this kind are taken into account in the template fits
by using the other POI measurements as nuisance parameters in the fit, with an uncertainty of
6\% for the $\tauh$ identification efficiency,
3\% for the $\tauh$ energy scale,
20\% for the jet $\to \tauh$ misidentification rate,
and 30\% for the $\Pe \to \tauh$ and $\Pgm \to \tauh$ misidentification rates.
The rate for $\Pe \to \tauh$ and $\Pgm \to \tauh$ instrumental background in the MC simulation
is corrected by the data-to-MC ratios given in Tables~\ref{tab:eToTauFakeRateResults} and~\ref{tab:muToTauFakeRateResults}.

\section{Measurement of the \texorpdfstring{$\tauh$}{hadronic tau} identification efficiency}
\label{sec:Tau_identification_efficiency}

The efficiency to reconstruct and identify $\tauh$ decays
is measured in $\cPZ/\Pggx \to \Pgt\Pgt \to \Pgm\tauh$ and
$\cPqt\cPaqt \to \cPqb\cPaqb\Pgm\tauh$ events.

\subsection{Tau identification efficiency in \texorpdfstring{$\cPZ/\Pggx \to \Pgt\Pgt$}{Z/gamma* to tau tau} events}
\label{sec:tauIdEfficiency_TnP_Ztautau}

The measurement of the $\tauh$ identification efficiency in $\cPZ/\Pggx \to \Pgt\Pgt$ events
is based on selecting a sample of $\cPZ/\Pggx \to \Pgt\Pgt \to \Pgm\tauh$ events without applying any $\tauh$ identification criteria
and determining the number of $\tauh$ decays passing and failing the $\tauh$ identification discriminant.

Following the event selection criteria described in Section~\ref{sec:validation_eventSelection_ZTT},
candidate events are required to pass a single-muon trigger, a higher \pt threshold for the muon with $\pt > 25$\GeV,
and, instead of requiring the event to contain a $\tauh$ candidate that passes the $\tauh$ identification discriminants,
a loose $\tauh$ candidate selection is applied as follows.
Reconstructed jets are required to satisfy the conditions $\pt^{\text{jet}} > 20$\GeV and $\lvert \eta_{\text{jet}} \rvert < 2.3$,
to be separated from the muon by $\Delta R > 0.5$, and to contain at least one track with $\pt > 5$\GeV.
The track of highest \pt within the jet is required to have a charge
opposite to that of the muon,
and to be compatible with originating from the same vertex.
When more than one jet passes the $\tauh$ candidate selection criteria,
the jet with largest \pt is used for this check.

In addition, tight kinematic criteria are applied to reduce contributions from background processes.
The $\mT$ criterion described in Section~\ref{sec:validation_eventSelection_ZTT}
is complemented by a requirement on a topological discriminant.
The topological discriminant~\cite{CDFrefPzeta} is based on the projections:
\begin{equation}
P_{\zeta} = \left( \ptvec^{\Pgm} + \ptvec^{\tauh} + \vecMET \right) \cdot \frac{\vec{\zeta}}{\abs{\vec{\zeta}}}
\qquad \text{and} \qquad
P_{\zeta}^{\text{vis}} = \left( \ptvec^{\Pgm} + \ptvec^{\tauh} \right) \cdot \frac{\vec{\zeta}}{\abs{\vec{\zeta}}}
\label{eq:PzetaDefinition}
\end{equation}
on the axis $\vec{\zeta}$, given by the bisector of the momenta in the transverse plane of the visible decay products of the two $\Pgt$ leptons.
The discriminant utilizes the fact that the angle between the neutrinos produced in $\Pgt$ decays and the visible $\Pgt$ decay products is typically small,
forcing the $\vecMET$ vector in $\cPZ/\gamma^{*} \to \Pgt\Pgt$ events to point in the direction of $\ptvec^{\Pgm} + \ptvec^{\tauh}$,
which is often not the case in $\PW$+jets and $\cPqt\cPaqt$ events.
Selected events are required to satisfy the condition
$P_{\zeta} - 1.85 P_{\zeta}^{\text{vis}} > -15$\GeV.
This reduces the sum of backgrounds passing the $\mT$ criterion by about a factor two.
Background from $\cPqt\cPaqt$ production is reduced
by vetoing events that contain jets of $\pt > 20$\GeV and $\abs{\eta} < 2.5$
that pass $\Pbottom$ tagging criteria.
The background contributions arising from $\PW\PW$, $\PW\cPZ$, and $\cPZ\cPZ$ production
are suppressed by rejecting events that contain an electron with $\pt > 15$\GeV and $\abs{\eta} < 2.4$
or a second muon with $\pt > 5$\GeV and $\abs{\eta} < 2.4$.
The electrons and muons considered for this veto are required to pass loose identification and isolation criteria.

The $\tauh$ identification efficiency, $\epsilon_{\tau}$, is obtained through
a simultaneous fit of the number of $\cPZ/\Pggx \to \Pgt\Pgt$ events, $N^{\tau}_{\text{pass}}$ and $N^{\tau}_{\text{fail}}$,
with $\tauh$ candidates passing (``pass'' region) and failing (``fail'' region) the $\tauh$ identification discriminant.
The fit is performed as described in Section~\ref{sec:validation_templateFits}.
The $\tauh$ identification efficiency is taken as the parameter of interest $\mu$ in the fit.
The number of $\cPZ/\Pggx \to \Pgt\Pgt$ events in the pass and fail regions as a function of $\mu$ are given by
$N^{\tau}_{\text{pass}} = \mu \, N^{\cPZ/\Pggx \to \Pgt\Pgt}$ and $N^{\tau}_{\text{fail}} = (1 - \mu) \, N^{\cPZ/\Pggx \to \Pgt\Pgt}$, respectively.
The normalization of the $\cPZ/\Pggx \to \Pgt\Pgt$ signal in the sum of pass and fail regions, $N^{\cPZ/\Pggx \to \Pgt\Pgt}$,
as well as the templates for signal in both regions are obtained from the MC simulation.
The systematic uncertainties discussed in Section~\ref{sec:validation_systematicUncertainties} are represented by nuisance parameters in the fit.
An additional nuisance parameter with an uncertainty of 3\% is included in the fit to account for the uncertainty in the energy scale of the $\tauh$ decays.

Contributions from background processes, especially to the fail region, are sizeable.
The distributions for $\cPZ/\Pggx \to \Pgm\Pgm$, $\PW$+jets, $\cPqt\cPaqt$, single top quark, and diboson backgrounds are obtained from MC simulation.
The uncertainty in the yield of $\cPZ/\Pggx \to \Pgm\Pgm$ and diboson ($\cPqt\cPaqt$ and single top quark) backgrounds is increased to 30\% (20\%),
to account for the uncertainty in the rate with which muons (light-quark and gluon jets) are misidentified as $\tauh$ decays.

The normalization of the $\PW$+jets background that is used as input to the fit is determined from data,
using a control region defined by inverting the $\mT < 40$\GeV selection and requiring $\mT > 70$\GeV instead.
The contributions of other backgrounds to this control region, referred to as high-$\mT$ sideband, are subtracted,
based on MC predictions,
before extrapolating the event yield observed in the control region into the signal region.
The extrapolation factor from $\mT > 70$\GeV to $\mT < 40$\GeV is obtained from MC simulation.
The uncertainty in the $\PW$+jets background in the signal region,
arising from the statistical uncertainty in the event yield in the control region, and from the uncertainty in the extrapolation factor,
amounts to 15\%, and is represented by a nuisance parameter in the fit.

The normalization and distribution of the multijet background is estimated from data,
using events in which the muon and loose $\tauh$ candidate have the same charge.
The extrapolation factor from the same-sign (SS) to the opposite-sign (OS) region
is measured in events in which the muon fails the isolation criterion.
The contributions from DY, $\cPqt\cPaqt$, single top quark, and diboson backgrounds to the OS and SS event samples with non-isolated muons,
and to the SS event sample with isolated muons,
are subtracted according to MC expectation.
The number of $\PW$+jets events subtracted is determined using a control region in which the muon and the $\tauh$ candidate have the same charge, and $\mT > 70$\GeV.
The procedure provides an estimate of the multijet background in the signal region with an uncertainty of 10\%.

Two alternative observables are used to perform the fit:
(i)~$\mVis$, the visible mass of the muon and the $\tauh$ candidate,
and (ii)~$\Ntracks$, the multiplicity of tracks within a cone of size $\Delta R < 0.5$ centred on the $\tauh$ direction.
The main results are obtained using $\mVis$.
Fits of the $\Ntracks$ distribution are used to measure the
$\tauh$ identification efficiency as function of \pt and $\eta$ of the $\tauh$ candidate,
and also as function of $\Nvtx$.

Two other uncertainties are considered when $\Ntracks$ is used in the fit.
The track reconstruction efficiency is measured with an uncertainty of 3.9\%~\cite{Chatrchyan:2014fea},
and an uncertainty of 10\% is attributed to the multiplicity of tracks associated with the $\tauh$ candidates
that are from $\text{jet} \to \tauh$ misidentifications.
The 10\% represents the uncertainty on the multiplicity of charged hadrons produced in the hadronization of quarks and gluons into jets.
The uncertainties in track reconstruction efficiency and hadronization affect the $\Ntracks$ distributions obtained from the MC simulation.
We account for these uncertainties by producing $\Ntracks$ distributions with means shifted by $\pm 3.9\%$ and $\pm 10\%$.
The shifted distributions are produced as follows:
for a given event, we set $\Ntracks^{\text{shifted}} = \Ntracks$.
We then iterate over the collection of reconstructed tracks.
For each track, we sample from a uniform distribution,
and when the random number thus selected
is below the magnitude of the shift (either 0.039 or 0.10) we reduce or increase $\Ntracks^{\text{shifted}}$ by one unit,
depending on whether we have, respectively, a downward- or upward-shifted template.

A closure test is performed using pseudo-data,
given by the sum of MC simulated signal and background events and the multijet background obtained from data.
Different pseudo-experiments are generated so as to be able to change signal yields and verify that the fit determines the $\tauh$ identification
efficiency without bias when the signal fraction differs from the nominal value.

An uncertainty of 3.9\% is added in quadrature to the uncertainty in $\epsilon_{\tau}$ determined in the fit.
The value of 3.9\% represents the uncertainty to pass the loose $\tauh$ candidate selections,
and in particular to reconstruct a track with $\pt > 5$\GeV.

\begin{figure}[htbp]
\centering
\includegraphics[width=0.48\textwidth]{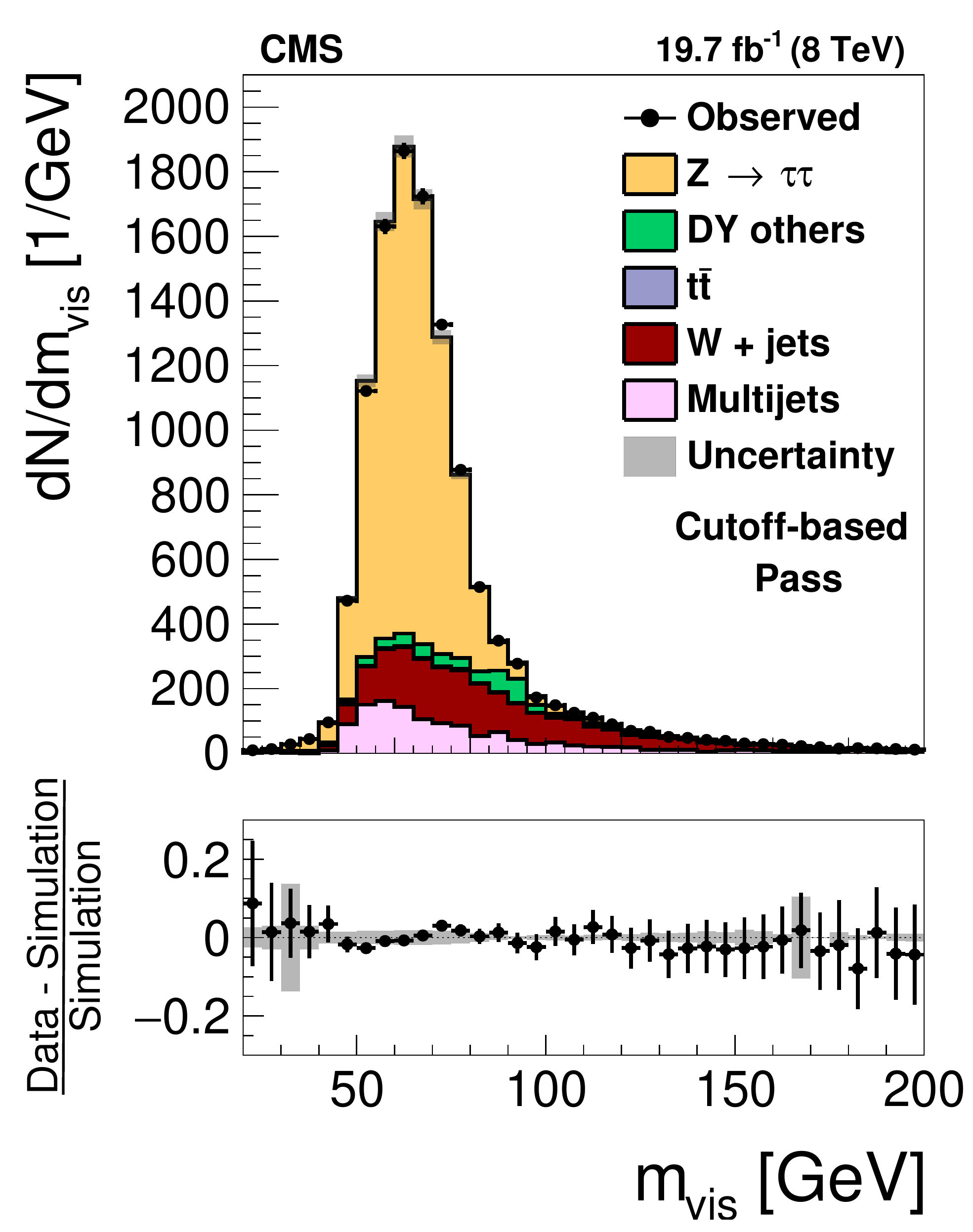}
\includegraphics[width=0.48\textwidth]{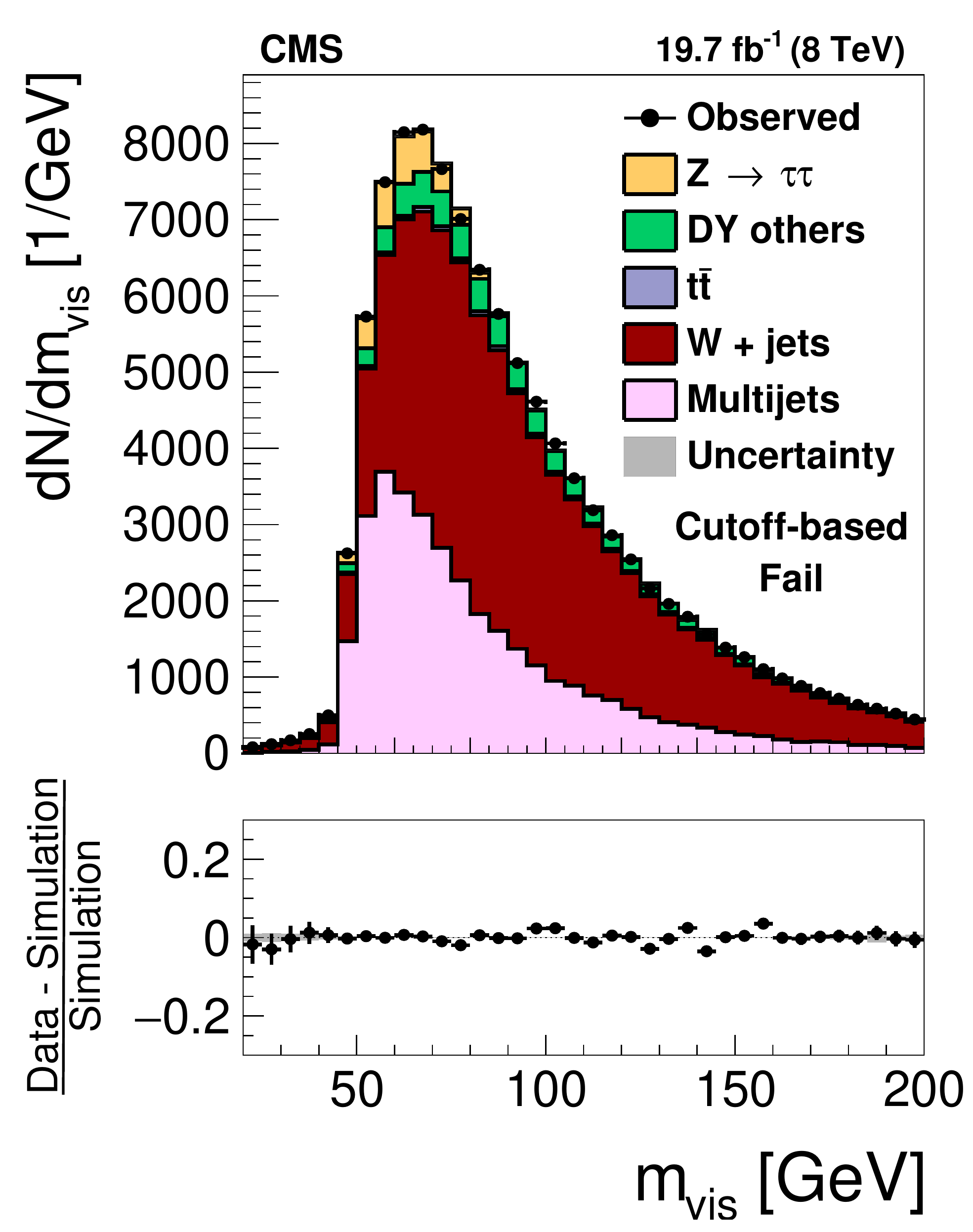}
\includegraphics[width=0.48\textwidth]{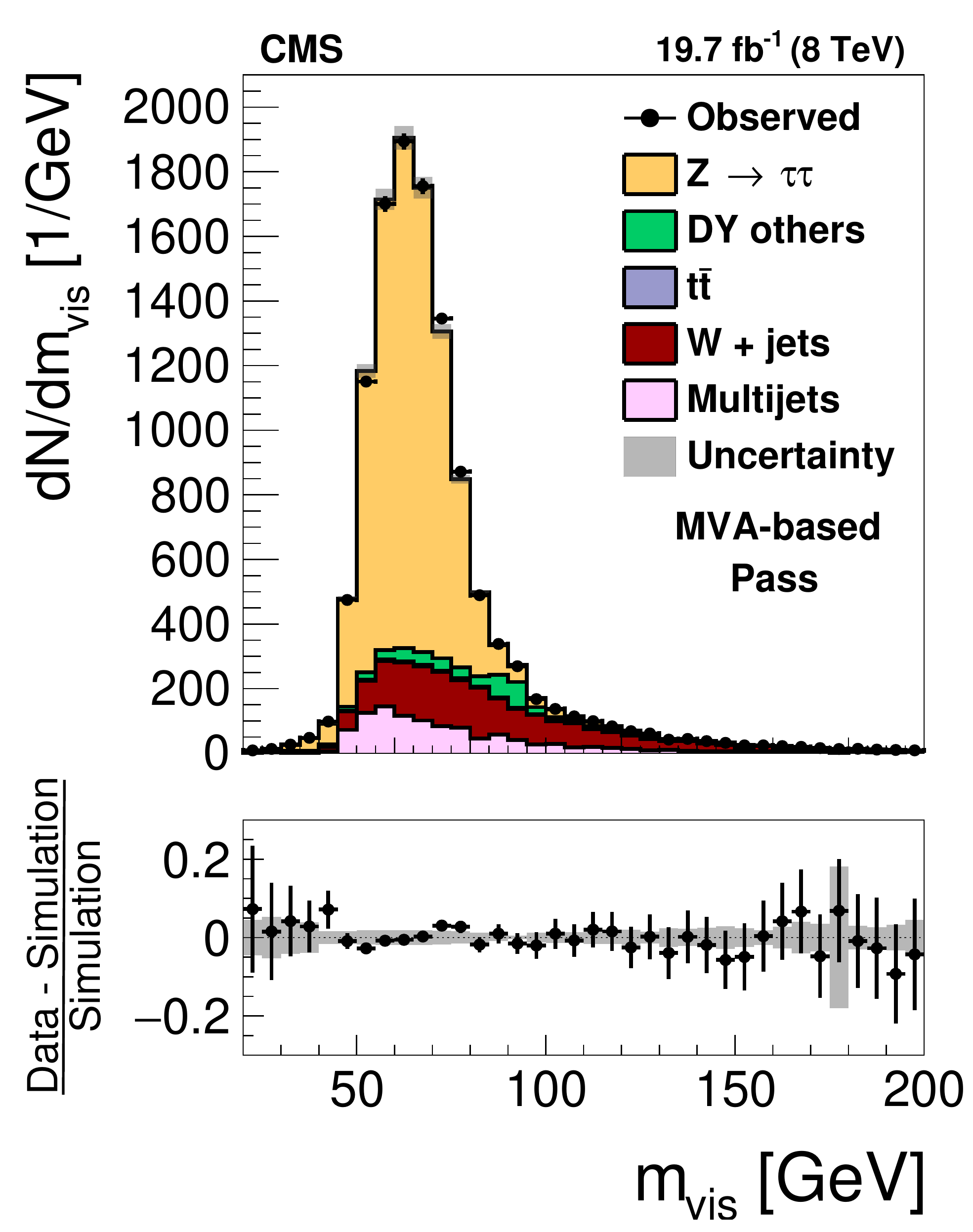}
\includegraphics[width=0.48\textwidth]{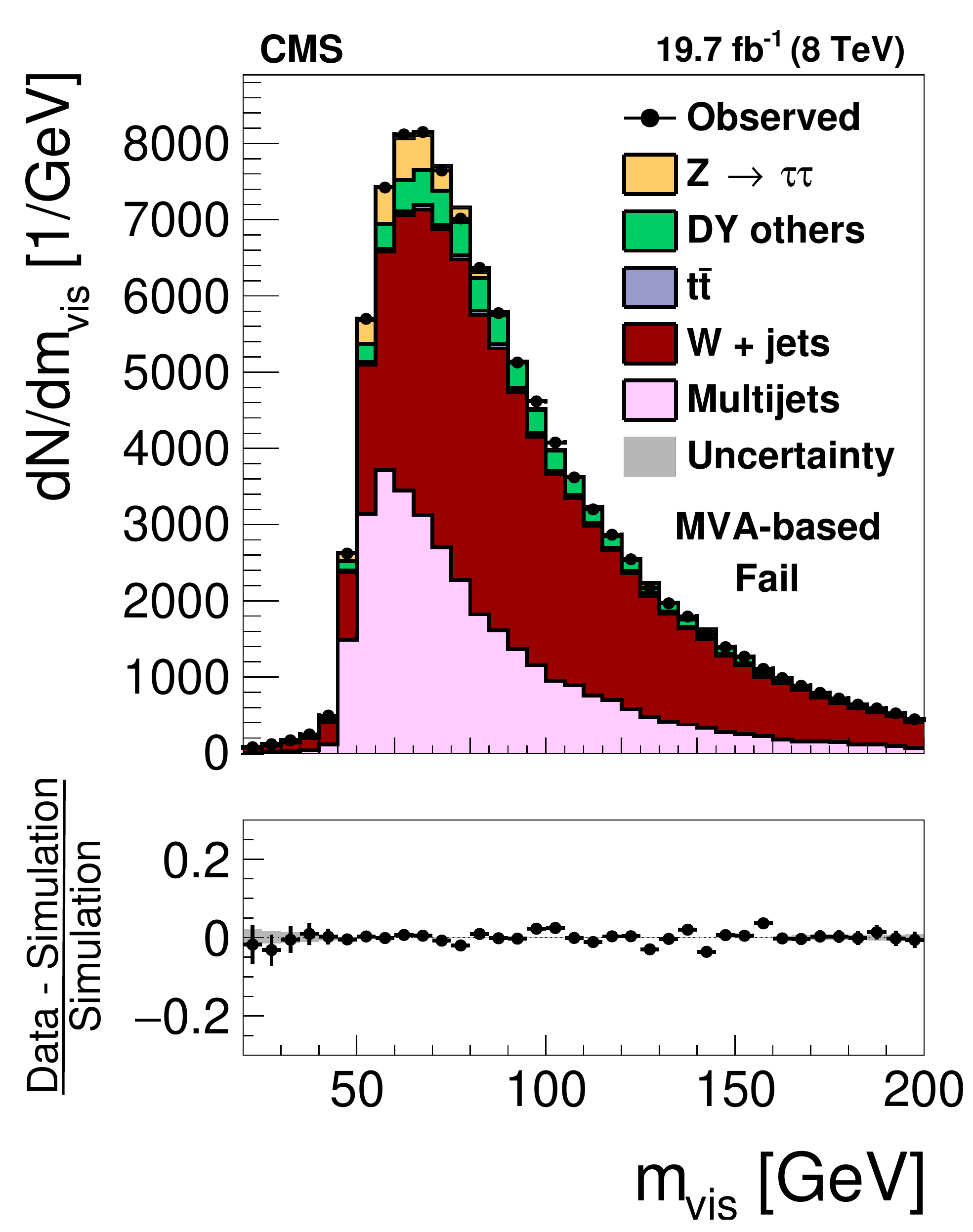}
\caption{
  Distribution in $\mVis$ observed in the pass (left) and fail (right) samples
  of $\cPZ/\Pggx \to \Pgt\Pgt$ candidate events used to measure the $\tauh$ identification efficiency,
  compared to the MC expectation,
  for the loose WP of the cutoff-based (top) and MVA-based (bottom) $\tauh$ isolation discriminants.
  $\cPZ/\Pggx \to \Plepton\Plepton$ ($\Plepton = \Pe$, $\Pgm$, $\Pgt$) events in which
  either the reconstructed muon or the reconstructed $\tauh$ candidate is due to a misidentification are denoted by  ``DY others''.
  The expected $\mVis$ distribution is shown for the values of nuisance parameters obtained from the likelihood fit to the data, described in Section~\ref{sec:validation_templateFits}.
  The ``Uncertainty'' bands represent the statistical and systematic uncertainties added in quadrature.
}
\label{fig:tauID_ZTauTau_mvis_Comb3_and_MVA3oldDMwLT}
\end{figure}

\begin{figure}[htbp]
\centering
\includegraphics[width=0.48\textwidth]{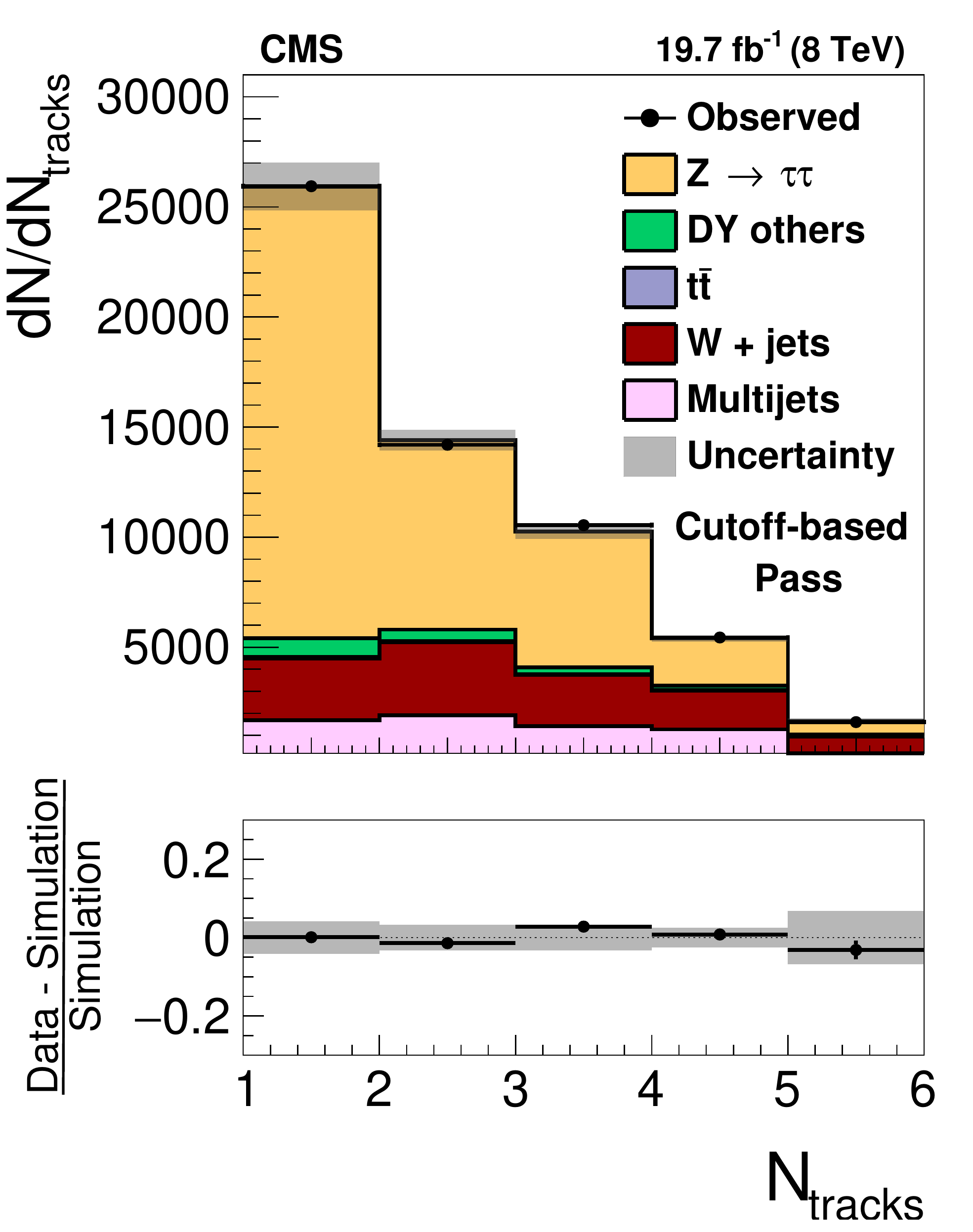}
\includegraphics[width=0.48\textwidth]{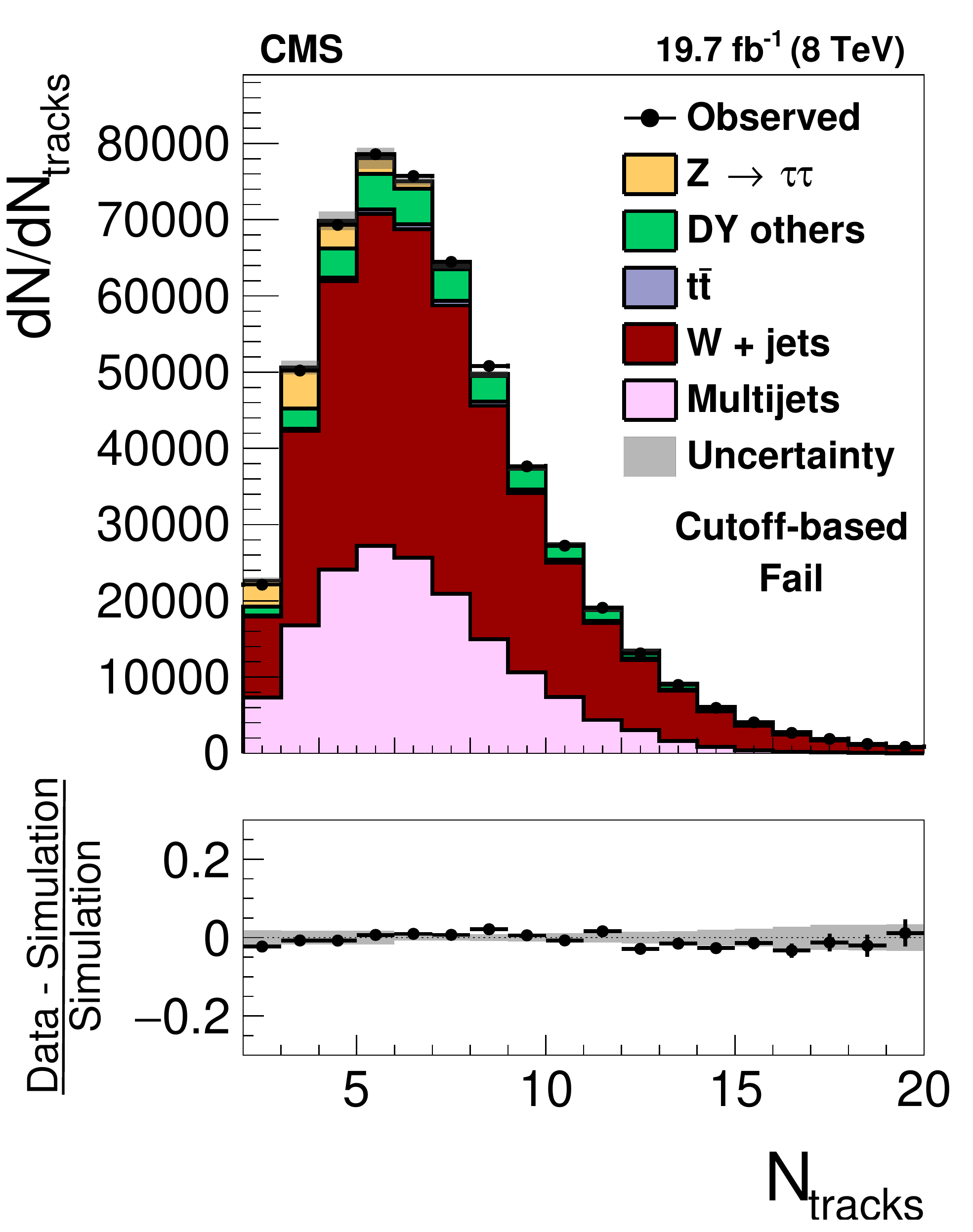}
\includegraphics[width=0.48\textwidth]{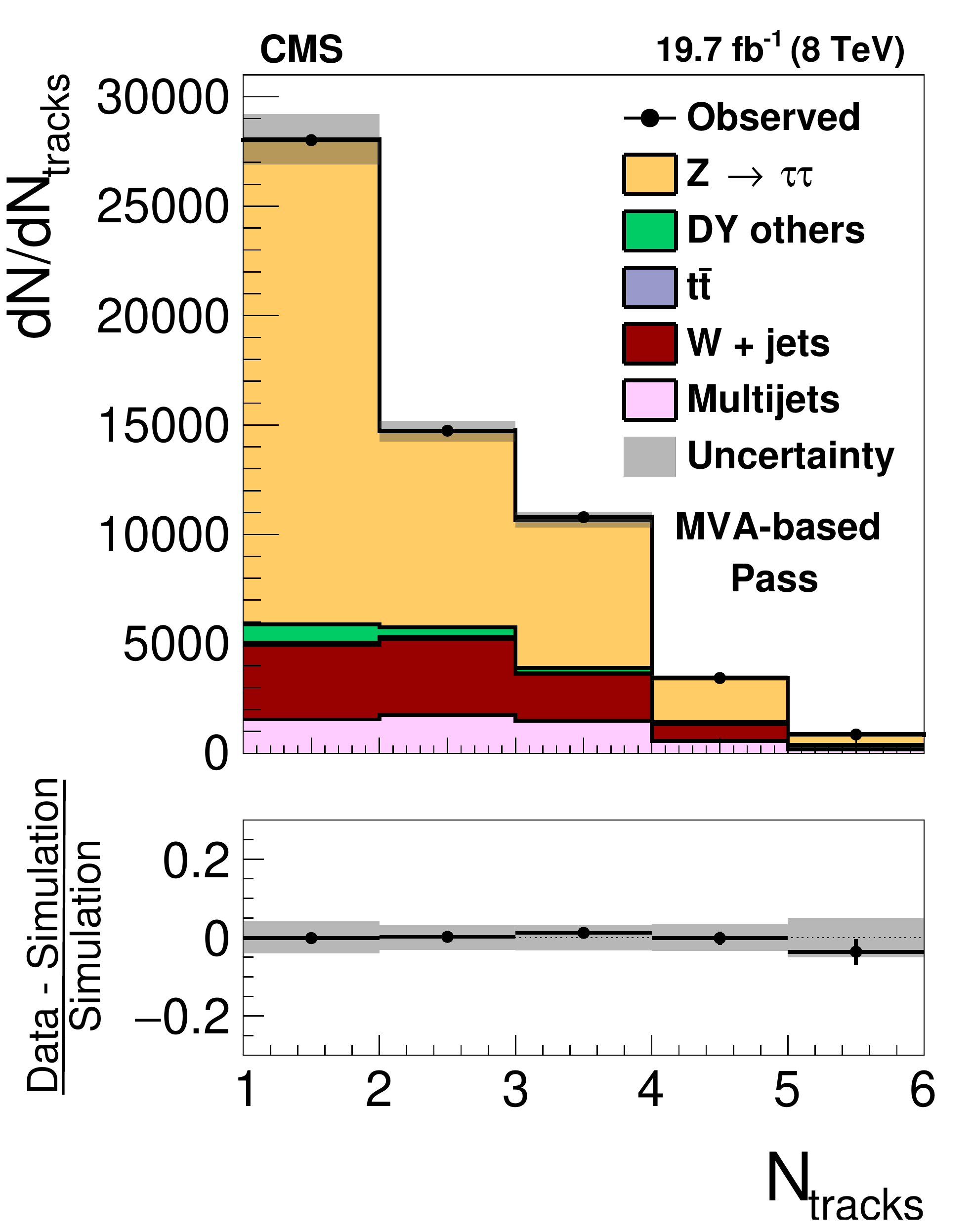}
\includegraphics[width=0.48\textwidth]{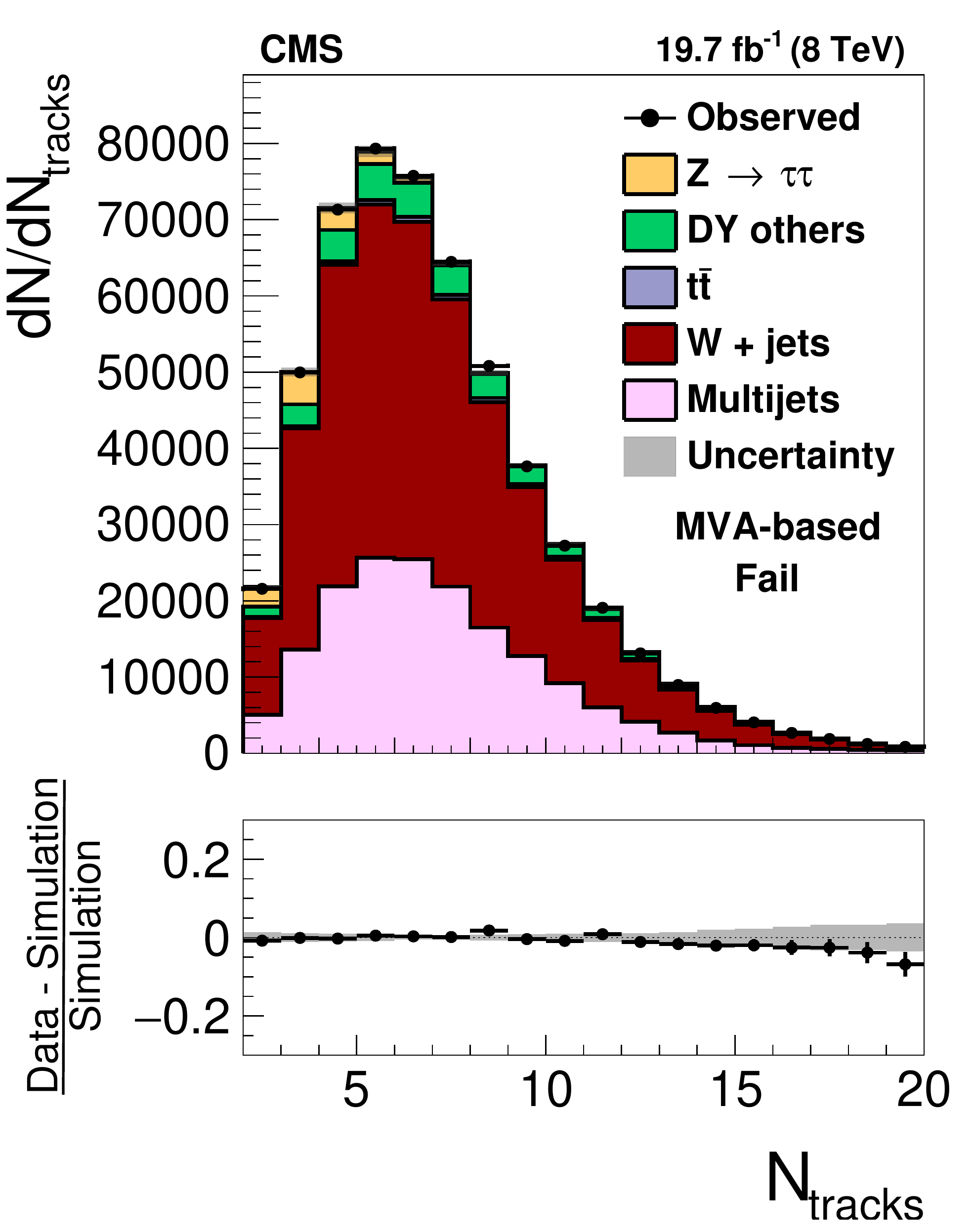}
\caption{
  Distribution in $\Ntracks$ observed in the pass (left) and fail (right) samples
  of $\cPZ/\Pggx \to \Pgt\Pgt$ candidate events used to measure the $\tauh$ identification efficiency,
  compared to the MC expectation,
  for the loose WP of the cutoff-based (top) and MVA-based (bottom) $\tauh$ isolation discriminants.
  $\cPZ/\Pggx \to \Plepton\Plepton$ ($\Plepton = \Pe$, $\Pgm$, $\Pgt$) events in which
  either the reconstructed muon or the reconstructed $\tauh$ candidate is due to a misidentification are denoted by  ``DY others''.
  The expected $\Ntracks$ distribution is shown for the values of nuisance parameters obtained from the likelihood fit to the data, described in Section~\ref{sec:validation_templateFits}.
  The ``Uncertainty'' bands represent the statistical and systematic uncertainties added in quadrature.
}
\label{fig:tauID_ZTauTau_ntracks_Comb3_and_MVA3oldDMwLT}
\end{figure}

The $\tauh$ identification efficiencies measured in the data are quoted relative to the MC expectation.
The results are given in Table~\ref{tab:results_ID_ZTT}.
The data-to-MC ratios obtained using $\mVis$ and $\Ntracks$ are compatible.
All ratios are compatible with unity within the estimated uncertainties of $\approx 4.5\%$.
Plots of the $\mVis$ and $\Ntracks$ distributions in the
pass and fail regions are presented in
Figs.~\ref{fig:tauID_ZTauTau_mvis_Comb3_and_MVA3oldDMwLT} and~\ref{fig:tauID_ZTauTau_ntracks_Comb3_and_MVA3oldDMwLT}.

\begin{table}[htbp]
\centering
\topcaption{
  Data-to-MC ratios of the efficiency for $\tauh$ decays to pass different identification discriminants,
  measured in $\cPZ/\Pggx \to \Pgt\Pgt \to \Pgm\tauh$ events.
  The results obtained using the observables $\mVis$ and $\Ntracks$ are quoted in separate columns.
}
\label{tab:results_ID_ZTT}
\begin{tabular}{lc|c}
\hline
\multirow{2}{10mm}{WP} & \multicolumn{2}{c}{Data/Simulation} \\
\cline{2-3}
 & $\mVis$ & $\Ntracks$ \\
\hline
\multicolumn{3}{c}{Cutoff-based} \\
\hline
Loose      & $1.006 \pm 0.044$ & $0.963 \pm 0.051$ \\
Medium     & $0.984 \pm 0.044$ & $0.982 \pm 0.048$ \\
Tight      & $0.982 \pm 0.044$ & $0.997 \pm 0.052$ \\
\hline
\multicolumn{3}{c}{MVA-based} \\
\hline
Very loose & $1.034 \pm 0.044$ & $0.940 \pm 0.086$ \\
Loose      & $1.017 \pm 0.044$ & $1.026 \pm 0.054$ \\
Medium     & $1.014 \pm 0.044$ & $0.992 \pm 0.057$ \\
Tight      & $1.015 \pm 0.045$ & $0.975 \pm 0.052$ \\
\hline
\end{tabular}
\end{table}

\begin{figure}[htbp]
\centering
\includegraphics[width=0.48\textwidth]{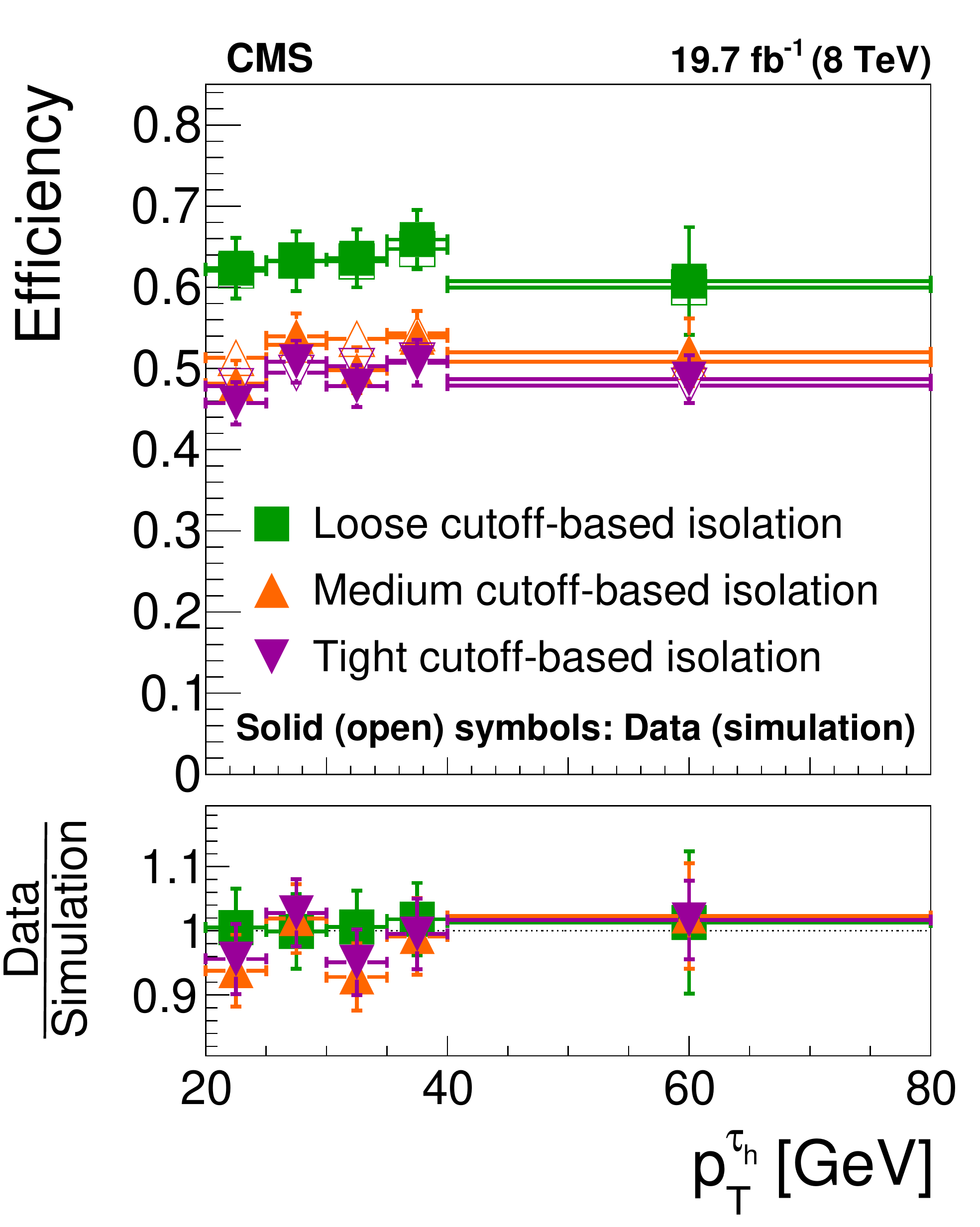}
\includegraphics[width=0.48\textwidth]{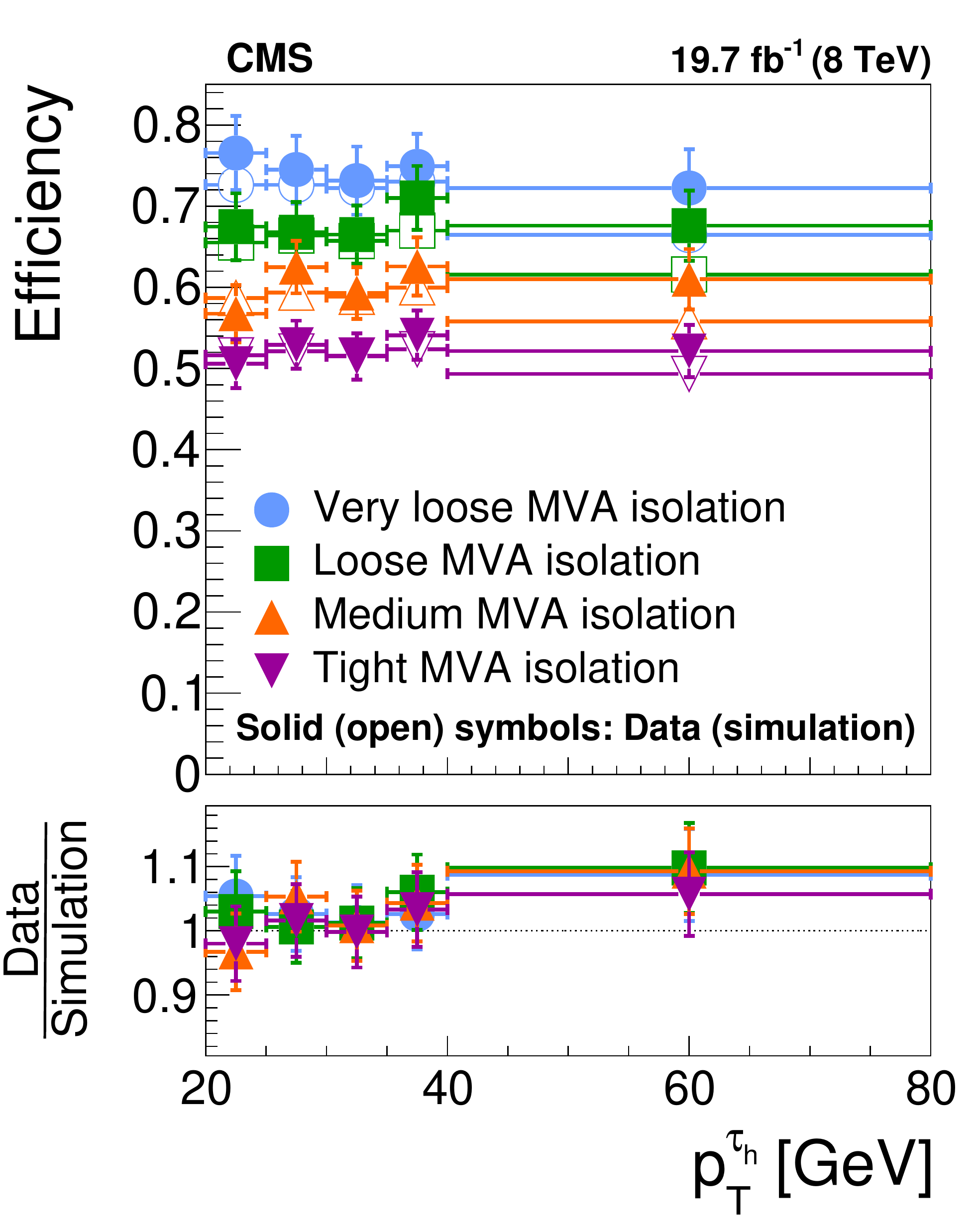}
\includegraphics[width=0.48\textwidth]{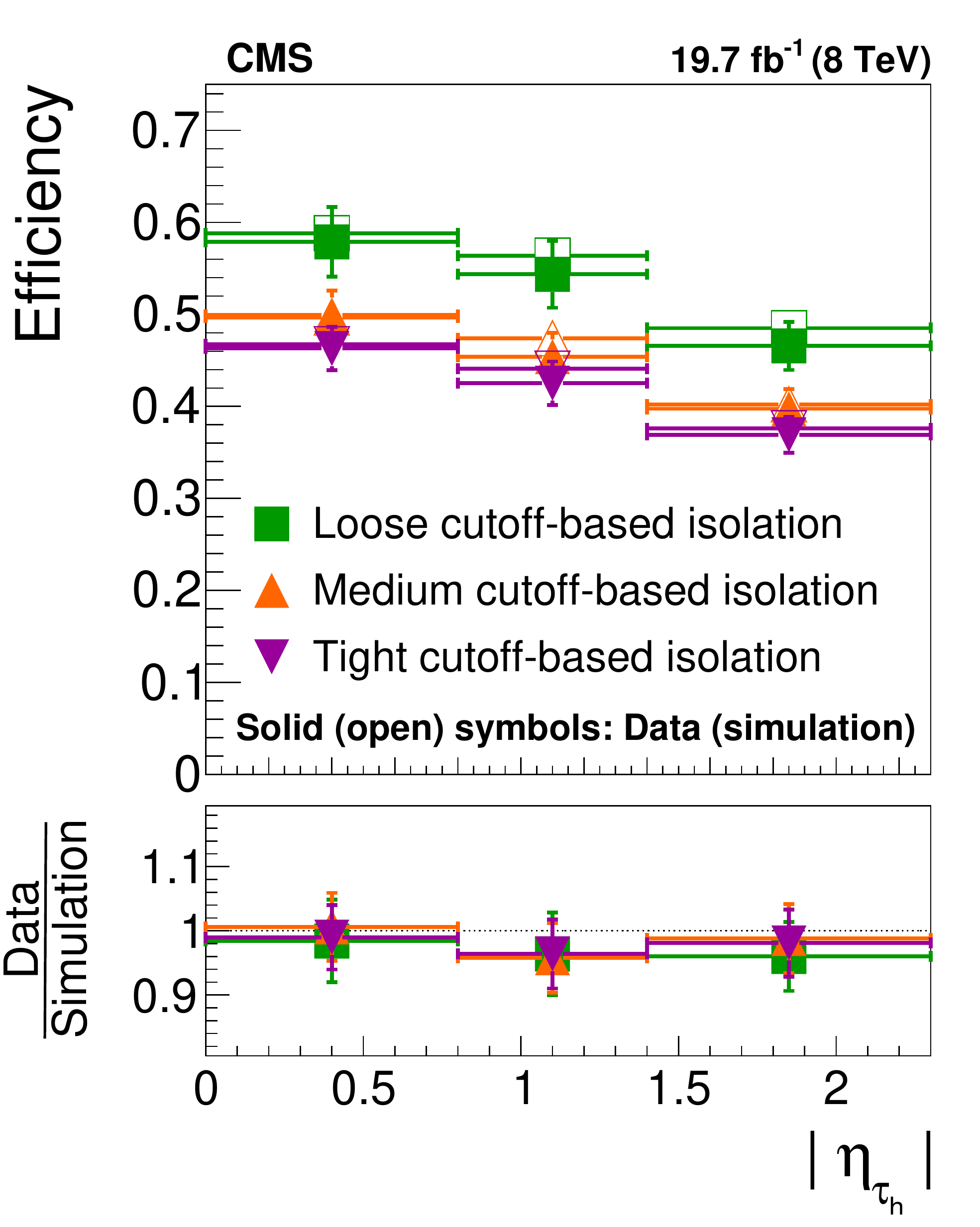}
\includegraphics[width=0.48\textwidth]{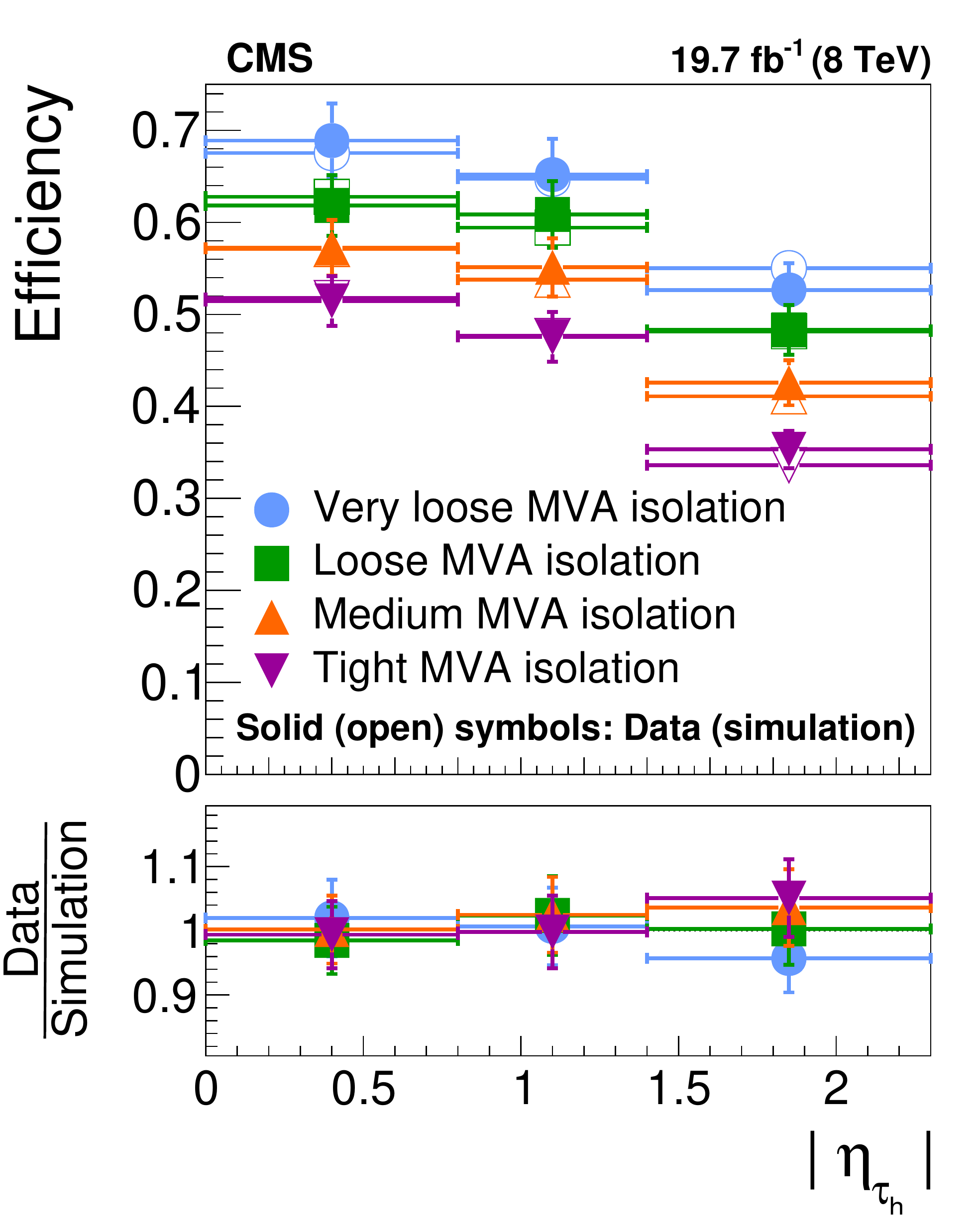}
\caption{
  Tau identification efficiency measured in $\cPZ/\Pggx \to \Pgt\Pgt \to \Pgm\tauh$ events
  as function of \pt and $\eta$,
  for the cutoff-based and MVA-based $\tauh$ isolation discriminants,
  compared to the MC expectation.
  The efficiency is computed relative to $\tauh$ candidates passing the loose $\tauh$ candidate selection described in Section~\ref{sec:tauIdEfficiency_TnP_Ztautau}.
}
\label{fig:tauID_ZTauTau_pT_and_eta}
\end{figure}

\begin{figure}[htb]
\centering
\includegraphics[width=0.48\textwidth]{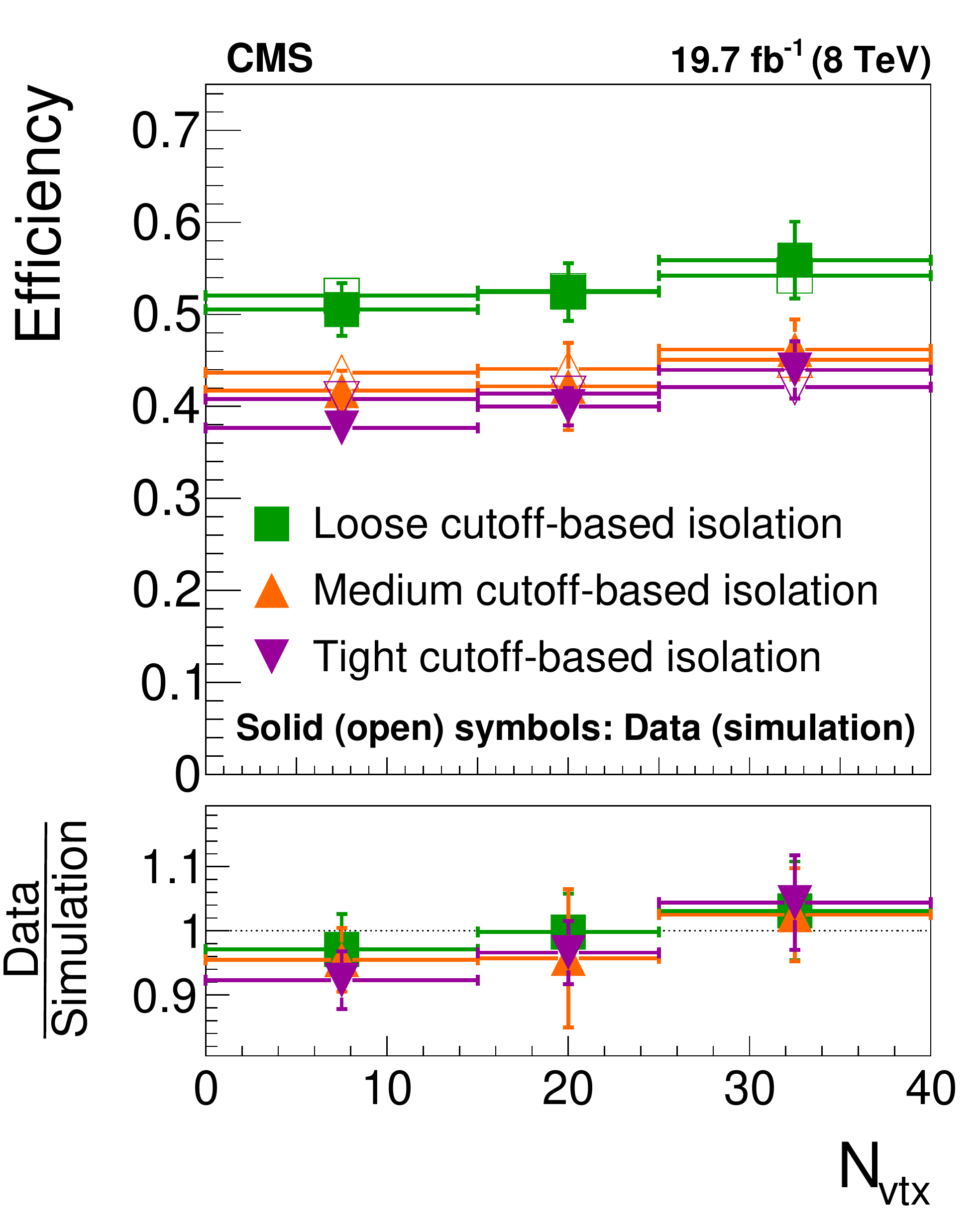}
\includegraphics[width=0.48\textwidth]{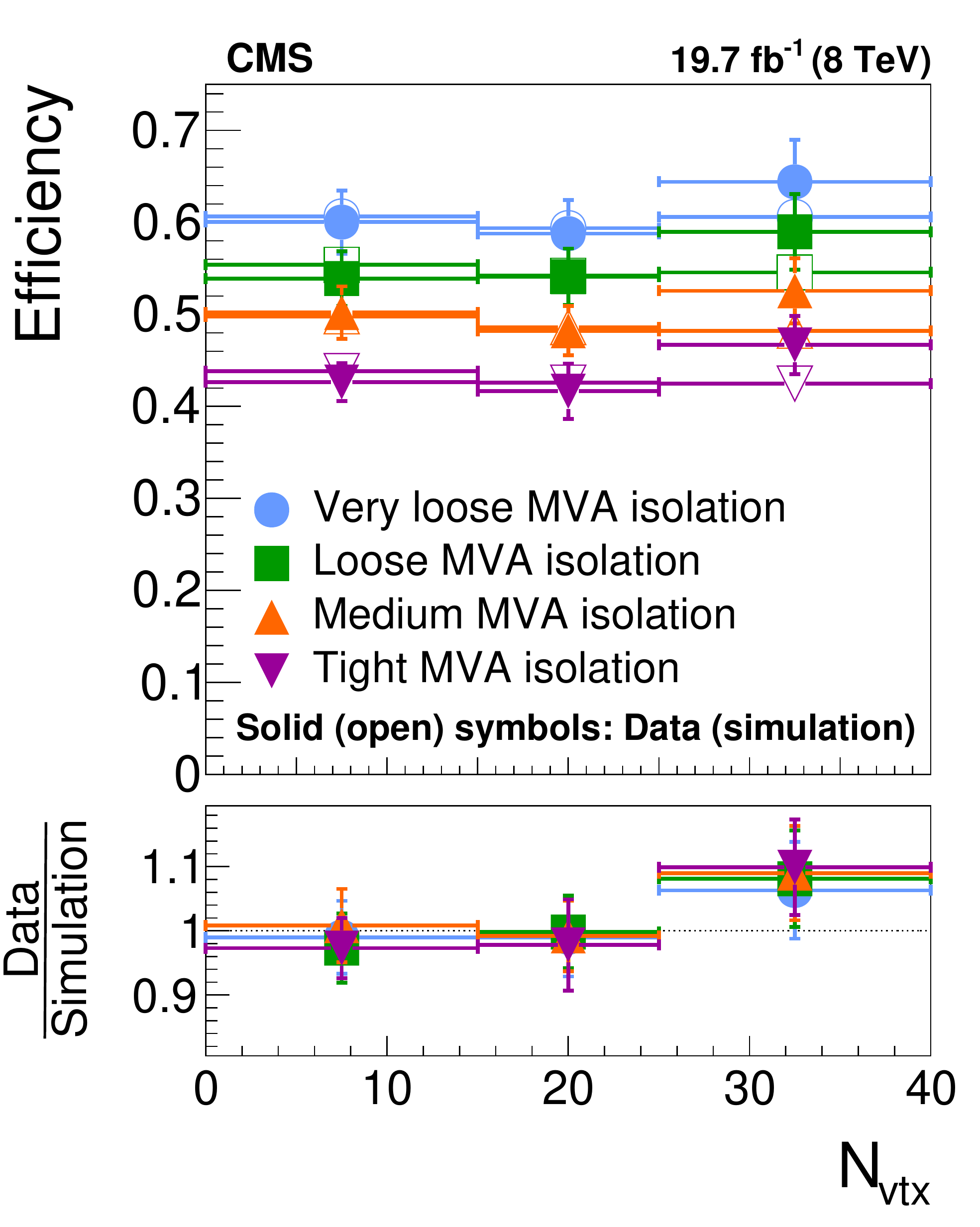}
\caption{
  Tau identification efficiency measured in $\cPZ/\Pggx \to \Pgt\Pgt \to \Pgm\tauh$ events
  as a function of the number of reconstructed vertices $\Nvtx$,
  for the cutoff-based and MVA-based $\tauh$ isolation discriminants,
  compared to the MC expectation.
  The efficiency is computed relative to $\tauh$ candidates passing the loose $\tauh$ candidate selection described in Section~\ref{sec:tauIdEfficiency_TnP_Ztautau}.
}
\label{fig:tauID_ZTauTau_PU}
\end{figure}

The fits of the $\Ntracks$ distribution are repeated for the pass and fail samples,
split into bins of \pt and $\eta$, and into bins of $\Nvtx$, to obtain the dependence of the
tau identification efficiency on \pt and $\eta$ of the $\tauh$ candidate, and on pileup, respectively.
The results are illustrated in Figs.~\ref{fig:tauID_ZTauTau_pT_and_eta} and~\ref{fig:tauID_ZTauTau_PU}.
Within uncertainties, amounting to $\approx 5\%$, the scale factors are compatible with unity.

The efficiency for $\tauh$ decays in $\cPZ/\Pggx \to \Pgt\Pgt$ events to pass the
discriminants for vetoing electrons and muons, described in Section~\ref{sec:discrAgainstElectronsAndMuons}, are also measured,
using a template fit to the $\mVis$ distribution.
Events passing the selection criteria described above,
and containing a $\tauh$ candidate with $\pt > 20$\GeV and $\abs{\eta} < 2.3$
reconstructed in one of the decay modes $\oneProngZeroPizero$, $\oneProngOnePizero$, $\oneProngTwoPizero$, or $\threeProngZeroPizero$,
and passing the loose WP of the cutoff-based tau isolation discriminant,
are divided into pass and fail samples,
depending on whether the $\tauh$ candidate passes or fails the electron or muon discriminants of Section~\ref{sec:discrAgainstElectronsAndMuons}, respectively.
The efficiencies measured in data are in agreement with the MC expectation
within the uncertainty of the measurement, amounting to less than 1\%.

\subsection{Tau identification efficiency in \texorpdfstring{$\cPqt\cPaqt \to \cPqb\cPaqb\Pgm\tauh$}{ttbar to bbbar muon hadronic tau} events}
\label{sec:tauIdEfficiency_TnP_TTbar}

The sample of $\cPqt\cPaqt \to \cPqb\cPaqb\Pgm\tauh$ candidate events is selected as described in Section~\ref{sec:validation_eventSelection_ttbar}.
The high level of background contamination in the $\cPqt\cPaqt \to \cPqb\cPaqb\Pgm\tauh$ event sample
impedes the measurement of the $\tauh$ identification efficiency
using the number of $\tauh$ decays that pass and fail $\Pgt$ identification criteria.
Instead, we determine the $\tauh$ identification efficiency $\varepsilon_{\Pgt}$
from the yield of $\cPqt\cPaqt \to \cPqb\cPaqb\Pgm\tauh$ signal events passing $\tauh$ identification criteria,
using the relation:
\begin{equation}
\varepsilon_{\Pgt}
 = \frac{N^{\Pgt}_{\text{pass}}}{\varepsilon_{\text{non}-\Pgt}  L  \sigma_{\cPqt\cPaqt}},
\label{eq:tauIdEffInTTbar}
\end{equation}
where $N^{\tau}_{\text{pass}}$ denotes the number of observed $\cPqt\cPaqt \to \cPqb\cPaqb\Pgm\tauh$ signal events, and
is obtained through a template fit that takes into account the contribution of background processes.
The symbol $L$ denotes the integrated luminosity of the analyzed data, and $\sigma_{\cPqt\cPaqt}$ the product of the $\cPqt\cPaqt$ production cross section and the branching fraction.
The efficiency of the event selection criteria other than the identification efficiency of $\tauh$ is denoted by $\varepsilon_{\text{non}-\Pgt}$, and is obtained from the MC simulation.
The MC-to-data corrections are applied for the muon trigger, identification, and isolation efficiencies, and for the $\MET$ resolution.
Residual differences between data and MC simulation that may affect $\varepsilon_{\text{non}-\Pgt}$ are considered as systematic uncertainties.
We refer to this sample as the pass region.

The number of $N^{\tau}_{\text{pass}}$ events,
as well as the contributions from background processes,
are determined by fitting the distribution in $\mT$ of Eq.~(\ref{eq:MtDefinition})
in the selected event sample, using templates for signal and background processes.

The templates for the $\cPqt\cPaqt \to \cPqb\cPaqb\Pgm\tauh$ signal and for DY, $\PW$+jets, single top quark, and diboson backgrounds are obtained from the MC simulation.
Due to the $\cPqt\cPaqt$ contribution in the high-$\mT$ sideband, the
normalization of the $\PW$+jets background cannot be determined from the data,
and is taken from the MC simulation, with an uncertainty of 30\%.
A substantial background arises from $\cPqt\cPaqt$ events in which the reconstructed $\tauh$ candidate corresponds to either a jet $\to \tauh$, $\Pe \to \tauh$, or $\Pgm \to \tauh$ misidentification.
The $\cPqt\cPaqt$ background with such $\tauh$ candidates is included in the fit as a separate contribution with an independent normalization.
The template for the $\cPqt\cPaqt$ background is obtained from the MC simulation.
The multijet template is obtained from a control region, by applying event selection criteria that are similar to the pass region,
except that the muon isolation requirement is changed to $I_{\Pgm} >
0.10 \, \pt^{\Pgm}$ and the jets are not required to pass $\Pbottom$ tagging criteria.
The contribution from $\cPqt\cPaqt$ and backgrounds from sources other than multijet events are subtracted according to MC predictions,
using the samples and cross sections described in Section~\ref{sec:datasamples_and_MonteCarloSimulation}.
Because of this subtraction, the template for the multijet background depends on systematic uncertainties that affect the $\cPqt\cPaqt$ signal and non-multijet backgrounds.
The dependence is taken into account through suitable changes in the template as function of the corresponding nuisance parameters in the fit.

Systematic uncertainties that can affect the yield of $\cPqt\cPaqt \to \cPqb\cPaqb\Pgm\tauh$ signal in the pass region,
as well as the rate for background processes,
are constrained using a control region dominated by $\cPqt\cPaqt \to \cPqb\cPaqb\Pgm\Pgm$ events,
which we refer to as the dimuon region.

Events in the dimuon region are selected by requiring two muons
with $\pt > 20$\GeV and $\abs{\eta} < 2.4$,
passing tight identification and isolation criteria.
The muons are required to be of opposite charge, and to be compatible with originating from the same vertex.
The mass of the muon pair is required to exceed $m_{\Pgm\Pgm} > 50$\GeV,
and not be within 10\GeV of the nominal $\cPZ$ boson mass, i.e. requiring $\lvert m_{\Pgm\Pgm} - m_{\cPZ} \rvert > 10$\GeV.
The event is also required to pass the single-muon trigger.
At least one of the muons is required to satisfy the conditions
$\pt > 25$\GeV and $\abs{\eta} < 2.1$,
to ensure that the single-muon trigger is fully efficient.
The event is further required to contain two jets with $\pt > 30$\GeV and $\abs{\eta} < 2.5$,
separated from each of the muons by $\Delta R > 0.5$.
At least one of the jets is required to pass $\Pbottom$ tagging criteria.
The $\MET$ in the event must be $> 40$\GeV.
Events containing additional electrons with $\pt > 15$\GeV and $\abs{\eta} < 2.3$,
or muons with $\pt > 10$\GeV and $\abs{\eta} < 2.4$ that pass loose identification and isolation criteria,
are rejected.

The trigger and event selection criteria that are applied to select $\cPqt\cPaqt \to \cPqb\cPaqb\Pgm\tauh$ and $\cPqt\cPaqt \to \cPqb\cPaqb\Pgm\Pgm$ events
are chosen to be as similar as possible.
This ensures that the systematic uncertainties affecting the yield of signal and background processes
are the same in the pass and in the dimuon regions.
The $\mT$ distributions observed in the two regions are fitted simultaneously.
In the dimuon control region, the transverse mass is computed by choosing one of the two muons at random.

\begin{figure}[htbp]
\centering
\includegraphics[width=0.48\textwidth]{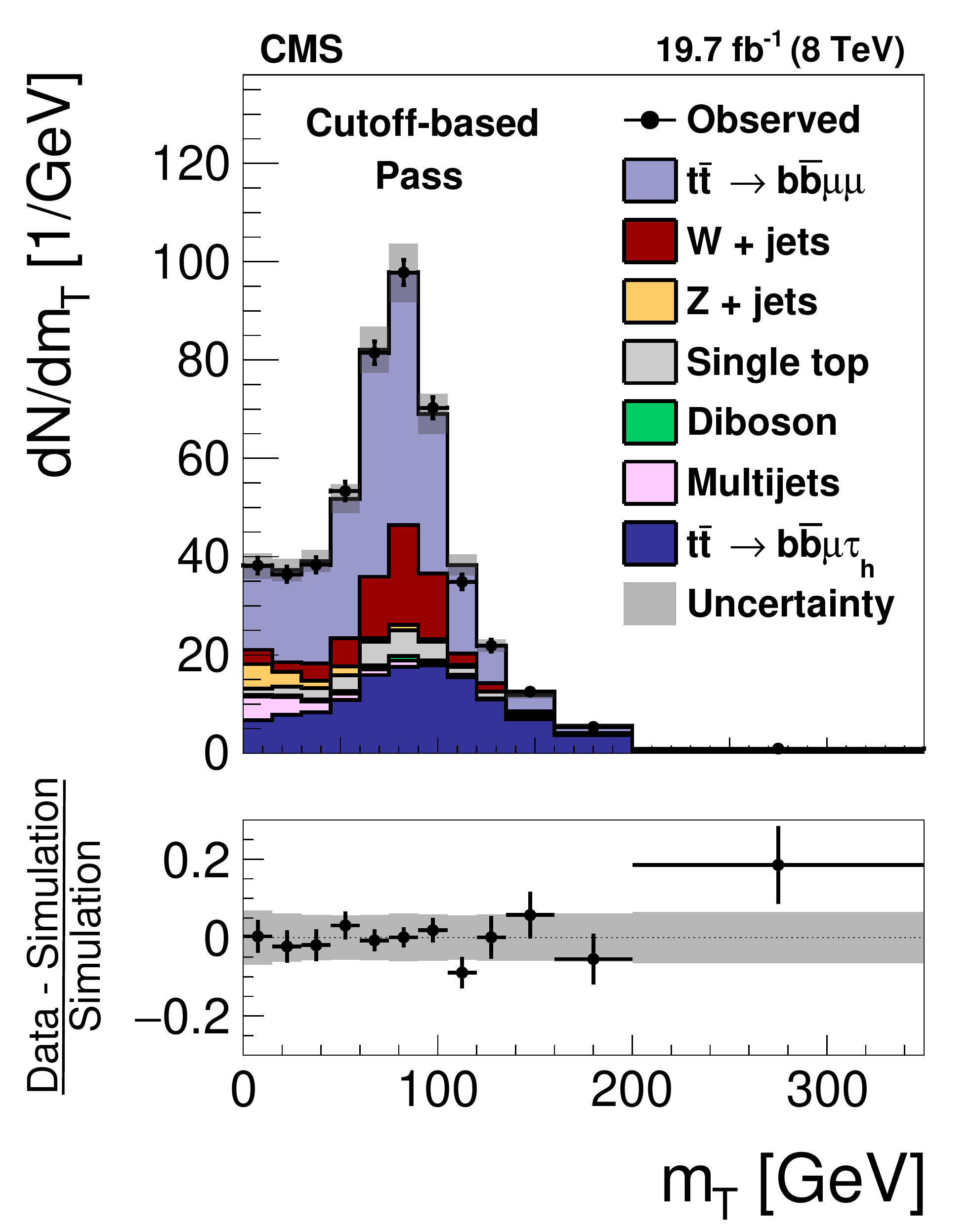}
\includegraphics[width=0.48\textwidth]{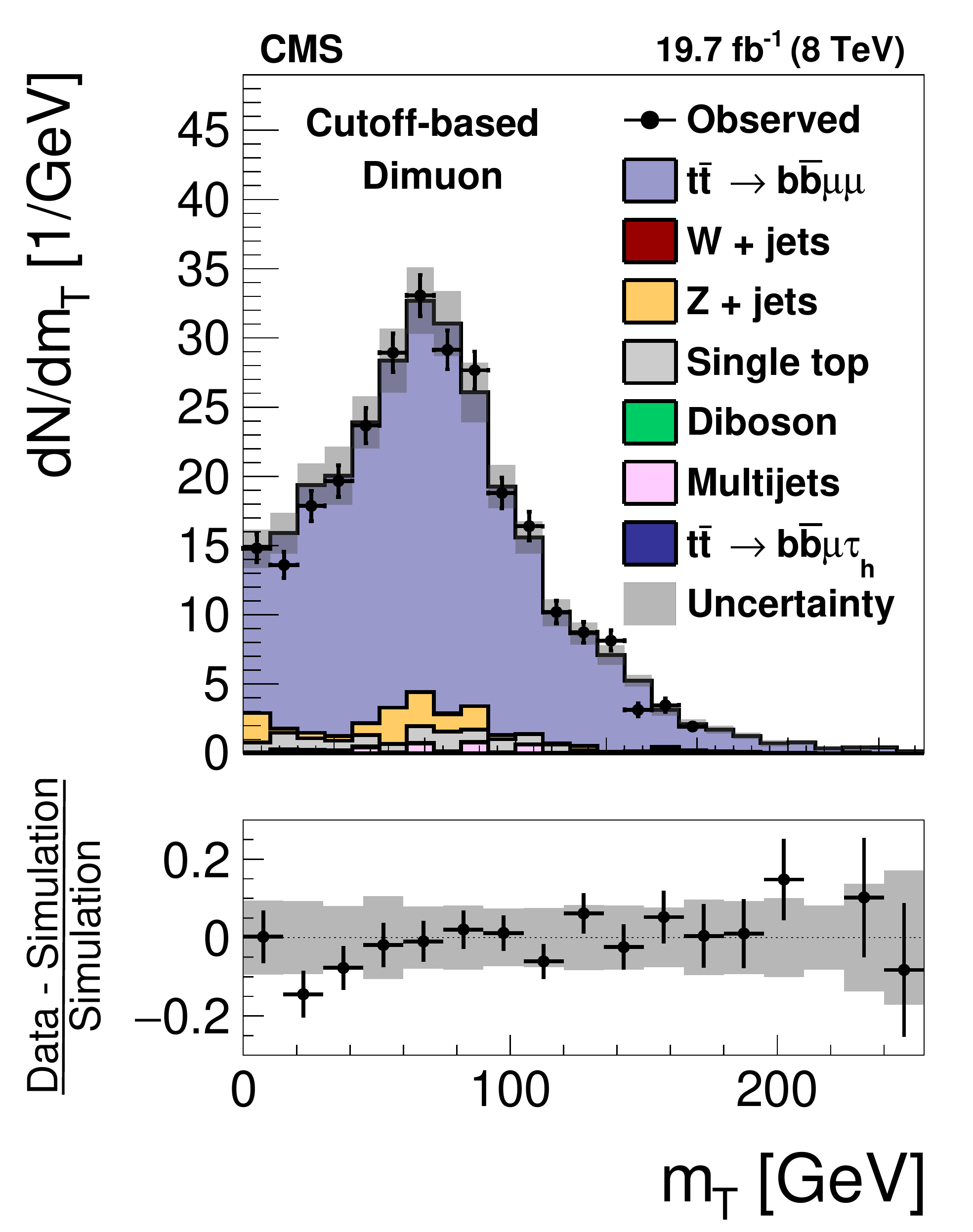}
\includegraphics[width=0.48\textwidth]{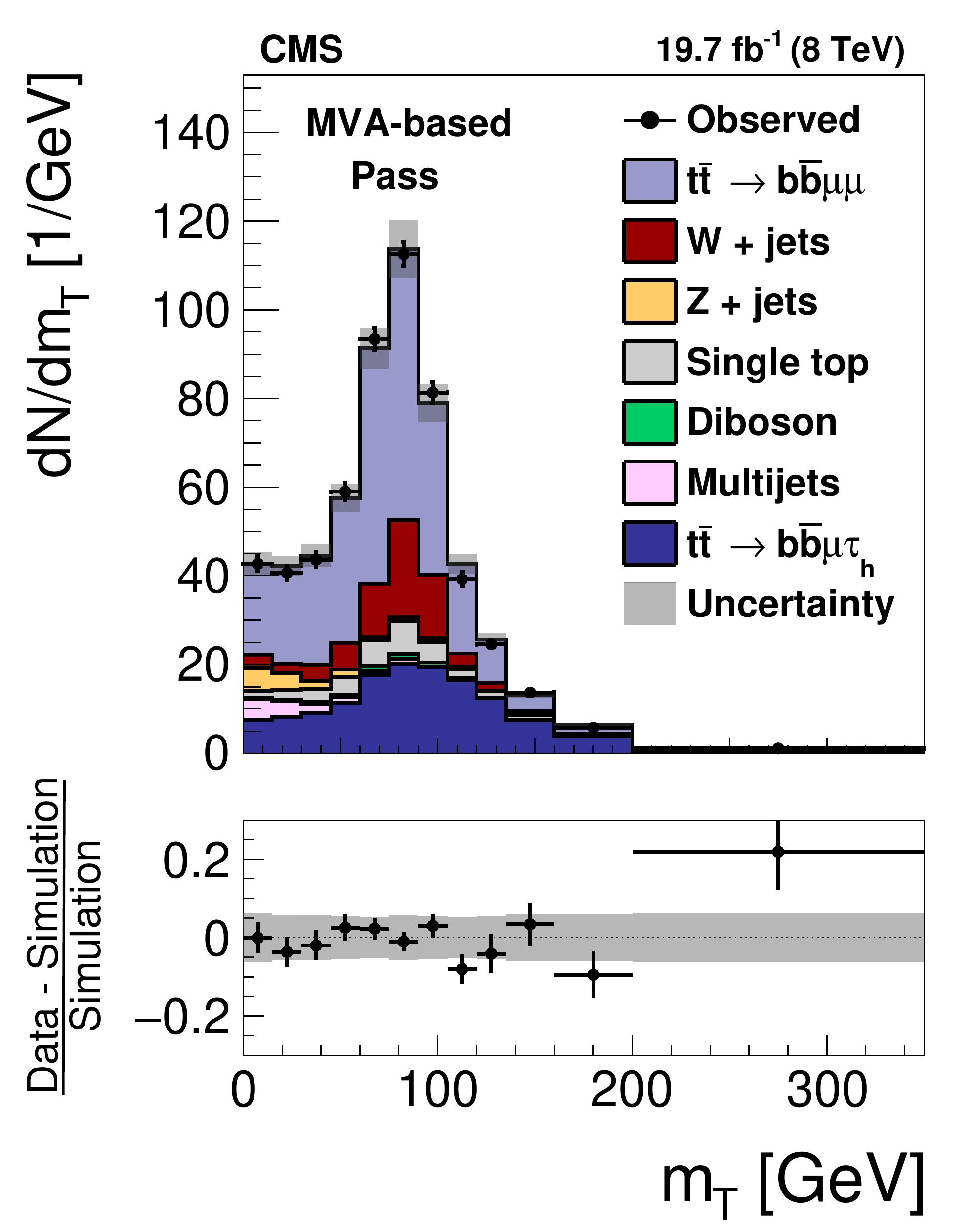}
\includegraphics[width=0.48\textwidth]{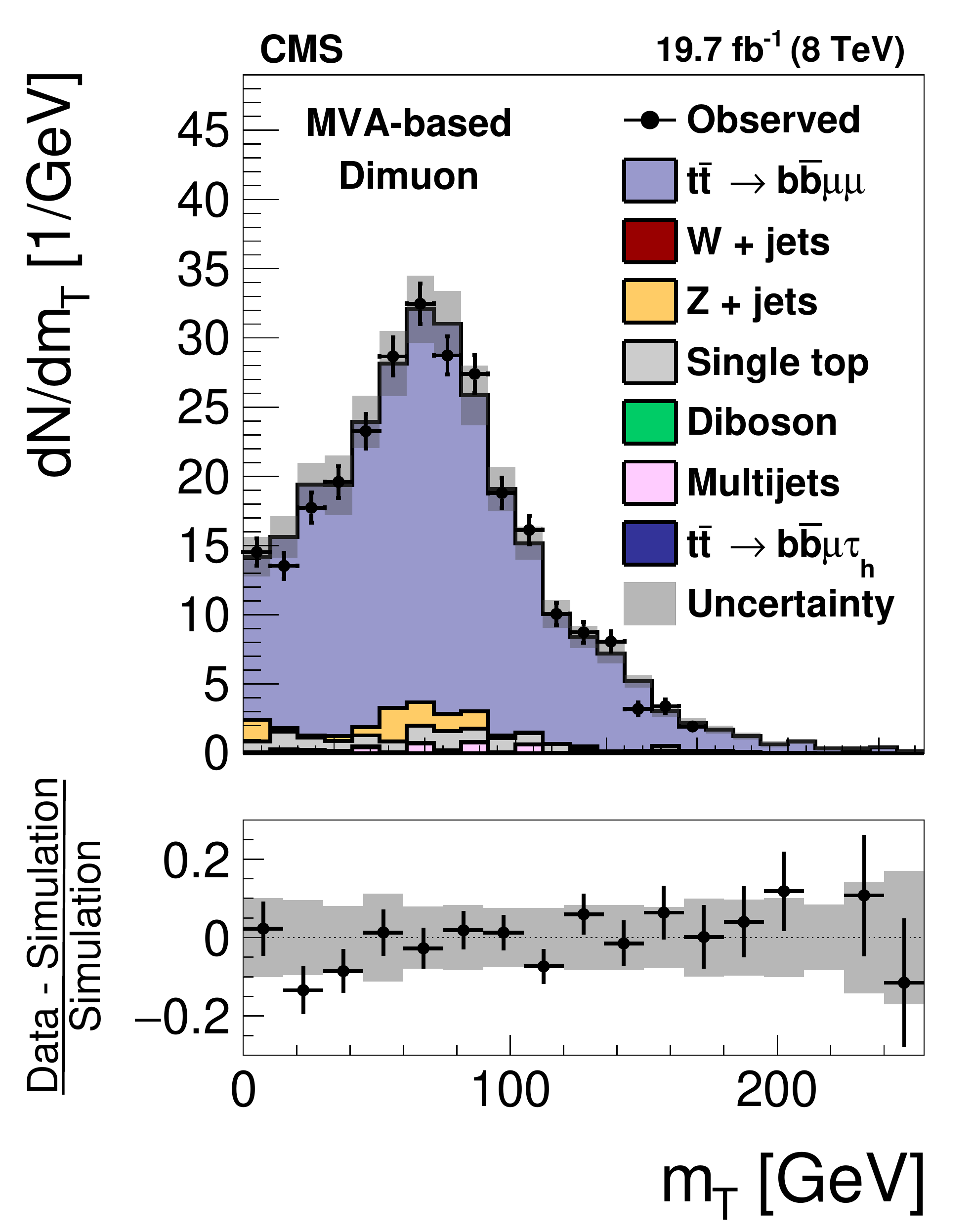}
\caption{
  Distribution in the transverse mass of the muon and $\MET$ in the
  pass region (left) and in the dimuon region (right) in $\cPqt\cPaqt$ events used to measure the $\tauh$ identification efficiency,
  for the loose WP of the cutoff-based (top) and MVA-based (bottom) $\tauh$ isolation discriminants, respectively.
  The $\cPqt\cPaqt$ events in which
  either the reconstructed muon or the reconstructed $\tauh$ candidate are misidentified are denoted by ``$\cPqt\cPaqt$ others''.
  The expected $\mT$ distribution is shown for the values of nuisance parameters obtained from the likelihood fit to the data,
  as described in Section~\ref{sec:validation_templateFits}.
  The ``Uncertainty'' band represents the statistical and systematic uncertainties added in quadrature.
}
\label{fig:tauID_ttbar_Comb3_and_MVA3oldDMwLT}
\end{figure}

The data-to-MC ratios of $\tauh$ identification efficiencies measured in $\cPqt\cPaqt \to \cPqb\cPaqb\Pgm\tauh$ events are given in
Table~\ref{tab:Results_tauID_ttbar}.
Within the uncertainty of the measurement of 9--11\%,
the efficiencies of all $\tauh$ identification discriminators are
compatible with the MC expectations.
Plots of the distribution in $\mT$ in the pass and dimuon control regions are shown in
Fig.~\ref{fig:tauID_ttbar_Comb3_and_MVA3oldDMwLT}.
Data and MC simulation agree within uncertainties after the fit.

\begin{table}[htbp]
\centering
\topcaption{
  Data-to-MC ratios of the efficiency
  for $\tauh$ decays in $\cPqt\cPaqt \to \cPqb\cPaqb\Pgm\tauh$ events
  to pass different $\tauh$ identification discriminants.
}
\label{tab:Results_tauID_ttbar}
\begin{tabular}{lc}
\hline
WP & Data/Simulation \\
\hline
\multicolumn{2}{c}{Cutoff-based} \\
\hline
Loose      & $1.037 \pm 0.097$ \\
Medium     & $1.050 \pm 0.107$ \\
Tight      & $1.047 \pm 0.108$ \\
\hline
\multicolumn{2}{c}{MVA-based} \\
\hline
Very loose & $0.927 \pm 0.097$ \\
Loose      & $1.009 \pm 0.097$ \\
Medium     & $0.956 \pm 0.118$ \\
Tight      & $1.080 \pm 0.117$ \\
\hline
\end{tabular}
\end{table}

\section{Measurement of the \texorpdfstring{$\tauh$}{hadronic tau} energy scale}
\label{sec:Tau_energy_scale}

The energy scale for $\tauh$ (referred to as $\Pgt$ES),
defined as the average reconstructed $\tauh$ energy relative to the generator level energy of the visible $\Pgt$ decay products,
is an important source of systematic uncertainty in many analyses with $\Pgt$ leptons in the final state.
In particular, $\Pgt$ES has a significant influence on the potential to discover a $\PHiggs \to \Pgt\Pgt$ signal
in the presence of the dominant irreducible background from DY $\cPZ/\Pggx \to \Pgt\Pgt$ production
in the $\Pgt\Pgt$ mass distribution~\cite{HIG-13-004}.

An MC-to-data $\Pgt$ES correction is determined
by fitting the distributions of observables sensitive to the energy scale,
using a sample of $\cPZ/\Pggx \to \Pgt\Pgt \to \Pgm\tauh$ events.
The events are selected as described in
Section~\ref{sec:validation_eventSelection_ZTT}, except that the $\tauh$
candidates are required to pass the medium WP of the MVA-based $\tauh$ isolation discriminant,
instead of the loose WP of the cutoff-based discriminant.

The $\Pgt$ES is measured separately for $\tauh$ candidates reconstructed in the decay modes
$\oneProngZeroPizero$, $\oneProngPizeros$, and $\threeProngZeroPizero$
in bins of $20 < \pt < 30$\GeV, $30 < \pt < 45$\GeV, and $\pt > 45$\GeV.

Two alternative observables are used to perform the fit:
the reconstructed mass of the $\tauh$ candidate $m_{\tauh}$,
and $\mVis$, the mass of muon and $\tauh$ candidate.
The $\mVis$ and $m_{\tauh}$ templates for the $\cPZ/\Pggx \to \Pgt\Pgt$ signal
are computed by changing the $\tauh$ four-momentum, reconstructed as described in Section~\ref{sec:decay_mode_reconstruction},
as a function of $\Pgt$ES, and recomputing $\mVis$ and $m_{\tauh}$ after each such change.
For $\tauh$ candidates reconstructed in the $\oneProngPizeros$ and $\threeProngZeroPizero$ modes,
all components of the $\tauh$ four-vector are scaled by the given $\Pgt$ES factor,
while for $\tauh$ candidates in the $\oneProngZeroPizero$ decay mode we scale the energy and adjust the momentum such that
$\eta$, $\phi$ and mass of the four-vector remain unchanged.
The observable $m_{\tauh}$ is defined only for $\tauh$ candidates reconstructed in the $\oneProngPizeros$ and $\threeProngZeroPizero$ modes,
and the energy scale of $\tauh$ candidates reconstructed in the $\oneProngZeroPizero$ decay mode is measured via $\mVis$.

The $\cPZ/\Pggx \to \Pgt\Pgt$ signal is modelled via the ``embedding'' technique~\cite{HIG-13-004}.
The method is based on selecting $\cPZ/\Pggx \to \Pgm\Pgm$ events in data,
and replacing the reconstructed muons by generator-level $\Pgt$ leptons.
The $\Pgt$ decays are simulated using {\TAUOLA},
and the {\sc geant4}-based detector simulation is used to model the detector response to the $\Pgt$ decay products.
The visible $\Pgt$ decay products are reconstructed with the PF algorithm,
and mixed with the remaining particles of the $\cPZ/\Pggx \to \Pgm\Pgm$ event,
after the two muons are removed.
Finally, $\tauh$ candidates, jets, and $\MET$ are reconstructed, the isolation of electrons and muons is computed,
and the event is analyzed as if it were data.
Embedded samples are produced for the entire data-taking period,
covering the same run ranges as the data used to measure the $\Pgt$ES correction.

The $\cPZ/\Pggx \to \Plepton\Plepton$ ($\Plepton$ = $\Pe$, $\Pgm$), $\PW$+jets, $\cPqt\cPaqt$, single top quark, and diboson backgrounds
are modelled using MC simulation.
The templates for background processes are kept unchanged as function of $\Pgt$ES.

The multijet background is obtained directly from data,
using events in which the muon is not isolated, and of the same charge as the $\tauh$ candidate,
as described in Section~\ref{sec:tauIdEfficiency_TnP_Ztautau}.

\begin{figure}[htb]
\centering
\includegraphics[width=0.32\textwidth]{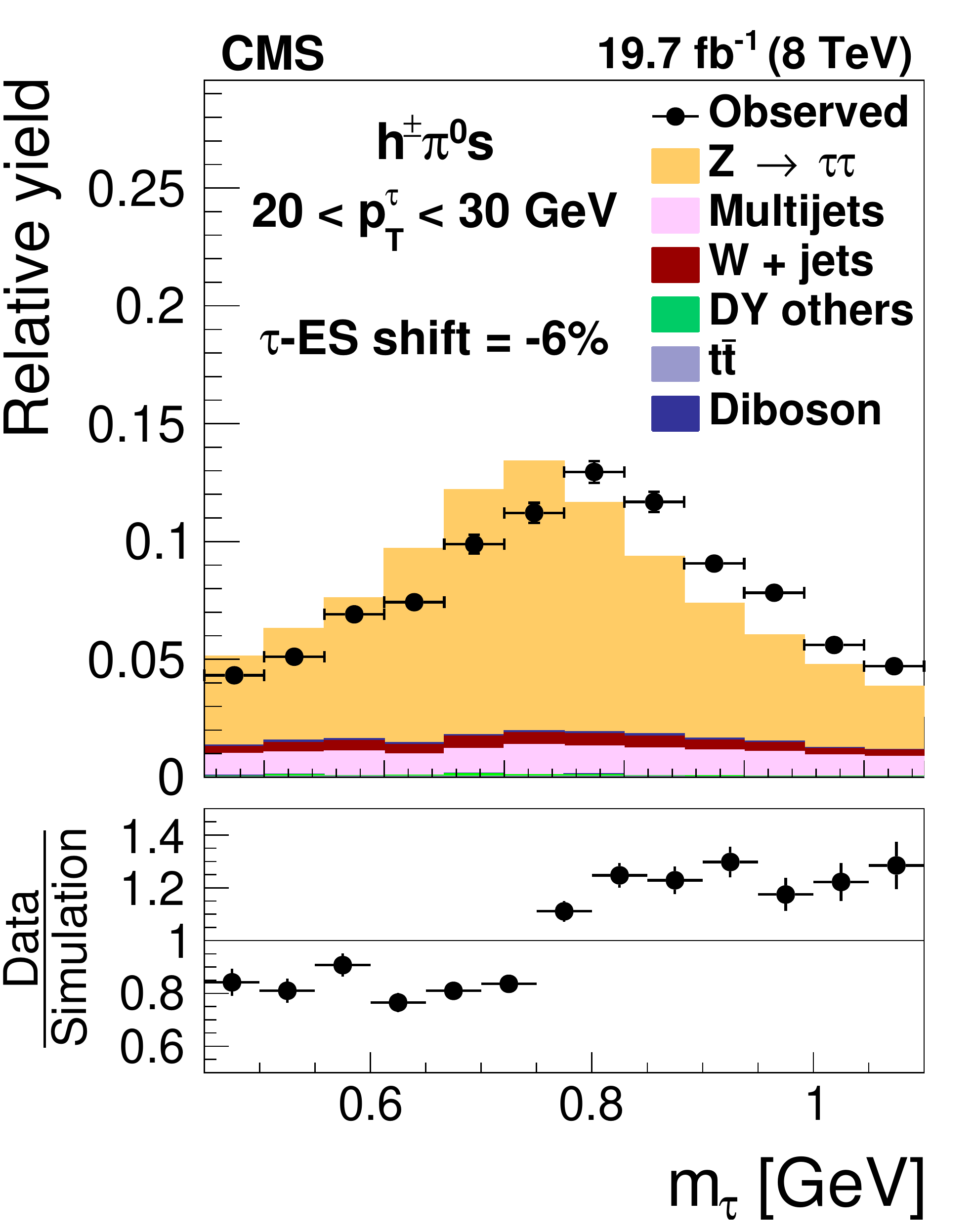}
\includegraphics[width=0.32\textwidth]{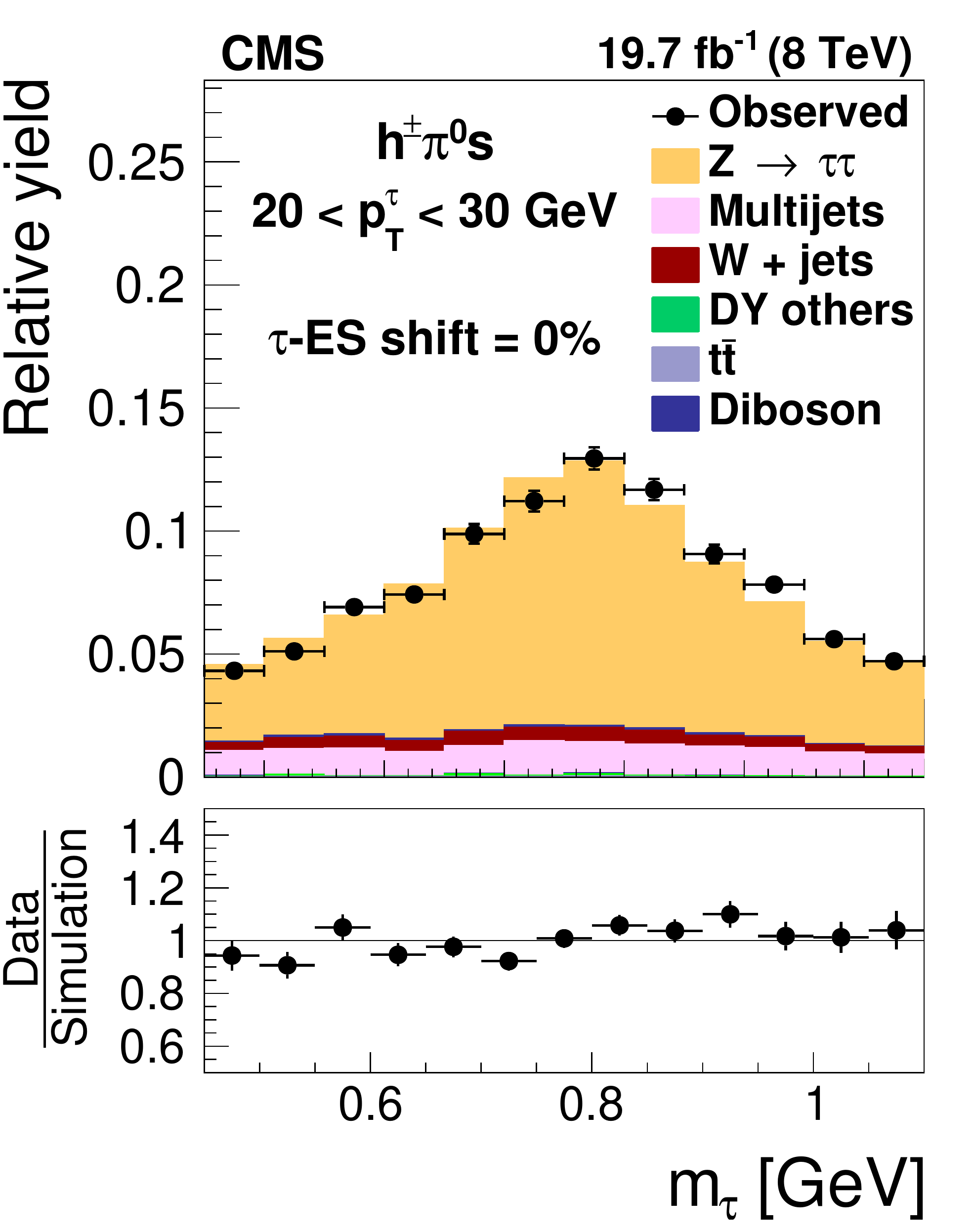}
\includegraphics[width=0.32\textwidth]{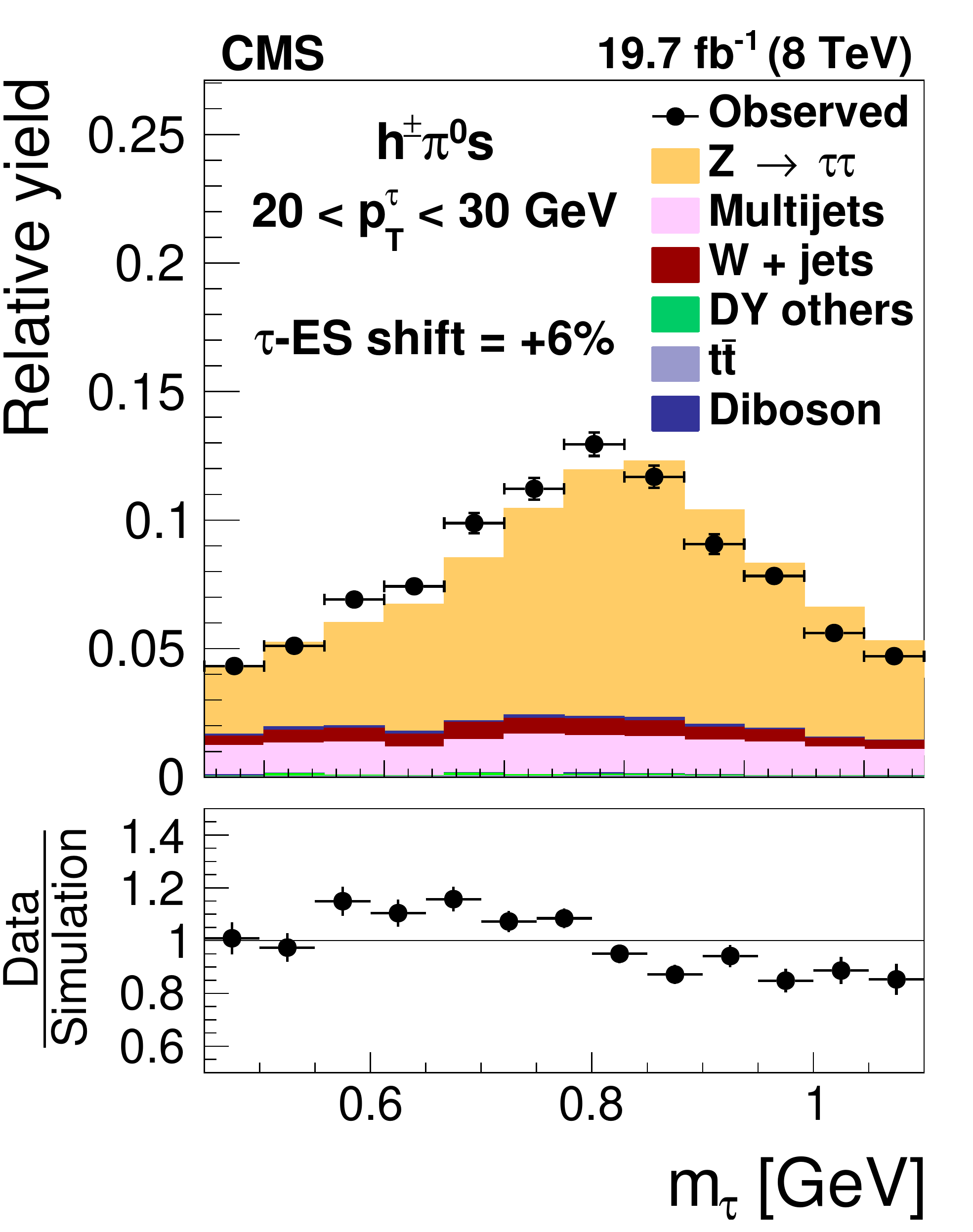}
\includegraphics[width=0.32\textwidth]{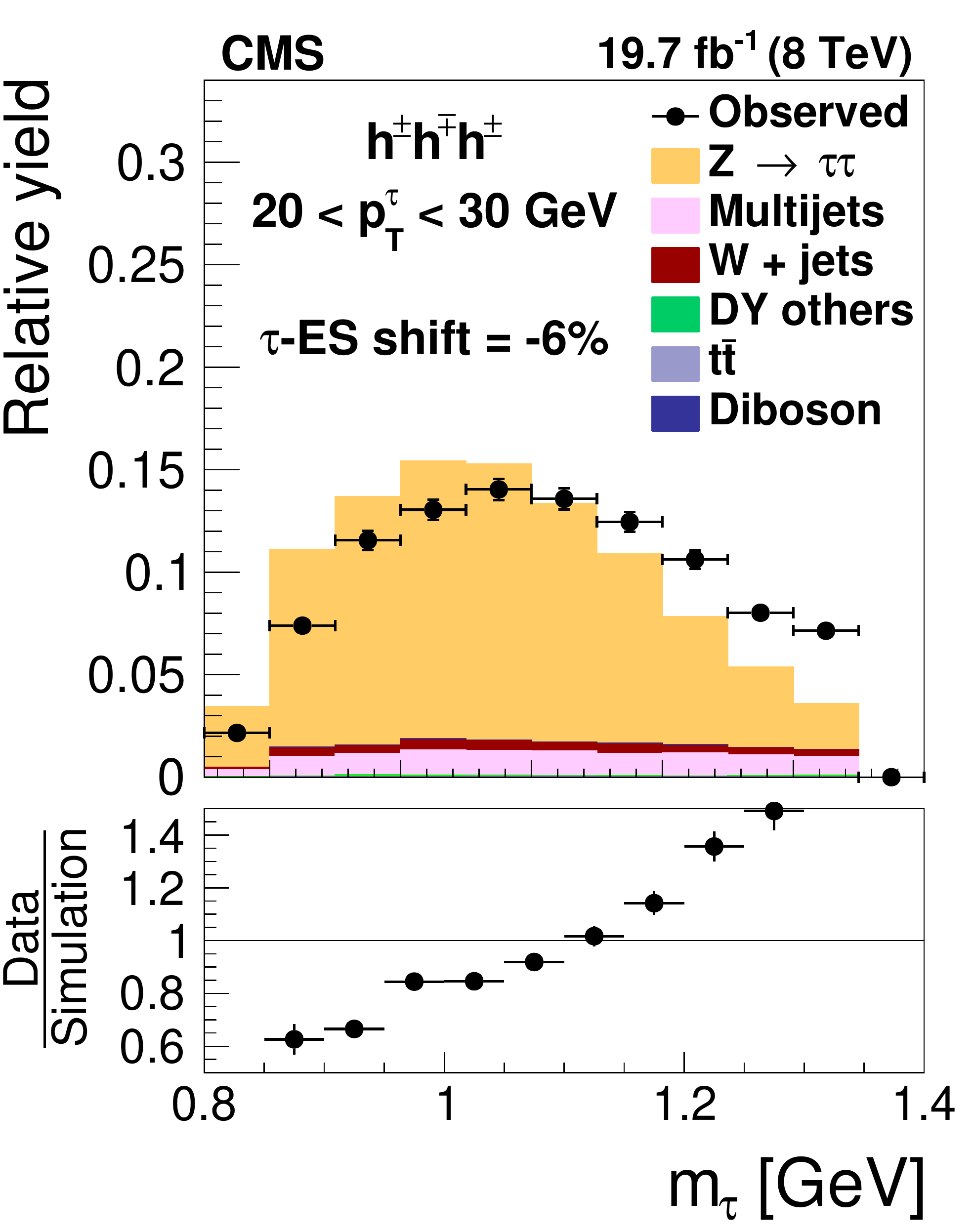}
\includegraphics[width=0.32\textwidth]{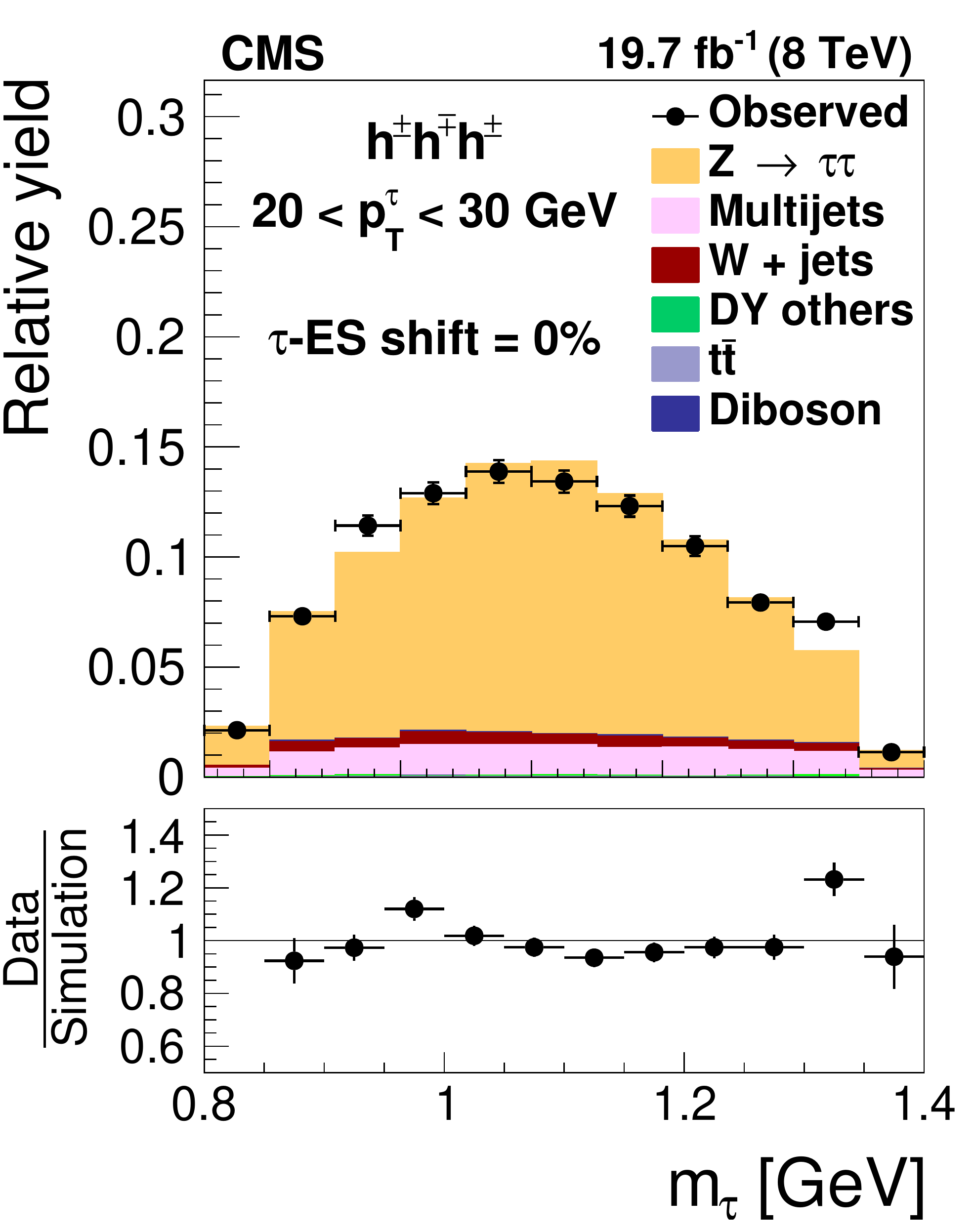}
\includegraphics[width=0.32\textwidth]{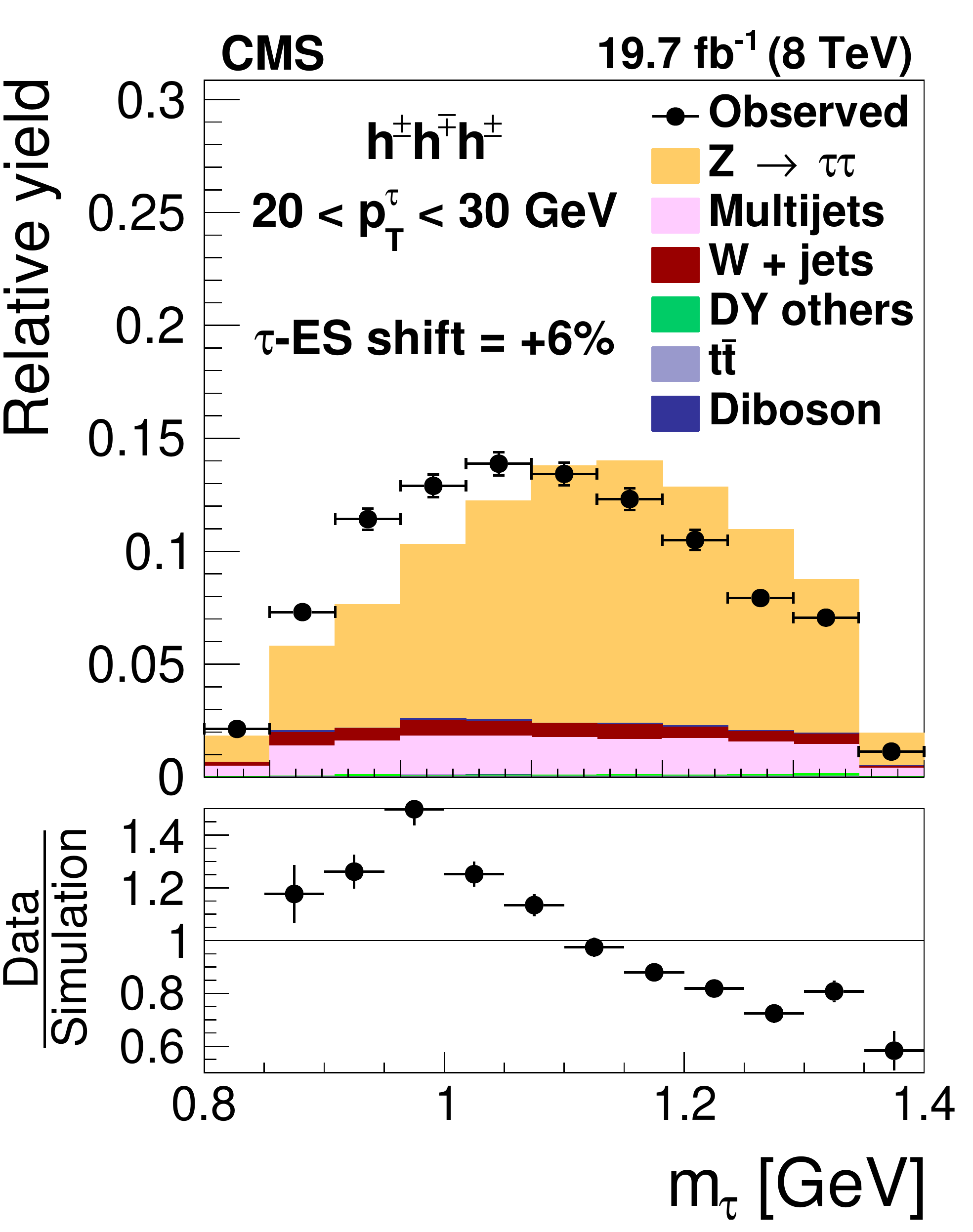}
\caption{
  Distribution in $m_{\tauh}$,
  observed in events containing $\tauh$ candidates of $20 < \pt < 30$\GeV,
  reconstructed in the decay modes $\oneProngPizeros$  (top) and $\threeProngZeroPizero$ (bottom),
  compared to the sum of $\cPZ/\Pggx \to \Pgt\Pgt$ signal plus background expectation.
  The $m_{\tauh}$ shape templates for the $\cPZ/\Pggx \to \Pgt\Pgt$ signal are shown for $\Pgt$ES variations of $-6\%$ (left), 0\% (centre) and $+6\%$ (right).
  For clarity, the symbols $\pt^{\Pgt}$ and $m_{\Pgt}$ are used instead of $\pt^{\tauh}$ and $m_{\tauh}$ in these plots.
}
\label{fig:ControPlots_mT}
\end{figure}

For illustration,
the $m_{\tauh}$ templates corresponding to $\Pgt$ES shifts of 0, $-6$, and $+6\%$
are shown for $\tauh$ candidates of $20 < \pt < 30$\GeV in Fig.~\ref{fig:ControPlots_mT}.
The data are compared to the sum of $\cPZ/\Pggx \to \Pgt\Pgt$ and expected background distributions.
A positive and negative slope in the data-to-MC ratio shown in the bottom parts of the figures
indicates that the best-fit values of the $\Pgt$ES correction are, respectively,
larger and smaller than the shift shown in the figure.

\begin{figure}[htb]
\centering
\includegraphics[width=0.48\textwidth]{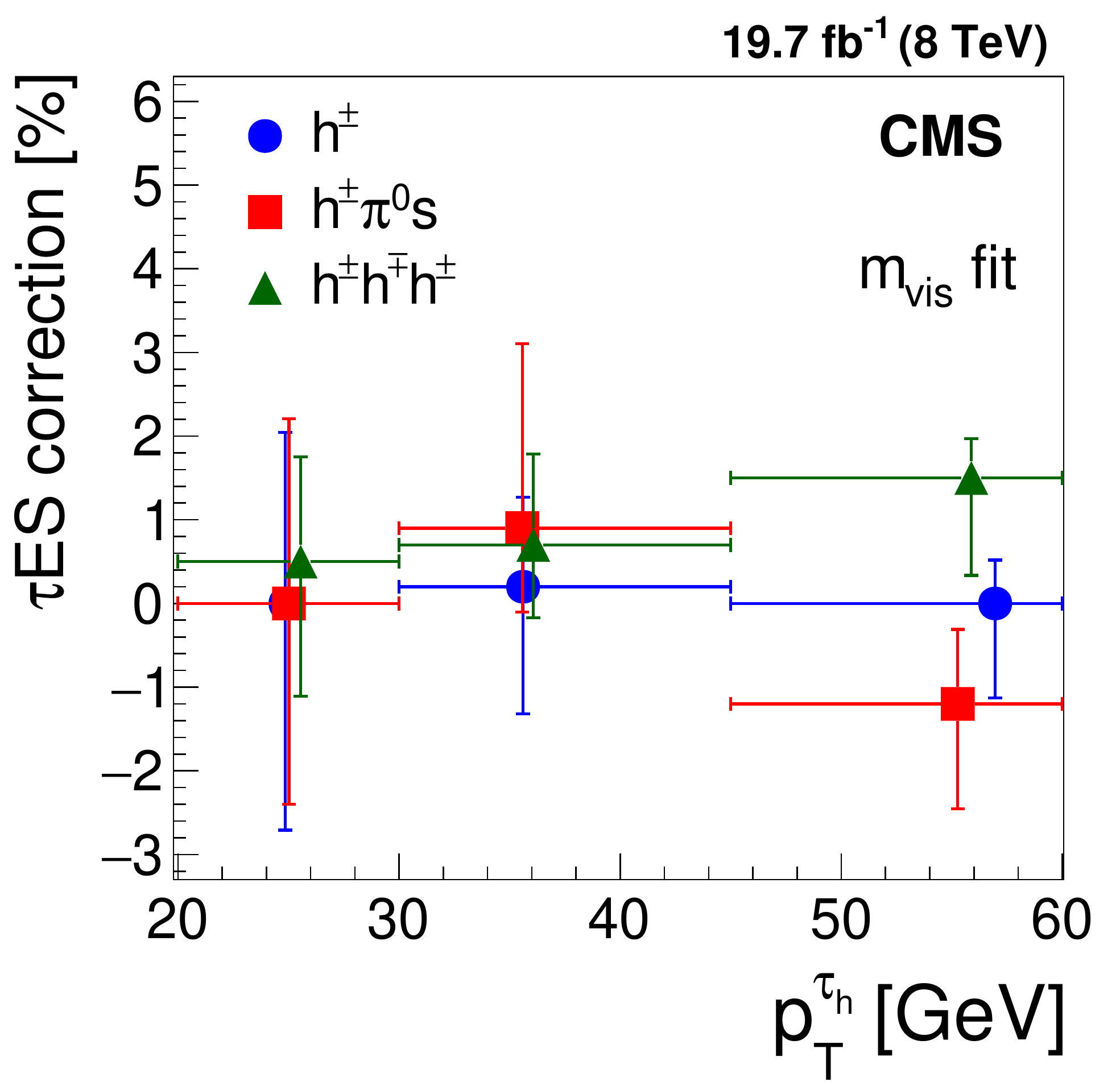}
\includegraphics[width=0.48\textwidth]{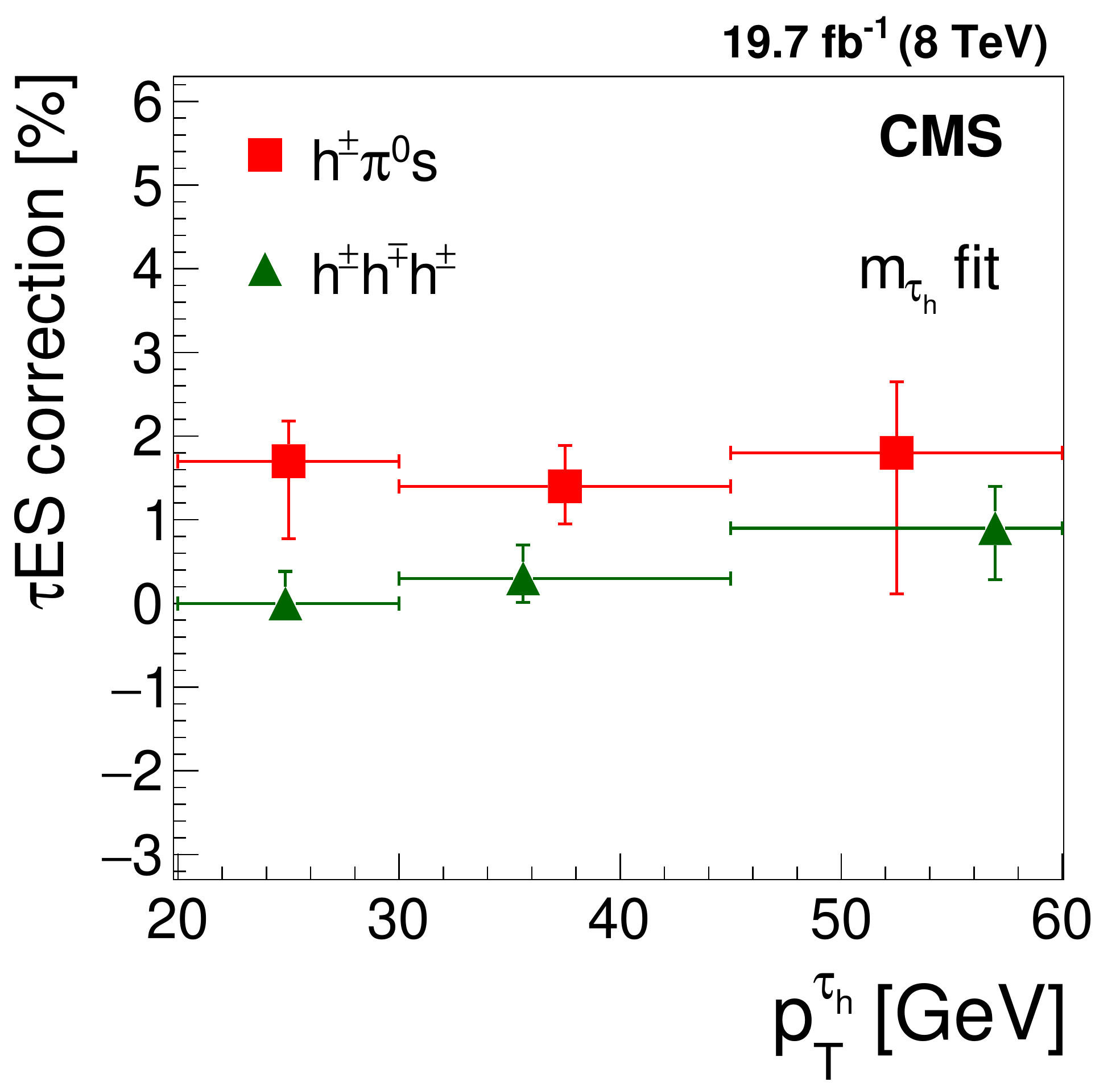}
\caption{
  Energy scale corrections for $\tauh$ measured in $\cPZ/\Pggx \to \Pgt\Pgt$ events,
  using the distribution in (left) visible mass of muon and $\tauh$ and (right) of the $\tauh$ candidate mass,
  for $\tauh$ reconstructed in different decay modes and in different ranges of $\tauh$ candidate \pt.
}
\label{fig:TESSummary_TT}
\end{figure}

The best-fit values for the $\Pgt$ES correction are presented in Fig.~\ref{fig:TESSummary_TT}.
The variable $m_{\tauh}$ is seen to be the more sensitive observable compared to $\mVis$,
as indicated by smaller uncertainties.

Numerical values of the measured $\Pgt$ES corrections are given in Table~\ref{tab:TESResults}.
The $\Pgt$ES corrections obtained using the observables $\mVis$ and $m_{\tauh}$ agree within their uncertainties.
For $\tauh$ reconstructed in the decay modes $\oneProngZeroPizero$ and $\threeProngZeroPizero$,
the energy scale measured in data agrees with the simulation.
The energy of $\tauh$ candidates reconstructed in the decay mode $\oneProngPizeros$
is lower by about 1\% in data than in simulation.
We do not find any indication of a dependence of the measured $\Pgt$ES corrections on $\pt^{\tauh}$.

\begin{table}[!ht]
\centering
\topcaption{
  Energy scale corrections for $\tauh$ measured in $\cPZ/\Pggx \to \Pgt\Pgt$ events,
  using the distribution in $\mVis$ and $m_{\tauh}$,
  for $\tauh$ reconstructed in different decay modes and $\tauh$ \pt bins.
  The $\Pgt$ES corrections measured for the combination of all $\tauh$ decay modes and \pt bins
  are also given in the table.
  It is obtained by means of an independent fit and hence may be
  different from the average of $\Pgt$ES corrections
  measured for individual decay modes.
}
\label{tab:TESResults}
\begin{tabular}{lcccc}
\hline
\multicolumn{5}{c}{$\Pgt$ES correction measured using $\mVis$\,[\%]}\\
\hline
Decay mode & $20 < \pt < 30$\GeV & $30 < \pt < 45$\GeV & $\pt > 45$\GeV & All \pt \\
\hline
$\oneProngZeroPizero$ & $0.0 \pm 2.3$ & $0.2 \pm 1.3$ & $0.0 \pm 0.8$ & $0.2 \pm 0.5$ \\
$\oneProngPizeros$ & $0.0 \pm 2.3$ & $0.9 \pm 1.6$ & $-1.2 \pm 1.0$ & $-0.3 \pm 0.6$ \\
$\threeProngZeroPizero$ & $0.5 \pm 1.4$ & $0.7 \pm 1.0$ & $1.5 \pm 0.8$ & $0.9 \pm 0.7$ \\
\hline
All decay modes       & $0.9 \pm 1.7$ & $0.7 \pm 0.8$ & $0.1 \pm 0.7$ & $0.5 \pm 0.5$ \\
\hline
\end{tabular}

\vspace{4mm}

\begin{tabular}{lcccc}
\hline
\multicolumn{5}{c}{$\Pgt$ES correction measured using $m_{\tauh}$\,[\%]}\\
\hline
Decay mode & $20 < \pt < 30$\GeV & $30 < \pt < 45$\GeV & $\pt > 45$\GeV & All \pt \\
\hline
$\oneProngZeroPizero$ & \NA & \NA & \NA & \NA \\
$\oneProngPizeros$ & $1.7 \pm 0.6$ & $1.4 \pm 0.4$ & $1.8 \pm 1.0$ & $1.5 \pm 0.4$\\
$\threeProngZeroPizero$ & $0.0 \pm 0.3$ & $0.3 \pm 0.3$ & $0.9 \pm 0.6$ & $0.3 \pm 0.2$\\
\hline
\end{tabular}

\end{table}

\section{Measurement of the misidentification rate for jets}
\label{sec:jetToTauFakeRate}

The rate for quark and gluon jets to be misidentified as $\tauh$ decays
is measured in $\PW$+jets and multijet events.
The events are selected as described in
Sections~\ref{sec:validation_eventSelection_Wjets}
and~\ref{sec:validation_eventSelection_QCD}, respectively.

The jet $\to \tauh$ misidentification rate is measured as a function of jet \pt and $\eta$,
and as a function of $\Nvtx$.
The rate is computed according to Eq.~(\ref{eq:jetToTauFakeRate}).
The jets considered in the denominator are required to pass a set of loose jet identification criteria~\cite{JME-10-003},
and to be compatible with originating from the primary collision vertex.

\begin{figure}[htbp]
\centering
\includegraphics[width=0.48\textwidth]{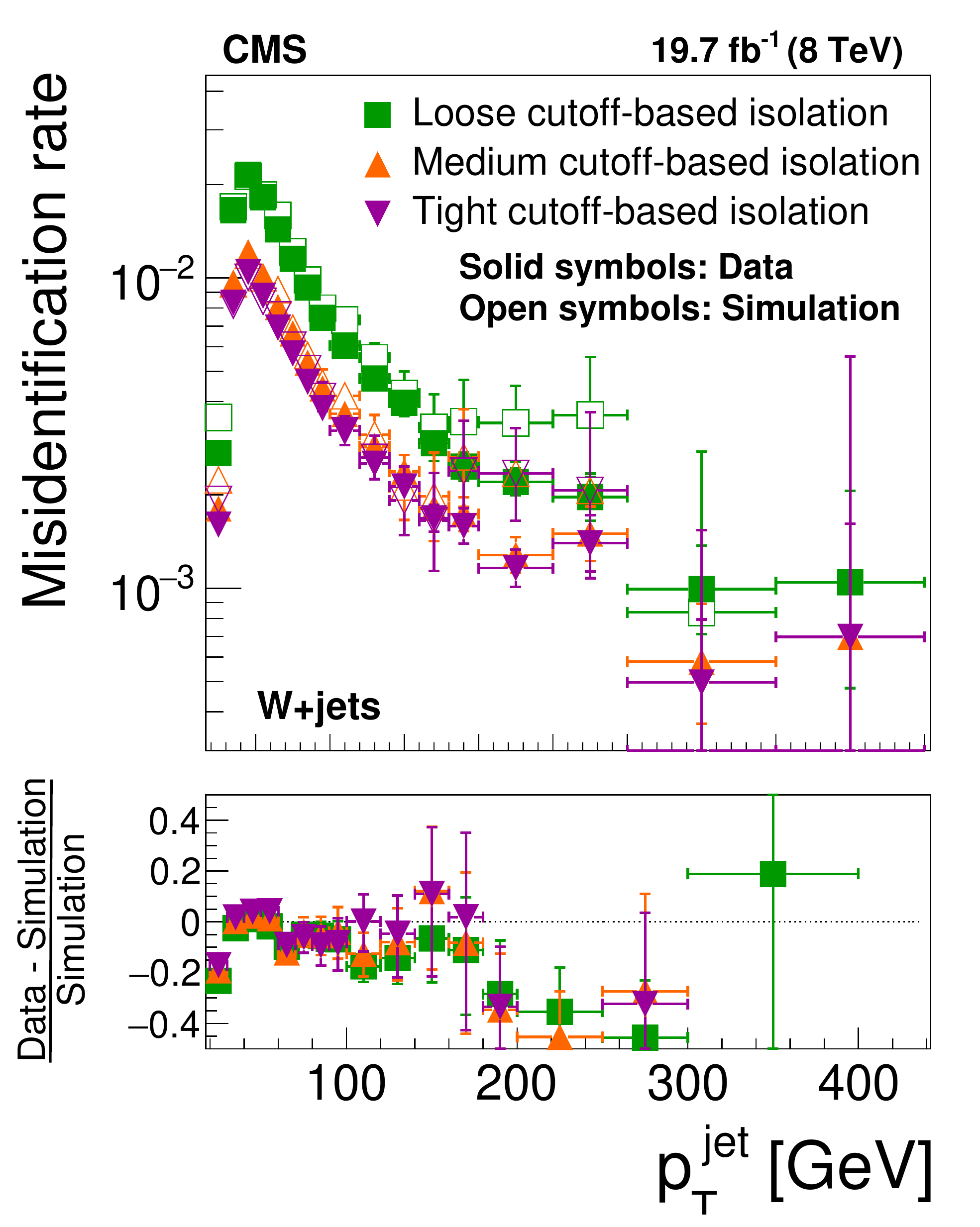}
\includegraphics[width=0.48\textwidth]{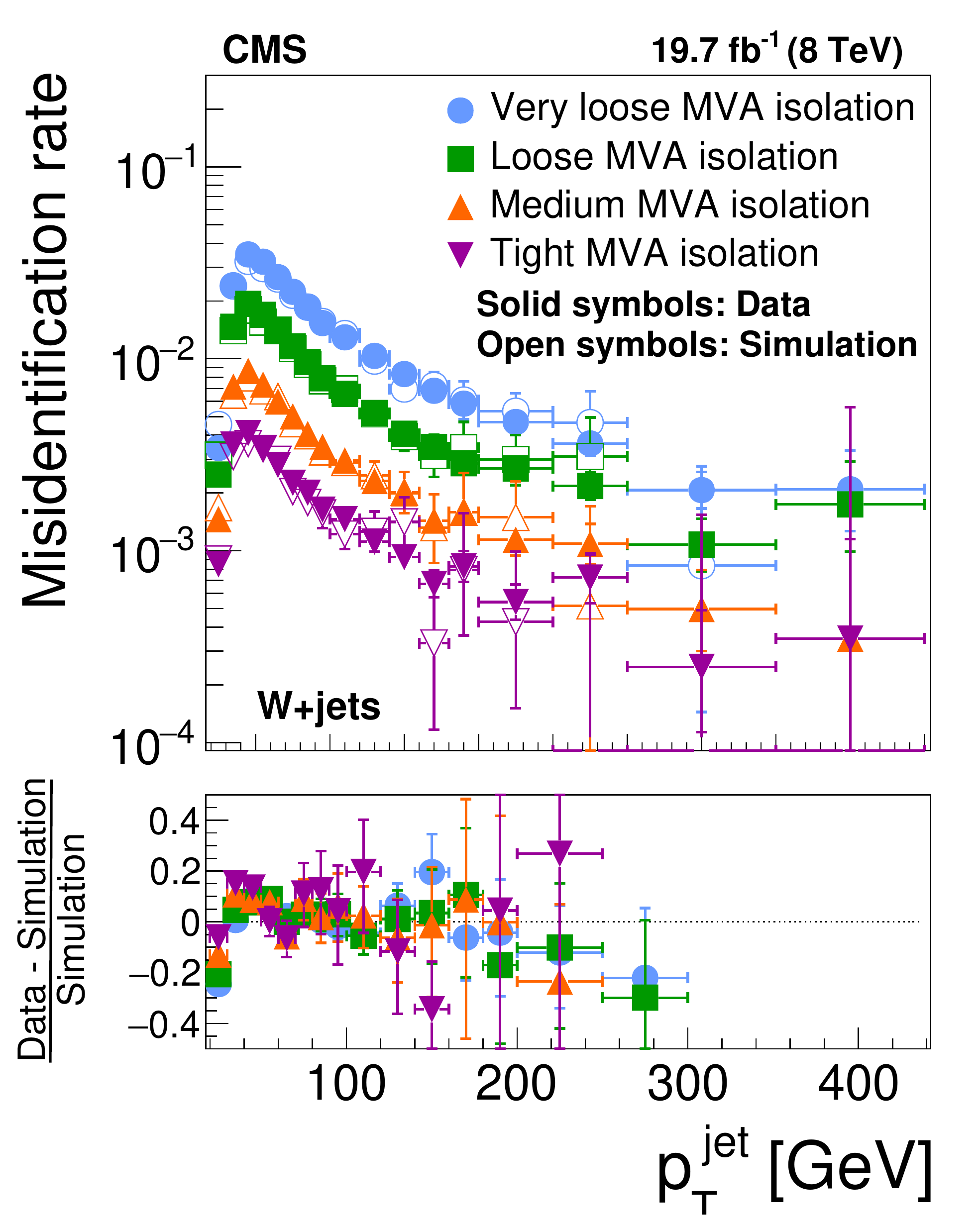}
\includegraphics[width=0.48\textwidth]{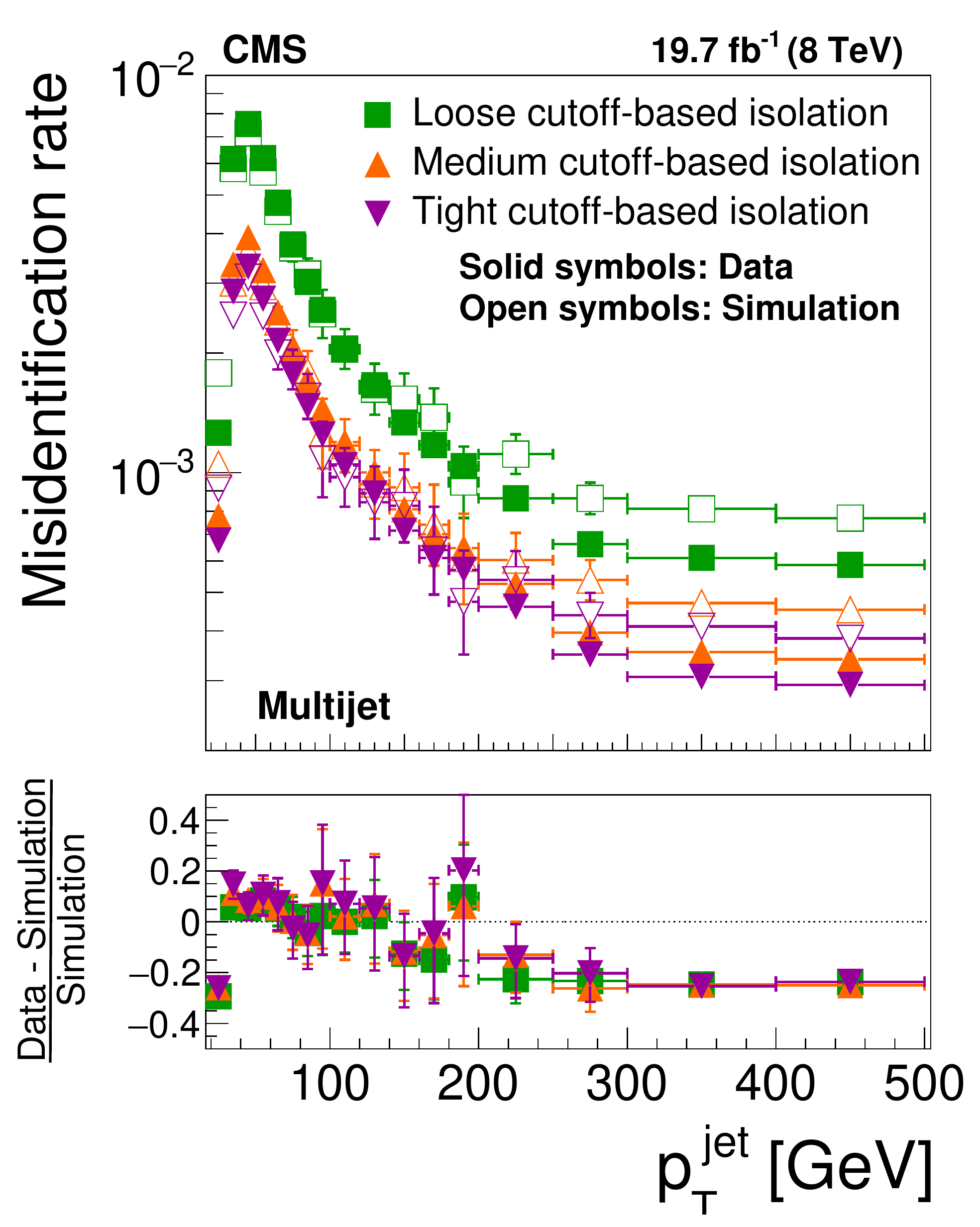}
\includegraphics[width=0.48\textwidth]{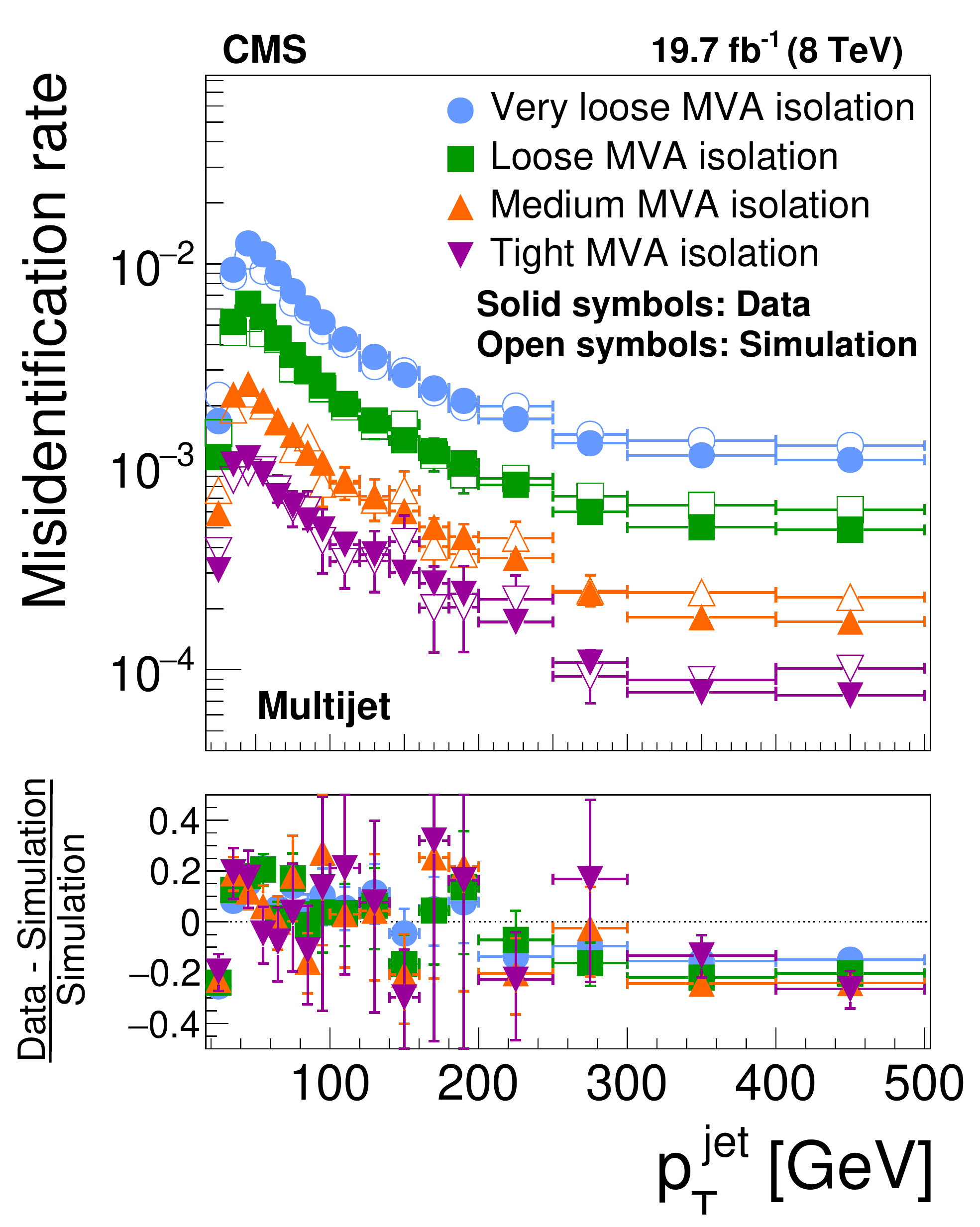}
\caption{
  Probabilities for quark and gluon jets in $\PW$+jets (top) and multijet (bottom) events
  to pass the cutoff-based (left) and MVA-based (right) $\tauh$ isolation discriminant,
  as a function of jet \pt.
  The misidentification rates measured in the data are compared to the
  MC expectation.
}
\label{fig:jetToTauFakeRate_Wjets_and_QCD_HPScombIso3Hit_and_MVAisoOldDMwLT_pt}
\end{figure}

\begin{figure}[htbp]
\centering
\includegraphics[width=0.48\textwidth]{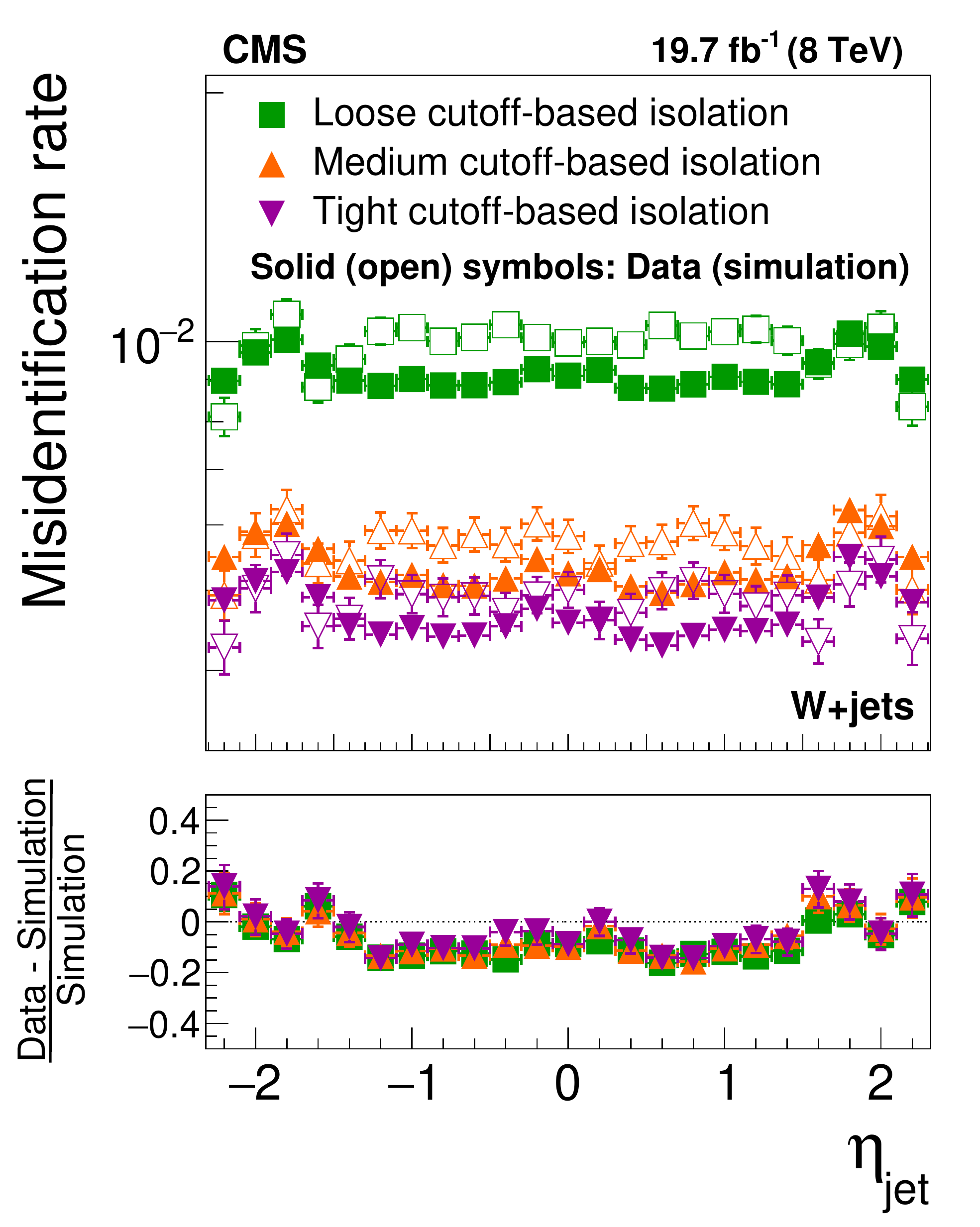}
\includegraphics[width=0.48\textwidth]{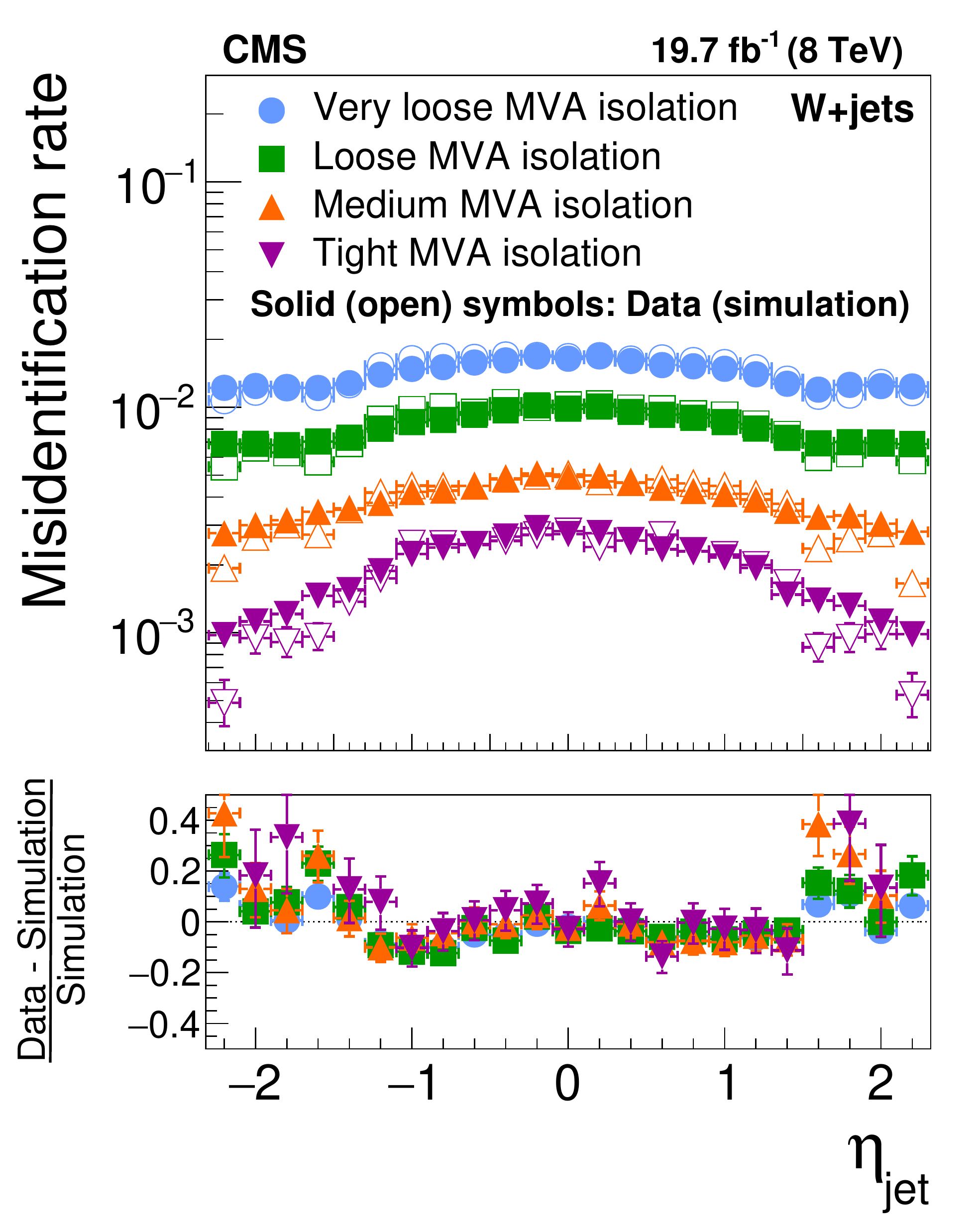}
\includegraphics[width=0.48\textwidth]{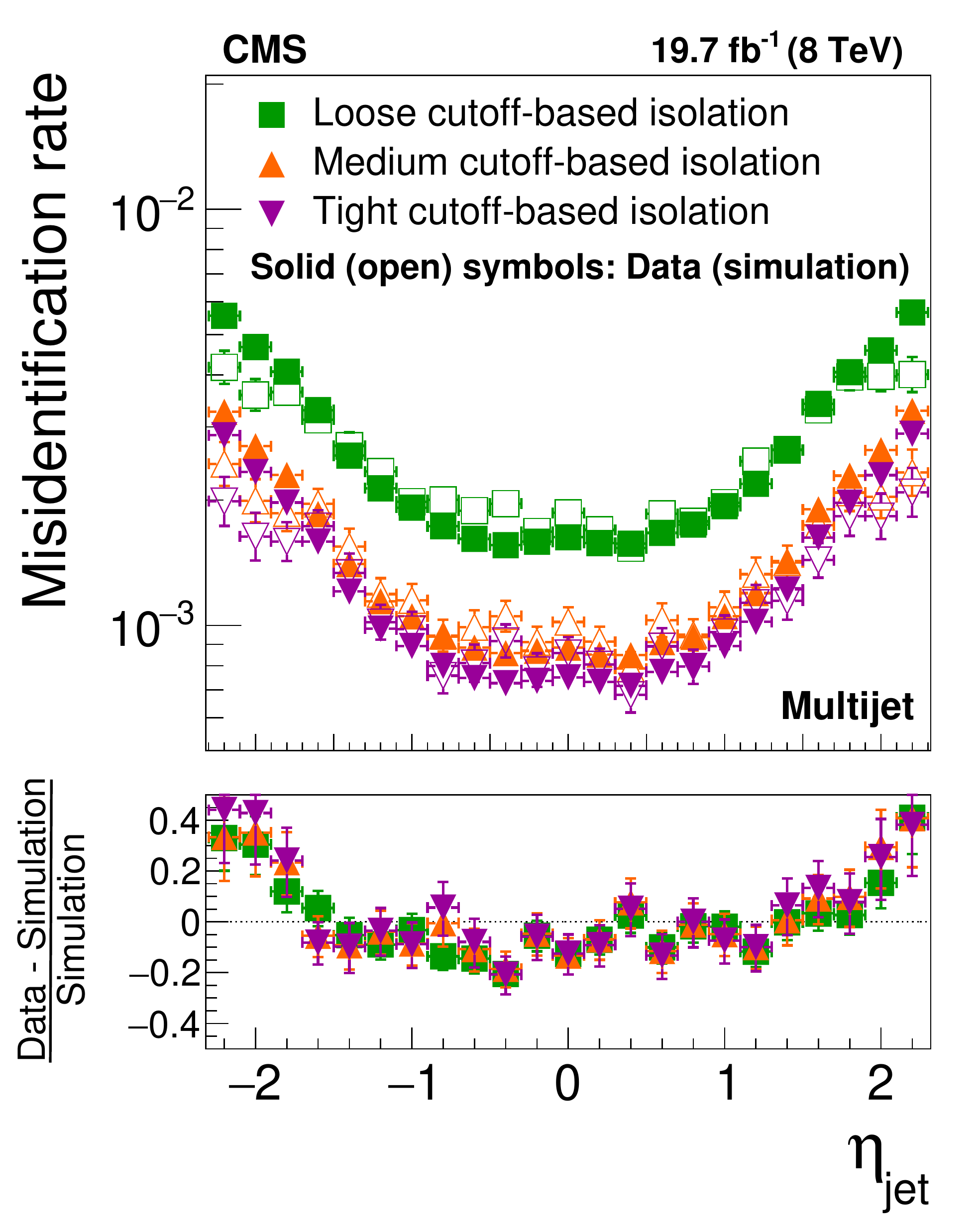}
\includegraphics[width=0.48\textwidth]{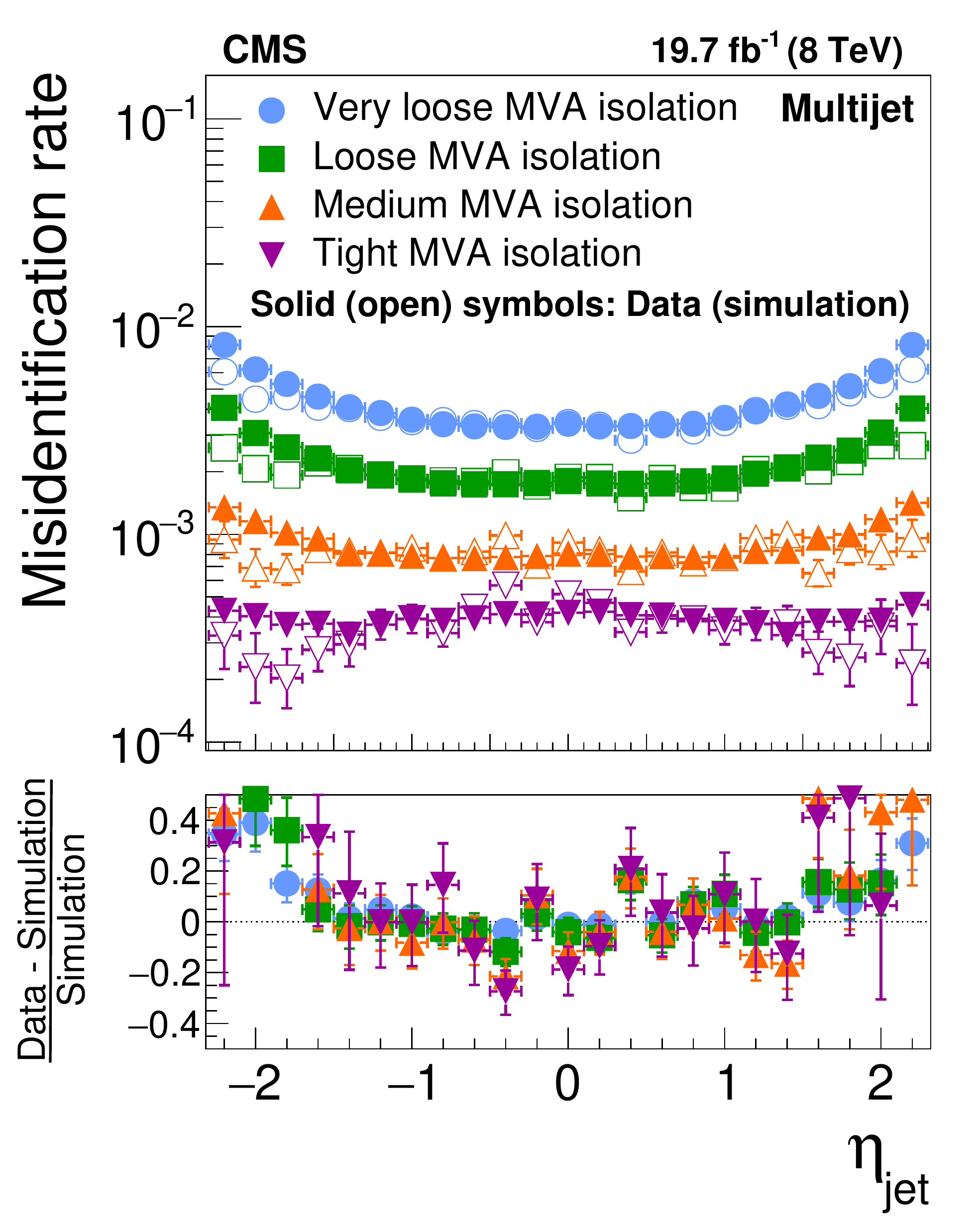}
\caption{
  Probabilities for quark and gluon jets in $\PW$+jets (top) and multijet (bottom) events
  to pass the cutoff-based (left) and MVA-based (right) $\tauh$ isolation discriminant,
  as a function of jet $\eta$.
  The misidentification rates measured in the data are compared to the MC expectation.
}
\label{fig:jetToTauFakeRate_Wjets_and_QCD_HPScombIso3Hit_and_MVAisoOldDMwLT_eta}
\end{figure}

\begin{figure}[htbp]
\centering
\includegraphics[width=0.48\textwidth]{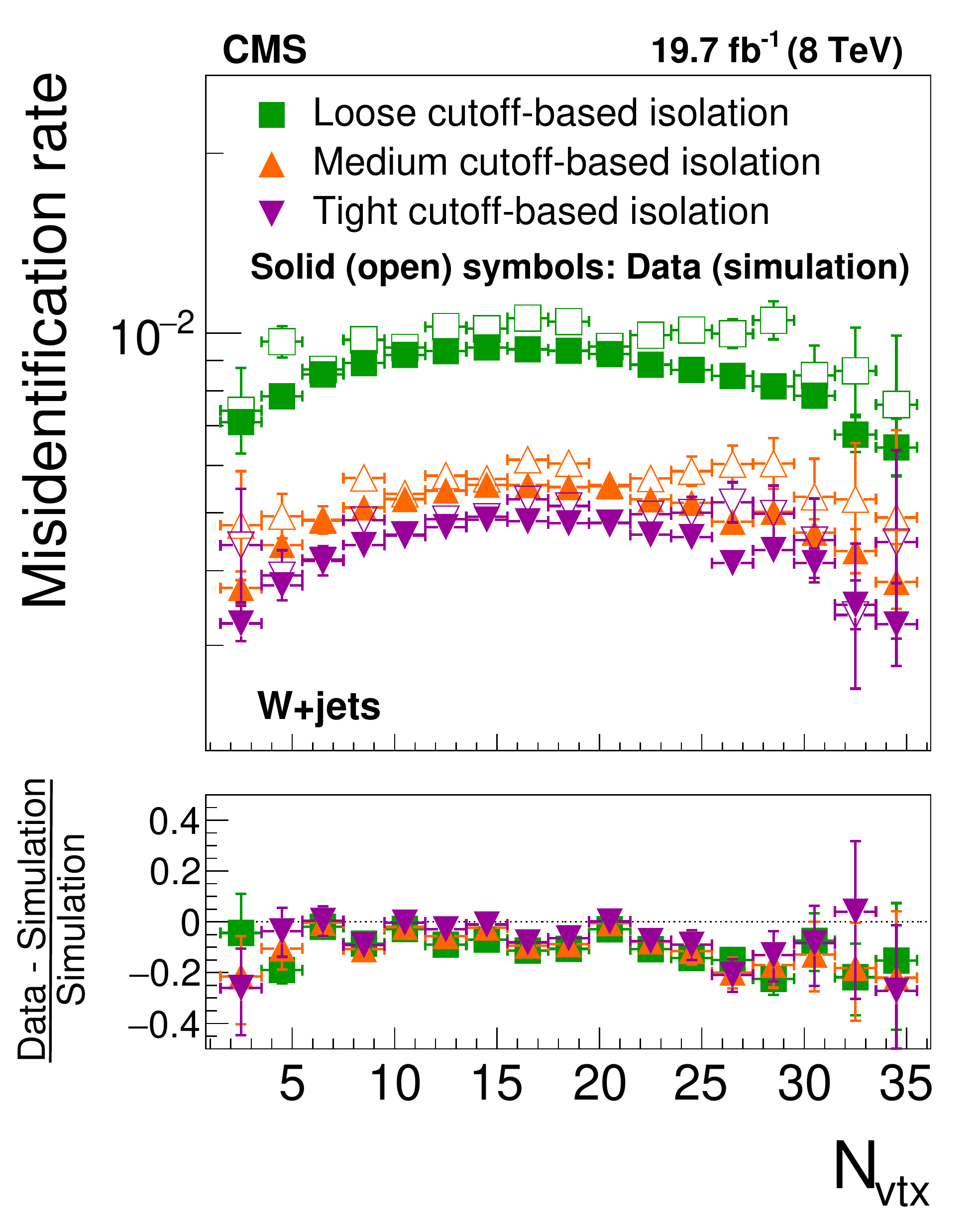}
\includegraphics[width=0.48\textwidth]{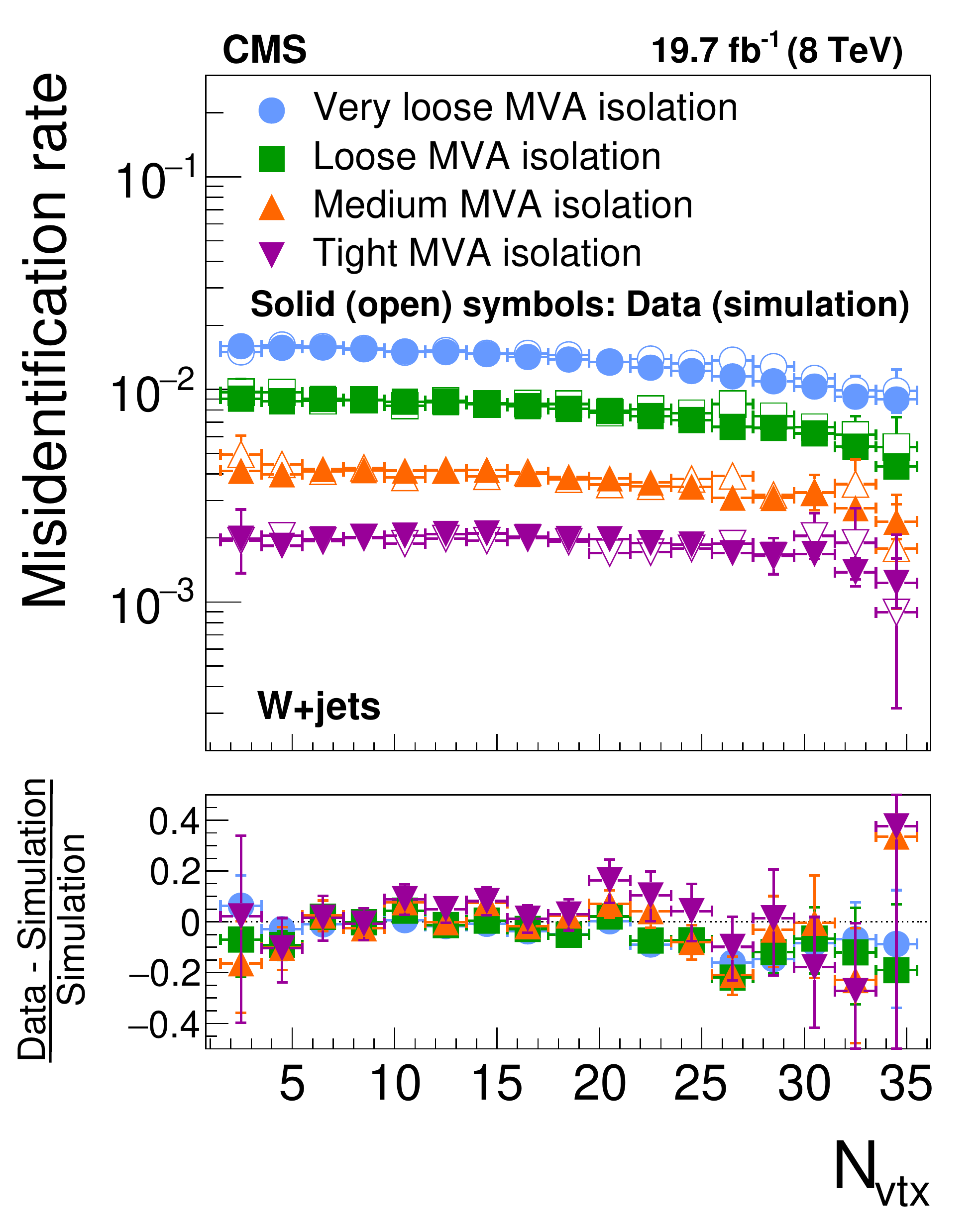}
\includegraphics[width=0.48\textwidth]{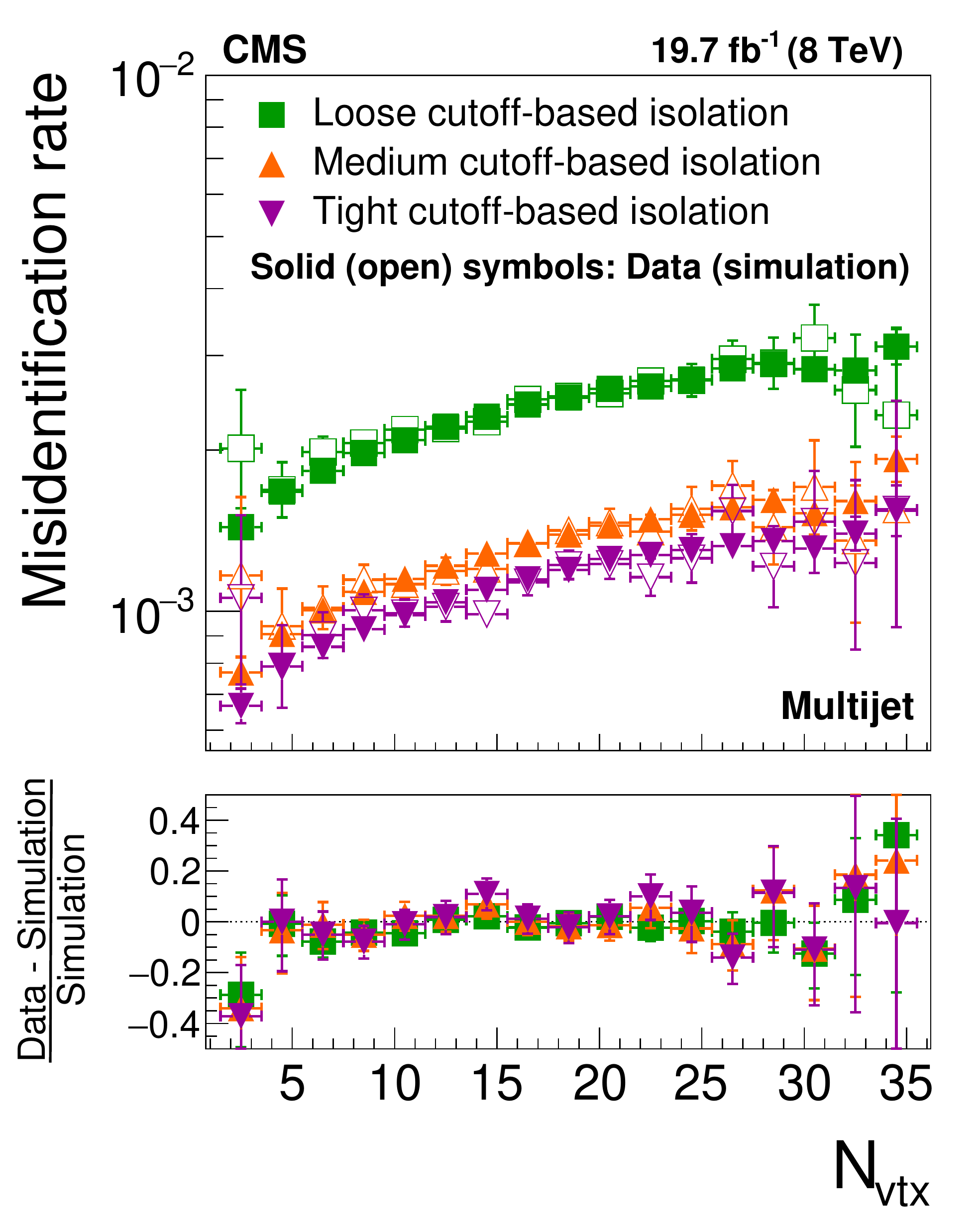}
\includegraphics[width=0.48\textwidth]{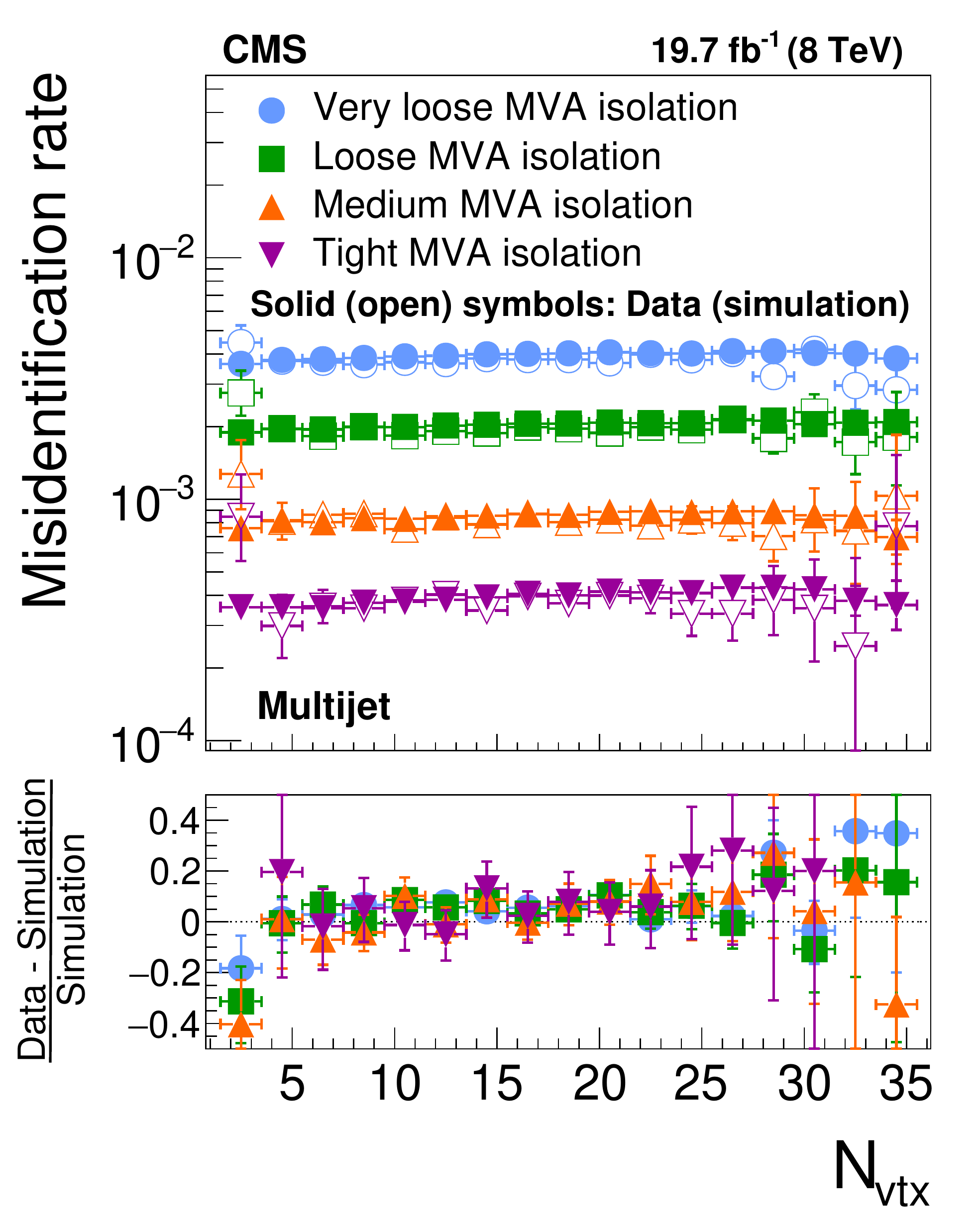}
\caption{
  Probabilities for quark and gluon jets in $\PW$+jets (top) and multijet (bottom) events
  to pass the cutoff-based (left) and MVA-based (right) $\tauh$ isolation discriminant,
  as a function of $\Nvtx$.
  The misidentification rates measured in the data are compared to the
  MC expectation.
}
\label{fig:jetToTauFakeRate_Wjets_and_QCD_HPScombIso3Hit_and_MVAisoOldDMwLT_Nvtx}
\end{figure}

The misidentification rates measured in $\PW$+jets and in multijet events
are shown in Figs.~\ref{fig:jetToTauFakeRate_Wjets_and_QCD_HPScombIso3Hit_and_MVAisoOldDMwLT_pt}--\ref{fig:jetToTauFakeRate_Wjets_and_QCD_HPScombIso3Hit_and_MVAisoOldDMwLT_Nvtx} and compared to MC expectation.
The contributions from background processes,
predominantly arising from $\cPqt\cPaqt$ and heavy-flavour jet production in the $\PW$+jets sample,
and from $\cPqt\cPaqt$ in the multijet sample, are included in the comparison.

In general, the misidentification rates are higher in $\PW$+jets than in multijet events.
The difference is due to the higher fraction of quark jets in $\PW$+jets events.
Quark jets typically have a lower particle multiplicity and are more collimated than gluon jets,
thereby increasing their probability to be misidentified as $\tauh$ decays.
The jet $\to \tauh$ misidentification rates for quark jets as well as gluon jets typically decrease as function of jet \pt,
as particle multiplicities increase for jets with larger \pt.

Moderate increases in the rate of jet $\to \tauh$ are observed at high pileup and at large $\abs{\eta}$.
The increase in the misidentification rate as a function of $\Nvtx$ is due to the $\Delta\beta$ correction described in Section~\ref{sec:tauIdDiscrCutBased}, which,
in events with high pileup, effectively relaxes the criteria on neutral-particle isolation,
as is necessary to maintain a high $\tauh$ identification efficiency.
The effect is reduced for the MVA-based $\tauh$ isolation discriminant.
The increase of the misidentification rate at high $\abs{\eta}$ results from a decrease in track reconstruction efficiency
near the edge of the geometric acceptance of the tracking detectors, which reduces the effectiveness of the isolation criteria.
The dependence of the misidentification rate on $\Nvtx$ and $\eta$ increases with jet \pt,
and is therefore more pronounced for multijet events compared to $\PW$+jets events.
Overall, the jet $\to \tauh$ misidentification rates vary between ${\approx}10^{-4}$ and ${\approx}4 \times 10^{-2}$.

Notable differences are observed in the
data/MC rate with which quark and gluon jets in $\PW$+jets and multijet events pass the
cutoff-based and MVA-based isolation discriminants at high $\abs{\eta}$.
Comparison with the rates for $\tauh$ identification discriminants based on charged-particle isolation
demonstrate that the difference is due to imprecise modelling of neutral particle isolation in the high $\abs{\eta}$ region in MC simulation.
The effect is caused by a restriction in detector simulation to a time window of $\pm$50\unit{ns} around the nominal bunch crossing,
while the ECAL electronics samples the signal amplitudes in 10 consecutive intervals of 25\unit{ns} within a time window of $-75$\unit{ns} to $+150$\unit{ns}
in order to correct, on an event-by-event basis, the energy reconstructed in the crystals for out-of-time pileup~\cite{Bruneliere:2006ra}.
The restriction in detector simulation to a time window of $\pm$50\unit{ns}
leads to a moderate mismodelling of the effect of out-of-time pileup on the isolation of $\tauh$ candidates reconstructed in the ECAL endcap
with respect to neutral particles.

A trend is observed in the ratio
of misidentification rates measured in multijet events relative to the MC simulation as a function of \pt.
While the rates for $\text{jet}\to \tauh$ measured in data exceed the MC expectation at low \pt,
the rates measured at high \pt fall short of the simulation.
The magnitude of the effect on the data/MC ratio is $\approx$20\%.
The trend is observed for the cutoff-based and for the MVA-based $\tauh$ identification discriminants,
and is of similar magnitude for jets in the central and forward regions.

\section{Measurement of misidentification rates for electrons and muons}
\label{sec:e_and_muToTauFakeRates}

The probability for electrons or muons to pass the $\tauh$ identification criteria,
and in particular to pass the dedicated discriminants against electrons or muons
described in Section~\ref{sec:discrAgainstElectronsAndMuons}
are measured through the tag-and-probe technique using $\cPZ/\Pggx \to \Pe\Pe$ or $\cPZ/\Pggx \to \Pgm\Pgm$ events~\cite{cmsTnP}.

The events are selected as described in Section~\ref{sec:validation_eventSelection_Zll}.
The probe is furthermore required to pass the loose WP of the cutoff-based $\tauh$ isolation discriminant.
Depending on whether the probe passes the lepton veto discriminator under study,
the event enters either the pass or the fail region.
When an event contains either two electron or two muon candidates that pass the tight selection criteria and qualify as tags,
both combinations of the tag and probe leptons are considered.

The $\Pe \to \tauh$ and $\Pgm \to \tauh$ misidentification rate, $P_{\text{misid}}$,
is measured using a simultaneous fit of the number of $\cPZ/\Pggx \to \Pe\Pe$ or $\cPZ/\Pggx \to \Pgm\Pgm$ events
in the pass and fail regions
($N^{\text{probe}}_{\text{pass}}$ and $N^{\text{probe}}_{\text{fail}}$).
The visible mass of the tag and probe pair is fitted using
templates for the $\cPZ/\Pggx \to \Pe\Pe$ or $\cPZ/\Pggx \to \Pgm\Pgm$ signal,
and $\cPZ/\Pggx \to \Pgt\Pgt$, $\PW$+jets, $\cPqt\cPaqt$, single top quark, diboson, and multijet backgrounds.
The templates for the $\cPZ/\Pggx \to \Pe\Pe$ and $\cPZ/\Pggx \to \Pgm\Pgm$ signal and for all background processes, except multijets, are obtained from simulation.
The normalization is performed according to the cross sections detailed in Section~\ref{sec:datasamples_and_MonteCarloSimulation},
except for the $\PW$+jets background,
the rate of which is determined from data, using the high-$\mT$ sideband method described in Section~\ref{sec:tauIdEfficiency_TnP_Ztautau}.
The distribution and normalization of the multijet background is determined from data,
using events in which tag and probe have the same charge.
Contributions from other backgrounds to the same-charge control region are subtracted, using the MC predictions.
The fit is performed as described in Section~\ref{sec:validation_templateFits},
taking the $\Pe \to \tauh$ or $\Pgm \to \tauh$ misidentification rate as the parameter of interest $\mu$.
The number of $\cPZ/\Pggx \to \Plepton\Plepton$ ($\Plepton = \Pe$, $\Pgm$) events in the pass and fail regions
are given, respectively, as a function of $\mu$ by
$N^{\text{probe}}_{\text{pass}} = \mu \, N^{\cPZ/\Pggx \to \Plepton\Plepton}$ and $N^{\text{probe}}_{\text{fail}} = (1 - \mu) \, N^{\cPZ/\Pggx \to \Plepton\Plepton}$.

Systematic uncertainties are represented by nuisance parameters in the template fits.
The uncertainties in the normalization of signal and background processes rescale the yield in the pass and fail region
by the same factor.
Uncertainties in the energy scale of the tag and probe leptons are represented by uncertainties in the fitted distributions.
The energy scales of tag electrons and muons are known with an uncertainty of 1\%.
Larger uncertainties of 5\% and 3\% are assigned to the energy scale of probe electrons and muons.

A correction is applied to account for the fact that not all probes in $\cPZ/\Pggx \to \Pe\Pe$ or $\cPZ/\Pggx \to \Pgm\Pgm$ events are electrons or muons.
In particular, in the pass region there is a few percent contamination from jet $\to \tauh$.
The contamination is corrected by subtracting from the number of $\cPZ/\Pggx \to \Pe\Pe$ or $\cPZ/\Pggx \to \Pgm\Pgm$ events in the fit
the expected number of jet $\to \tauh$ misidentifications, obtained from MC simulation.
A 20\% systematic uncertainty is assigned to the small number of jet $\to \tauh$ events subtracted,
motivated by the level of agreement of the jet $\to \tauh$ misidentification rates observed between data and simulation
presented in Section~\ref{sec:jetToTauFakeRate}.

\subsection{Misidentification rate for electrons}
\label{sec:eToTauFakeRate}

\begin{figure}[htbp]
\centering
\includegraphics[width=\figWidth]{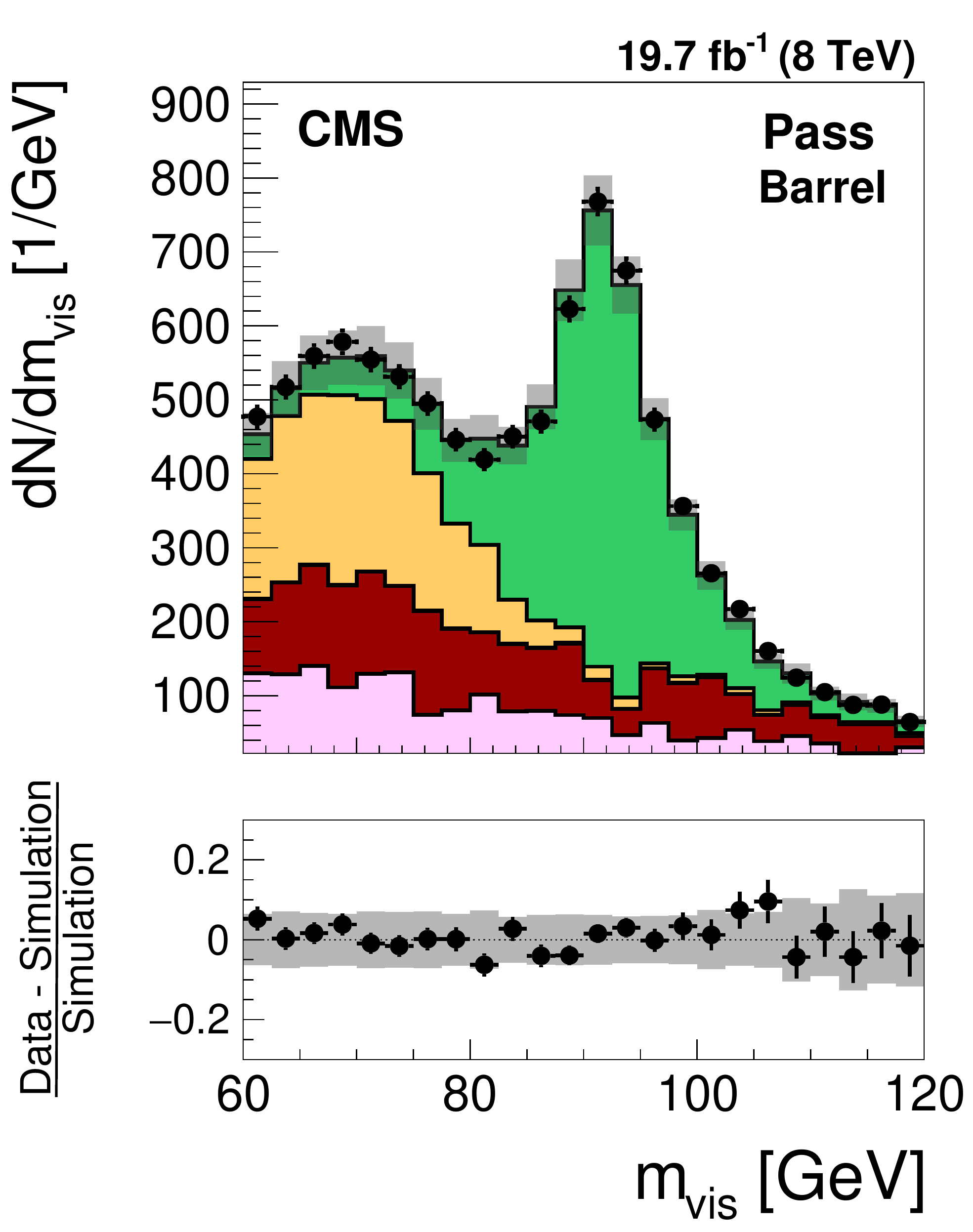}
\includegraphics[width=\figWidth]{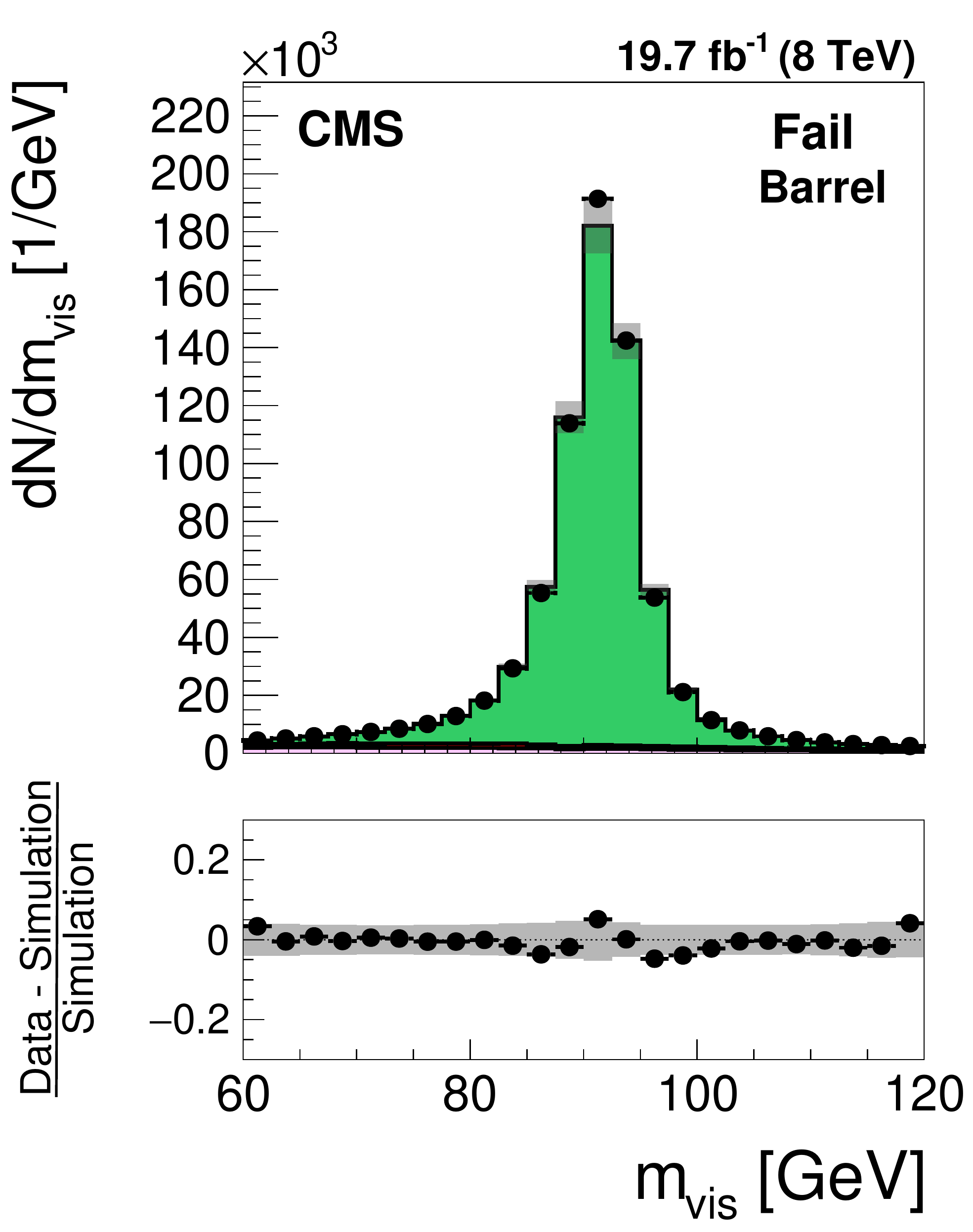}
\parbox[b][18em][t]{\figWidthScale}{\includegraphics[width=\figWidthScale]{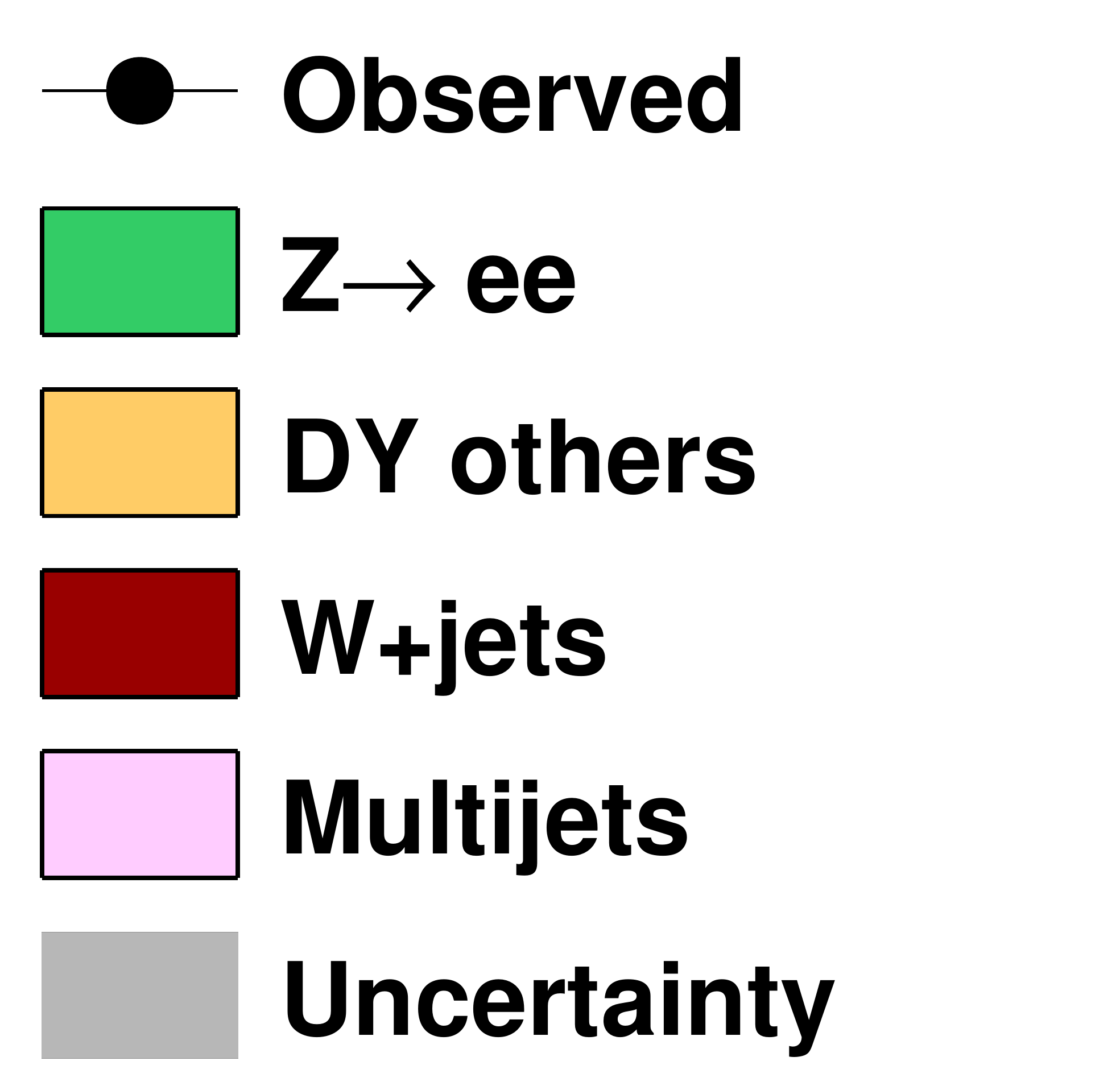}}
\includegraphics[width=\figWidth]{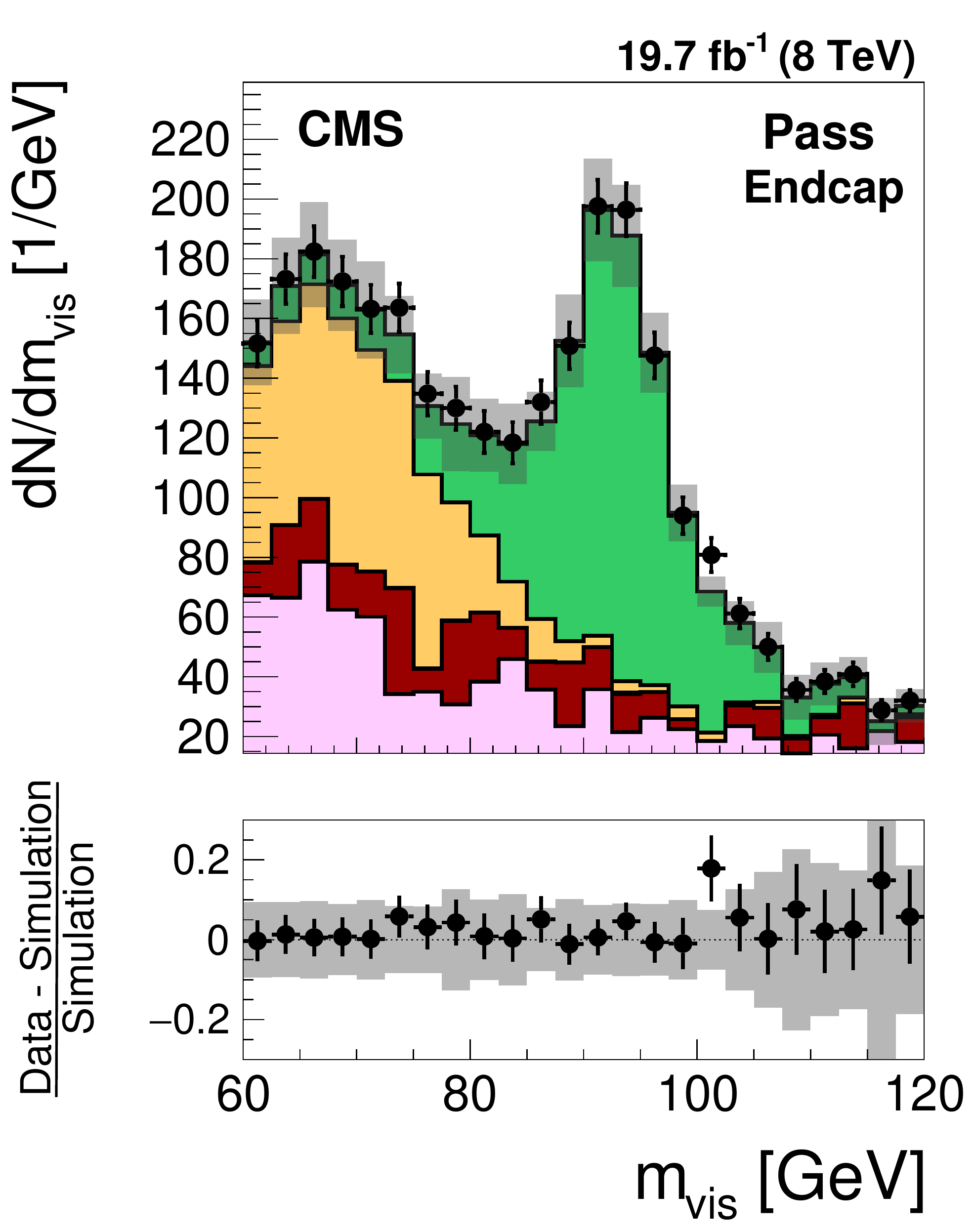}
\includegraphics[width=\figWidth]{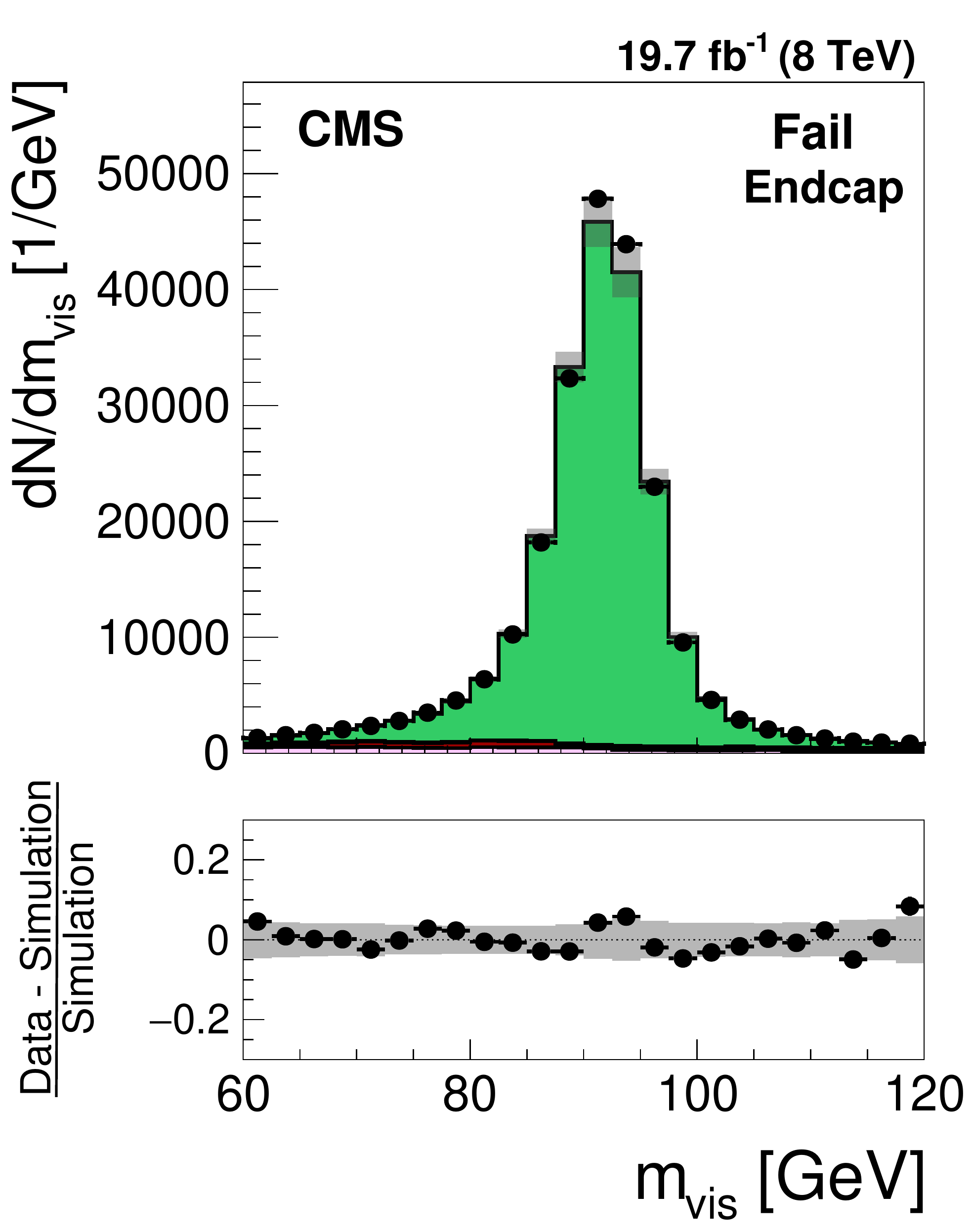}
\parbox[b][18em][t]{\figWidthScale}{\includegraphics[width=\figWidthScale]{plots/paper/eToTauFakeRate2/makeElecToTauFakeRateControlPlots_fromRaman_legend.pdf}}
\caption{
  Distribution in the visible mass of the tag and probe pair in the pass (left) and fail (right) regions,
  for the loose WP of the electron discriminant in the barrel (top) and endcap (bottom) regions.
  The distributions observed in $\cPZ/\Pggx \to \Pe\Pe$ candidate events selected in data
  are compared to the MC expectation, shown for the values of nuisance parameters obtained from the likelihood fit to the data,
  as described in Section~\ref{sec:validation_templateFits}.
  The contributions from $\cPZ/\Pggx \to \Pgt\Pgt$ background are denoted by ``DY others''.
  The $\cPqt\cPaqt$, single top quark, and diboson backgrounds yield a negligible contribution to the selected event sample and,
  while present in the fit, are omitted from the legend.
  The ``Uncertainty'' band represents the statistical and systematic uncertainties added in quadrature.
}
\label{fig:eToTauFakeRatePostfitPlots}
\end{figure}

In the measurement of the $\Pe \to \tauh$ misidentification rate,
the distribution in $\mVis$ is fitted within the range $60 < \mVis < 120$\GeV.
Separate fits are performed for probes in the barrel ($\abs{\eta} < 1.46$) and in the endcap ($\abs{\eta} > 1.56$) regions of ECAL.
Plots of the $\mVis$ distributions in the pass and fail regions
are presented for the loose WP of the electron discriminant in Fig.~\ref{fig:eToTauFakeRatePostfitPlots}.
In $\cPZ/\Pggx \to \Pgt\Pgt$ events that enter the pass region, 
the tag electrons are mainly due to $\Pgt^{-} \to \Pe^{-} \, \Pagne \, \Pnut$ decays,
while the probes are typically due to hadronic $\Pgt$ decays.

\begin{figure}[htb]
\centering
\includegraphics[width=0.6\textwidth]{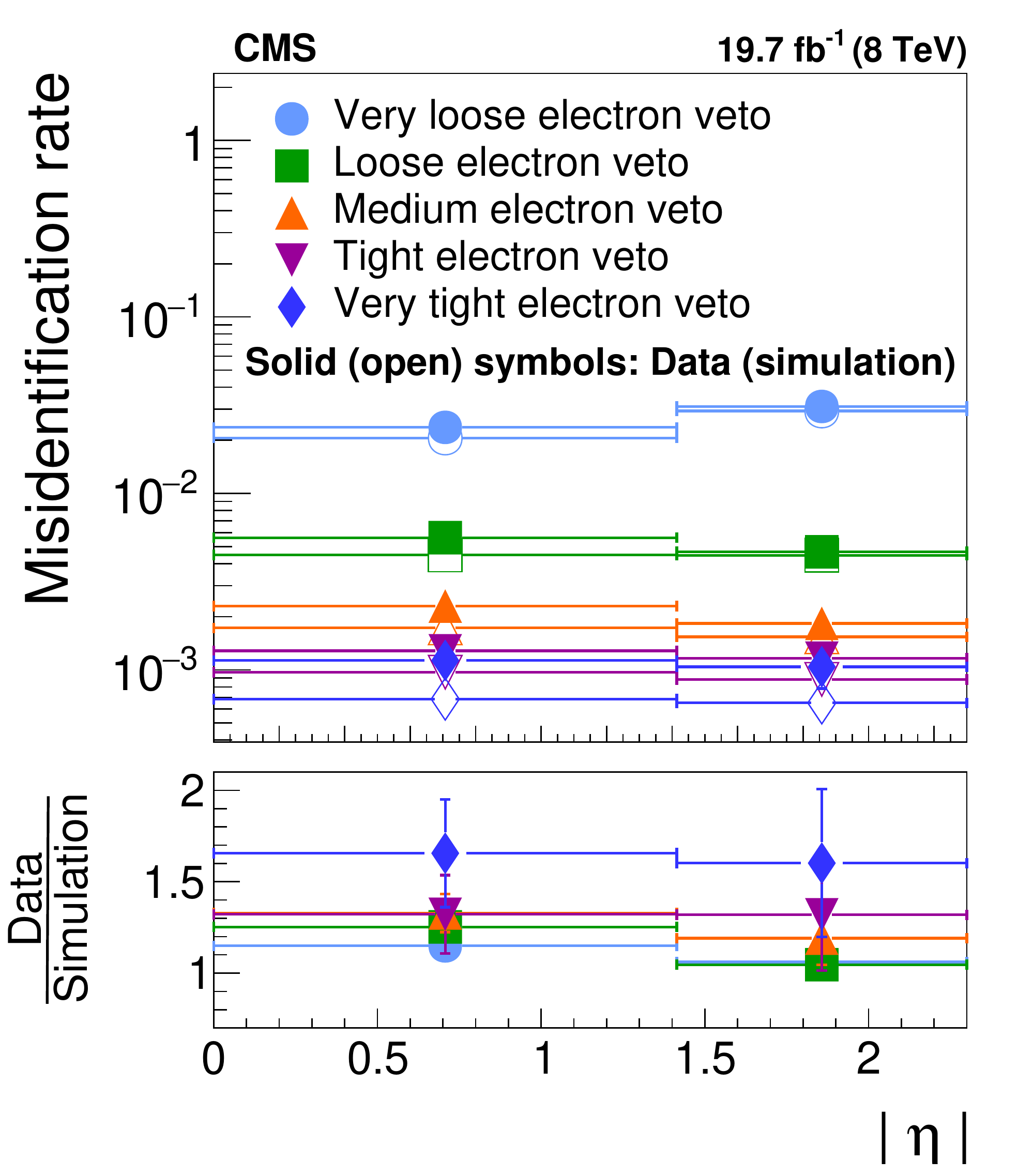}
\caption{
  Probability for electrons in $\cPZ/\Pggx \to \Pe\Pe$ events to pass different WP of the discriminant against electrons.
  The $\Pe \to \tauh$ misidentification rates measured in data are compared to the MC expectation,
  separately for electrons in the barrel ($\abs{\eta} < 1.46$) and in the endcap ($\abs{\eta} > 1.56$) regions of the electromagnetic calorimeter.
}
\label{fig:eToTauFakeRateResults}
\end{figure}

The $\Pe \to \tauh$ misidentification rates measured for different WP of the electron discriminant are given in Table~\ref{tab:eToTauFakeRateResults}.
The measured misidentification rates exceed the MC prediction by up to a factor of 1.7.
The difference between data and MC simulation,
quantified by the deviation in the ratio data/simulation from unity,
increases for tight and very tight WP.
Figure~\ref{fig:eToTauFakeRateResults} shows a graphical comparison of the misidentification rates measured in data to the MC expectation.
The measured data/simulation ratios are taken into account in physics analyses
by applying suitable MC-to-data correction factors.

\begin{table}[!ht]
\centering
\topcaption{
  Probability for electrons to pass different WP of the discriminant against electrons.
  The $\Pe \to \tauh$ misidentification rates measured in $\cPZ/\Pggx \to \Pe\Pe$ events are compared to the MC expectation,
  separately for electrons in the ECAL barrel and endcap regions.
}
\label{tab:eToTauFakeRateResults}
\begin{tabular}{lccc}
\hline
WP & Simulation & Data & Data/Simulation  \\
\hline
\multicolumn{4}{c}{ECAL barrel ($\abs{\eta} < 1.46$)} \\
\hline
Very loose & $(2.06 \pm 0.01) \times 10^{-2}$ & $(2.37 \pm 0.06) \times 10^{-2}$ & $1.15 \pm 0.03$ \\
Loose      & $(4.48 \pm 0.05) \times 10^{-3}$ & $(5.61 \pm 0.17) \times 10^{-3}$ & $1.25 \pm 0.04$ \\
Medium     & $(1.73 \pm 0.03) \times 10^{-3}$ & $(2.30 \pm 0.18) \times 10^{-3}$ & $1.33 \pm 0.10$ \\
Tight      & $(9.70 \pm 0.02) \times 10^{-4}$ & $(1.28 \pm 0.21) \times 10^{-3}$ & $1.32 \pm 0.21$ \\
Very tight & $(6.83 \pm 0.02) \times 10^{-4}$ & $(1.13 \pm 0.20) \times 10^{-3}$ & $1.66 \pm 0.30$ \\
\hline
\multicolumn{4}{c}{ECAL endcap ($\abs{\eta} > 1.56$)} \\
\hline
Very loose & $(2.93 \pm 0.02) \times 10^{-2}$ & $(3.11 \pm 0.09) \times 10^{-2}$ & $1.06 \pm 0.03$ \\
Loose      & $(4.46 \pm 0.09) \times 10^{-3}$ & $(4.67 \pm 0.22) \times 10^{-3}$ & $1.05 \pm 0.05$ \\
Medium     & $(1.54 \pm 0.05) \times 10^{-3}$ & $(1.83 \pm 0.22) \times 10^{-3}$ & $1.19 \pm 0.15$ \\
Tight      & $(8.83 \pm 0.38) \times 10^{-4}$ & $(1.16 \pm 0.26) \times 10^{-3}$ & $1.32 \pm 0.31$ \\
Very tight & $(6.50 \pm 0.33) \times 10^{-4}$ & $(1.04 \pm 0.26) \times 10^{-3}$ & $1.60 \pm 0.40$ \\
\hline
\end{tabular}
\end{table}

\subsection{Misidentification rate for muons}
\label{sec:muToTauFakeRate}

\begin{figure}[htbp]
\centering
\includegraphics[width=\figWidth]{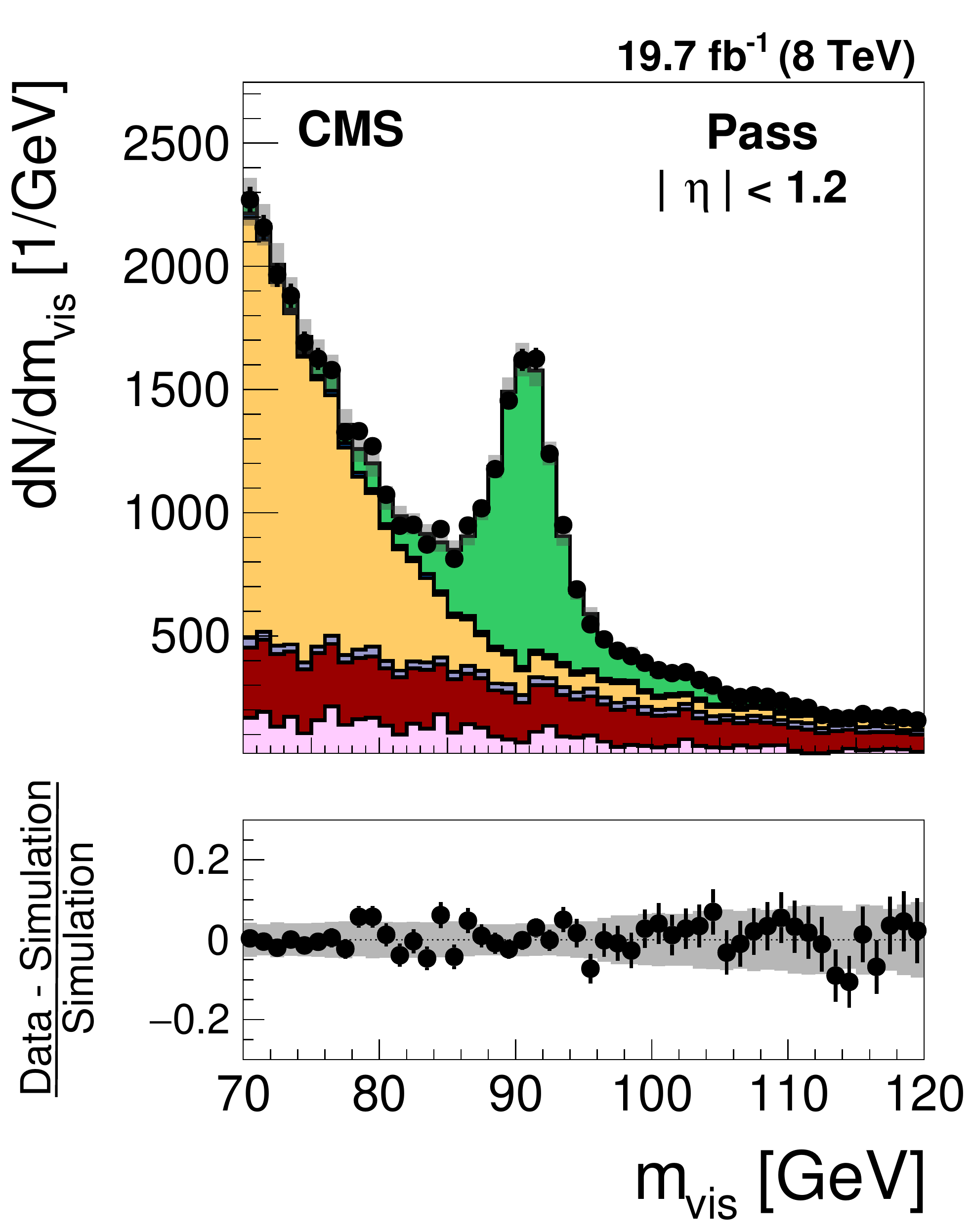}
\includegraphics[width=\figWidth]{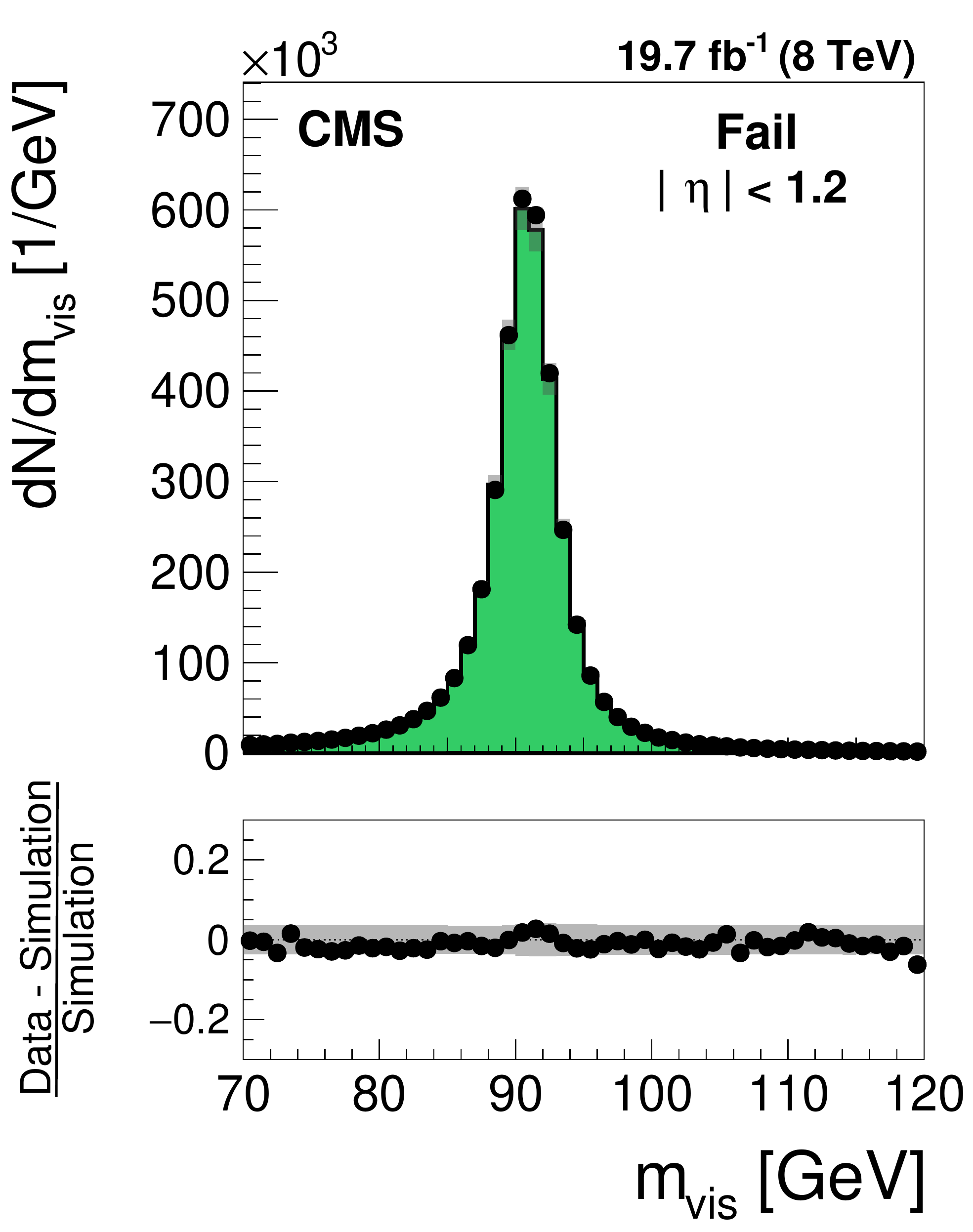}
\parbox[b][18em][t]{\figWidthScale}{\includegraphics[width=\figWidthScale]{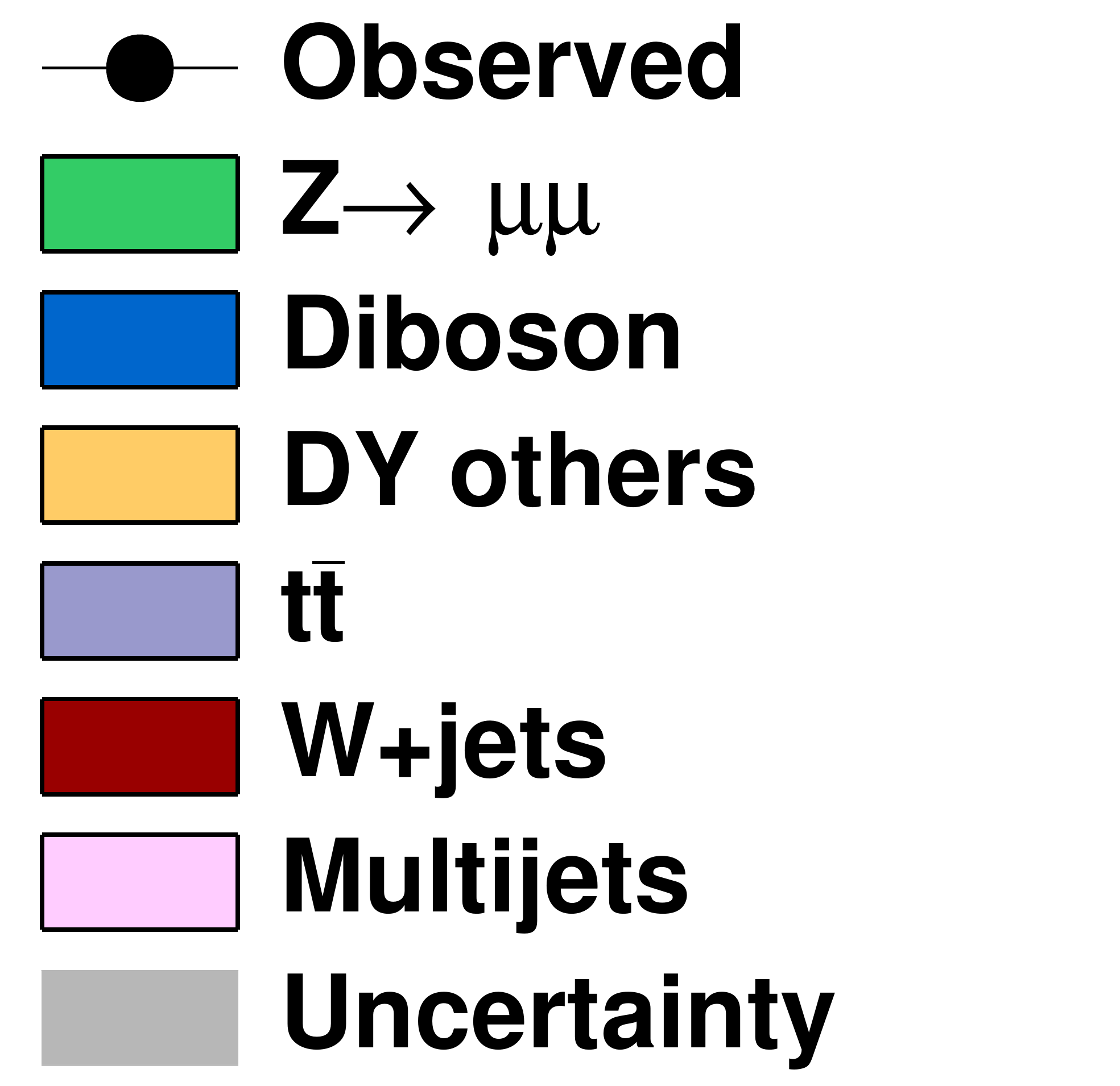}}
\includegraphics[width=\figWidth]{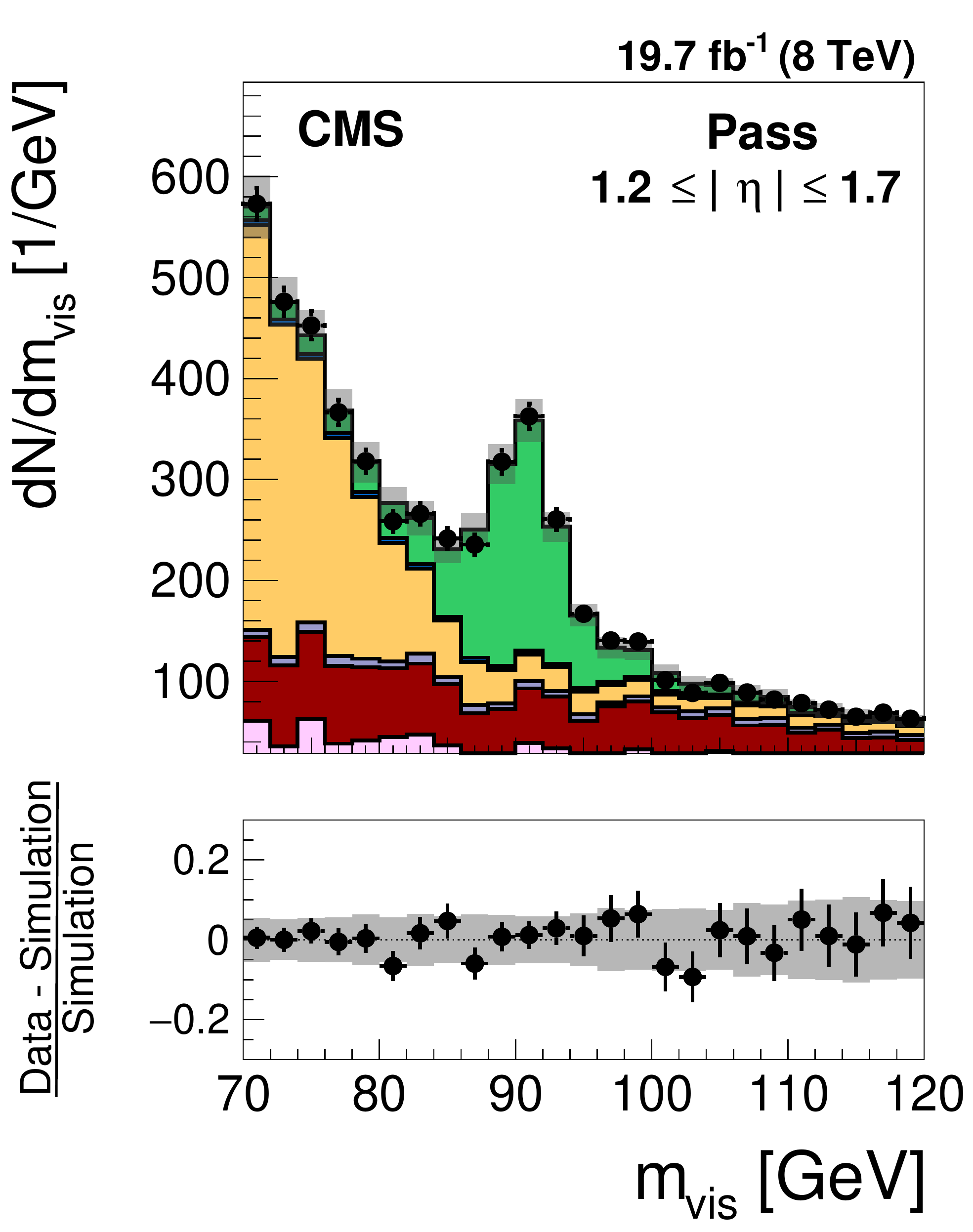}
\includegraphics[width=\figWidth]{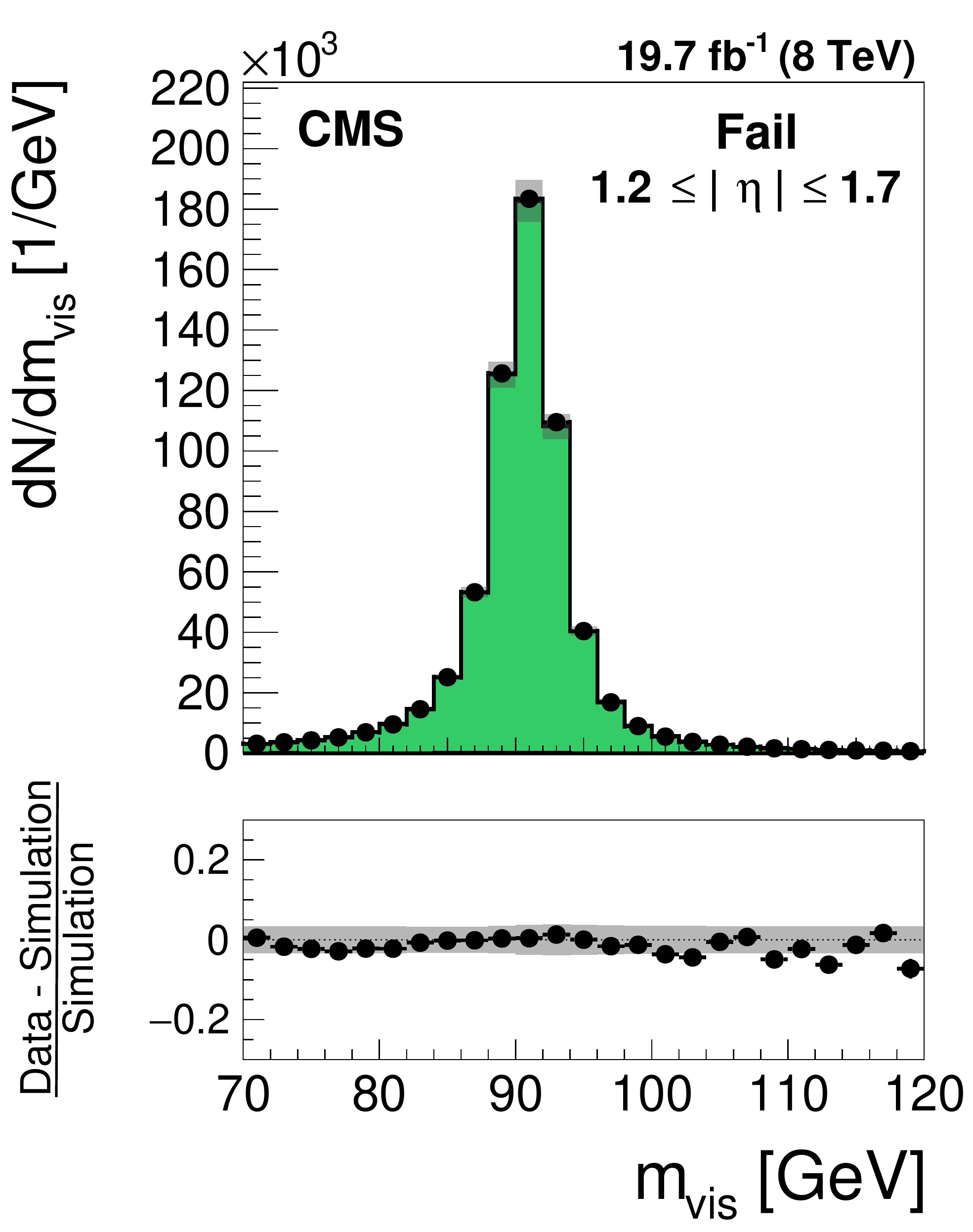}
\parbox[b][18em][t]{\figWidthScale}{\includegraphics[width=\figWidthScale]{plots/paper/muToTauFakeRate/makeMuToTauFakeRateControlPlots_fromCesare_legend.pdf}}
\caption{
  Distribution in the mass of the tag and probe pair in the pass (left) and fail (right) regions,
  for the loose WP of the cutoff-based muon discriminant in the region
  $\abs{\eta} < 1.2$ (top) and $1.2 \leq \abs{\eta} \leq 1.7$ (bottom).
  The distributions in $\cPZ/\Pggx \to \Pgm\Pgm$ candidate events selected in data
  are compared to the MC expectation, shown for the values of nuisance parameters obtained from the likelihood fit to the data,
  as described in Section~\ref{sec:validation_templateFits}.
  The contributions from $\cPZ/\Pggx \to \Pgt\Pgt$ background are denoted by ``DY others''.
  The ``Uncertainty'' band represents the statistical and systematic uncertainties added in quadrature.
}
\label{fig:muToTauFakeRatePostfitPlots1}
\end{figure}

\begin{figure}[htb]
\centering
\includegraphics[width=\figWidth]{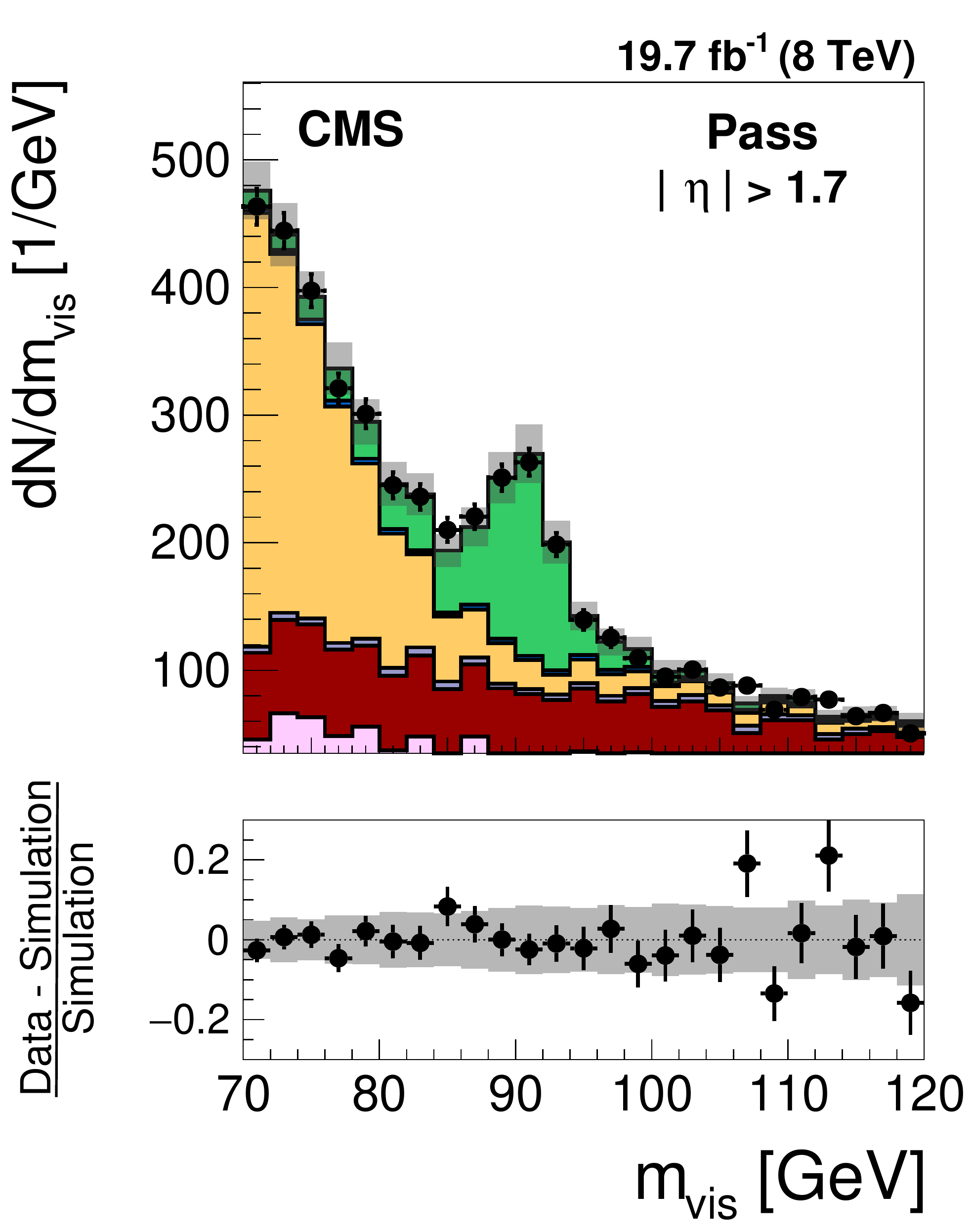}
\includegraphics[width=\figWidth]{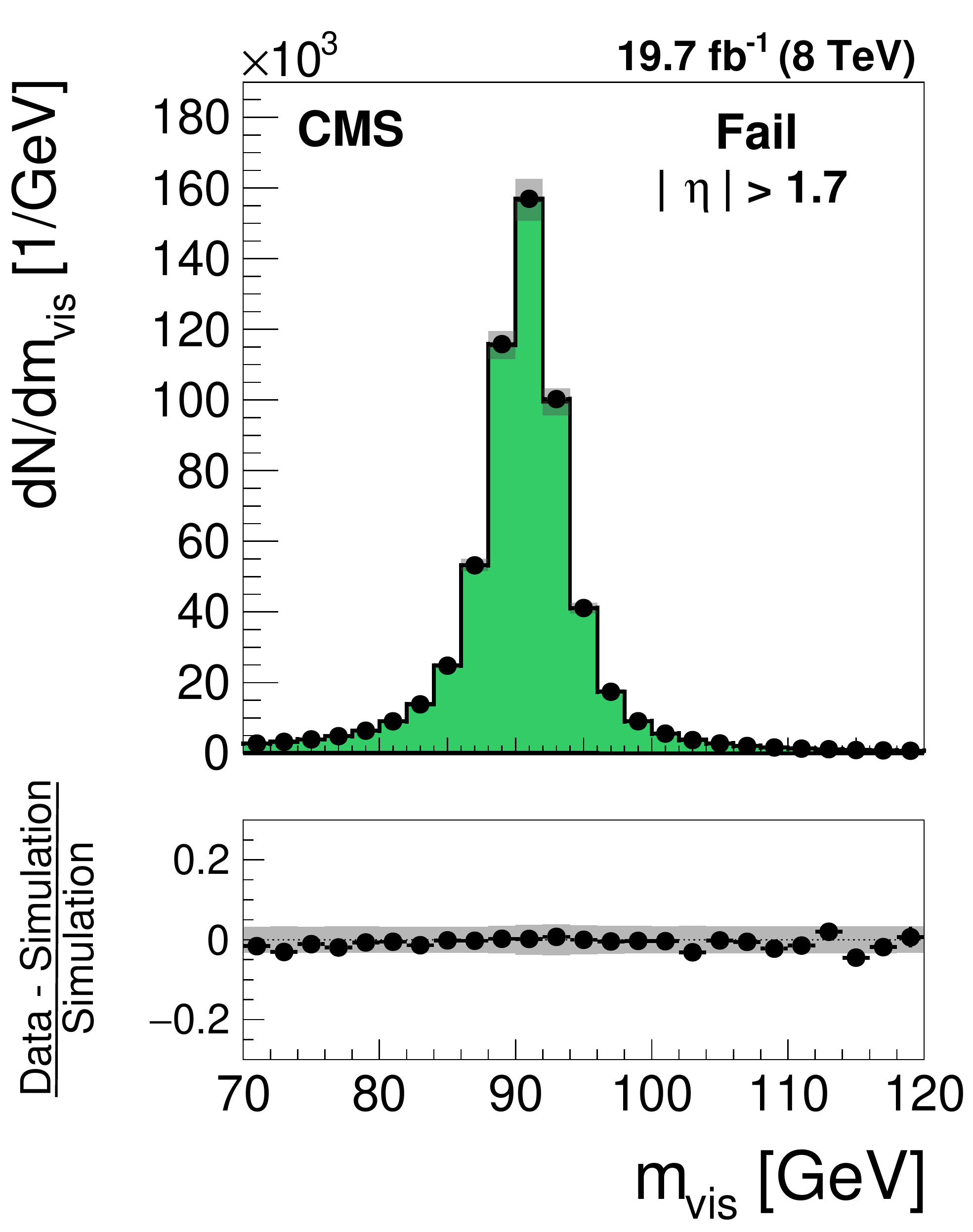}
\parbox[b][18em][t]{\figWidthScale}{\includegraphics[width=\figWidthScale]{plots/paper/muToTauFakeRate/makeMuToTauFakeRateControlPlots_fromCesare_legend.pdf}}
\caption{
  Distribution in the mass of the tag and probe pair in the pass (left) and fail (right) regions,
  for the loose WP of the cutoff-based muon discriminant in the region
  $\abs{\eta} > 1.7$.
  The distributions in $\cPZ/\Pggx \to \Pgm\Pgm$ candidate events selected in data
  are compared to the MC expectation, shown for the values of nuisance parameters obtained from the likelihood fit to the data,
  as described in Section~\ref{sec:validation_templateFits}.
  The contributions from $\cPZ/\Pggx \to \Pgt\Pgt$ background are denoted by ``DY others''.
  The ``Uncertainty'' band represents the statistical and systematic uncertainties added in quadrature.
}
\label{fig:muToTauFakeRatePostfitPlots2}
\end{figure}

In the measurement of the $\Pgm \to \tauh$ misidentification rate,
the distribution in $\mVis$ is fitted within the range $60 < \mVis < 120$\GeV.
The fit is performed separately in the regions $\abs{\eta} < 1.2$, $1.2 \leq \abs{\eta} \leq 1.7$, and $\abs{\eta} > 1.7$.
Plots of the $\mVis$ distributions in the pass and fail regions
are presented for the loose WP of the cutoff-based muon discriminant in Figs.~\ref{fig:muToTauFakeRatePostfitPlots1} and~\ref{fig:muToTauFakeRatePostfitPlots2}.
In $\cPZ/\Pggx \to \Pgt\Pgt$ events that enter the pass region, 
the tag muons are mainly due to $\Pgt^{-} \to \Pgm^{-} \, \Pagngm \, \Pnut$ decays,
while the probes are typically due to hadronic $\Pgt$ decays.

\begin{figure}[htb]
\centering
\includegraphics[width=0.48\textwidth]{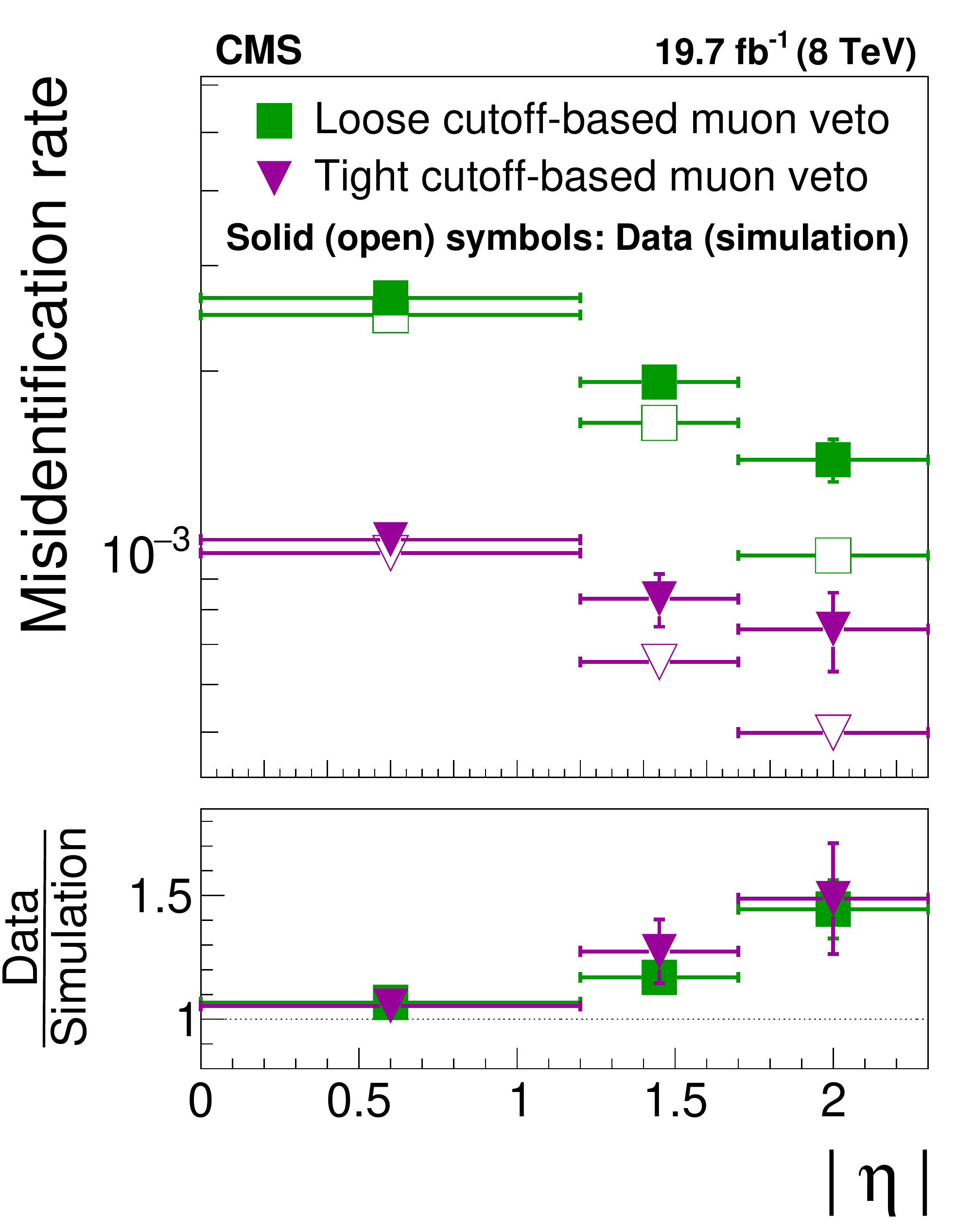}
\includegraphics[width=0.48\textwidth]{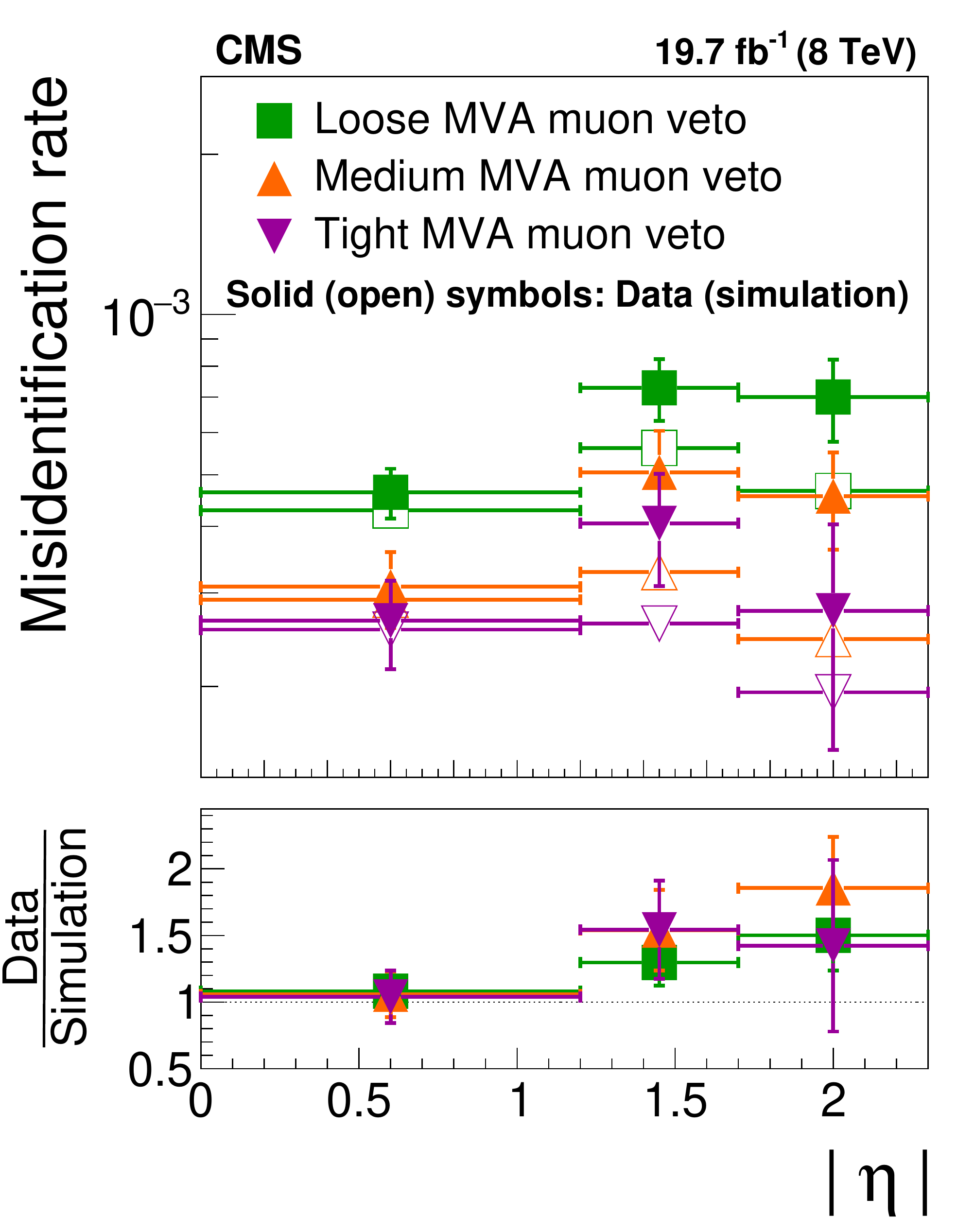}
\caption{
  Probability for muons in $\cPZ/\Pggx \to \Pgm\Pgm$ events to pass different WP of the (left) cut-based and (right) MVA-based discriminants against muons.
  The $\Pgm \to \tauh$ misidentification rates measured in data are compared to the MC simulation in the regions
  $\abs{\eta} < 1.2$, $1.2 \leq \abs{\eta} \leq 1.7$, and $\abs{\eta} > 1.7$.
}
\label{fig:muToTauFakeRateResults}
\end{figure}

The $\Pgm \to \tauh$ misidentification rates measured for different WP of the cutoff-based and MVA-based muon discriminants are given in Table~\ref{tab:muToTauFakeRateResults}.
The rates measured in the data exceed the MC prediction.
The difference between data and MC simulation is higher in the forward
than in the central region,
and increases as the muon rejection criteria are tightened.
Figure~\ref{fig:muToTauFakeRateResults} illustrates the results given in Table~\ref{tab:muToTauFakeRateResults}.
The observed differences between data and simulation have little effect on most analyses,
as the background due to muons that get misidentified as $\tauh$ decays is typically very small compared to other backgrounds.

\begin{table}[htb]
\centering
\topcaption{
  Probability for muons to pass different WP of the cutoff-based and MVA-based discriminants against muons.
  The $\Pgm \to \tauh$ misidentification rates measured in $\cPZ/\Pggx \to \Pgm\Pgm$ events are compared to the MC predictions in the regions
  $\abs{\eta} < 1.2$, $1.2 \leq \abs{\eta} \leq 1.7$, and $\abs{\eta} > 1.7$.
}
\label{tab:muToTauFakeRateResults}
\begin{tabular}{lccc}
\hline
WP & Simulation & Data & Data/Simulation  \\
\hline
\multicolumn{4}{c}{$\abs{\eta} < 1.2$} \\
\hline
Cutoff-based loose & $(2.48 \pm 0.02) \times 10^{-3}$ & $(2.65 \pm 0.06) \times 10^{-3}$ & $1.07 \pm 0.03$ \\
Cutoff-based tight & $(9.94 \pm 0.10) \times 10^{-4}$ & $(1.05 \pm 0.05) \times 10^{-3}$ & $1.05 \pm 0.05$ \\
\hline
MVA loose        & $(4.28 \pm 0.09) \times 10^{-4}$ & $(4.63 \pm 0.49) \times 10^{-4}$ & $1.08 \pm 0.12$ \\
MVA medium       & $(2.91 \pm 0.07) \times 10^{-4}$ & $(3.08 \pm 0.50) \times 10^{-4}$ & $1.06 \pm 0.17$ \\
MVA tight        & $(2.56 \pm 0.07) \times 10^{-4}$ & $(2.66 \pm 0.50) \times 10^{-4}$ & $1.04 \pm 0.20$ \\
\hline
\multicolumn{4}{c}{$1.2 \leq \abs{\eta} \leq 1.7$} \\
\hline
Cutoff-based loose & $(1.64 \pm 0.03) \times 10^{-3}$ & $(1.92 \pm 0.10) \times 10^{-3}$ & $1.17 \pm 0.07$ \\
Cutoff-based tight & $(6.54 \pm 0.19) \times 10^{-4}$ & $(8.33 \pm 0.81) \times 10^{-4}$ & $1.27 \pm 0.13$ \\
\hline
MVA loose        & $(5.61 \pm 0.18) \times 10^{-4}$ & $(7.28 \pm 0.94) \times 10^{-4}$ & $1.30 \pm 0.17$ \\
MVA medium       & $(3.28 \pm 0.14) \times 10^{-4}$ & $(5.05 \pm 0.97) \times 10^{-4}$ & $1.54 \pm 0.30$ \\
MVA tight        & $(2.63 \pm 0.12) \times 10^{-4}$ & $(4.06 \pm 0.95) \times 10^{-4}$ & $1.54 \pm 0.37$ \\
\hline
\multicolumn{4}{c}{$\abs{\eta} > 1.7$} \\
\hline
Cutoff-based loose & $(9.85 \pm 0.30) \times 10^{-4}$ & $(1.42 \pm 0.11) \times 10^{-3}$ & $1.45 \pm 0.12$ \\
Cutoff-based tight & $(4.99 \pm 0.18) \times 10^{-4}$ & $(7.42 \pm 1.09) \times 10^{-4}$ & $1.49 \pm 0.22$ \\
\hline
MVA loose        & $(4.66 \pm 0.17) \times 10^{-4}$ & $(6.99 \pm 1.20) \times 10^{-4}$ & $1.50 \pm 0.26$ \\
MVA medium       & $(2.46 \pm 0.12) \times 10^{-4}$ & $(4.57 \pm 0.92) \times 10^{-4}$ & $1.86 \pm 0.38$ \\
MVA tight        & $(1.95 \pm 0.11) \times 10^{-4}$ & $(2.77 \pm 1.25) \times 10^{-4}$ & $1.42 \pm 0.64$ \\
\hline
\end{tabular}
\end{table}

\section{Summary}
\label{sec:summary}

The algorithms used by the CMS experiment for reconstruction and identification of hadronic $\Pgt$ decays
in Run 1 data from the LHC have been presented,
and their performance validated with proton-proton collision data
recorded at $\sqrt{s} = 8$\TeV,
corresponding to an integrated luminosity of 19.7\fbinv.

The algorithms achieve a $\tauh$ identification efficiency of typically 50--60\%,
and misidentification rates for quark and gluon jets, electrons, and muons
that vary between the per mille and per cent level.
The reconstruction of different $\tauh$ decay modes and their energies
is demonstrated to be robust against pileup.

The $\tauh$ identification efficiency measured in the data agrees with the MC expectation
within the uncertainty of the measurement of about 4.5\%.
The measured jet $\to \tauh$ misidentification rates are
about 20\% higher than predicted for low-\pt jets and 20\% lower for high-\pt jets.
The probabilities for electrons and muons to pass the $\tauh$ identification criteria,
including dedicated discriminants that were developed to reduce the $\Pe \to \tauh$ and $\Pgm \to \tauh$ misidentification rates,
have been measured with a precision that ranges from a few \% and 25\%,
for loose and tight working points, respectively.
The measured misidentification rate for electrons exceeds the MC expectation by up to a factor 1.7.

The differences observed between data and MC simulation
in the probabilities for jets, electrons, and muons to be misidentified as $\tauh$ decays
have been taken into account in physics analyses by applying appropriate MC-to-data correction factors.

The procedures developed for studying $\tauh$ decays have provided a powerful tool
for precision measurements as well as for the search for new phenomena beyond the standard model in Run 2 of the LHC.

\begin{acknowledgments}
\hyphenation{Bundes-ministerium Forschungs-gemeinschaft Forschungs-zentren} We congratulate our colleagues in the CERN accelerator departments for the excellent performance of the LHC and thank the technical and administrative staffs at CERN and at other CMS institutes for their contributions to the success of the CMS effort. In addition, we gratefully acknowledge the computing centres and personnel of the Worldwide LHC Computing Grid for delivering so effectively the computing infrastructure essential to our analyses. Finally, we acknowledge the enduring support for the construction and operation of the LHC and the CMS detector provided by the following funding agencies: the Austrian Federal Ministry of Science, Research and Economy and the Austrian Science Fund; the Belgian Fonds de la Recherche Scientifique, and Fonds voor Wetenschappelijk Onderzoek; the Brazilian Funding Agencies (CNPq, CAPES, FAPERJ, and FAPESP); the Bulgarian Ministry of Education and Science; CERN; the Chinese Academy of Sciences, Ministry of Science and Technology, and National Natural Science Foundation of China; the Colombian Funding Agency (COLCIENCIAS); the Croatian Ministry of Science, Education and Sport, and the Croatian Science Foundation; the Research Promotion Foundation, Cyprus; the Ministry of Education and Research, Estonian Research Council via IUT23-4 and IUT23-6 and European Regional Development Fund, Estonia; the Academy of Finland, Finnish Ministry of Education and Culture, and Helsinki Institute of Physics; the Institut National de Physique Nucl\'eaire et de Physique des Particules~/~CNRS, and Commissariat \`a l'\'Energie Atomique et aux \'Energies Alternatives~/~CEA, France; the Bundesministerium f\"ur Bildung und Forschung, Deutsche Forschungsgemeinschaft, and Helmholtz-Gemeinschaft Deutscher Forschungszentren, Germany; the General Secretariat for Research and Technology, Greece; the National Scientific Research Foundation, and National Innovation Office, Hungary; the Department of Atomic Energy and the Department of Science and Technology, India; the Institute for Studies in Theoretical Physics and Mathematics, Iran; the Science Foundation, Ireland; the Istituto Nazionale di Fisica Nucleare, Italy; the Ministry of Science, ICT and Future Planning, and National Research Foundation (NRF), Republic of Korea; the Lithuanian Academy of Sciences; the Ministry of Education, and University of Malaya (Malaysia); the Mexican Funding Agencies (CINVESTAV, CONACYT, SEP, and UASLP-FAI); the Ministry of Business, Innovation and Employment, New Zealand; the Pakistan Atomic Energy Commission; the Ministry of Science and Higher Education and the National Science Centre, Poland; the Funda\c{c}\~ao para a Ci\^encia e a Tecnologia, Portugal; JINR, Dubna; the Ministry of Education and Science of the Russian Federation, the Federal Agency of Atomic Energy of the Russian Federation, Russian Academy of Sciences, and the Russian Foundation for Basic Research; the Ministry of Education, Science and Technological Development of Serbia; the Secretar\'{\i}a de Estado de Investigaci\'on, Desarrollo e Innovaci\'on and Programa Consolider-Ingenio 2010, Spain; the Swiss Funding Agencies (ETH Board, ETH Zurich, PSI, SNF, UniZH, Canton Zurich, and SER); the Ministry of Science and Technology, Taipei; the Thailand Center of Excellence in Physics, the Institute for the Promotion of Teaching Science and Technology of Thailand, Special Task Force for Activating Research and the National Science and Technology Development Agency of Thailand; the Scientific and Technical Research Council of Turkey, and Turkish Atomic Energy Authority; the National Academy of Sciences of Ukraine, and State Fund for Fundamental Researches, Ukraine; the Science and Technology Facilities Council, UK; the US Department of Energy, and the US National Science Foundation.

Individuals have received support from the Marie-Curie programme and the European Research Council and EPLANET (European Union); the Leventis Foundation; the A. P. Sloan Foundation; the Alexander von Humboldt Foundation; the Belgian Federal Science Policy Office; the Fonds pour la Formation \`a la Recherche dans l'Industrie et dans l'Agriculture (FRIA-Belgium); the Agentschap voor Innovatie door Wetenschap en Technologie (IWT-Belgium); the Ministry of Education, Youth and Sports (MEYS) of the Czech Republic; the Council of Science and Industrial Research, India; the HOMING PLUS programme of the Foundation for Polish Science, cofinanced from European Union, Regional Development Fund; the OPUS programme of the National Science Center (Poland); the Compagnia di San Paolo (Torino); the Consorzio per la Fisica (Trieste); MIUR project 20108T4XTM (Italy); the Thalis and Aristeia programmes cofinanced by EU-ESF and the Greek NSRF; the National Priorities Research Program by Qatar National Research Fund; the Rachadapisek Sompot Fund for Postdoctoral Fellowship, Chulalongkorn University (Thailand); and the Welch Foundation, contract C-1845.
\end{acknowledgments}

\bibliography{auto_generated}

\cleardoublepage \appendix\section{The CMS Collaboration \label{app:collab}}\begin{sloppypar}\hyphenpenalty=5000\widowpenalty=500\clubpenalty=5000\textbf{Yerevan Physics Institute,  Yerevan,  Armenia}\\*[0pt]
V.~Khachatryan, A.M.~Sirunyan, A.~Tumasyan
\vskip\cmsinstskip
\textbf{Institut f\"{u}r Hochenergiephysik der OeAW,  Wien,  Austria}\\*[0pt]
W.~Adam, E.~Asilar, T.~Bergauer, J.~Brandstetter, E.~Brondolin, M.~Dragicevic, J.~Er\"{o}, M.~Flechl, M.~Friedl, R.~Fr\"{u}hwirth\cmsAuthorMark{1}, V.M.~Ghete, C.~Hartl, N.~H\"{o}rmann, J.~Hrubec, M.~Jeitler\cmsAuthorMark{1}, V.~Kn\"{u}nz, A.~K\"{o}nig, M.~Krammer\cmsAuthorMark{1}, I.~Kr\"{a}tschmer, D.~Liko, T.~Matsushita, I.~Mikulec, D.~Rabady\cmsAuthorMark{2}, B.~Rahbaran, H.~Rohringer, J.~Schieck\cmsAuthorMark{1}, R.~Sch\"{o}fbeck, J.~Strauss, W.~Treberer-Treberspurg, W.~Waltenberger, C.-E.~Wulz\cmsAuthorMark{1}
% ...
\vskip\cmsinstskip
% ...
1:~~Also at Vienna University of Technology, Vienna, Austria\\
2:~~Also at CERN, European Organization for Nuclear Research, Geneva, Switzerland\\
% ...

\end{sloppypar}
\end{document}